%% file: main.tex
\documentclass[aps,pra,10pt,a4paper,superscriptaddress,onecolumn,preprintnumbers,floatfix,nofootinbib]{revtex4-1}

\usepackage{amsmath,amssymb,amsfonts}
\usepackage{graphicx,graphics,latexsym,placeins}
\usepackage{epsfig}
\usepackage{bm}
\usepackage{footmisc}
\usepackage{slashed}
\usepackage{natbib}
\usepackage{url}
\usepackage{dcolumn}
\usepackage{color}
\usepackage{psfrag}
\usepackage{subfigure}
\usepackage{tabularx}
\usepackage{physics}
\usepackage{hyperref}
\hypersetup{
	colorlinks=true,       
	linkcolor=blue,        
	citecolor=blue,        
	filecolor=blue,     
	urlcolor=blue         
}
\usepackage{orcidlink}
\usepackage[left=3cm,right=3cm,top=3cm,bottom=3cm]{geometry}
\newcommand{\be}{\begin{equation}}
	\newcommand{\ee}{\end{equation}}
\newcommand{\ba}{\begin{eqnarray}}
	\newcommand{\ea}{\end{eqnarray}}
\newcommand\bea{\begin{eqnarray}}
	\newcommand\eea{\end{eqnarray}}

\newcommand{\D}{{\rm d}}
\renewcommand{\vec}[1]{\boldsymbol{\mathbf{#1}}}

\usepackage[export]{adjustbox}
\newcommand{\eqfig}[1]{\includegraphics[valign=c]{#1}}

\usepackage{multirow}
\newcommand{\tablestyle}{\displaystyle\rule[-20pt]{0pt}{42pt}}

\usepackage{microtype}
\usepackage[defaultlines=2,all]{nowidow}
\newcommand{\vph}[1]{{\vphantom{#1}}}

\newlength\customcitespacing
\setcitestyle{citesep={,\kern-\customcitespacing}}

\usepackage{mathtools}
\newcommand{\coloneq}{\coloneqq}

\newcommand{\pp}{\mathrm{pp}}
\newcommand{\mpl}{m_\mathrm{Pl}} 
\newcommand{\M}{\mathcal{M}} 
\newcommand{\avg}[1]{{\langle #1 \rangle}} 
\newcommand{\w}{\omega}
\newcommand{\vecell}{\hat{\bm\ell}}
\newcommand{\projell}[1]{\vecell\!\hspace{0.5pt}\cdot\!\bm{#1}}
\newcommand{\yl}{\mathfrak{l}}
\newcommand{\ym}{\mathfrak{m}}
\newcommand{\x}{\mathtt{x}}

\newcommand{\GN}{G_\mathrm{N}}
\newcommand{\Geff}{G_{12}}
\newcommand{\gammaeff}{\gamma_{12}}
\newcommand{\gammapeff}{\gamma'\!{}_{12}}
\newcommand{\alphaprime}{\alpha'\!{}}
\newcommand{\comboA}[2]{\mathfrak a_{#1}^{(#2)}}
\newcommand{\comboAp}[2]{{\mathfrak a}_{#1}^{\prime\,(#2)}}
\newcommand{\comboBS}[2]{\tilde{\mathfrak b}_{#1}^{(#2)}}

\newcommand{\K}{\kappa}

\newcommand{\epsLadder}{\epsilon_\mathrm{L}} 
\newcommand{\epsSpin}{\epsilon_\mathrm{S}} 

\newcommand{\dx}{\mathrm{d}}
\newcommand{\diff}[2]{\frac{\dx #1}{\dx #2}}

\newcommand{\pdiff}[2]{\frac{\partial #1}{\partial #2}}

\newcommand{\scount}[1]{{\small\textsc{[\MakeLowercase{#1}]}}}
\newcommand{\sector}[1]{{(\mathrm{#1})}}
\newcommand{\scriptsector}[1]{{\scriptscriptstyle\sector{#1}}}
\newcommand{\dip}{\mathrm{d}}
\newcommand{\nondip}{\mathrm{nd}}

\newcommand{\QD}{\mathrm{QD}}

\newcommand{\CEA}{\affiliation{Universit{\'e} Paris-Saclay, CNRS, CEA, Institut de physique th{\'e}orique, 91191, Gif-sur-Yvette, France}}
\newcommand{\DAMTP}{\affiliation{Department of Applied Mathematics and Theoretical Physics (DAMTP),\\Center for Mathematical Sciences, University of Cambridge, CB3 0WA, United Kingdom}}
\newcommand{\KICC}{\affiliation{Kavli Institute for Cosmology (KICC), University of Cambridge,\\Madingley Road, Cambridge CB3 0HA, United Kingdom}}
\newcommand{\MPI}{\affiliation{Max Planck Institute for Gravitational Physics (Albert Einstein Institute), Am M{\"u}hlenberg 1, 14476 Potsdam, Germany}}
\newcommand{\UNISA}{\affiliation{Dipartimento di Fisica ``E.R.\ Caianiello'', Universit\`a degli Studi di Salerno,\\ Via Giovanni Paolo II, 132 - 84084 Fisciano (SA), Italy}}
\newcommand{\INFNSA}{\affiliation{Istituto Nazionale di Fisica Nucleare - Gruppo Collegato di Salerno - Sezione di Napoli,\\ Via Giovanni Paolo II, 132 - 84084 Fisciano (SA), Italy}}
\newcommand{\IPM}{\affiliation{School of Physics, Institute for Research in Fundamental Sciences (IPM),	P.\ O.\ Box 19395-5531, Tehran, Iran}}
\newcommand{\Damghan}{\affiliation{School of Physics, Damghan University, Damghan 3671641167, Iran}}
\newcommand{\Az}{\affiliation{Center for Theoretical Physics, Khazar University, 41 Mehseti Str., AZ1096 Baku, Azerbaijan}}
\newcommand{\UF}{\affiliation{Institut f\"ur Theoretische Physik, Goethe Universit\"at, Max-von-Laue-Stra{\ss}e 1, 60438 Frankfurt am Main, Germany}}
\newcommand{\ICTP}{\affiliation{International Centre for Theoretical Physics Asia-Pacific, University of Chinese Academy of Sciences, Beijing 100190, China}}
\newcommand{\UCAS}{\affiliation{Taiji Laboratory for Gravitational Wave Universe (Beijing/Hangzhou), University of Chinese Academy of Sciences, Beijing 100049, China}}
\newcommand{\TDLI}{\affiliation{Tsung-Dao Lee Institute $\&$ School of Physics and Astronomy, Shanghai Jiao Tong University,
Shanghai 201210, China}}

\begin{document}

\preprint{IPPP/26/23}

{
\makeatletter
\def\frontmatter@thefootnote{%
 \altaffilletter@sw{\@fnsymbol}{\@fnsymbol}{\csname c@\@mpfn\endcsname}%
}%
\makeatother

\title{Illuminating the dark universe in the multi-messenger era}

\author{Philippe Brax \orcidlink{0000-0003-0727-3186}}
\email{philippe.brax@cea.fr}
\CEA

\author{Anne-Christine Davis \orcidlink{0000-0002-3708-8283}}
\email{ad107@cam.ac.uk}
\DAMTP \KICC

\author{Md Riajul Haque \orcidlink{0000-0002-6618-4899}}
\email{riaj.0009@gmail.com}
\TDLI

\author{C{\'e}dric Jockel \orcidlink{0009-0007-7617-7178}}
\email{cedric.jockel@aei.mpg.de}
\MPI

\author{Gaetano Lambiase \orcidlink{0000-0001-7574-2330}}
\email{lambiase@sa.infn.it}
\UNISA \INFNSA

\author{Michiru Uwabo-Niibo \orcidlink{0000-0003-3733-1834}}
\email{michiru@ibs.re.kr}
\affiliation{Cosmology, Gravity, and Astroparticle Physics Group, Center for Theoretical Physics of the Universe, Institute for Basic Science (IBS), Daejeon, 34126, Korea}

\author{Mohsen Khodadi \orcidlink{0000-0001-7949-0441}}
\email{khodadi@kntu.ac.ir}
\IPM \Damghan  \Az

\author{Tanmay Kumar Poddar \orcidlink{0000-0002-9078-6224}}
\email{tanmay.k.poddar@durham.ac.uk}
\affiliation{Institute for Particle Physics Phenomenology (IPPP), Department of Physics, Durham University, Durham DH1 3LE, United Kingdom}

\author{Laura Sagunski \orcidlink{0000-0002-3506-3306}}
\email{sagunski@itp.uni-frankfurt.de}
\UF

\author{Luca Visinelli \orcidlink{0000-0001-7958-8940}}
\email{lvisinelli@unisa.it}
\UNISA \INFNSA

\author{Jun Zhang \orcidlink{0000-0001-5314-3505}}
\email{zhangjun@ucas.ac.cn}
\ICTP \UCAS

\begin{abstract}

The precision era of multi-messenger astronomy, together with modern astrophysical, cosmological, and gravitational wave observations, increasingly points toward the existence of a ``dark" sector that cannot be explained within the framework of the Standard Model of particle physics and General Relativity. In this review, we explore extensions of standard physics and examine how observational data can be used to probe new particles and interactions. We consider a wide range of scales, from Solar System tests to galactic and cosmological observations, and investigate both conventional dark matter candidates, such as weakly interacting massive particles, and alternative scenarios including ultralight fields and primordial black holes. We discuss constraints derived from compact objects such as neutron stars, black holes, pulsars, and magnetars observations as well as from high-energy astrophysical phenomena. In addition, we analyze extensions of General Relativity involving additional scalar fields and their impact on gravitational wave signals and stochastic backgrounds from primordial black holes. We also study the capture and accumulation of dark matter in compact objects, which can alter properties such as mass, radius, and tidal deformability, and consider scenarios in which dark matter decays into Standard Model particles. While current observations already place significant limits on dark matter and modified-gravity models, upcoming experiments and observatories are expected to further probe or discover such new physics by improving constraints on particle masses and interaction strengths. 

\end{abstract}
 
\maketitle
\tableofcontents

\newpage

\input{1_Introduction}
\input{2_Compact_objects}
\input{3_Theory}
\input{4_StochasticGW}
\input{5_Radiation_particles}
\input{6_Superradiance_and_search_for_bosonic_DM}
\input{7_Compact_stars_properties}
\input{8_Discussion}

\input{9_Acknowledgements}
\input{10_Appendix}
\input{11_Acronyms_Conventions}

\bibliographystyle{utphys}
\bibliography{all_references}

\end{document}

%% file: 1_Introduction.tex
\section{Introduction}
\label{sec1}

For over a century, General Relativity (GR) has provided the foundation of our understanding of gravity. Its predictions have been confirmed with impressive precision, from Solar System experiments to observations of binary pulsars~\cite{Will:1994fb, Weisberg:2010zz}. More recently, the direct detection of gravitational waves (GWs) via interferometry efforts has pushed these tests into the highly dynamical, strong-field regime~\cite{LIGOScientific:2016aoc, LIGOScientific:2017bnn}. Together, these results establish GR as a successful effective description of gravity, while also opening a new window to probe possible deviations where they are most likely to emerge.

Despite its triumphant record, GR faces conceptual and empirical puzzles that call for extensions of the theory. On the largest scales, cosmic acceleration suggests that either a new energy component such as dark energy permeates spacetime~\cite{SupernovaCosmologyProject:1998vns, SupernovaSearchTeam:1998fmf} or that gravity itself must be modified in the infrared~\cite{Clifton:2011jh, Joyce:2016vqv}. On galactic and sub-galactic scales, several anomalies persist in the cold dark matter (CDM) framework, including small-scale structure tensions~\cite{Bullock:2017xww} and the unknown particle nature of dark matter (DM)~\cite{Bertone:2004pz}. On the smallest scales, GR remains incompatible with quantum field theory, motivating ultraviolet completions such as string theory and quantum gravity~\cite{Burgess:2003jk}. Together, these considerations strongly suggest the presence of new degrees of freedom beyond those contained in GR and the Standard Model (SM).

From the perspective of particle physics, extensions of gravity and of the dark sector generically predict new light fields. Scalars, pseudoscalars, vectors, and tensor degrees of freedom arise naturally in a wide range of scenarios beyond the SM, including axion and axion-like particle models, dark photons, dilatons, moduli fields, and massive gravity.
Well-studied realizations include scalar-tensor gravity~\cite{Brans:1961sx, Horndeski:1974wa}, Einstein-dilaton-Gauss-Bonnet gravity~\cite{Kanti:1995vq, Pani:2011xm}, dynamical Chern-Simons gravity~\cite{Alexander:2009tp}, screened fifth-force theories~\cite{Khoury:2003aq, Brax:2004qh, Hinterbichler:2010es, Joyce:2014kja}, and massive gravity~\cite{deRham:2010kj, deRham:2014zqa}. 
These theories predict striking observational signatures such as dipolar radiation in compact binaries~\cite{Shao:2014wja, Shao:2017gwu}, scalar or vector hair for black holes (BH)~\cite{Herdeiro:2014goa}, BH superradiance triggered by ultralight bosons~\cite{Arvanitaki:2009fg, Arvanitaki:2010sy, Brito:2015oca}, modified dispersion and luminosity distance of GWs~\cite{Lombriser:2015sxa, Baker:2017hug}, and environmental phenomenology including DM-induced gravitational drag~\cite{Eda:2014kra}. Compact objects play a central role in amplifying these effects. BHs and neutron stars (NSs) probe curvature, density, and timescale regimes that are beyond the reach of laboratory experiments, effectively acting as natural detectors for light degrees of freedom. In such environments, even feeble couplings between new particles and matter or gravity can lead to observable signatures. GWs emitted by compact binaries are particularly powerful probes, as they describe the dissipative dynamics of the system, allowing small deviations from GR to accumulate into measurable phase shifts over many orbital cycles.

The multi-messenger observation of GW170817 and its electromagnetic (EM) counterpart has already excluded broad classes of modified gravity theories that predict an anomalous GW propagation speed~\cite{LIGOScientific:2017vwq}. Yet enormous discovery potential remains available in upcoming detectors. Space-based interferometers such as the Laser Interferometer Space Antenna (LISA)~\cite{Amaro-Seoane:2012aqc, LISA:2017pwj}, Taiji~\cite{Hu:2017mde}, and the DECi-hertz Interferometer Gravitational Wave Observatory (DECIGO)~\cite{Kawamura:2006up, Kawamura:2008zza, Kawamura:2011zz}, together with third-generation terrestrial instruments like the Einstein Telescope~\cite{Punturo:2010zz, ET:2025xjr} and Cosmic Explorer~\cite{Reitze:2019iox}, will probe extreme mass-ratio inspiral (EMRI), intermediate-mass BHs, and early-universe GWs. These systems provide pristine access to the strong-gravity regime, where even subtle deviations from GR accumulate into measurable phase shifts across hundreds of thousands of orbital cycles~\cite{Gair:2017ynp, Babak:2017tow}.

The dark sector emerges as a particularly compelling arena linking modified gravity and astrophysics. Ultralight DM fields can cluster around compact objects~\cite{Hui:2016ltb}, drive BH superradiant instabilities~\cite{Brito:2015oca}, or form dense spikes whose gravitational imprint modulates inspiral waveforms~\cite{Gondolo:1999ef, Eda:2014kra, Kavanagh:2020cfn}. Non-annihilating and annihilating DM models offer complementary signatures, from fifth-force-like effects to exotic radiation channels~\cite{Barausse:2014pra, Cardoso:2019rvt}. Meanwhile, primordial BHs (PBHs) and GW backgrounds from phase transitions, cosmic strings, or inflation yield probes of new physics inaccessible to laboratory experiments~\cite{Caprini:2018mtu, Aggarwal:2025noe}.

This review provides a coherent roadmap through the rapidly expanding landscape of gravity beyond GR. The accelerating pace of observational breakthroughs, especially in GW astronomy, has begun to blur traditional boundaries between cosmology, astrophysics, and particle physics. As new data probe deeper into the strong-field regime and the dark sector, theoretical developments have diversified into a wide range of modified gravity frameworks, each predicting distinct multi-messenger signatures. This review emphasizes where deviations from GR may realistically emerge and which experiments are most suitable to reveal them. The goal is to clarify the connections, stress the complementarity of observational channels, and illuminate the scientific opportunities that define the coming decade of gravitational physics.

\newpage

%% file: 2_Compact_objects.tex
\section{Dark sector probes with compact objects}

In this section, we provide a brief overview of dark matter (DM) and its possible realizations in the Universe. We then review gravitational wave (GW) searches for DM-induced structures, such as DM spikes, as well as GW signatures arising in alternative theories of gravity. In this context, we discuss how additional degrees of freedom beyond general relativity (GR), or DM itself, can mediate effective fifth forces, potentially pointing to new fundamental interactions of nature.
\label{sec:GW}

\subsection{Dark matter}

There is significantly more non-luminous matter in the universe than visible matter. This unseen component, known as dark matter (DM), interacts primarily through gravity. The concept of DM was first proposed by Fritz Zwicky in 1933 while studying the Coma cluster. By analyzing the motion of galaxies within the cluster, Zwicky estimated its total mass and compared it to the mass inferred from the brightness and number of galaxies. He found a discrepancy, a total mass roughly $400$ times greater than what visible matter could account for. This became known as the ``missing mass'' problem~\cite{Zwicky:1933gu}. Further evidence came in the 1970s, when Vera Rubin studied the rotation curves of spiral galaxies. According to Newtonian mechanics, stars orbiting the galactic center should exhibit decreasing rotational velocities with increasing distance much like planets in our solar system. However, Rubin's observations showed that stellar velocities remain roughly constant beyond a certain radius. This flat rotation curve suggests the presence of a large amount of unseen mass distributed in a halo-like structure around the galaxy, extending far beyond the luminous disk~\cite{Rubin:1970zza, Rubin:1980zd}. Today, it is widely accepted that most DM in the universe is non-relativistic, meaning it moves slowly compared to the speed of light, and is therefore classified as cold DM, with a velocity distribution following a Maxwellian profile~\cite{Drukier:1986tm, Lewin:1995rx}.

Compelling evidence for the collisionless nature of DM came from the Chandra X-ray observations of the Bullet Cluster, a system formed by the merger of two massive galaxy clusters. The X-ray data revealed that the hot baryonic gas slowed down due to hydrodynamical interactions, while the dominant mass component passed through almost unaffected, behaving as if it were collisionless. Gravitational lensing maps revealed that the majority of mass in the system was concentrated around regions devoid of luminous matter, hinting at the dominant presence of DM~\cite{Markevitch:2004qk, Clowe:2006eq}. On even larger scales, the case for DM is reinforced by the remarkable agreement between the standard model of cosmology--$\Lambda$-cold dark matter ($\Lambda$CDM) and precision cosmological observations. The temperature anisotropies of the cosmic microwave background (CMB), the detailed shape of its angular power spectrum, and the formation and distribution of the large-scale structure of the universe all require a substantial cold DM component~\cite{Planck:2018vyg}.

The Planck satellite provides precise measurements of the universe's energy composition within the framework of $\Lambda$CDM. Within this framework, the energy density fractions of baryonic matter and cold DM are determined to be
\begin{equation}
    \Omega_b h^2 = 0.02237 \pm 0.00015, \qquad 
    \Omega_{\rm CDM} h^2 = 0.1200 \pm 0.0012 \,,
\end{equation}
where $h$ denotes the reduced Hubble constant. These measurement refer to the marginalized means and 68\% confidence level (CL) for the base-$\Lambda$CDM cosmological parameters from \textit{Planck} TT,TE,EE+lowE+lensing, using the \texttt{Plik} likelihood~\cite{Planck:2018vyg}.

For a particle to qualify as a viable DM candidate, it must be massive, electrically neutral, stable on cosmological timescales, and cold at the time of structure formation~\cite{Jungman:1995df, Bertone:2004pz, Bertone:2016nfn, Cirelli:2024ssz}. The prototypical candidate is the Weakly Interacting Massive Particle (WIMP), which arises naturally in extensions of the Standard Model (SM) such as supersymmetry. Efforts to detect DM via non-gravitational interactions have thus far yielded null results~\cite{XENON:2025vwd, PandaX:2024qfu, LZ:2022lsv, AMS:2019rhg, Fermi-LAT:2015att}, motivating the development of complementary approaches to probe its nature. Direct searches for DM focus on three complementary strategies. Direct detection experiments such as CDMS~\cite{CDMS-II:2009ktb}, LUX~\cite{LUX:2016ggv}, XENON~\cite{XENON:2018voc, XENON:2025vwd}, and LZ~\cite{LZ:2022lsv} aim to observe the tiny nuclear recoils produced when DM particles scatter off detector nuclei~\cite{Goodman:1984dc, Drukier:1986tm, Lewin:1995rx}. Indirect detection experiments, including PAMELA~\cite{PAMELA:2010kea}, AMS-02~\cite{AMS:2019rhg}, and the Fermi Large Area Telescope~\cite{Fermi-LAT:2015att}, search for secondary cosmic rays, gamma rays, or neutrinos generated by DM annihilation or decay in regions of high DM density, such as the Galactic center or dwarf spheroidal galaxies~\cite{Baltz:2008wd,Bergstrom:2012fi}. Collider searches, most notably at the Large Hadron Collider (LHC), attempt to produce DM particles in high-energy collisions, where their presence would be inferred from missing transverse momentum in association with SM final states~\cite{Birkedal:2004xn,Goodman:2010yf,Abdallah:2015ter}. Gamma-ray searches have been a central focus of indirect detection, using instruments such as the Fermi Large Area Telescope (Fermi-LAT), H.E.S.S.~\cite{HESS:2016pst}, VERITAS~\cite{VERITAS:2006lyc}, and LHAASO~\cite{Cao:2010zz}. These telescopes aim to detect high-energy photon fluxes originating from dense DM environments, thereby offering potential insight into DM interactions. Since the annihilation rate scales with $\rho_{\rm DM}^2$, regions with density enhancements such as galactic centers, dwarf spheroidals, or spikes around compact objects are especially promising targets. Despite decades of effort, no conclusive evidence has emerged. The null results have imposed stringent bounds on WIMP models, particularly for masses above the GeV scale, as illustrated by the most recent constraints on the WIMP--nucleon scattering cross section~\cite{LZ:2022lsv, PandaX:2024qfu, XENON:2025vwd}.

This persistent absence of WIMP signals has motivated a broad exploration of alternative scenarios. Among these are feebly interacting massive particles (FIMPs), which never reach thermal equilibrium in the early universe~\cite{Hall:2009bx, Bernal:2017kxu}; strongly interacting massive particles (SIMPs), whose relic density is set by $3 \to 2$ self-annihilation processes~\cite{Hochberg:2014dra,Hochberg:2014kqa}; and fuzzy dark matter (FDM), consisting of ultralight bosons with masses around $10^{-22}$\,eV that suppress structure on kiloparsec scales~\cite{Hu:2000ke,Hui:2016ltb}. Additional well-motivated candidates include sterile neutrinos, which may leave imprints in the form of unexplained X-ray line features~\cite{Dodelson:1993je,Abazajian:2001vt,Boyarsky:2009ix}; axions and axion-like particles, originally proposed to solve the strong CP problem~\cite{Peccei:1977hh,Weinberg:1977ma,Wilczek:1977pj}, and now widely studied as DM candidates~\cite{Preskill:1982cy,Abbott:1982af,Dine:1982ah,Sikivie:2006ni, DiLuzio:2020wdo}; and ultralight scalar fields, which can lead to novel signatures in cosmology and astrophysics~\cite{Arvanitaki:2009fg, Marsh:2015xka}. Non-particle candidates are also actively investigated, most notably black holes (BHs) of primordial origin, which could account for part or all of the DM abundance depending on their mass spectrum~\cite{Carr:2016drx,Carr:2020gox,Green:2020jor}. While cosmological observations firmly establish the existence of DM, they do not uniquely determine its microscopic nature. Astrophysical structure formation on galactic and sub-galactic scales offers a sensitive way to discriminate among competing models.

\subsection{Sub-Galactic Tensions in $\Lambda$CDM}

On galactic and sub-galactic scales, several long-standing tensions arise between the predictions of the $\Lambda$CDM paradigm and astronomical observations. These most prominent are the core--cusp problem, the missing satellites problem, and the too--big--to--fail problem~\cite{Weinberg:2013aya, Bullock:2017xww}. On scales below $\sim 1$\,kpc, observations of dwarf galaxies reveal a core-like DM density profile, with a nearly constant central density~\cite{Flores:1994gz, deBlok:2009sp}. This contrasts sharply with predictions from $\Lambda$CDM simulations, which favor a steep, cuspy profile~\cite{Navarro:1996gj, Moore:1999gc}. This inconsistency is referred to as the core--cusp problem. For instance, data from the dwarf galaxy survey indicate that within the inner one--third kiloparsec, the density remains flat--contrary to the cuspy profile expected from collisionless CDM simulations. One possible resolution is baryonic feedback, where energy injection from star formation or supernovae can smooth out the inner density peak~\cite{Governato:2009bg, Pontzen:2011ty}. Alternatively, self--interacting dark matter (SIDM)~\cite{Spergel:1999mh, Tulin:2017ara} and fuzzy dark matter (FDM)~\cite{Hu:2000ke, Hui:2016ltb} models naturally produce cored profiles without invoking complex baryonic physics.

Another issue, known as the missing satellite problem, stems from the discrepancy between the number of satellite galaxies predicted by CDM simulations and those actually observed in the Milky Way. CDM predicts a much higher number of satellite galaxies with halo masses $>10^7~ \mathrm{M}_\odot$ than are observed, even down to masses of $\sim 300\,\mathrm{M}_\odot$ within $300$\,kpc of the Milky Way. A common explanation is that galaxy formation becomes inefficient in low-mass halos, possibly due to insufficient baryonic infall. However, more recent analyses from the Sloan Digital Sky Survey (SDSS), which account for observational selection effects, suggest that the number of satellite galaxies may indeed match CDM predictions when corrected for detection efficiency~\cite{Koposov:2007ni, Walsh:2007hc}.

A third problem, known as the too-big-to-fail problem, refers to the presence of massive subhalos (with central densities $\gtrsim  10^{10}~\mathrm{M}_\odot$) in CDM simulations that are too dense to have failed in forming stars, yet are not observed in the local universe. This suggests a mismatch between simulation predictions and the real galaxy population. Proposed solutions include modifications to the nature of DM itself, with models such as Warm Dark Matter (WDM), SIDM, and FDM offering potential resolutions by suppressing small-scale structure formation or altering halo density profiles.

\subsection{Dark matter spikes}

Since the annihilation rate scales with the square of the DM density, $\rho_{\rm DM}^2$, regions with strong overdensity are especially promising for indirect and gravitational probes. One particularly important mechanism arises in the vicinity of compact objects. In the adiabatic growth of a BH at the center of a DM halo, the surrounding DM can undergo contraction, forming a steep overdensity known as a \textit{spike}~\cite{Gondolo:1999ef}. In such environments, GW astronomy provides a promising new avenue for probing DM. In particular, the inspiral of an extreme mass ratio inspiral (EMRI) composed of a stellar-mass compact object into a more massive BH, is sensitive to the gravitational environment surrounding the central object. If an intermediate-mass black hole (IMBH) resides within a dense DM distribution, the inspiral dynamics and resulting GW signal can be measurably altered, offering an indirect probe of the DM density profile and its properties~\cite{Behnke:2014tma, Chan:2022gqd}. The resulting DM density is typically modeled as a power-law, $\rho(r) \propto r^{-\gamma_{\text{sp}}}$, where the slope $\gamma_{\text{sp}}$ depends on the initial halo profile and the BH growth history. If DM is self-annihilating, the innermost region of the spike can flatten into a central plateau where annihilation balances the DM inflow~\cite{Vasiliev:2007vh, Fields:2014pia}. As the compact object spirals into the IMBH, its motion is influenced not only by GW emission but also by dynamical friction from the surrounding DM and energy loss due to annihilation processes. These effects modify the inspiral trajectory and leave characteristic imprints on the GW waveform~\cite{Eda:2014kra}. In particular, annihilation-induced flattening of the spike reduces dynamical friction in the innermost region, leading to distinctive waveform deformations that encode information about the DM profile and annihilation cross section. We develop a semi-analytic model of the DM spike that incorporates annihilation effects, derive the resulting modifications to the GW signal, and explore the observational prospects for space-based detectors such as LISA~\cite{Amaro-Seoane:2012aqc, LISA:2017pwj}, Taiji~\cite{Hu:2017mde}, and DECIGO~\cite{Kawamura:2006up, Kawamura:2008zza, Kawamura:2011zz}. The results suggest that EMRI systems in DM-rich environments can serve as sensitive probes of DM annihilation near BHs, opening a new window on the nature of particle DM.

Massive compact objects, such as BHs, are expected to accrete surrounding material, with baryonic matter, primarily interstellar gas, providing the most immediate and well-understood contribution. In astrophysical settings, this gas often forms an accretion disk around the BH, enabled by mechanisms that transport angular momentum and dissipate energy through radiation. These processes allow the gas to gradually spiral inward, feeding the BH and influencing its spin and mass. In binary BH systems, the presence of a circumbinary gas disk can substantially affect the inspiral dynamics. First, the gravitational potential of the disk perturbs the binary orbit, either accelerating or delaying the inspiral depending on the disk's structure and mass distribution~\cite{Barausse:2006vt, Macedo:2013qea, Barausse:2014tra}. Second, each BH experiences dynamical friction as it moves through the gaseous medium, generating a gravitational wake that slows its motion~\cite{Barausse:2007dy, Barausse:2007ph}. Third, accretion of gas can alter the BHs' masses and spins over time~\cite{Barausse:2006vt, Barausse:2014tra}. In contrast, the accretion of DM is more subtle due to its lack of electromagnetic (EM) interactions. Without the ability to radiate energy, DM cannot efficiently shed angular momentum, making it incapable of forming accretion disks or collapsing into compact baryon-like structures. Nonetheless, DM particles passing sufficiently close to a BH can still be gravitationally captured, especially in regions of high density such as within a halo or a steep central \textit{spike} formed through adiabatic contraction or other dynamical mechanisms. Over cosmological timescales, such captured DM can contribute to the BH's mass growth. In the case of non-annihilating DM, the accreted material is expected to form an approximately spherical distribution around the BH on scales comparable to the gravitational radius, $r_g \equiv GM/c^2$, reflecting the collisionless and pressureless nature of DM, unlike the flattened morphology of baryonic disks. The DM profile can be parameterized as:
\begin{equation}
    \rho(r) = \begin{cases}
        0\,, & \hbox{for $r < 2r_g$}\,,\\
        \left(\frac{r_{\rm sp}}{r}\right)^{\gamma_{\rm sp}}\,, & \hbox{for $2r_g \leq r < r_{\rm halo}$}\,,\\
        \rho_{\rm halo}(r)\,, & \hbox{for $r \geq r_{\rm halo}$}\,,
    \end{cases}
\end{equation}
where $\rho_{\rm halo}(r)$ is the galactic DM distribution. Although DM accretion is less efficient than baryonic accretion, its long-term contribution may be significant, particularly for primordial or supermassive BHs, and could offer observational signatures relevant to indirect DM detection efforts. This result originates from the reconstruction of the final density after adiabatic growth of the BH, 
\begin{equation}
    \rho(r) = \frac{4 \pi}{r^2}\,\int_{\mathcal{E}_m}^0 {\rm d}\mathcal{E}_f \int_{L_c}^{L_m} {\rm d}L_f  \,\frac{L_f}{v_r}\,f_f(\mathcal{E}_f, L_f)\,,
\end{equation}
where $f(\mathcal{E}, L)$ is the phase space distribution of DM particles, as a function of the energy $\mathcal{E}$ and angular momentum $L$, and the suffix ``$f$'' denotes the evaluation after accretion. For a DM distribution $\rho_{\rm halo}(r) \propto r^{-\gamma}$, the index for the spike slope is $(9 - 2 \gamma)/(4 - \gamma)$~\cite{Gondolo:1999ef}. For instance, the Navarro-Frenk-White (NFW) DM profile predicts a slope with an index $\gamma = 1$ in the inner region hosting a SMBH, $\rho_{\rm halo} \propto 1/r$, leading to the slope for the DM spike $\gamma_{\rm sp} = 7/3$.

The relation with the initial phase space distribution $f_i = f_i(r, v)$ is found by considering the probability of finding a DM particle at a radius $r$ for an orbit specified by energy $\mathcal{E}$ and angular momentum $L$ as~\cite{Ullio:2001fb}
\begin{equation}
    P(r|\mathcal{E}, L)\,{\rm d}r = \frac{2}{T_{\rm orb}}\frac{1}{v_r}\,{\rm d}r\,,
\end{equation}
where $T_{\rm orb}$ is the orbital period. The integration of the probability distribution over the phase space leads to~\cite{Bertone:2024wbn}
\begin{equation}
    \rho(r) = \frac{1}{4\pi r^2}\,\int {\rm d}^3 {\bf r}_i\, {\rm d}^3 {\bf v}_i \,\frac{2\,f_i}{T_{\rm orb}\,v_r}\,.
\end{equation}
When the phase-space distribution can be decomposed in terms of the spatial and velocity distributions as $f_i = \rho_i(r_i)\,f_v({\bf v}_i)$, the final result for a spherically-symmetric distribution is~\cite{Eroshenko:2016yve}
\begin{equation}
    \label{eq:rhof}
    \rho(r) = \frac{1}{r^2}\,\int {\rm d}r_i\,r_i^2\,\rho_i(r_i)\,\int {\rm d}^3 {\bf v}_i \,\frac{2}{T_{\rm orb}\,v_r}\,f_v({\bf v}_i)\,.
\end{equation}
The properties of this integral form have been explored in Ref.~\cite{Carr:2020mqm} for the accretion of primordial BHs in a cosmological context. For example, the accretion in a radiation-dominated Universe, for spikes of cosmological origin, equating the cosmological expansion with the BH field gives the slope $\gamma_{\rm sp} = 9/4$. This slope arises from analyzing the balance between the gravitational field of a primordial black hole (PBH) and the cosmological expansion during the radiation-dominated epoch. In this setting, DM spikes can form around PBHs due to the adiabatic contraction of surrounding DM, which is gravitationally pulled inward as the PBH grows. The structure of such spikes depends on the initial DM density distribution and the background cosmological environment. To estimate the spike profile, one considers the equilibrium between the PBH's gravitational attraction and the Hubble expansion. This balance defines a characteristic turnaround radius $r_{\rm ta}$~\cite{Adamek:2019gns}, beyond which the Hubble flow dominates, while matter within $r_{\rm ta}$ can become gravitationally bound to the PBH. Under the assumptions of spherical symmetry, pressureless DM, and radial infall, a self-similar solution can be derived. During the radiation-dominated era, the background density scales as $\rho(t) \propto t^{-2}$, and the resulting infall leads to a power-law density profile given by $\rho(r) \propto r^{-9/4}$, corresponding to a spike with slope $\gamma_{\rm sp} = 9/4$. As discussed in Ref.~\cite{Carr:2020mqm}, incorporating a Maxwell-Boltzmann distribution for the DM velocity dispersion and integrating over the phase-space distribution maintains the same radial dependence, $\rho(r) \propto r^{-9/4}$, but introduces a multiplicative normalization factor. This enhancement arises from the velocity distribution of DM particles during infall and results in an overall concentration factor of approximately $1.53$ in the final density profile. Note however, that the decomposition of the phase-space properties leading to Eq.~\eqref{eq:rhof} are only an approximation given that the velocity distribution might not be Maxwellian, but it is related to the density distribution through the continuity equation. The proper method, as outlined in Ref.~\cite{Bertone:2024wbn}, is to evaluate the initial distribution function following the Eddington inversion procedure, assuming a spherically symmetric density profile~\cite{2008gady.book.....B, Lacroix:2018qqh}.

In scenarios involving self-annihilating DM, the central density of a DM spike surrounding a BH cannot grow arbitrarily large. At sufficiently high densities, the annihilation rate becomes significant and can efficiently deplete the innermost region of the spike. This results in the formation of a density plateau, where the annihilation timescale becomes comparable to the age of the BH. The maximal DM density in this inner region is obtained by requiring that the annihilation rate per unit volume, $\Gamma_{\rm ann} \sim \langle\sigma v\rangle \rho^2 / m_{\rm DM}^2$, does not exceed the inverse age of the BH, $1/t_{\rm BH}$~\cite{Gondolo:1999ef, Ullio:2001fb, Fields:2014pia}. This leads to an approximate expression for the maximum density:
\begin{equation}
	\rho_{\rm max} \sim \frac{m_{\rm DM}}{\langle\sigma v\rangle\,t_{\rm BH}}\,,
\end{equation}
where $m_{\rm DM}$ is the mass of the DM particle, $\langle\sigma v\rangle$ is the velocity-averaged annihilation cross section, and $t_{\rm BH}$ is the age of the central BH. The radius $r_{\rm cut}$ within which this plateau forms can be estimated by solving $\rho(r_{\rm cut}) = \rho_{\rm max}$, using the unperturbed spike profile, $\rho \propto r^{-\gamma_{\rm sp}}$. Inside this radius, the density flattens to $\rho_{\rm max}$, while outside it retains the steep power-law behavior associated with the spike. This annihilation-induced flattening has important implications for indirect detection of DM via gamma rays or other annihilation products, as it suppresses the signal from the innermost region while still allowing for potentially detectable emission from the surrounding spike. Detailed models incorporating relativistic effects and dynamical evolution further refine this picture~\cite{Eda:2013gg, Shapiro:2016ypb, Ferrer:2017xwm}.

In the cold, collisionless DM paradigm, density spikes are expected to form around BHs across a wide range of masses. These include supermassive black holes (SMBHs)~\cite{Gondolo:1999ef, Ullio:2001fb}, IMBHs~\cite{Bertone:2005hw, Zhao:2005zr}, and stellar-mass or even primordial BHs~\cite{Kohri:2014lza, Eroshenko:2016yve, Boucenna:2017ghj}. If a SMBH grows adiabatically within a DM halo exhibiting a central cusp, a high-density spike can develop with a power-law profile steepening to $\rho_{\rm spike}(r) \propto r^{-\alpha}$, where $2.25 \lesssim \alpha \lesssim 2.5$. These enhancements greatly increase the DM annihilation rate, potentially producing detectable gamma-ray fluxes~\cite{Bringmann:2011ut, Sandick:2016zeg, Kalaja:2019uju, Gow:2020bzo}. However, this idealized spike is subject to dynamical erosion due to galaxy mergers, baryonic feedback, and subhalo interactions~\cite{Macedo:2013qea, Barausse:2014pra}, raising questions about spike survival in realistic galactic environments~\cite{Bertone:2024wbn, Jangra:2024sif}.

\subsection{Gravitational-wave searches for DM spikes}

GW astronomy is a groundbreaking field that uses the direct detection of GWs to explore astrophysical phenomena and test Einstein's GR. GWs are faint ripples in spacetime, predicted over a century ago, that encode information about some of the most energetic events in the Universe, such as mergers of BHs and NSs. The first indirect evidence for GWs came from observing the orbital decay of binary pulsars, notably the Hulse-Taylor pulsar~\cite{Hulse:1974eb}. For decades, direct detection remained elusive despite early experimental efforts, including Explorer~\cite{Astone:1993ur} at CERN and Nautilus~\cite{Astone:1997gi} at INFN. A confirmation occurred in 2015, when the Advanced Laser Interferometer Gravitational-Wave Observatory (aLIGO) and VIRGO collaborations observed the GW strain emitted from the merger of two inspiraling BHs~\cite{LIGOScientific:2016aoc}. This detection not only confirmed a key prediction of GR but also opened an unprecedented observational window to the Universe, enabling tests of GR, measurements of BH and NS properties, and searches for phenomena beyond the SM. Nowadays, GW astronomy is revolutionizing our view of the cosmos, enabling the study of inaccessible phenomena, from the dynamics of compact objects to fundamental physics. Recent research has focused on using GWs to probe the dark sector. Gravitational radiation may carry subtle imprints from an interacting dark sector~\cite{Califano:2024xzt}, non-standard GW polarizations such as those sourced by a dark photon~\cite{Nomura:2024cku}, or deviations from GR captured in post-Einstein frameworks~\cite{Wilcox:2024sqs}. These studies remark the growing synergy between GW astronomy and the search for new physics.

Beyond direct signals from compact binary merger, GWs can also leave subtle, indirect imprints that are accessible through the search for DM. In particular, ultralight bosonic fields, such as axions, may form bound states with rotating BHs, leading to superradiant instabilities that emit long-duration, nearly monochromatic GWs~\cite{Arvanitaki:2014wva, Baryakhtar:2020gao}. The absence of these signals in GW data places constrains on axion-like particle masses and their coupling to gravity. Additionally, dense DM structures such as axion miniclusters or boson stars could interact with stellar remnants, inducing GW bursts via tidal disruption, collisions, or transit-induced perturbations~\cite{Tkachev:1991ka, Bai:2017feq}. These indirect GW signatures offer a novel probe of the small-scale structure and composition of DM. In this way, the search for DM is intertwined with GW observations.

Besides the DM spikes around massive BHs discussed earlier, an alternative formation process involves IMBHs, which may develop surrounding DM mini-halos that could exhibit a spiky density profile within a parsec of the IMBH, forming DM mini-spikes. These systems are of particular interest for GW astronomy~\cite{Eda:2014kra}. When a stellar-mass object orbits an IMBH embedded in a DM mini-spike, the GW signals emitted during the inspiral phase are influenced by the DM distribution~\cite{Amaro-Seoane:2007osp, MacLeod:2015bpa}. This effect has been explored in studies proposing that upcoming space-based GW detectors, such as LISA~\cite{Amaro-Seoane:2012aqc} and DECIGO~\cite{Kawamura:2008zza, Kawamura:2011zz}, could measure DM mini-spike parameters, such as the density profile's power-law index, with remarkable precision~\cite{Kavanagh:2020cfn, Coogan:2021uqv}.

Recent advancements aim to probe DM complementary to gamma-ray observations. Unlike EM waves, GWs pass through matter with minimal interaction, preserving pristine information about DM structures. This makes GW signals from IMBH-stellar binaries uniquely suited to exploring DM properties, independent of the particle's annihilation characteristics. Furthermore, matched filtering techniques and Fisher matrix analyses indicate that DM mini-spikes can have a measurable impact on GW waveforms, offering a new avenue for DM detection. These methods provide sensitivity to a wide range of DM candidates, extending beyond WIMPs, and may reveal insights into both the structure of DM mini-spikes and the underlying nature of DM particles.

In particular, matched filtering techniques and Fisher matrix analyses suggest that future detectors could resolve DM mini-spike parameters such as the power-law index $\alpha$ and normalization of the density profile~\cite{Kavanagh:2020cfn, Coogan:2021uqv}. Unlike gamma-ray searches, which rely on assumptions about DM annihilation, GW probes are sensitive to gravitational effects alone, making them agnostic to the DM particle model. This opens a new avenue for exploring DM physics using waveform distortions induced by gravitational interactions with surrounding DM. Importantly, these effects could be measurable even in the absence of self-annihilation or decay, making GWs a truly complementary probe of the dark sector.

\subsection{Gravitational waves from compact objects in modified gravity theories}
\label{sec:gw_modified_gravity}

Compact objects such as BHs and NSs probe the strong-field, highly dynamical regime of gravity that is inaccessible to traditional astrophysical observations. As they inspiral and merge, they emit GWs carrying detailed information about the underlying gravitational dynamics. These waveforms can deviate from GR predictions if gravity is governed by a modified theory at high energies or large distances. Several classes of modified gravity theories predict alterations to the structure and dynamics of compact objects, as well as to the generation and propagation of GWs. Examples include scalar-tensor theories such as Jordan-Brans-Dicke gravity~\cite{Brans:1961sx, Dicke:1961gz}, Einstein-dilaton-Gauss-Bonnet gravity~\cite{Kanti:1995vq, Pani:2009wy}, dynamical Chern-Simons gravity~\cite{Alexander:2009tp}, and massive gravity~\cite{deRham:2010kj}. These models often introduce new degrees of freedom, such as scalar or vector fields, that can modify the inspiral rate, alter the quasi-normal mode spectrum, or introduce dipole radiation.

In scalar-tensor theories, for example, BHs in GR remain indistinguishable from those in the modified theory due to no-hair theorems, but NSs can acquire scalar charges, leading to dipole radiation that accelerates the inspiral~\cite{Damour:1993hw, Damour:1995kt, Will:1994fb}. This mechanism affects the GW phase evolution and can be constrained using binary pulsar observations and LIGO/Virgo data~\cite{Shao:2017gwu}. In Einstein-dilaton-Gauss-Bonnet gravity, BHs can acquire scalar hair~\cite{Kanti:1995vq, Pani:2009wy}, resulting in observable deviations in the ringdown spectrum following merger events. GW propagation can also be affected by modifications to GR. In some massive gravity models or theories with Lorentz symmetry violation, the GW speed can deviate from the speed of light or exhibit dispersion~\cite{Lombriser:2015sxa, Ezquiaga:2017ekz, Copeland:2018yuh}. The near-simultaneous detection of GWs and EM counterparts from the binary NS merger GW170817~\cite{LIGOScientific:2017zic} placed stringent bounds on such deviations, excluding broad classes of scalar-tensor and Horndeski theories. Future space-based detectors~\cite{Amaro-Seoane:2012aqc, Kawamura:2008zza, Kawamura:2011zz} will dramatically enhance our ability to test modified gravity. These observatories will access lower frequency bands and probe inspirals involving intermediate-mass BHs and EMRIs, which are particularly sensitive to deviations from GR due to their long-duration signals and strong-field dynamics~\cite{Barack:2006pq, Babak:2017tow}.

We now turn to binary BH mergers and DM effects. Ref.~\cite{Fakhry:2024kjj} explores the merger rates of compact binaries within DM spikes surrounding SMBHs, in the context of two widely studied massive gravity (MG) models: Hu-Sawicki $f(R)$ gravity~\cite{Hu:2007nk} and the normal branch of the Dvali-Gabadadze-Porrati (nDGP) gravity~\cite{Dvali:2000hr}. By employing three SMBH mass functions--Benson, Vika, and Shankar--Ref.~\cite{Fakhry:2024kjj} analyzes how these MG models and mass functions influence the predicted merger rates of binaries involving PBHs and NSs. The findings indicate that both MG models predict significantly higher merger rates for PBH-PBH and PBH-NS binaries compared to GR, particularly in environments with high DM densities, such as those surrounding SMBHs. This enhancement is more pronounced at lower SMBH masses and for lower values of the power-law index, suggesting that MG models could play a crucial role in shaping compact binary mergers. Moreover, the comparison of different SMBH mass functions reveals that the choice of mass function significantly affects the predicted merger rates, with the Shankar and Vika mass functions yielding the lowest PBH abundances. These results emphasize the importance of accurately modeling the SMBH mass distribution and the structure of DM spikes, which, when influenced by the growth of SMBHs and DM halo models, can drive substantial variations in merger rates.

Additionally, the Hu-Sawicki $f(R)$ models generally predict lower PBH abundances than the nDGP models, especially when using the Vika and Shankar mass functions. This suggests that Hu-Sawicki $f(R)$ gravity may provide a more effective framework for explaining observed merger rates and constraining the properties of DM spikes. The study also observes that within the Hu-Sawicki $f(R)$ model, different orders of enhancement, denoted by $f_4$, $f_5$, $f_6$, highlight the role of the field strength $f_{R0}$, whereas in the nDGP models, the crossover distance $r_c$ significantly influences deviations from GR. Importantly, the findings confirm that the merger rates predicted by MG models, particularly at lower power-law indices, fall within the sensitivity range of current GW detectors such as the LIGO-Virgo-KAGRA (LVK) collaboration~\cite{LIGOScientific:2007fwp, Aso:2013eba, VIRGO:2014yos}. This suggests that ongoing and future GW observations could potentially distinguish between GR and MG scenarios by analyzing compact binary mergers in these extreme environments. By comparing the results with observational constraints on PBH abundances and LVK merger event rates, the study demonstrates that MG models can provide valuable insights into the formation and evolution of compact binaries. These findings underscore the potential of GW observations as a tool for testing and constraining alternative theories of gravity.

Beyond their impact on binary merger rates, MG models may also significantly influence the dynamics of PBHs near SMBHs. The high DM density in spikes around SMBHs can create a gravitational well that enhances PBH capture, leading to an increased rate of EMRIs, where a smaller BH spirals into a much larger SMBH. MG models, by introducing an effective fifth force, can further amplify these interactions, driving higher merger rates compared to GR. The energy loss through gravitational radiation in such systems accelerates inspiral processes, making EMRIs promising targets for space-based GW observatories like LISA, which could provide crucial insights into fundamental physics and the nature of DM. However, while LVK observations could help differentiate between GR and MG scenarios, interpreting merger rates is challenging. Multiple astrophysical channels, including isolated binary evolution, dynamical mergers in star clusters, and PBH mergers, contribute to the observed population. The complexity of these merger channels, along with uncertainties in the fraction of PBH-driven events, may limit the ability to distinguish between different MG models solely based on merger rates. Thus, while GW observations offer a promising avenue for testing alternative gravity theories, astrophysical uncertainties necessitate a careful approach when drawing conclusions about the nature of gravity in DM-rich environments.

\subsection{Fifth Force}

In nature, four fundamental forces are currently known: gravity, electromagnetism, the strong nuclear force, and the weak nuclear force. However, theoretical models and certain experimental results have suggested the possible existence of an additional interaction, commonly referred to as the \textit{fifth force}. The first indication of such a force emerged in 1986~\cite{Fischbach:1985tk}, following a reanalysis of the E{\"o}tv{\"o}s experiment, which measured the acceleration of objects with different compositions in Earth's gravitational field. Subsequent experiments have provided further hints, though no direct or conclusive evidence for a fifth force has yet been observed~\cite{Adelberger:1990ye, Fischbach:2020kwt}. A notable development occurred in 2015 when the ATOMKI research group reported evidence for a potential new particle, termed X17, with a mass of approximately 17\,MeV~\cite{Alves:2023ree, Denton:2023gat}. This particle, observed in anomalous nuclear transitions, is hypothesized to be a light boson capable of mediating a short-range fifth force. Similarly, in 2021, results from the muon $g$--2 experiment at Fermilab suggested the possibility of BSM physics, potentially linked to a new fundamental force~\cite{Muong-2:2021vma, Sorbara:2025mtx, Karuza:2024tju}.

The fifth force is generally assumed to be mediated by ultra-light scalar or vector particles and to follow a Yukawa-type potential. Any deviation from the Newtonian inverse-square law in experimental measurements could point to the presence of such a force. The fifth force is typically characterized by two key parameters: its strength and its range. The strength must be comparable to or weaker than gravity--otherwise, it would likely have been detected already. Its range may vary from sub-millimeter scales to cosmological distances~\cite{Moody:1984ba}.

One of the most intriguing aspects of the fifth force hypothesis is its potential connection to the dark sector, including DM and dark energy. Some theories suggest that the so called ``missing mass" problem in the universe may be attributable to a fifth force. Similarly, the observed accelerated expansion of the universe could be driven by a form of quintessence--a dynamic field behaving like dark energy, which may itself represent a fifth force~\cite{Dvali:2001dd}. Due to these far-reaching implications, the search for the fifth force has become an active area of research. Various forms of fifth forces have been proposed, including long-range and short-range types, as well as composition-dependent and composition-independent variants. Moreover, theoretical frameworks such as Kaluza-Klein theories, string theory, and supergravity predict fifth forces with Yukawa-like behavior arising from additional dimensions or fields~\cite{Mashhoon:1998tp}.

\newpage

%% file: 3_Theory.tex
\section{GW theory framework and fifth force}

In this section, we discuss tests of modified gravity, focusing on Brans-Dicke theory and screened scalar-field models, and examine their implications for Solar System experiments and GW observations. We also study the dynamics of spinning compact objects in the presence of scalar couplings using the worldline effective field theory approach, including the associated energy loss due to scalar radiation. Such scalar fields can modify GW phases and waveforms and lead to observable fifth-force effects. Finally, we consider ultralight bosons that generate long-range potentials, affecting perihelion precession, gravitational light bending, and the Shapiro time delay. Using Solar System tests, we derive constraints on the mass and coupling strength of these bosons.

\subsection{From Brans-Dicke to scalar-tensors}

Very light scalars on the scales of binary system can radiate away energy and therefore induce a change in the phase of the observed GW signal. This can take place as long as matter couples to the scalars. In the simple single-scalar case as is presented below, the most general coupling of a scalar to matter is associated to the conformal-disformal metric~\cite{Bekenstein:1992pj}  playing the role of a Jordan metric. 
In the following, we will develop the theory of these nearly-massless scalars coupled to matter and their influence on the dynamics of binary systems. We will also include the effect of spin which can both lead to a modification of trajectories and add new extra radiation. 

Brans-Dicke (BD) theory~\cite{Brans:1961sx, Dicke:1961gz} is the earliest attempt to introduce a dynamical gravitational coupling within the GR framework. As a scalar-tensor theory, BD represents a subset of the more general Horndeski models~\cite{Horndeski:1974wa}. Alongside the metric tensor, BD theory incorporates an additional scalar degree of freedom, $\varphi$, which modulates the gravitational interaction strength as $G = G_N / \varphi$, where $G_N$ denotes the local measurement of the Newtonian gravitational constant. The scalar field $\varphi$ interacts non-minimally with gravity through the Ricci scalar, and its dynamics are governed by a new dimensionless parameter, $\omega_{\rm BD}$ (or alternatively $\epsilon_{\rm BD} \equiv 1 / \omega_{\rm BD}$). The action of the theory, expressed in the Jordan frame, is
\begin{equation}
S = \int {\rm d}^4x \sqrt{-g} \left[ \frac{1}{16\pi G_N} \left( \varphi R - \frac{\omega_{\rm BD}}{\varphi} g^{\mu \nu} \partial_\mu \varphi \partial_\nu \varphi \right) - \rho_\Lambda \right] + S_m (\chi_i, g_{\mu \nu})\,,
\end{equation}
where the matter action $S_m$ arises from the Lagrangian density of matter fields, denoted by $\chi_i$ and $\rho_\Lambda$ denotes cosmological constant energy density. The equation of motion for the dimension-less $\varphi$ takes the form
\begin{equation}
    \Box \varphi = 8\pi G_N \,(3 + 2\omega_{\rm BD})^{-1}\,T\,,
\end{equation}
where $T$ represents the trace of the total energy-momentum tensor. GR is recovered in the limits $\varphi \to 1$ and $\epsilon_{\rm BD} \to 0$, which suppress $\varphi$'s dynamics and set its constant value to $G_N$.
As we will see below, this is nothing but a scalar-tensor theory with the choices
\begin{equation}
    \varphi=A^{-2}(\phi),\ \quad 3+2\omega_{\rm BD}=m_{\rm Pl}^2\left(\frac{d\phi}{d\ln A}\right)^2.
\end{equation}

GW observations offer a powerful means to probe potential deviations from GR. The GW170817 event, the first observed binary NS merger in NGC 4993 and announced by the LIGO-Virgo network, marked a significant milestone in testing GR through its multimessenger nature: GWs were detected almost simultaneously with gamma rays~\cite{LIGOScientific:2017zic, LIGOScientific:2017vwq}. The time delay between the detection of the GW and the gamma-ray burst counterpart, measured at $\delta t = (1.734\pm0.054)$\,s, constrains the relative speed difference between GWs and light to less than one part in $10^{15}$, ruling out many modified gravity theories that predict deviations in the GW propagation speed~\cite{Tasinato:2021wol}. The event also provided independent constraints on the GW luminosity distance, allowing for an analysis of the distance-redshift relation without relying solely on EM observations~\cite{Holz:2005df, Dalal:2006qt, MacLeod:2007jd, Nissanke:2009kt, Cutler:2009qv}. By comparing the Hubble constant inferred from GW170817 with measurements obtained using other methods, potential deviations from GR and the standard cosmological model are constrained. For instance, recent detections have excluded modified gravity models that predict a GW speed differing from the speed of light. The size of extra dimensions is also tested this way~\cite{Visinelli:2017bny}. Measurements of GW luminosity distance can also reveal friction effects caused by modified gravity as GWs traverse cosmological backgrounds. Precision measurements of GW at high redshifts by future observatories like LISA and the Einstein Telescope will further investigate discrepancies with GR.

Such models give rise to fifth forces that are severely constrained by solar system tests. Such tests constrain deviations from GR to be less than $10^{-5}$ unless the fifth force is screened. Here we review a class of scalar-tensor models with the usual quadratic kinetic terms and where the fifth forces are screened either with the chameleon~\cite{Khoury:2003rn}, symmetron~\cite{Hinterbichler:2010es}or the Damour-Polyakov mechanism~\cite{Damour:1994zq,Brax:2010gi}. We will not discuss the K-mouflage~\cite{Babichev:2009ee, Brax:2012jr} and Vainshtein~\cite{Horndeski:1974wa, Vainshtein:1972sx,Nicolis:2008in} screenings here. 

\subsection{Screened Scalar-tensor Models}

In the following, we will introduce a class of screened modified gravity models. In Einstein frame, we can write the action as 
\begin{equation}
   S = \int {\rm d}^4 x \sqrt{-g} \left( \frac{M_\mathrm{Pl}^2}{2} R - \frac{1}{2} (\partial \phi)^2 - V(\phi) \right) + S_\mathrm{matter}[A^2(\phi) g_{\mu_\nu}; \psi]~,
\end{equation}
with potential $V(\phi)$ and a coupling to matter $A(\phi)$. The matter action has not been written down explicitly, but note that in this class of models matter moves in the Jordan frame. For the chameleon model, the potential and coupling function are 
\begin{equation}
    V(\phi) = \Lambda^4 \left(1 + \frac{\Lambda^n}{\phi^n} \right)~, \quad\quad\quad\quad A(\phi) = e^{\phi/M}~,
\end{equation}
where $n$ is an integer, $\Lambda$ and $M$ scales specifying the model. This was originally proposed in Ref.~\cite{Khoury:2003rn}. When computing the equations of motion, the coupling to matter needs to be taken into account. This results in the field responding to both the potential and the matter coupling, becoming density dependent in the presence of matter of density $\rho_\mathrm{m}$. We can write this as an ``effective potential'' 
\begin{equation}
    V_\mathrm{eff} = \Lambda^4 \left(1 + \frac{\Lambda^n}{\phi^n} \right) + \frac{\phi}{M} \rho_\mathrm{m}~,
    \label{Veff-chameleon}
\end{equation}
The minimum of the effective potential, and hence the mass of the scalar particle, is density dependent. Therefore, the  particle is massive in a dense environment, evading all solar system tests, but almost massless in sparse environments; for instance cosmologically and thus giving rise to interesting physical effects. 

Another such model is the symmetron which has  effective potential
\begin{equation}
    V_\mathrm{eff}(\phi) = \frac{1}{2} \left( \frac{\rho}{M^2} - \mu^2 \right) \phi^2 + \frac{1}{4} \lambda \phi^4~.
\end{equation}
meaning that in regions where the ambient matter density is large, $\rho > \mu^2 M^2$, with $\mu$ the mass parameter and $\lambda$ the self interaction strength, the effective potential is positive and is minimized when $\phi = 0$ but symmetry can be spontaneously broken in regions with sufficiently small matter density $\rho$. Thus the coupling to matter `switches off' in dense environments,  but not in sparse ones below a critical density. 

One might expect a runaway dilaton field  to be another such model. This has potential and coupling to matter of 
\begin{equation}
     V(\phi) = V_{0} e^{-\alpha\phi}~,
     \quad\quad\quad A(\phi) = e^{\beta\phi}~.
\end{equation}
Whilst the mass derived from  the effective potential is density dependent, there is not a thin shell around a dense body where the field changes. As a consequence fifth forces are not screened so the model will not pass solar system tests. Instead one can use the environmentally dependent dilaton model~\cite{Brax:2010gi} but this will not be discussed further.  It should be emphasized that when considering this class of models, it is important to test the thin shell mechanism. This happens when the field inside an objects $\phi_{\rm in}$ and outside $\phi_{\rm out}$ are such that the effective coupling to matter is smaller than the coupling outside 
\begin{equation} 
\beta_{\rm eff}\equiv \frac{\vert \phi_{\rm out}-\phi_{\rm in}\vert }{2m_{\rm Pl} \Phi_N}\le \beta_{\rm out}
\end{equation}
where $\beta= m_{\rm Pl} \frac{{\rm d}\ln A(\phi)}{{\rm d}\phi}$, $m_{\rm Pl}$ is the reduced Planck scale and $\Phi_N$ the Newtonian potential at the surface of the object.

\subsection{GWs and fifth-force}

The absence of strong deviations from GR in the solar system may indicate that light scalar on large scales may be screened locally in the solar system for instance. 
It is possible to study the effects of a scalar field on NSs and BHs in models of screened modified gravity, such as chameleon theories~\cite{Khoury:2003aq,Brax:2004qh}. For example the no-hair theorem is based on a static BH in vacuum. As soon as the conditions of the no-hair theorem are violated it is possible for a BH to carry \textit{scalar hair}~\cite{Jacobson:1999vr}. This is the case in a realistic, astrophysical BH with an accretion disk~\cite{Davis:2014tea, Davis:2016avf}. The class of models considered have Lagrangian 
\begin{equation}
   S = \int {\rm d}^4 x \sqrt{-g} \left( \frac{m_\mathrm{Pl}^2}{2} R - \frac{1}{2} (\partial \phi)^2 - V(\phi) \right) + S_\mathrm{matter}[A^2(\phi) g_{\mu_\nu}; \psi]\,,
\end{equation}
with potential $V(\phi)$ and a coupling to matter $A(\phi)$. This is written in the Einstein frame, while matter moves in the Jordan frame. In the chameleon case, this reads 
\begin{equation}
    V(\phi) = \Lambda^4 \left(1 + \frac{\Lambda^n}{\phi^n} \right)~, \quad\quad\quad\quad A(\phi) = e^{\beta\phi/m_\mathrm{Pl}}\,,
\end{equation}
where $n$ is an integer, while $\Lambda$ and $M$ are model parameters. In this model the fifth force is screened since the minimum of the effective potential, and hence the mass of the scalar, is density dependent. There is also a thin shell about a dense object resulting in the fifth force only coming from a very small region at the surface of the object. 

In this framework, the effect of an accretion disk around a massive, Schwarzschild  BH was investigated. In Ref.~\cite{Davis:2016avf} Weyl geometry was used to model a realistic, astrophysical accretion disk. Both Schwarzschild and  Kerr geometry were considered. After analysis it was found that the effect of the chameleon scalar was a shift of the scalar towards the ISCO, which is pushed closer to the event horizon of the BH. The scalar profile has potential astrophysical implications, particularly for SMBHs. Whilst the fifth force was found to be small in comparison to the Newtonian force for realistic coupling parameters, even for SMBHs, the scalar radiation could be potentially observable. Taking an extreme mass ratio inspiral, e.g. a stellar mass object orbiting a SMBH, then the scalar radiation compared to the gravitational radiation was found to be~\cite{Davis:2016avf} 
\begin{equation}
    \frac{\dot E_{\phi}} {\dot E_{\rm GR}} \approx \beta \left(\frac{R_0}{R_S}\right)^{9/2} \left(\frac{\delta\phi}{\delta r}\right) \frac{M_{\rm BH}}{M_{\rm Pl}^3} [\frac{M_{\rm BH}}{m_t}]\,,
\end{equation}
with $m_t$ the test mass, which could be taken as a solar mass object, $R_0$ the ISCO and $R_s$ the Schwarzschild radius and $\delta\phi/\delta r$ the scalar displacement on the accretion disk. For a solar mass test mass taking 
\begin{equation}
    10^6 \le \frac{M_{\rm BH}}{M_{\odot}} \le 10^{10}\,,
\end{equation}
giving
\begin{equation}
    \beta^2 10^{-3} \le \frac {E_{\phi}}{E_{\rm GR}} \le \beta^2 10^7\,.
\end{equation}
For ``ultramassive'' BHs of mass above $10^{11}\,M_\odot$~\cite{Natarajan:2008ks, Carr:2020erq}, this can be very large and potentially observable.

Another important case is the effect of screened scalar models on NSs, which could lead to observational signatures in LVK data. Two models have been considered in detail: the chameleon theory discussed earlier, and a dilaton model~\cite{Brax:2010gi}, which is motivated by string theory and screens via the Damour-Polyakov mechanism~\cite{Damour:1994zq}. Both models gave similar physical effects. A Schwarzschild metric was used to model the relativistic star, or NS with the Tolman–Oppenheimer–Volkoff (TOV) equation. This was used in conjunction with the modified Einstein equations to include the scalar field and the Klein-Gordon equation. The system was studied numerically. The scalar field settled into its value in the interior of the star and took its exterior value, or value in low densities, outside. The change was over a very small region close to the surface of the star, indicating the screening of fifth forces. 

\begin{figure}
\centering
\includegraphics[width=6cm]{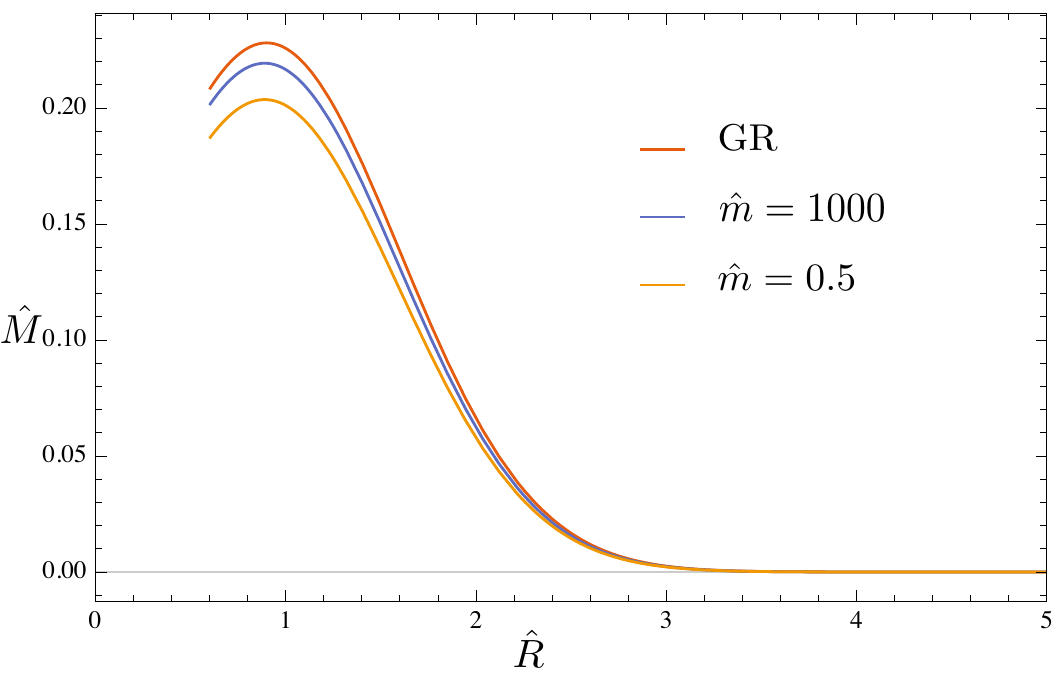}
\includegraphics[width=6cm]{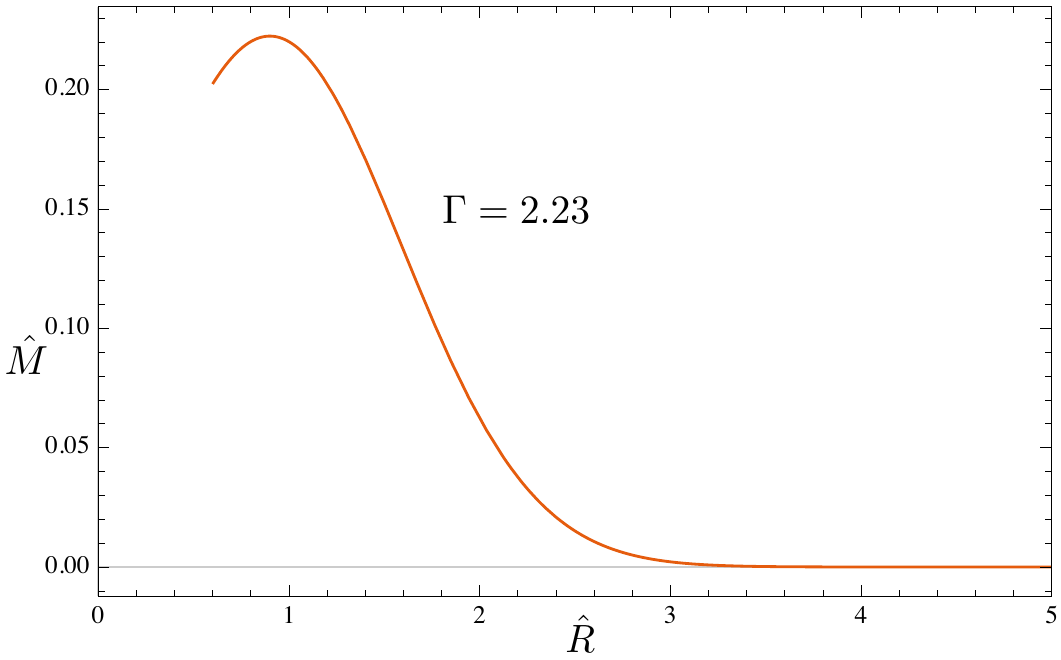}
\caption{Left: scaled mass--radius relation of a polytropic star ($\Gamma = 2$) for different chameleon masses. Right: the same relation in GR with $\Gamma = 2.23$, which mimics the $\hat{m} = 1000$ chameleon case. Adapted from Ref.~\cite{Brax:2017wcj}.}
\label{chamstar1}
\end{figure}

To solve the system one needs an equation of state (EoS) for the NS. A polytropic EoS of the form
\begin{equation}
    \tilde\rho = \left(\frac{\tilde P}{K}\right)^{1/\Gamma} + \frac{\tilde P}{\Gamma -1}\,,
\end{equation}
was used, with $\Gamma$ the polytropic index and $K$ a scale parameter. The pressure and matter density are specified in Jordan frame. Realistic NSs have $\Gamma \approx 1-3$. In our analysis we use values close to 2. The GR equation for a relativistic star such as a NS was solved with the TOV equation. This constrains the structure of a spherically symmetric body of isotropic material which is in static gravitational equilibrium. The equation was derived by solving Einstein equations for a spherically symmetric metric, such as the Schwarzschild metric. The TOV equation is 
\begin{equation}
\tilde P^{\prime} = -(\tilde\rho + \tilde P)\,\frac{m + r^3 \tilde P /(2m_\mathrm{Pl}^2)}{r^2(1- 2m/r)}\,.
\end{equation}
One solves this in conjunction with equations for the mass parameter. The full Einstein equations for the scalar model was solved with the TOV equation and the polytropic EoS. It was found that in the EoS $K=0.39$, both chameleon and dilaton gravity models have stellar configurations which were very similar. Solving the full equations it was found there was a degeneracy between the EoS for the NS and the scalar model parameters. This is shown in Fig.~\ref{chamstar1}. It is still unclear whether or not this could be detected in NS binary collisions or whether it would give observable consequences. 

There are more general possibilities one can study with scalar fields. For example, the most general coupling of the scalar field to matter involves both the conformal and disformal coupling, as discussed in Ref.~\cite{Bekenstein:1992pj}. In this case the metric between moving bodies is given by
\begin{equation}
g_{\mu\nu} = A^2(\phi)g^E_{\mu\nu} + B^2(\phi)\partial_{\mu}\phi\partial_{\nu}\phi\,,
\end{equation}
where $g^E_{\mu\nu}$ is the Einstein frame metric and the conformal and disformal couplings are respectively
\begin{equation}
    A(\phi) = e^{\beta\phi/m_\mathrm{Pl}}\,,\qquad B^2(\phi) = \frac{2}{m_{\rm Pl}^2 \Lambda^2}= \frac{1}{{\cal M}^4}\,,
\end{equation}
where the constant $\Lambda$ characterizes the strength of the disformal interaction. This was used to consider the effects of the disformal coupling on binary orbits in Refs.~\cite{Brax:2018bow, Brax:2019tcy} using perturbative techniques. For perturbation theory to be valid, and to recover the Newtonian limit we require the disformal term to be subdominant. The system is solved by using the Einstein and Klein-Gordon equations where the disformal term is treated perturbatively. At leading order, the Klein-Gordon equation is 
\begin{equation}
    \Box\phi = -\beta\frac{T}{m_\mathrm{Pl}} + \frac{1}{{\cal M}^4}D_{\mu} \partial_{\nu}\phi T^{\mu\nu}\,,
\end{equation}
where T is the trace of the matter energy-momentum tensor and $D_{\mu}$ is the covariant derivative for the Einstein metric. This expression can be solved iteratively using $\phi = \phi^{0} + \delta\phi$, where $\phi^{0}$ solves the Klein-Gordon equation with just the conformal coupling and the fluctuation follows
\begin{equation}
    \Box\delta\phi = \frac{1}{{\cal M}^4}D_{\mu} \partial_{\nu}\phi^{0} T^{\mu\nu}\,.
\end{equation}
Solving this expression, we find a series representation corresponding to an expansion in ladder diagrams 
\begin{equation}
    \partial\phi = \sum\limits_{n\ge 0}\delta\phi^{(n)}\,.
\end{equation}
The system is now solved iteratively with each iteration bringing in another insertion of the energy-momentum tensor and is suppressed by higher powers of $\Lambda^4$. The formalism can be applied to binary systems with the effect of the disformal term being obtained in two steps. Firstly one solves
\begin{equation}
    \Box\phi^{(0)} = -\beta\frac{T^A + T^B}{m_\mathrm{Pl}}\,,
\end{equation}
where the energy-momentum tensor contains both parts from particle A and B. The solution for $\phi^{(0)}$ is given by a linear combination. This solution sources $\delta\phi^{(0)}$. We can now work to leading order in $G_N$ and ${\cal M}^4$ and in the low velocity limit to consider the interaction of bodies A and B that are both conformally and disformally coupled. Essentially, we did a post-Newtonian (PN) expansion but including the conformal and disformal coupling terms. First one computes the full action consisting of the gravitational, matter and scalar parts by consistently expanding to leading and first non-leading order and then going to the centre of mass frame. One finally considers physical effects. The disformal term gives a negligible contribution to the Shapiro time delay but does affect the perihelion advance of orbits. After calculation one finds the perihelion advance for an orbit 
\begin{equation}
    \Delta\theta = 2\pi \frac{G_N m_A}{p}\Big[(3-2\beta^2) + 5\frac{\beta^2m_A}{2\pi {\cal M}^4 p^3}\Big]\,,
\end{equation}
for particle B orbiting particle A and 
\begin{equation}
    p = a(1-e^2)\,,
\end{equation}
where $e$ is the eccentricity of the orbit and $a$ the semi-major axis. Taking the Cassini bound for $\beta^2$ of $10^{-5}$, the perihelion advance of Mercury gives
\begin{equation}
    {\cal M}\ge {10^{-4}}{\rm \,MeV}\,,
\end{equation}
which is weaker than some laboratory constraints. 

\begin{figure}
\centering
\includegraphics[width=12cm]{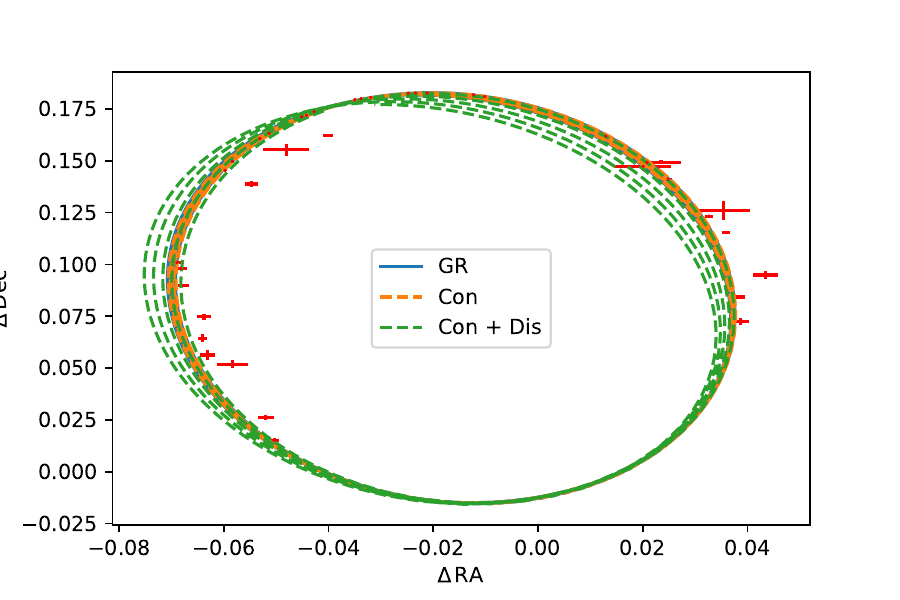}
\caption{The best fit of the S2 motion with conformal and disformal interactions. The evolution of the numerical simulation is for 60 years. Adapted from Ref.~\cite{Benisty:2021cmq}.}
\label{S2fit}
\end{figure}

One can however consider the effect of such coupled scalar particles in other astrophysical or cosmological situations, for example to two inspiralling NSs or stars orbiting a SMBH. If a BH had an accretion disk then there could be an effect on the orbits of stars around the BH. This situation was considered in Ref.~\cite{Benisty:2021cmq}. The S-stars are known to orbit the SMBH Sagittarius A$^\star$ (SgrA$^\star$), with the orbits being highly eccentric so reaching large velocities. The two stars whose orbits are closest to SgrA$^\star$ are $S2$ and $S38$ and there are $30$ year observational data on $S2$ in particular. Using the techniques outlined above and a MCMC numerical analysis predictions were made for the orbits of $S2$ and $S38$ over a $60$ year period and combined constraints on the disformal parameter made after inputting the Cassini bound for $\beta^2$. The $60$ year prediction for $S2$ is given in Fig.~\ref{S2fit}. The combined constraint of ${\cal M} \approx 0.08$\,MeV was obtained from the known orbits, taking the Cassini constraint into account. This is rather more stringent than that found for the perihelion advance of Mercury and close to that obtained in laboratory experiments. 

The method outlined above treated the disformal coupling in a perturbative expansion. In Ref.~\cite{Davis:2019ltc} it was shown there were two regimes for the disformal coupling, the perturbative and non-perturbative regime, with the ladder expansion breaking down when
\begin{equation}
    \frac{\Delta_{\mu}\phi\Delta_{\nu}\phi}{{\cal M}^4} \ge \frac{\phi}{m_\mathrm{Pl}}g_{\mu\nu}.    
\end{equation}
The ladder parameter is defined to be
\begin{equation}
    {\cal{L}} = \frac{v^2 (m_1m_2)^{\frac {1}{2}}}{r^3{\cal M}^4},
\end{equation}
where $m_1$ and $m_2$ are the particle masses and $v$ their typical relative velocity. For large separations, ${\cal{L}}\ll 1$ the effects of the disformal term are small and can be treated perturbatively, but when ${\cal{L}} \gg 1$, at small separations and large relative velocities, the effect of the disformal interactions are large and the ladder expansion breaks down. Instead a resummation of the ladder series was found. It was found that this resummation led to a new screening mechanism of fifth forces with fifth forces screened as $1/\cal{L}$, appropriate for binaries systems.  

\begin{figure}
    \centering
    \includegraphics[width=0.8\linewidth]{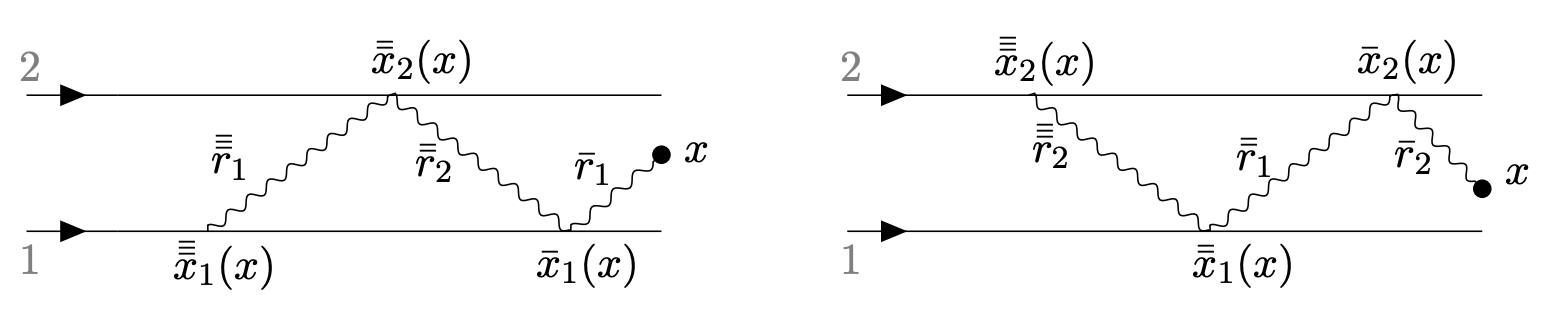}
    \caption{Iterated retarded positions on the worldline for the two bodies. The disformal coupling always has two legs out of a vertex. The exchange could be a scalar or a graviton and propagates along geodesics between the two bodies. Adapted from Ref.~\cite{Davis:2019ltc}.}
    \label{fig:ladder}
\end{figure}

The ladder approximation breaks down when corrections from subsequent ladder diagrams become $O(1)$. When this happens a different approach needs to be taken. In the ladder approximation, we have
\begin{equation}
    \Box\phi^0 = -\frac{\beta T}{m_\mathrm{Pl}}\,,
\end{equation}
which gives the series of fields in the expansion as $\phi^0$, $\phi^1$, $\phi^2$, and so on. For two bodies we can represent this diagrammatically as in Fig~\ref{fig:ladder}.

\begin{figure}
    \centering
    \includegraphics[width=0.8\linewidth]{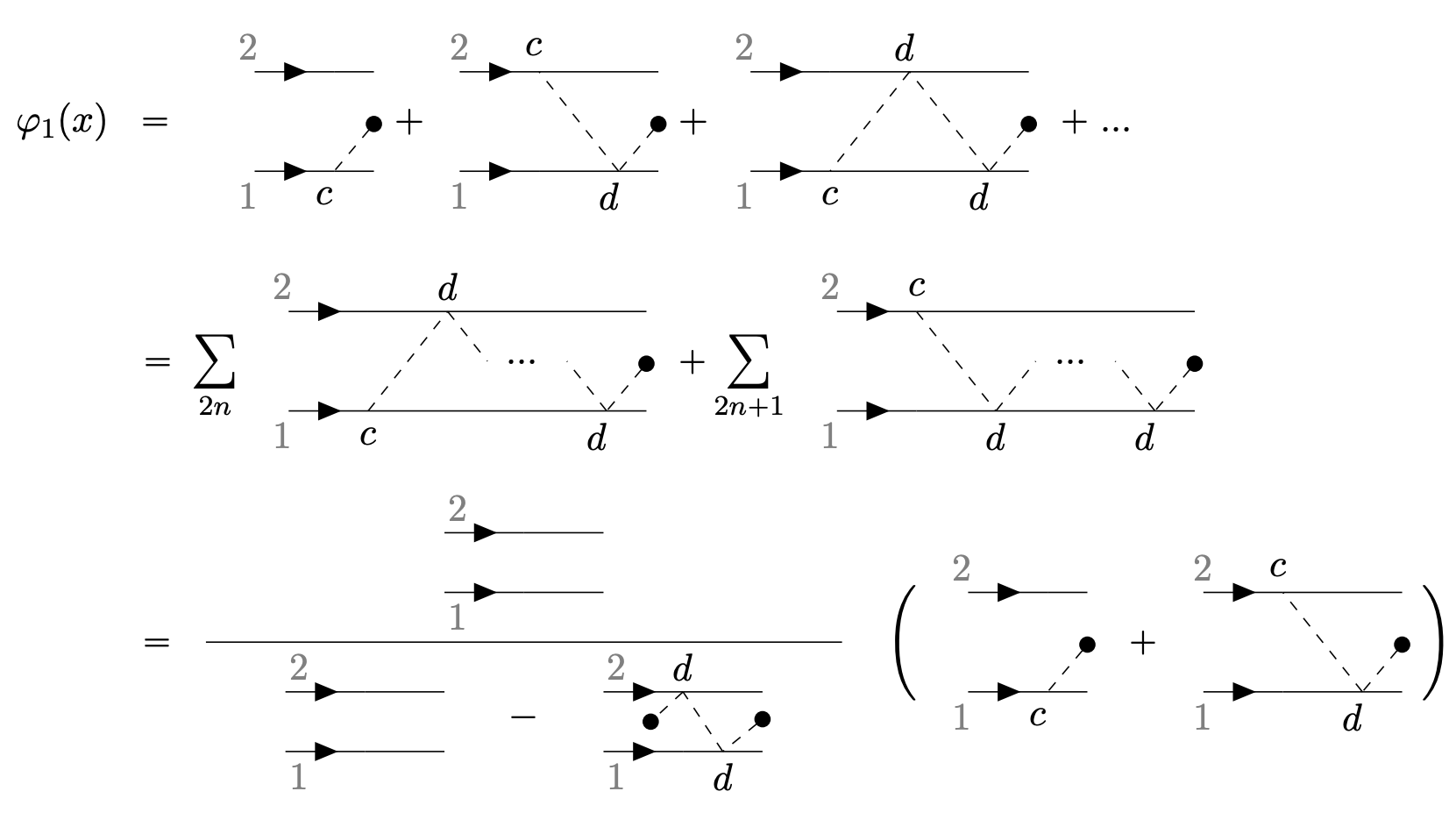}
    \caption{Resummation of the ladder, where $c$ and $d$ stand for conformal and disformal. Adapted from Ref.~\cite{Davis:2019ltc}.}
    \label{fig:ladder_resum}
\end{figure}

Formally, the ladder series takes the form
\begin{equation}
    \phi_1 = \sum\limits_{n\ge 0}\phi_1^{(n)}\,,
\end{equation}
for body $1$ and the $n^{\rm th}$ term can be expressed recursively using the preceding terms. 
\begin{equation}
    \phi_1^{(n)} =  \hat D_1 \left[ \phi_2^{(n-1)} \right]  = \hat D_1 \hat D_2 \left[ \phi_1^{(n-2)} \right] \,,
\end{equation}
using the two operators $\hat{D_A}$ which we can ascertain,
\begin{equation}
    \phi_1 = \frac{1}{1- \hat D_1 \hat D_2} \left[  \phi_1^{(0)} (x) + \phi_1^{(1)} (x) \right]\,.
\end{equation}
This is shown diagrammatically in Fig~\ref{fig:ladder_resum}. In the diagram, the first line corresponds to $\phi_1^0$ and $\phi_1^1$, i.e., the field sourced by body 1 at the first two orders in the expansion. The second line represents even and odd insertions of the disformal term and the final term gives the resummation.  

In Ref.~\cite{Davis:2019ltc}, this was applied to different physical binary systems to display the effect of the disformal screening uncovered by the resummation. In particular, it was applied to circular and elliptical motion. In all cases considered, it was found that the ladder resummation led to an efficient screening mechanism with the potential fifth force being screened by $1/{\cal{L}}$ in the ${\cal{L}} \gg 1$ regime. This was tested for planetary orbits in particular. One could consider such disformal theories in BH or NS collisions. Given the efficiency of the disformal screening there is unlikely to be a significant effect in the GW signal from the late time inspiral since that is clearly in the ladder resummation regime. Whether or not it is possible to detect such theories when the bodies are orbiting each other in the ${\cal{L}} \ll 1$ regime remains to be seen but could be quite hard to detect experimentally.

\subsection{Coupling spinning particles to scalars}
\label{sec:review}
\subsubsection{From the coupled action to the point-particle dynamics}
Below we will be  interested in light scalars of nearly vanishing masses and their influence of the dynamics of compact objects as captured by 
the general action\footnote{Our metric signature is~$(-,+,+,+)$. Units in which ${\hbar = c = 1}$ are used throughout and, as is common in the NRGR literature, we define the reduced Planck mass by ${\mpl \equiv (8\pi\GN)^{-1/2}}$. This section follows the line of Ref.~\cite{Brax:2021qqo}}
\begin{equation}
	S = S_\text{fields}[g,\phi]
	+
	S_\text{matter}[\tilde g(g,\phi)],
\label{eq:review_S_full}
\end{equation}
where matter couples to an effective Jordan-frame metric~$\tilde g_{\mu\nu}$, which we take to be~\cite{Bekenstein:1992pj}
\begin{equation}
	\tilde g_{\mu\nu}
	=
	A^2(\phi ,X) [g_{\mu\nu}
	+
	C^2(\phi,X) \partial_\mu\phi\,\partial_\nu\phi\,]
	\qquad
	(X \coloneq g^{\mu\nu}\partial_\mu\phi\,\partial_\nu\phi),
\label{eq:review_Jordan_frame}
\end{equation}
where $B=AC$.
For spinning objects such as pulsars, it is more convenient to work with the vielbein 
\begin{equation}
	\tilde e_\mu^a = A(\phi,X)[ e_\mu^a + D(\phi,X) \nabla^a\phi\nabla_\mu\phi]
\label{eq:pp_disformal_transformation}
\end{equation}
with $\tilde g_{\mu\nu}=\eta_{ab} \tilde e^a_\mu \tilde  e^b_\nu$.
Notice that the conformal coupling function~$C$ is dimensionless by construction, and so must depend on $\phi$ and $X$ only via the dimensionless combinations $\phi/\mpl$ and $X/\M^4$, where $\mpl$ is the reduced Planck mass and $\M$ is some strong-coupling scale.
In practice, we use the expansion
\begin{equation}
	A_\K = 1 - \alpha_\K \bigg(\frac{\varphi}{\mpl}\bigg)
	- \frac{\alphaprime_\K}{2} \bigg(\frac{\varphi}{\mpl}\bigg)^{\!2}
	+ \cdots,
	\quad
	D_\K = \frac{\beta_\K}{\M^4} + \cdots
\end{equation}
where we allow for a breaking of universality and different couplings for each particle. The parameters in this expansion will appear in the scalar-induced interactions between objects and the radiation by scalars.

In this action, the matter objects can in principle be of finite size. We will be interested in the long distance regime where they can effectively be treated as point-like to leading order. In this regime the disformal term can be treated in the ladder approximation. In the context of GWs emitted by such a system, the finite-size corrections induced by the Love numbers appear at higher order in the Post-Newtonian approximation (PN) than what we will consider. The resulting point-particle action is thus
\begin{equation}
	S_\text{eff} = S_\text{fields}[g,\phi]
	+
	\sum_{\K=1}^2 S_{\pp,\K}[x_\K,S_\K; g,\phi],
\label{eq:review_S_eff}
\end{equation}
where the point-like nature of the objects has been emphasized. The positions of the objects have been denoted by $x_\K$ and their spins $S_\K$.
Far away from the compact objects, the effective description can be summarised by first integrating out the short distance dynamics of the fields, i.e. the scalars and the graviton, corresponding to high frequency modes. One can then  separate the  effective action into a conservative part describing the energy-conserving dynamics of the point particles and a radiative sector involving the long wavelength radiative fields $\bar h$ for the gravitons and $\bar\varphi $ for the scalar
\begin{equation}
	S'_\text{eff}
	=
	S_\text{con}[x_\K,S_\K]
	+
	S_\text{rad}[x_\K,S_\K;\,\bar h,\bar\varphi].
\label{eq:review_S_eff'}
\end{equation} 
The conservative dynamics do not include radiation-reaction forces which lead to a loss of energy and are taken into account in the balance equation between the energy of the point-particles and the radiated power. The conservative action is defined by 
the Routhian\footnote{The Routhian mixes the Lagrangian coordinates $x_\K$ and $\dot x_\K$ with the Hamiltonian spin $S_\K$.}
\begin{equation}
	\mathcal R
	=
	-\sum_\K^\vph{\K} m_\K^{} \sqrt{1-\bm v_\K^2}
	-
	V(x_\K,v_\K,S_\K),
\label{eq:review_def_Routhian}
\end{equation}
defined by the integral over the coordinate time~$t$ where  ${S_\text{con} = \int\dx t\, \mathcal R}$. The point-particles have masses $m_\K$.  The conservative potential~$V$ arises from summing over all Feynman diagrams involving internal short-range scalar and graviton  lines and no external~ones:\looseness=-1%
\begin{equation}
	-\int\dx t\;V
	\eqfig{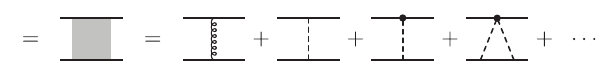}.
\label{eq:review_def_V}
\end{equation}
The Feynman rules involve the graviton (twiddly line) and the scalar (dashed line) and depend on the interactions between the short distance scalars and gravitons modes with matter. 
The equations of motions for the point particles and their spins are a mixed of the Euler-Lagrange equations for the Lagrangian coordinates and the Hamilton equations for the spins
\looseness=-1
\begin{equation}
	\diff{}{t}\bigg( \pdiff{\mathcal R}{\bm v_\K} \bigg)
	=
	\pdiff{\mathcal R}{\bm x_\K},
	\quad
	\diff{\bm S_\K}{t}
	=
	\{ \bm S_\K, \mathcal R \},
\label{eq:review_consv_eom}
\end{equation}
where the Poisson brackets are
\begin{equation}
	\{ S_{ab}, x^\mu \} = 0,
	\quad
	\{ S_{ab}, S_{cd} \}
	=
	- \eta_{ac \vph{d}} S_{bd} + \eta_{ad} S_{bc}
	- \eta_{bd} S_{ac \vph{d}} + \eta_{bc} S_{ad}.
\label{eq:pp_S_PB}
\end{equation}
This determines the trajectories of particles and their spins after taking into account all the interactions between objects mediated by scalars and gravitons. 

The radiative sector is characterized by a set of multipoles capturing the short-distance dynamics. In a nutshell, one considers the emission of a single radiative model and define the effective source $J$ obtained by summing Feynman diagrams with one open line.
\begin{equation}
	\int\dx^4x\, J(x) \bar\varphi(x)
	\eqfig{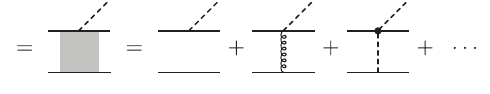},
\label{eq:review_def_J}
\end{equation}
from which $J(x)$ can be deduced. Any internal lines in these diagrams correspond to short-distance modes. Similarly, the effective  energy-momentum tensor~$T^{\mu\nu}(x)$ follows from summing over all Feynman diagrams with one external radiation-mode graviton,
\begin{equation}
	\frac{1}{2\mpl}\int\dx^4x\, T^{\mu\nu}(x) \bar h_{\mu\nu}(x)
	\eqfig{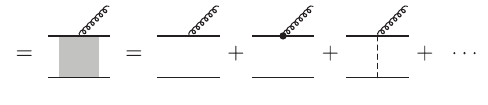}.
\label{eq:review_def_T}
\end{equation}

\begin{table}[t]
\centering
\def\arraystretch{1.3}
\begin{tabular}{|cl|cl|}
\hline
Symbol & Definition & Symbol & Definition \\
\hline
\multicolumn{2}{|l|}{Relative coordinates}		& \multirow{2}{*}{$\gammaeff$}
												& \multirow{2}{*}{$\displaystyle 1 -
													\frac{4\alpha_1\alpha_2}{1+2\alpha_1\alpha_2}$} \\
$\bm r$	& $\bm x_1 - \bm x_2$ 					&& \\
$\bm n$ 	& $\bm r/r \quad (r=|\bm r|)$     	& \multirow{2}{*}{$\gammapeff$}
												& \multirow{2}{*}{$\displaystyle 1 -
														\frac{2(m \alpha_1^2\alpha_2^2
														+ m_1^\vph{2}\alpha_1^2\alphaprime_2^\vph{2}
														+ m_2^\vph{2}\alpha_2^2\alphaprime_1^\vph{2})%
														}{ m (1+2\alpha_1\alpha_2)^2}$}\\
$\bm v$	& $\bm v_1 - \bm v_2$ 					&&\\
$\bm a$	& $\bm a_1 - \bm a_2$					& \multicolumn{2}{l|}{Radiative sector}\\
\multicolumn{2}{|l|}{Mass combinations}			& $\Delta\alpha$ & $\alpha_1 - \alpha_2$\\
$m$			& $m_1 + m_2$ 						& \multirow{2}{*}{$\comboA{\pm}{\ell}$}
												& \multirow{2}{*}{$\displaystyle\frac{
												\alpha_1^\vph{\ell} m_1^\vph{\ell} m_2^{\ell}
												\pm (-1)^\ell
												\alpha_2^\vph{\ell} m_2^\vph{\ell} m_1^{\ell}
												}{ m^{\ell+1} }$}\\
$\Delta m$	& $m_1 - m_2$						&&\\
$\nu$		& $m_1 m_2 / m^2$ 					& \multirow{2}{*}{$\comboAp{\pm}{\ell}$}
												& \multirow{2}{*}{$\displaystyle\frac{
												\alphaprime_1^\vph{2}\alpha_2^\vph{\ell} m_1^\vph{\ell} m_2^{\ell}
												\pm (-1)^\ell
												\alphaprime_2^\vph{2}\alpha_1^\vph{\ell} m_2^\vph{\ell} m_1^{\ell}
												}{ m^{\ell+1} }$}\\
\multicolumn{2}{|l|}{Spin combinations}			&&\\
$S$			& $S_1 + S_2$ 						& \multirow{2}{*}{$\comboBS{\pm}{\ell}$}
												& \multirow{2}{*}{$\displaystyle\frac{
												\tilde\beta_1^\vph{\ell}\alpha_2^\vph{\ell}
												m_1^\vph{\ell} m_2^{\ell}
												\pm (-1)^\ell
												\tilde\beta_2^\vph{\ell}\alpha_1^\vph{\ell}
												m_2^\vph{\ell} m_1^{\ell}
												}{ m^{\ell+1} } $}\\
$\Sigma$	& $m \, (S_2/m_2 - S_1/m_1)$ 		&&\\
$S_\pm$		& $(m_2/m_1)S_1 \pm (m_1/m_2)S_2$~~	& \multirow{2}{*}{$\comboA{\text{NLO}}{1}$}
												& \multirow{2}{*}{$\displaystyle
												  \frac{ 2\comboAp{-}{2} - \comboA{-}{2} }{1+2\alpha_1\alpha_2}
												  - \frac{3}{5}\comboA{+}{3} $}\\
$\chi_\K^{}$ & $S_\K^{}/(\GN m_\K^2)$ &&\\
\multicolumn{2}{|l|}{Conservative sector}		& \multirow{2}{*}{$\zeta$}
												& \multirow{2}{*}{$\displaystyle
												  1 + \frac{1}{3}\bigg(\frac{ \comboA{+}{2} }{\nu}\bigg)^2 $}\\
 $\Geff$ 	& $\GN(1+2\alpha_1\alpha_2)$ 		&&\\[5pt]
\hline
\end{tabular}
\caption{Definitions for all of the variables and parameter combinations used in this work. Both the antisymmetric tensor~$S_\K^{ij}$ and the 3-vector~$S_\K^i$ encode the same information about the bodies' spins, and one can easily switch between the two by using the definition ${S_\K^{ij} \equiv \epsilon^{ijk} S_\K^k}$. Adapted from Ref.~\cite{Brax:2021qqo}.}
\label{table:definitions}
\end{table}

Next, one Taylor expands in the radiation fields, scalars and gravitons,  about the centre of energy of the system taken to coincide with the origin.  Then using as basis the  symmetric and trace-free~(STF) operators, the result in the scalar sector~is%

\begin{equation}
	\int\dx^4x\, J(x)\,\bar\varphi(x)
	=
	\int\dx^4x\, J(x)
	\bigg(
		\sum_{\ell=0}^\infty
		\frac{1}{\ell!} \bm x^L \partial_L \bar\varphi(t,\bm 0)
	\bigg)
	=
	\sum_{\ell=0}^\infty\frac{1}{\ell!}\int\dx t\,
	\mathcal Q^L(t)\partial_L\bar\varphi(t,\bm 0),
\label{eq:review_Q_multipole_expansion}
\end{equation}
where the scalar multipole moments $\mathcal Q^L(t)$ are given by~\cite{Ross:2012fc}
\begin{equation}
	\mathcal Q^L(t)
	=
	\sum_{p=0}^\infty \frac{(2\ell+1)!!}{(2p)!!(2\ell+2p+1)!!}
	\left( \diff{}{t} \right)^{2p}\!\int\dx^3x\,
	J(t,\bm x)\, \bm x^{2p} \bm x^\avg{L}.
\label{eq:review_def_Q}
\end{equation}
This allows one to calculate the scalar radiated power by the system 
\begin{subequations}
\label{eq:review_rad_power}
\begin{equation}
	P_\phi
	=
	\sum_{\ell=0}^\infty
	\frac{ \langle ( {}^{(\ell+1)}\!\mathcal Q^L )^2 \rangle }%
		{4\pi\ell!(2\ell+1)!!},
\label{eq:review_rad_power_scalar}
\end{equation}
while the power radiated into GWs is~\cite{Ross:2012fc, Thorne:1980ru}
\begin{equation}
	P_g =
	\sum_{\ell=2}^\infty
	\frac{\GN}{\ell!(2\ell+1)!!}
	\bigg( \frac{ (\ell+2)(\ell+1) }{\ell(\ell-1)}
	\langle ( {}^{(\ell+1)}\!\mathcal I^L )^2 \rangle
	+
	\frac{ 4\ell(\ell+2) }{(\ell+1)(\ell-1)}
	\langle ( {}^{(\ell+1)}\!\mathcal J^L )^2 \rangle
	\bigg).
\label{eq:review_rad_power_gw}
\end{equation}
\end{subequations}
The total power ${P = P_\phi + P_g}$ computed in the radiative sector\,---\,can now be combined into the balance equation
\begin{equation}
	\bigg\langle \diff{E}{t} \bigg\rangle
	=
	-\avg{P}.
\label{eq:review_balance}
\end{equation}
This balance equation gives how the point-particles are subject to radiation-reaction forces which are not included in the conservative action.

In the case of a binary system and for circular non-precessing orbits,  both $E$ and $P$ depend on time only through the orbital frequency~$\Omega$, and so Eq.~\eqref{eq:review_balance} can in this case be recast into a differential equation for the binary's orbital phase ${\psi = \int\dx t\,\Omega}$. We will make this explicit below. The notation and parameters corresponding to the two-body problem are expounded in Table \ref{table:definitions}.

\subsubsection{Spinning point particles in scalar-tensor theories}
\label{sec:pp}

A spinning point particle can be described by a worldline $x^\mu(\lambda)$ and an orthonormal tetrad~$e^\mu_A(\lambda)$. The first specifies its trajectory  and is  the Jacobian ${e^\mu_A \equiv \partial x^\mu/\partial y^A}$ transforming between a general coordinate chart~$x^\mu$ and the particle's body-fixed frame~$y^A$. This transformation encodes the intrinsic rotation of the particle defined by an  angular velocity given by ${\Omega^{\mu\nu} \coloneq \eta^{AB} e^\mu_A D e^{\nu \vph{\mu}}_B/D\lambda}$, where ${D/D\lambda\equiv \dot x^\mu\nabla_\mu}$ is the covariant derivative along the tangent $\dot x^\mu$~($\equiv \dx x^\mu/\dx\lambda$) to the worldline. The conjugate momentum variables
\begin{equation}
	p_\mu
	=
	\bigg(\frac{\delta S_\pp}{\delta\dot x^\mu}\bigg)_\Omega
	\quad\text{and}\quad
	S_{\mu\nu}
	=
	\bigg(\frac{\delta S_\pp}{\delta\Omega^{\mu\nu}}\bigg)_{\dot x},
\end{equation}
are the momentum and the spin of the point-particle. The action can then be written as 
\begin{equation}
	S_\pp
	=
	\int \mathcal L_\pp\,\dx\lambda,
	\quad
	\mathcal L_\pp
	=
	p_\mu \dot x^\mu
	+
	\frac{1}{2} S_{\mu\nu}\Omega^{\mu\nu}
	-
	\mathcal H_\pp.
\label{eq:pp_general_action}
\end{equation}
where the Hamiltonian will be given below.

The transformation from a general set of coordinates~$x^\mu$ to the body-fixed coordinates~$y^A$ is obtained by going to the locally flat frame and then a Lorentz transformation to go into the body-fixed frame, i.e. ${e^\mu_A = \Lambda^a{}^\vph{\mu}_A e^\mu_{a\vph{A}} }$, where $e^\mu_a$ is the vierbein to the locally flat frame, while $\Lambda^a{}_A$ is the  Lorentz transformation. This implies that
\begin{equation}
	S_{\mu\nu}\Omega^{\mu\nu}
	=
	S_{ab} (\Omega^{ab}_\Lambda +  \w_\mu^{ab} \dot x^\mu),
\label{eq:pp_SOmega}
\end{equation}
where ${\Omega^{ab}_\Lambda = \eta^{AB} \Lambda^a{}_A \dot\Lambda^b{}_B}$ is the angular velocity relative to the locally flat frame and 
${S_{ab} = e^\mu_{a \vph{bb}} e^{\nu \vph{\mu}}_b S_{\mu\nu} }$
is the corresponding spin tensor in this frame. The minimal coupling between gravity and spin appears via the spin connection
\begin{equation}
	\w_\mu^{ab}
	\coloneq
	g^{\rho\sigma} e^b_\sigma \nabla_\mu^\vph{b} e^{a \vph{b}}_\rho.
\end{equation}
The generalised coordinates $(x^\mu,\Lambda^a{}_A)$ and their conjugate momenta $(p_\mu, S_{ab})$ together give us a total of 20 phase-space variables when only 12 are needed to describe a spinning point particle\,---\,three coordinates for its position and three angles to describe its orientation. This is reduced to 12 by imposing extra conditions in the Hamiltonian with associated Lagrange multipliers~\cite{Steinhoff:2009ei}
\begin{equation}
	\mathcal H_\pp
	=
	\frac{e}{2m}(p^2 + m^2)
	+
	e\chi_a ({\textstyle\sqrt{-p^2}}\Lambda^a{}_0 - p^a)
	+
	e\,\xi^a S_{ab} p^b,
\label{eq:pp_H}
\end{equation}
where the fields~$e(\lambda)$, $\chi_a(\lambda)$, and $\xi^a(\lambda)$ serve as Lagrange multipliers. They impose
${ S_{ab} p^b \approx 0}$~\cite{Tulczyjew:1959ssc}
in the Dirac sense of constraints. This is the covariant spin supplementary condition~(SSC) which  removes the three unphysical degrees of freedom contained in the spin tensor~$S_{ab}$. The other  constraint is~${\Lambda^a{}_0 \approx p^a/\sqrt{-p^2}}$~\cite{Hanson:1974qy}, which also removes just three degrees of freedom. 
The mass-shell constraint ${p^2 + m^2 \approx 0}$ is also imposed. The last remaining degree of freedom amounts to choosing the right parameter $\lambda$  for instance by demanding ${\dot x_\mu \dot x^\mu = -1}$.
The previous action must now be written in the Jordan frame. Then the conjugate momentum must be integrated out and the constraints must be applied~\cite{Brax:2021qqo}. This eventually leads to the Routhian\footnote{Defined as \begin{equation}
	\mathcal R_\pp
	=
	\mathcal L_\pp - \frac{1}{2} S_{ab}\Omega_\Lambda^{ab}.
\label{eq:pp_def_Routhian}
\end{equation}.}
\begin{align}
	\mathcal R_\pp
	=&
	-mA
	\sqrt{-\dot x^2}
	\bigg(
		1 + \frac{B(\dot x\cdot\!\nabla\phi)^2}{\dot x^2}
	\bigg)
	+
	\frac{1}{2}S_{ab\,} \w_\mu^{ab}\dot x^\mu
	-
	S_{\mu\nu}
	\dot x^{\mu}\nabla^{\nu}\log A
	\nonumber\\&
	-
	 D S_{\mu\nu}
	\dot x^\rho \nabla^{\mu}\phi \nabla^{\nu}\nabla_\rho\phi
	+D
	\frac{\dot x^\sigma \dot x^\mu}{\dot x^2}
	S_{\mu\nu}\dot x^\rho
	\big[
	   \nabla_\rho\!\nabla_\sigma \nabla^\nu \phi+
	  \nabla_\rho\phi\nabla_\sigma\nabla^\nu\phi
	\big].
\label{eq:pp_Routhian}
\end{align}
In the non-relativistic case, we have  ${v_\K^\mu \coloneq \dx x^\mu_\K/\dx t = (1,\bm v_\K)}$ as the corresponding gauge-fixed version of the tangent vector~$\dot x^\mu$. Figuratively, the point-particle action contains multiple interaction vertices and reads
\begin{equation}
	S_{\pp,\K} = -m_\K^\vph{2} \int\dx t\,\sqrt{1-\bm v_\K^2}
	\;\;+\;
	\bigg(
		\eqfig{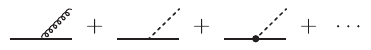}
	\bigg).
\end{equation}
Explicit expressions for the four main worldline vertices are presented here in Table~\ref{table:feyn_1}. 

\input{table_feyn_1}

In the case of a binary system with velocity $v$ in the centre of energy frame, perturbation theory depends on three dimensionless parameters. These are the orbital velocity~$v$, the ladder parameter ${\epsLadder \sim mv^2/(2\pi \M^4 r^3)}$, and the ratio ${\epsSpin \sim S/L}$, where ${L \sim mrv}$ is the binary's orbital~angular~momentum. Typically, we will limit ourselves to the first spin correction and the lowest orders in the disformal term. We will also quote results at the lowest non-trivial PN order.

\subsubsection{The conservative potential}

The conservative dynamics depend on the potential which can be split up in orbital (O) and spin (SO) parts. Similarly, contributions depend on the conformal interaction (C) and the disformal one (D). The spin-independent conformal part is given by~\cite{Damour:1992we}
\begin{subequations}
\label{eq:consv_V}
\begin{align}
	V^\sector{C}\scount{o}
	=&
	-
	\frac{\Geff m_1 m_2}{r}
	\bigg(
		1
		-
		\frac{\Geff m}{2r} (2\gammapeff - 1)
		\nonumber\\&
		+
		\frac{1}{2}
		\big[
			\bm v_1^2 + \bm v_2^2
			-
			3 \bm v_1\cdot\bm v_2
			-
			(\bm n\cdot\bm v_1)(\bm n\cdot\bm v_2)
			+
			2\gammaeff \bm v^2
		\big]
	\bigg),
\label{eq:consv_V_O_C}
\end{align}
where ${\bm v = \bm v_1 - \bm v_2}$ is the relative velocity between the two bodies, ${\bm r = \bm x_1 - \bm x_2}$ is their relative displacement, and ${\bm n = \bm r/r}$ is the unit vector pointing in this direction. The disformal potential is~\cite{Brax:2019tcy}
\begin{equation}
	V^\sector{D}\scount{o}
	=
	- 
	\frac{\GN m_1 m_2}{2\pi\M^4 r^4}
	(m_1^\vph{2}\alpha_1^2\beta_2^\vph{2} 
	+
	m_2^\vph{2}\alpha_2^2\beta_1^\vph{2})
	(\bm n \cdot \bm v)^2.
\label{eq:consv_V_O_D}
\end{equation} 
Now for the spin-orbit part of the potential, we have that~\cite{Brax:2021qqo}
\begin{align}
	V^\sector{C}\scount{so}
	&=
	\frac{\GN m_2}{r^2} n^i
	\big[
		2 S_1^{ij} v^j 
		+
		(1 + 2\alpha_1\alpha_2)
		(S^{i0}_1 - S_1^{ij} v_1^j)
	\big]
	+
	(1\leftrightarrow 2),
	\label{eq:consv_V_SO_C}
	\\
	V^\sector{D}\scount{so}
	&=
	\frac{\GN m_2^2 \alpha_2^2 }{2\pi\M^4 r^5} 
	\tilde\beta_1^\vph{j} S^{ij}_1 n^i v^j
	+
	(1\leftrightarrow 2).
\label{eq:consv_V_SO_D}
\end{align}
\end{subequations}
These interactions correct the motion of the spinning particles which ``wobble" along their trajectories.

\subsubsection{Radiated power and the gravitational-wave phase}
\label{sec:rad}
 
A two-body system eventually coalesces by emitting  gravitational and scalar waves. Here we focus on circular orbits only. The scalar  power can be split into a dipolar~($\dip$) and non-dipolar~($\nondip$) part
\begin{equation}
	P_\phi = P_{\phi,\dip} + P_{\phi,\nondip}.
\label{eq:rad_power_dipolar_split}
\end{equation}
The non-dipolar part is defined as the set of terms that survive in the limit ${\Delta\alpha \to 0}$. We can 
further subdivide each piece into a spin-independent and spin-orbit part. A subsequent decomposition into a conformal and disformal part  will also be made in the spin-orbit sector. The end result is thus a division of $P_\phi$ into six qualitatively distinct parts:
\begin{equation}
	P_\phi
	=
	\big(
		P_{\phi,\dip}^\vph{(}\scount{o}
		+
		P_{\phi,\dip}^\sector{C}\scount{so}
		+
		P_{\phi,\dip}^\sector{D}\scount{so}
	\big)
	+
	\big(
		P_{\phi,\nondip}^\vph{(}\scount{o}
		+
		P_{\phi,\nondip}^\sector{C}\scount{so}
		+
		P_{\phi,\nondip}^\sector{D}\scount{so}
	\big).
\end{equation}
It is convenient to define the Newtonian variable
\begin{equation}
	\x \coloneq (\Geff m \Omega)^{2/3},
\label{eq:consv_def_x}
\end{equation}
which behaves like $v^2$.
The power  in the spin-independent sector~is
\begin{subequations}
\label{eq:rad_power_phi}	
\begin{align}
	P_{\phi,\dip}\scount{o}
	&=
	\frac{2\Delta\alpha^2\nu^2}{3\Geff(1+2\alpha_1\alpha_2)} 
	\,\x^4,
\label{eq:rad_power_phi_O_dip}
	\\
	P_{\phi,\nondip}\scount{o}
	&=
	\frac{32 (\comboA{+}{2})^{2\vph{)}}_\vph{+} }%
		{15 \Geff(1+2\alpha_1\alpha_2)}
	\,\x^5,
\label{eq:rad_power_phi_O_nondip}
\end{align}
see, e.g., refs.~\cite{Damour:1992we, Kuntz:2019zef, Sennett:2016klh}. In the conformal spin-orbit sector, the leading terms are~\cite{Brax:2021qqo}
\begin{align}
	P_{\phi,\dip}^\sector{C}\scount{so}
	=&\:
	\frac{2\Delta\alpha^2\nu^2}{3\Geff(1+2\alpha_1\alpha_2)}
	\frac{2}{3}
	\bigg[
		{-}
		\frac{ \projell{S} }{\Geff m^2}
		\frac{5+2\alpha_1\alpha_2}{1+2\alpha_1\alpha_2}
		\nonumber\\[-3pt]&
		+
		\frac{ \projell{\Sigma} }{\Geff m^2}
		\bigg(
			\frac{3 \comboA{+}{0} }{\Delta\alpha}
			-
			\frac{3+2\alpha_1\alpha_2}{1+2\alpha_1\alpha_2}
			\frac{\Delta m}{m}
		\bigg)
	\bigg]
	\,\x^{11/2},
\label{eq:rad_power_phi_SO_C_dip}	
\\[3pt]
	P_{\phi,\nondip}^\sector{C}\scount{so}
	=&\:
	\frac{32 (\comboA{+}{2})^{2\vph{)}}_\vph{+} }%
		{15 \Geff(1+2\alpha_1\alpha_2)}
	\frac{4}{3}
	\bigg[
		{-}
		\frac{ \projell{S} }{\Geff m^2}
		\frac{5+2\alpha_1\alpha_2}{1+2\alpha_1\alpha_2}
		\nonumber\\[-3pt]&
		+
		\frac{ \projell{\Sigma} }{\Geff m^2}
		\bigg(
			\frac{ 15\nu \comboA{+}{0}\comboA{\text{NLO}}{1} }%
				{ 32 (\comboA{+}{2})^{2\vph{)}}_\vph{+} }
			-
			\frac{3+2\alpha_1\alpha_2}{1+2\alpha_1\alpha_2}
			\frac{\Delta m}{m}
		\bigg)
	\bigg]
	\,\x^{13/2},
\label{eq:rad_power_phi_SO_C_nondip}
\end{align}
while in the disformal spin-orbit sector, we have~\cite{Brax:2021qqo}
\label{eq:rad_power_phi_SO_D}
\begin{align}
	P_{\phi,\dip}^\sector{D}\scount{so}
	=&\:
	\frac{m}{2\pi\M^4(\Geff m)^3}
	\frac{ 2\Delta\alpha^2\nu^2 }{3\Geff (1+2\alpha_1\alpha_2)}
	\sum_{\sigma=\pm}
	\frac{ \projell{S}_\sigma }{\Geff m^2}
	\nonumber\\[-5pt]&
	\times\!
	\bigg(
		\frac{\comboBS{-\sigma}{0}}{\Delta\alpha\nu}
		-
		\frac{4}{3}
		\frac{\alpha_2^2\tilde\beta_1^\vph{2} + 	
			\sigma\alpha_1^2\tilde\beta_2^\vph{2}}%
		{1+2\alpha_1\alpha_2}
	\bigg)
	\,\x^{17/2},
\label{eq:rad_power_phi_SO_D_dip}
\\
	P_{\phi,\nondip}^\sector{D}\scount{so}
	=&\:
	\frac{m}{2\pi\M^4(\Geff m)^3}
	\frac{32 (\comboA{+}{2})^{2\vph{)}}_\vph{+} }%
		{15 \Geff(1+2\alpha_1\alpha_2)}
	\sum_{\sigma=\pm}
	\frac{ \projell{S}_\sigma }{\Geff m^2}
	\nonumber\\[-5pt]&
	\times\!
	\bigg(
		\frac{ 5\comboA{\text{NLO}}{1} \comboBS{-\sigma}{0} }%
			{ 16 (\comboA{+}{2})^{2\vph{)}}_\vph{+} }
		+
		\frac{ 2\comboA{+}{2} \comboBS{-\sigma}{1} }%
			{ (\comboA{+}{2})^{2\vph{)}}_\vph{+} }
		-
		\frac{8}{3}
		\frac{\alpha_2^2\tilde\beta_1^\vph{2} + 	
				\sigma\alpha_1^2\tilde\beta_2^\vph{2}}%
			{1+2\alpha_1\alpha_2}
	\bigg)
	\,\x^{19/2}.
\label{eq:rad_power_phi_SO_D_nondip}
\end{align}
\end{subequations}
These emitted powers correspond to radiation-reaction forces which make the orbits spiral inwards. 

\subsubsection{Inspiraling and the balance equation}

The total power $P$~(${ = P_\phi + P_g }$) enters in the balance equations which specifies the way energy is radiated in an adiabatic way and influences the orbits of the two inspiraling objects
\looseness=-1%
\begin{equation}
	\diff{t}{\x} = -\frac{E'}{P},
\label{eq:rad_dtdx}
\end{equation}
where~${E' \equiv \dx E/\dx\x}$. Using the relation  ${\Omega \equiv \dx\psi/\dx t}$ to  the orbital phase~$\psi$ of the emitted orbits in the quasi-circular motion of the reduced particle in the centre of energy frame, we have 
\begin{equation}
	\diff{\psi}{\x} = -\frac{\x^{3/2}}{\Geff m}\frac{E'}{P}.
\label{eq:rad_phase_master_formula}
\vphantom{\sum_|}
\end{equation}
In practice, this leads to the parametric definition of the phase ${t \equiv t(\x)}$ and~${\psi \equiv \psi(\x)}$. The evolution of the orbital separation~$r$ can be obtained by writing the equations of motion for a circular orbit subject to the potential $V$ leading to~\cite{Brax:2021qqo}

\begin{align}
	\frac{\Geff m}{r}
	=
	\x\,&
	\bigg[
		1
		+
		\frac{1}{3} (2\gammapeff + \gammaeff - \nu) \,\x
		+
		\frac{1}{3}
		\bigg(
			\frac{ \projell{S} }{\Geff m^2}
			\frac{5+2\alpha_1\alpha_2}{1+2\alpha_1\alpha_2}
			+
			\frac{ \projell{\Sigma} }{\Geff m^2}
			\frac{3+2\alpha_1\alpha_2}{1+2\alpha_1\alpha_2}
			\frac{\Delta m}{m}
		\bigg)
		\,\x^{3/2}
		\nonumber\\&
		+
		\frac{m}{2\pi\M^4 (\Geff m)^3}
		\sum_{\sigma=\pm}
		\frac{2}{3}
		\frac{ \projell{S}_\sigma }{\Geff m^2}
		\frac{\alpha_2^2\tilde\beta_1^\vph{2} + 	
				\sigma\alpha_1^2\tilde\beta_2^\vph{2}}%
				{1+2\alpha_1\alpha_2}
		\,\x^{9/2}
	\bigg],
\label{eq:consv_circles_eom_2}
\end{align}
which corrects the Newtonian expression.
A simplifying assumption is whether the inspiraling phase is due to the dipolar or non-dipolar parts in the scalar sector. 
Dipolar radiation occurs when 
\begin{equation}
	\x \ll \frac{5\Delta\alpha^2}{48\zeta}.
\label{eq:rad_DD_QD_boundary}
\end{equation}
We will focus on the non-dipolar emission dominated by the quadrupole emission.

\input{table_coefficients}

It is convenient to introduce the dimensionless function
\begin{equation}
	\rho(\x) \coloneq - \frac{1}{\Geff m}\frac{E'}{P},
\label{eq:rad_def_rho}
\end{equation}
which is proportional to both Eqs.~\eqref{eq:rad_dtdx} and~\eqref{eq:rad_phase_master_formula}.
In the quadrupole-driven case, we expand 
\begin{equation}
	\frac{E'}{P}
	\simeq
	\frac{E'}{P_\nondip}
	\bigg(
		1 - \frac{P_\dip}{P_\nondip}
	\bigg),
\label{eq:rad_QD_first_order_approximation}
\end{equation}
 where
\begin{equation}
	P_\dip = P_{\phi,\dip}
	\quad\text{and}\quad
	P_\nondip = P_{\phi,\nondip} + P_g.
\end{equation}
The approximation in Eq.~\eqref{eq:rad_QD_first_order_approximation} naturally causes $\rho(\x)$ to split into a dipolar and nondipolar part as well, and this motivates writing ${\rho_\QD(\x) = \rho_{\QD,\nondip}(\x) + \rho_{\QD,\dip}(\x) }$, with
\begin{subequations}
\begin{align}
	\rho_{\QD,\nondip}(\x)
	&=
	-\frac{1}{\Geff m}\frac{E'}{P_\nondip},
\label{eq:rad_QD_rho_nondip}
\\
	\rho_{\QD,\dip}(\x)
	&=
	\frac{1}{\Geff m}\frac{E'}{P_\nondip}\frac{P_\dip}{P_\nondip}.
\label{eq:rad_QD_rho_dip}
\end{align}
\end{subequations}
Both parts of $\rho_\QD$ can be expanded in powers of $\x$ leading to 
\begin{subequations}
\label{eq:rad_QD_rho}
\begin{align}
	&&
	\rho_{\QD,\nondip}^\vph{()}\scount{o}
	&=
	\frac{1+2\alpha_1\alpha_2}{32\zeta\nu}
	\frac{5}{2}\x^{-5},
&\qquad
	\rho_{\QD,\dip}^\vph{()}\scount{o}
	&=
	-\frac{1+2\alpha_1\alpha_2}{32\zeta\nu}
	\frac{25\Delta\alpha^2}{336\zeta}
	\frac{7}{2}\x^{-6},
&
\label{eq:rad_QD_rho_O}
\\
	&&
	\rho_{\QD,\nondip}^\sector{C}\scount{so}
	&=
	\mathcal{S}_{\QD,\nondip}^\sector{C}
	\,\x^{-7/2},
	&
	\rho_{\QD,\dip}^\sector{C}\scount{so}
	&=
	\mathcal{S}_{\QD,\dip}^\sector{C}
	\,\x^{-9/2},
&
\label{eq:rad_QD_rho_SO_C}
\\[3pt]
	&&
	\rho_{\QD,\nondip}^\sector{D}\scount{so}
	&=
	\mathcal{S}_{\QD,\nondip}^\sector{D}
	\,\x^{-1/2},
	&
	\rho_{\QD,\dip}^\sector{D}\scount{so}
	&=
	\mathcal{S}_{\QD,\dip}^\sector{D}
	\,\x^{-3/2},
&
\label{eq:rad_QD_rho_SO_D}
\end{align}
\end{subequations}
 to leading order in~$\x$. The relation 
Eq.~\eqref{eq:rad_phase_master_formula} can be integrated to yield 
\begin{align}
	\psi^\vph{()}_\QD
	=
	\psi_0^\vph{()}
	+
	\big(
		&
		\psi_{\QD,\nondip}^\vph{()}\scount{o}
		+
		\psi_{\QD,\nondip}^\sector{C}\scount{so}
		+
		\psi_{\QD,\nondip}^\sector{D}\scount{so}
	\big)
	\nonumber\\
	+\:
	\big(
		&
		\psi_{\QD,\dip}^\vph{()}\scount{o}
		\:+\:
		\psi_{\QD,\dip}^\sector{C}\scount{so}
		\:+\:
		\psi_{\QD,\dip}^\sector{D}\scount{so}
	\big).
\end{align}
The integration constant~$\psi_0$ is determined by the  initial conditions and the remaining terms are to leading order
\looseness=-1
\begin{subequations}
\label{eq:rad_QD_phase}
\begin{align}
	&&
	\psi_{\QD,\nondip}^\vph{()}\scount{o}
	&=
	-\frac{1+2\alpha_1\alpha_2}{32\zeta\nu}
	\x^{-5/2},
&\qquad
	\psi_{\QD,\dip}^\vph{()}\scount{o}
	&=
	\frac{1+2\alpha_1\alpha_2}{32\zeta\nu}
	\frac{25\Delta\alpha^2}{336\zeta}
	\x^{-7/2},
&
\label{eq:rad_QD_phase_O}
\\
	&&
	\psi_{\QD,\nondip}^\sector{C}\scount{so}
	&=
	-\mathcal{S}_{\QD,\nondip}^\sector{C}
	\,\x^{-1},
	&
	\psi_{\QD,\dip}^\sector{C}\scount{so}
	&=
	-\frac{1}{2}\mathcal{S}_{\QD,\dip}^\sector{C}
	\,\x^{-2},
&
\label{eq:rad_QD_phase_SO_C}
\\[3pt]
	&&
	\psi_{\QD,\nondip}^\sector{D}\scount{so}
	&=
	\frac{1}{2}
	\mathcal{S}_{\QD,\nondip}^\sector{D}
	\,\x^{2},
	&
	\psi_{\QD,\dip}^\sector{D}\scount{so}
	&=
	\mathcal{S}_{\QD,\dip}^\sector{D}
	\,\x.
&
\label{eq:rad_QD_phase_SO_D}
\end{align}
\end{subequations}
The phase~$\Psi$ of GWs derives directly from the binary's orbital phase~$\psi$. After decomposing the signal into spin-weighted spherical harmonics, the phase of the ${(\yl,\ym})$~mode is given by ${\Psi_{\yl\ym} = \ym \psi}$. If
${
	h_{\yl\ym}(t) = A_{\yl\ym}(t) e^{-i\ym\psi(t)}
}$
is the waveform for a particular oscillatory mode in the time domain, its counterpart in the frequency domain is given by the~Fourier~transform
\begin{equation}
	\tilde h_{\yl\ym}(f)
	=
	\int_{-\infty}^{\infty}\dx t\;e^{2i\pi ft} h_{\yl\ym}(t).
\label{eq:rad_freq_domain}
\end{equation}
The integral in Eq.~\eqref{eq:rad_freq_domain} can be evaluated using the stationary-phase approximation, i.e.
\[
	\tilde h_{\yl\ym}(f)
	=
	\tilde A_{\yl\ym}(f)
	e^{-i\tilde\Psi_{\yl\ym}(f)-i\pi/4},
\]
where~\cite{Sennett:2016klh}
\begin{equation}
	\tilde\Psi_{\yl\ym}(f)
	=
	\ym
	\bigg(
		 \psi(\x) - \frac{\x^{3/2}}{\Geff m} \,t(\x)
	\bigg)_{\x=\x_f},
	\quad
	\x_f = ( 2\pi\Geff m f/\ym)^{2/3}.
\label{eq:rad_gw_phase_Fourier}
\end{equation}
This can be used to forecast the effects of the disformal interactions in  experimental situations such as double NS encounters.

\subsubsection{Double neutron stars and observational prospects}
\label{sec:obsv}

 Let $\mathcal N$ be the total number of GW cycles that accumulate in the detector as a signal passes through. Following Ref.~\cite{Will:1994fb}, an interaction will be considered to  be undetectable if its contribution to~$\mathcal N$\,---\,integrated over the entire time that the signal spends in the detector's sensitivity band\,---\,is less than~$1$. We will focus on NS binaries as they can be coupled to matter contrary to BHs with no hair. The evolution of the binary can be split broadly into three phases:  the perturbative inspiral phase, the nonperturbative inspiral phase, and the merger phase. 
 \begin{table}
\centering
\def\arraystretch{1.5}
\begin{tabular}{|l l l l |}
\hline
& $m_\K$ & $\alpha_\K$ & $\chi_\K$ \\[2pt]
\hline
Weakly scalarised, slowly spinning (W)
& $1.25~M_\odot$
& $5\times 10^{-4}$
& $2.0\times 10^{-5}$

\\
Strongly scalarised, slowly spinning (S)
& $1.70~M_\odot$
& $7\times 10^{-3}$
& $2.0\times 10^{-5}$
\\
Strongly scalarised, rapidly spinning (S${}^{\textstyle\ast}$)
& $1.70~M_\odot$
& $7\times 10^{-3}$
& $1.4\times 10^{-2}$~
\\[2pt]
\hline
\end{tabular}
\caption{Fiducial parameters for several types of neutron stars.}
\label{table:neutron_star_parameters}
\end{table}
We have
\begin{equation}
	\mathcal N
	=
	\mathcal N_\text{pert}
	+
	\mathcal N_\text{nonpert}
	+
	\mathcal N_\text{merger}.
\label{eq:obsv_N}
\end{equation}
The first of these phases corresponds to the early portion of the inspiral, during which all three expansion parameters\,---\,$v$, $\epsSpin$, and $\epsLadder\epsSpin$\,---\, are small, where these parameters are defined after Table I. The second phase is present when $\epsLadder\epsSpin$ is large enough to invalidate our perturbative expansion.  Finally, the merger phase begins when the two NSs come into sufficiently close contact~\cite{Radice:2020ddv, Baiotti:2016qnr, Faber:2012rw}.
Perturbation theory stands as long as 
\begin{equation}
	\frac{m}{2\pi\M^4(\Geff m)^3}
	\frac{\max(\tilde\beta_1\chi_1, \tilde\beta_2\chi_2)}{1+2\alpha_1\alpha_2} 
	(\Geff m\pi f)^3
	\ll 1,
\end{equation}
where ${\chi_\K \coloneq \projell{S}_\K/(\GN m_\K^2)}$.
This defines the frequency when the non-perturbative regime starts as
\begin{equation}
	f_\text{nonpert}
	=
	\bigg|
		\frac{2\epsilon \mathcal{M}^4(1+2\alpha_1\alpha_2)}%
			{m\pi^2 \max(\tilde\beta_1\chi_1, \tilde\beta_2\chi_2) }
	\bigg|^{1/3},
\label{eq:obsv_f_nonpert}
\end{equation}
where $\epsilon \sim 0.1$ is a fudge factor.

Focusing on the dominant $(2,2)$ mode of the GW signal,  the number of cycles is given by
\begin{equation}
	\mathcal N_\text{pert}
	=
	\frac{\Psi_{2,2}(\x_2) - \Psi_{2,2}(\x_1)}{2\pi},
\label{eq:obsv_N_pert}
\end{equation}
where the initial time ${t_1 \equiv t(\x_1)}$ is the time at which the signal enters the detector's sensitivity band, while the final time ${t_2 \equiv t(\x_2)}$ marks the end of the perturbative inspiral phase (or the merging when no non-perturbative phase is present). For the former, we have ${\x_1 = (\Geff m \pi f_1)^{2/3}}$, and following ref.~\cite{Shao:2017gwu}, we consider three possible values for~$f_1$:
\begin{equation}
	f_1 \in \{ 10~\text{Hz},\, 5~\text{Hz},\, 1~\text{Hz} \}.
\label{eq:obsv_f1}
\end{equation}

\begin{figure}[t]
\centering\includegraphics[width=\textwidth]{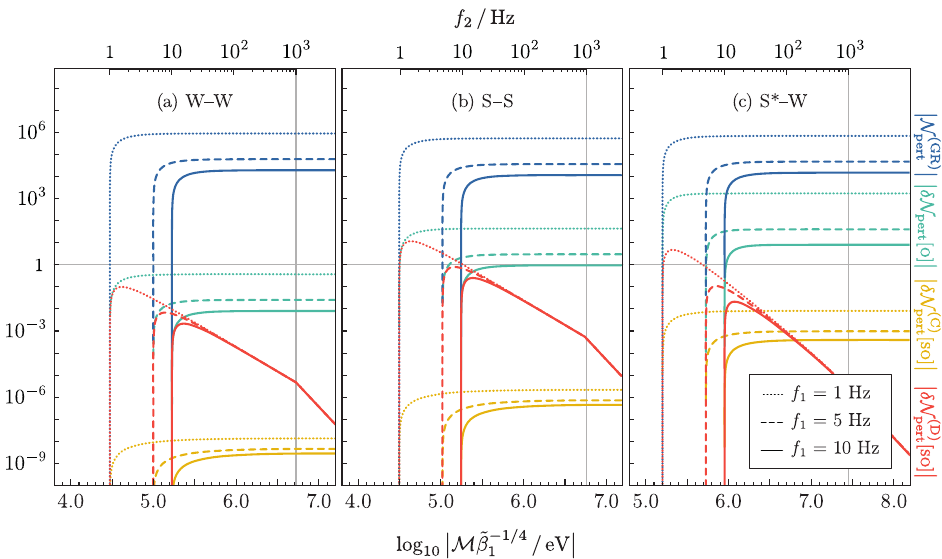}
\caption{Number of GW cycles~$\mathcal N_\text{pert}$  within the frequency band ${f \in [f_1,f_2]}$ during the inspiral of a double NS binary for different values of the disformal coupling scale~$\M$. Each type of NS (labeled W, S, or~S${}^{\textstyle\ast}$) are listed in Table~\ref{table:neutron_star_parameters}.
~$\mathcal N_\text{pert}$ splits into four parts: $\mathcal N_\text{pert}^\scriptsector{GR}$ is the prediction of GR, while $\delta\mathcal N_\text{pert}^{}\scount{o}$, $\delta\mathcal N_\text{pert}^\scriptsector{C}\scount{so}$, and $\delta\mathcal N_\text{pert}^\scriptsector{D}\scount{so}$ are the scalar  spin-independent, conformal spin-orbit, and disformal spin-orbit contributions.  The three values used for the lower frequency bound~$f_1$ give the beginning of the sensitivity bands for Advanced LIGO, Cosmic Explorer, and the Einstein Telescope. The upper frequency bound~$f_2$ corresponds to the end of the perturbative inspiral phase. For values of~$\M$ above a threshold shown in each panel as a vertical grey line, the binary goes directly from its perturbative inspiral phase into the merger phase, implying that ${f_2 = f_\text{contact}}$, with~${ f_\text{contact} \sim 1~\text{kHz} }$. For lower values of~$\M$, the binary system goes into a nonperturbative inspiral phase and ${f_2 = f_\text{nonpert}}$, with $f_\text{nonpert}$ given by Eq.~\eqref{eq:obsv_f_nonpert} as noted in the main text. The values of $f_2$ are shown along the top axis, and ${\mathcal N_\text{pert} = 0}$ when ${ f_2 \leq f_1}$. In each panel, there is a horizontal grey line giving the threshold for at least one GW cycle. Adapted from Ref.~\cite{Brax:2021qqo}.
\looseness=-1}
\label{fig:dephasing}
\end{figure}

The largest of these frequencies corresponds to the beginning of the sensitivity band for Advanced LIGO~\cite{Martynov:2016fzi}, while the remaining two are the forecasts for  third-generation ground-based detectors like Cosmic Explorer~(CE) and the Einstein Telescope~(ET)~\cite{Hild:2010id}. As for the end of the perturbative inspiral phase, we take ${\x_2 = (\Geff m \pi f_\text{nonpert})^{2/3}}$.
Only the formula for~$\psi_\QD$ in Eq.~\eqref{eq:rad_QD_phase} is needed to evaluate Eq.~\eqref{eq:obsv_N_pert}. The result can be split into four parts,
\begin{equation}
	\mathcal N_\text{pert}^{}
	=
	\mathcal N^\sector{GR}_\text{pert}
	+
	\delta\mathcal N_\text{pert}^{}\scount{o}
	+
	\delta\mathcal N^\sector{C}_\text{pert}\scount{so}
	+
	\delta\mathcal N^\sector{D}_\text{pert}\scount{so},
\label{eq:obsv_N_pert_split}
\end{equation}
where the different contributions from the non-dipolar part are
\begin{subequations}
\label{eq:obsv_N_pert_explicit}
\begin{align}
	\mathcal N_\text{pert}^\sector{GR}
	=&\:
	\bigg\{
		{-} \frac{1}{32\nu}
		(\GN m \pi f)^{-5/3}
	\bigg\}
	\Big._{f_1}^{f_2},
\allowdisplaybreaks\\[5pt]
	\delta\mathcal N_\text{pert}^{}\scount{o}
	=&\:
	\bigg\{
		-
		\frac{1}{32\nu}
		\bigg[
			\frac{1+2\alpha_1\alpha_2}{\zeta}
			-
			\bigg( \frac{\Geff}{\GN} \bigg)^{\! 5/3\,}
		\bigg]
		(\Geff m \pi f)^{-5/3}
	\bigg\}
	\Big._{f_1}^{f_2},
\label{eq:obsv_dN_pert_O}
\allowdisplaybreaks\\[5pt]
	\delta\mathcal N^\sector{C}_\text{pert}\scount{so}
	=&\:
	\bigg\{
		-
		\bigg[
			\mathcal{S}^\sector{C}_{\QD,\nondip}
			-
			\frac{1}{32\nu}
			\bigg(
				\frac{235}{6}
				\frac{\projell{S}}{\GN m^2}
				+
				\frac{125}{8}
				\frac{\Delta m}{m}
				\frac{\projell{\Sigma}}{\GN m^2}
			\bigg)
			\bigg( \frac{G_{12}}{G_N} \bigg)^{ 2/3}
		\bigg]
		(\Geff m \pi f)^{-2/3}
	\bigg\}
	\Big._{f_1}^{f_2},
\label{eq:obsv_dN_pert_SO_C}
\\[5pt]
	\delta\mathcal N^\sector{D}_\text{pert}\scount{so}
	=&\:
	\bigg\{
		\frac{1}{2}
		\mathcal{S}^\sector{D}_{\QD,\nondip}
		(G_{12} m \pi f)^{4/3}
	\bigg\}
	\Big._{f_1}^{f_2}.
\label{eq:obsv_dN_pert_SO_D}
\end{align} 
\end{subequations}
Typical NSs, are shown in Table~\ref{table:neutron_star_parameters}. 
For each, the magnitudes of $\mathcal N_\text{pert}^\scriptsector{GR}$, $\delta\mathcal N_\text{pert}^{}\scount{o}$, $\delta\mathcal N_\text{pert}^\scriptsector{C}\scount{so}$, and $\delta\mathcal N_\text{pert}^\scriptsector{D}\scount{so}$ are plotted as~functions~of~$\M$ in Fig. \ref{fig:dephasing}.
As can be seen, the spin-orbit effects on the number of cycles of the disformal interactions could become observable in a narrow band of values of $\M$ of the order $10^5$ eV. At this energy scale, classical tests of disformal couplings are weak, leaving this window open for future tests~\cite{Brax:2020vgg}. On the other hand, the conformal spin-orbit contribution will certainly remain unobservable. 

\subsection{Solar system tests of ultralight boson-induced fifth force}
An ultralight electrophilic scalar can interact with electrons through a scalar-type interaction. For instance, such a light scalar particle may act as a mediator between electrons within the Earth (or any astrophysical object) and those in a nearby gyroscope orbiting the object. The resulting scalar-mediated Yukawa potential modifies the geodetic precession of the gyroscope, leading to a shifted geodetic angle~\cite{Aliberti:2024udm}
\begin{equation}
\alpha = \frac{3\pi M}{R}
+ \frac{g^2 Q q}{4 R M_{\mathrm{gy}}}
  e^{-m_\phi R}(1+m_\phi R),
  \label{geodetic1}
  \end{equation}
where $M$ is the mass of the central object, $R$ is the orbital radius, $M_{\mathrm{gy}}$ is the gyroscope mass, $g$ denotes the scalar-electron coupling, $m_\phi$ is the scalar mass, and $Q$ and $q$ represent the total electron charge of the astrophysical object and the gyroscope, respectively.

This scalar-mediated interaction effectively induces a long-range fifth force in addition to the standard gravitational interaction between the gyroscope and the central body. The sensitivity to the scalar-electron coupling for a Sun-Earth system from the geodetic precession can be obtained as $g\lesssim 8\times 10^{-23}$ for the mass of the scalar $m_\phi\lesssim 10^{-18}\,\mathrm{eV}$~\cite{Aliberti:2024udm}. 

Furthermore, the presence of DM can also influence the gyroscope's motion. Assuming a constant local DM density $\rho_0$ along the gyroscope's orbit, the corresponding geodetic shift is modified to~\cite{Aliberti:2024udm}
\begin{equation}
\alpha = 3\pi M + 4\pi^2\rho_0 R^2.
\label{geodetic2}
\end{equation}

The same Sun-Earth system results the sensitivity on the DM overdensity as $\rho_0/\bar{\rho}\lesssim 7\times 10^{9}$~\cite{Aliberti:2024udm} from the geodetic precession, where $\bar{\rho}=0.4\,\mathrm{GeV}/\mathrm{cm^3}$.

In addition to scalar fields, ultralight pseudoscalars such as axions can also be probed through measurements of geodetic and frame-dragging effects. However, since axions couple to matter via spin-dependent interactions with nucleons in the gyroscope and in the central body, the resulting bounds are relatively weak~\cite{Poddar:2021ose}.

Similarly, ultralight dark photons, other vector gauge bosons, and even unparticle scenarios can, in principle, be constrained using such precision measurements~\cite{Poddar:2021ose}.

Ultralight vector bosons associated with gauged $L_e-L_{\mu,\tau}$ symmetries can also act as mediators between the Sun and the planets, generating a Yukawa-type potential in addition to the Newtonian gravitational potential. Such an interaction leads to an extra contribution to the perihelion precession of planetary orbits. The total shift in the perihelion angle, including both the general relativistic and vector-mediated effects, can be expressed as~\cite{KumarPoddar:2020kdz}
\begin{equation}
\Delta\phi \approx
\frac{6\pi GM}{a(1-e^2)}
+\frac{g^2 N_1 N_2 M_{Z^\prime}^2 a^2 (1-e^2)}
{4 M_p \left(GM + \dfrac{g^2 N_1 N_2}{4\pi M_p}\right)(1+e)}
\exp\big[-M_{Z^\prime} a (1-e^2)\big],
\label{perihelionm}
\end{equation}
where $a$ and $e$ denote the semi-major axis and eccentricity of the planetary orbit, respectively; $M$ is the solar mass; $M_p$ is the planetary mass; $g$ is the vector coupling constant; $N_1$ and $N_2$ correspond to the respective lepton-number charges of the Sun and the planet; and $M_{Z^\prime}$ is the mass of the ultralight vector mediator.

The first term in Eq.~\eqref{perihelionm} represents the general relativistic contribution, while the second term arises from the $L_e-L_{\mu,\tau}$ ultralight vector-mediated Yukawa potential. The perihelion precession of planets puts limit on the gauge coupling as $g\lesssim 3\times 10^{-25}$ for $M_{Z^\prime}\sim 10^{-18}\,\mathrm{eV}$~\cite{KumarPoddar:2020kdz}. Similar study has been done for asteroids as well~\cite{Tsai:2021irw,Tsai:2022jnv,Tsai:2023zza}.

The finite density corrections to the axion potential in celestial objects result scalar interaction of axions with nucleons which result a fifth force between objects in the solar system along with the gravitational force and affect gravitational light bending and Shapiro time delay. The axion-mediated long-range Yukawa potential modifies the gravitational light bending and hence the total bending is~\cite{Poddar:2021sbc}
\begin{equation}
\Delta \phi=\frac{\frac{4M}{b^2}+\frac{q_1 q_2}{2\pi M_p L^2}(1-0.347m^2_a b^2)\exp[-\frac{m_a L^2}{M}]}{\frac{1}{b}+\frac{q_1 q_2 m^2_a b^2}{8\pi M_p L^2}\exp[-\frac{m_a L^2}{M}]}  
\label{light-bending}
\end{equation}

where $q_1=4\pi f_a R_\odot$, $q_2=4\pi f_a R_\oplus$ are the axion charges of the Sun and the Earth respectively, $L^2=MD(1-e^2)$. The impact parameter $b\sim R_\odot$ is the solar radius and $R_\oplus$ denotes the Earth's radius. $D$ is the semi-major axis of the Earth's orbit and $e$ is the orbital eccentricity, $M$ and $M_p$ denote the masses of the Sun and Earth respectively, and $m_a$ denotes the mass of the axion with the decay constant $f_a$. In absence of axion-mediated Yukawa type force $(q_1=0=q_2)$, Eq.~\eqref{light-bending} reduces to the standard GR result 
\begin{equation}
\Delta \phi|_\mathrm{GR}=\frac{4M}{b}=\frac{4GM}{c^2R_\odot}=1.75\,\mathrm{arc~second}.    
\end{equation}

The axion-mediated fifth force can also modify the Shapiro time delay experienced by light signals propagating between the Earth and Venus. When a radar beam travels from the Earth to Venus and is reflected back, the presence of the Sun causes both a deflection and a delay in the light's propagation due to the gravitational potential along its trajectory. If the Earth and Venus possess axion-induced charges, then the axion-mediated potential contributes an additional correction to this delay on top of the usual gravitational one.

Accordingly, the total Shapiro time delay in the presence of both gravitational and axion-mediated interactions is given by~\cite{Poddar:2021sbc}
\begin{equation}
\begin{split}
\Delta T=4M\Big[\ln \Big(\frac{4r_er_v}{r^2_0}\Big)+1\Big]+(2b_0c_0(-1+c_0M)(r_e+r_v)+\frac{b_0c^2_0}{2}(r^2_e+r^2_v)+2b_0-4c_0Mb_0+\\
2a_0(r_e+r_v)+\frac{b_0}{24}(48+36 c^2_0r^2_0)[{\rm Ei}(-c_0 r_e)+{\rm Ei}(-c_0r_v)])\exp\Big[-\frac{m_a L^2}{M}\Big],
\end{split}
\label{shapiro}
\end{equation}
where 
\begin{equation}
a_0=\frac{q_1q_2 e^{-m_a r_0}}{4\pi M_p E^2 r_0},~~~ b_0=\frac{q_1q_2}{4\pi M_p E^2},~~~ c_0=m_a\,.
\end{equation}
Here, ${\rm Ei}(x)$ denotes the exponential integral function, $r_0$ is the solar radius, and $r_e$ and $r_v$ are the distances from the Sun to the Earth and Venus, respectively. $M_p$ represents the Earth's mass, while $q_1$ and $q_2$ are the axion-induced effective charges of the Sun and the Earth. The energy parameter is approximated by

\begin{equation}
E^2\approx \frac{L^2}{r^2_0}\Big(1-\frac{2M}{r_0}\Big)\Big(1-\frac{q_1q_2 e^{-m_a r_0}}{2\pi M_p E^2 r_0}\Big).
\end{equation}
In the absence of axion-mediated effects $(q_1=q_2=0)$, Eq.~\eqref{shapiro} reduces to the standard GR prediction for the Shapiro delay
\begin{equation}
\Delta T|_{\rm GR}=4M\Big[\ln \Big(\frac{4r_er_v}{r^2_0}\Big)+1\Big]=200\,\mu s,
\end{equation}
which corresponds to the well-known time delay measured for radar signals exchanged between the Earth and Venus in the Sun’s gravitational field. Precision measurements of the light bending and Shapiro time delay put upper limits on the axion decay constant as $f_a\lesssim 10^{10}\,\mathrm{GeV}$ and $f_a\lesssim 10^{7}\,\mathrm{GeV}$ for $m_a\lesssim 10^{-18}\,\mathrm{eV}$ respectively~\cite{Poddar:2021sbc}. Astrophysical observations can also be used to constrain mixed couplings when such ultralight particles act as mediators~\cite{OHare:2020wah,Poddar:2023bgk}. 

Beyond the Solar System, these solar system tests can also be investigated in astrophysical systems such as binary pulsars or in the orbit of the S2 star around Sgr A*, which has a semi-major axis of approximately $1000\,\mathrm{AU}$.

\newpage

%% file: table_feyn_1.tex
\begin{table}
\centering
\begin{tabular}{|@{\hspace{10pt}} l @{\hspace{25pt}} l @{\hspace{25pt}} l @{\hspace{10pt}}|}
\hline
Diagram & Interaction vertex & Scaling \\
\hline
\includegraphics[valign=c]{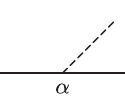}
&
$\tablestyle
\frac{\alpha_\K m_\K}{2\mpl}
\int\dx t\: \varphi$
&
$\sqrt{L}$
\\
\includegraphics[valign=c]{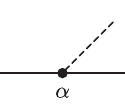}
&
$\tablestyle
\frac{\alpha_\K}{2\mpl}
\int\dx t\:
(S_\K^{i0} - S_\K^{ij}v_\K^j)\,\partial_i\varphi $
&
$\sqrt{L}v^2\epsSpin$
\\
\includegraphics[valign=c]{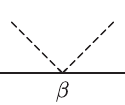}
&
$\tablestyle
\frac{\beta_\K m_\K}{\M^4} \int\dx t\:
\dot\varphi^2$
&
$\tablestyle
\epsLadder e^2$
\\
\includegraphics[valign=c]{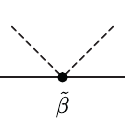}
&
$\tablestyle
\frac{\beta_\K}{\M^4}
\int\dx t\: S^{ij}_\K
(\partial_i\dot\varphi)\,\partial_j\varphi $
&
$\epsLadder\epsSpin$
\\[25pt]
\hline
\end{tabular}
\caption{Feynman rules for the four main worldline vertices. The scalar field is evaluated along the worldline; i.e., ${\varphi \equiv \varphi(t,\bm x_\K)}$, and  $\dot\varphi \equiv v_\K^\mu\partial_\mu\varphi$ for a total time derivative. Each factor of~$\varphi$ is a dashed line, while the worldline is a solid line. Vertices without black dots are spin-independent, while those with a black dot are coupled to one power of spin. The right column lists the power-counting rules for each vertex scales, assuming all factors of~$\varphi$ are taken to be short-distance modes. If $\varphi$  is taken  a radiation mode, include an extra factor of $\sqrt{v}$. Adapted from Ref. \cite{Brax:2021qqo}.}
\label{table:feyn_1}
\end{table}

%% file: table_coefficients.tex
\begin{table}
\centering
\def\arraystretch{1.4}
\begin{tabular}{| @{\hspace{10pt}} l @{\hspace{15pt}} l @{\hspace{10pt}} |}
\hline

\multicolumn{2}{|c|}{\small Quadrupole-driven regime}\\[2pt]
\hline
$\begin{aligned}
\tablestyle
	\mathcal{S}_{\QD,\nondip}^\sector{C}
\\[5pt]~
\end{aligned}$
&
$\begin{aligned}
\tablestyle
	&
	\frac{1+2\alpha_1\alpha_2}{32\zeta\nu}
	\frac{5}{6}
	\bigg\{
		\frac{ \projell{S} }{\Geff m^2}
		\bigg(
			\frac{55-2\alpha_1\alpha_2}
				{1+2\alpha_1\alpha_2}
			-
			\frac{8}{\zeta}
		\bigg)
		\nonumber\\[-10pt]&
		-
		\frac{\projell\Sigma}{\Geff m^2}
		\bigg[
			\frac{ 5\comboA{+}{0}\comboA{\text{NLO}}{1} }
				{8\zeta\nu}
			-
			\bigg(
				\frac{27-2\alpha_1\alpha_2}
					{1+2\alpha_1\alpha_2}
				-
				\frac{33}{4\zeta}
			\bigg)
			\frac{\Delta m}{m}
		\bigg]
	\bigg\}
\end{aligned}$
\\[35pt]
$\begin{aligned}
\tablestyle
	\mathcal{S}_{\QD,\dip}^\sector{C}
\\[5pt]~
\end{aligned}$
&
$\begin{aligned}
\tablestyle
	&
	\frac{1+2\alpha_1\alpha_2}{32\zeta\nu}
	\frac{25\Delta\alpha^2}{336\zeta}
	\frac{7}{6}
	\bigg\{
		{-}
		\frac{ \projell{S} }{\Geff m^2}
		\bigg(
			\frac{65+2\alpha_1\alpha_2}{1+2\alpha_1\alpha_2}
			-
			\frac{48}{3\zeta}
		\bigg)
		\nonumber\\[-10pt]&
		-
		\frac{\projell{\Sigma}}{\Geff m^2}
		\bigg[
			\frac{ 6\comboA{+}{0} }{\Delta\alpha}
			-
			\frac{ 5\comboA{+}{0}\comboA{\text{NLO}}{1} }
				{4\zeta\nu}
			+
			\bigg(
				\frac{33 + 2\alpha_1\alpha_2}{1+2\alpha_1\alpha_2}
				-
				\frac{33}{2\zeta}
			\bigg)
			\frac{\Delta m}{m}
			\bigg)
		\bigg]
	\bigg\}
\end{aligned}$
\\[33pt]
$\begin{aligned}
\tablestyle
	\mathcal{S}_{\QD,\nondip}^\sector{D}
\\[5pt]~
\end{aligned}$
&
$\begin{aligned}
\tablestyle
	&
	\frac{m}{2\pi\M^4(\Geff m)^3}
	\frac{1+2\alpha_1\alpha_2}{32\zeta\nu}
	\frac{5}{6}
	\sum_{\sigma=\pm}
	\frac{ \projell{S}_\sigma }{\Geff m^2}
	\nonumber\\[-10pt]&
	\times\!
	\bigg(
		\!-
		\frac{ 5\comboA{\text{NLO}}{1}\comboBS{-\sigma}{0} }
			{16\zeta\nu^2}
		-
		\frac{2\comboA{+}{2}\comboBS{-\sigma}{1}}
			{\zeta\nu^2}
		+
		\frac{52( \alpha_2^2\tilde\beta_1^\vph{2}
			+ \sigma\alpha_1^2\tilde\beta_2^\vph{2} ) }%
			{ 1 + 2\alpha_1\alpha_2 }
	\bigg)
\end{aligned}$
\\[33pt]
$\begin{aligned}
\tablestyle
	\mathcal{S}_{\QD,\dip}^\sector{D}
\\[5pt]~
\end{aligned}$
&
$\begin{aligned}
\tablestyle
	&
	\frac{m}{2\pi\M^4(\Geff m)^3}
	\frac{1+2\alpha_1\alpha_2}{32\zeta\nu}
	\frac{25\Delta\alpha^2}{336\zeta}
	\frac{7}{6}
	\sum_{\sigma=\pm}
	\frac{ \projell{S}_\sigma }{\Geff m^2}
	\nonumber\\[-10pt]&
	\times\!
	\bigg(
		\!-
		\frac{3\comboBS{-\sigma\,}{0}}{\Delta\alpha\nu}
		+
		\frac{ 5\comboA{\text{NLO}}{1}\comboBS{-\sigma}{0} }
			{8\zeta\nu^2}
		+
		\frac{4\comboA{+}{2}\comboBS{-\sigma}{1}}
			{\zeta\nu^2}
		-
		\frac{56( \alpha_2^2\tilde\beta_1^\vph{2}
			+ \sigma\alpha_1^2\tilde\beta_2^\vph{2} ) }%
			{ 1 + 2\alpha_1\alpha_2 }
	\bigg)
\end{aligned}$
\\[37pt]
\hline
\end{tabular}
\caption{Explicit expressions for the spin-orbit coefficients that appear in our solutions to the orbital and GW phase. All of the symbols used above are defined in Table~\ref{table:definitions}. Adapted from Ref. \cite{Brax:2021qqo}.}
\label{table:coefficients}
\end{table}

%% file: 4_StochasticGW.tex
\section{Stochastic Gravitational Wave Background from the Early Universe}
\label{sec:SGW}

The stochastic GW background (SGWB) offers a unique window into the physical processes that shaped the early Universe. A primary contribution arises from tensor perturbations generated during cosmic inflation, where quantum fluctuations of the metric were stretched to cosmological scales, producing a nearly scale-invariant spectrum of primordial GWs~\cite{Starobinsky:1979ty, Mukhanov:1981xt, Rubakov:1982df}. Beyond these primary tensor modes, a variety of secondary and post-inflationary mechanisms can enrich the GW spectrum. These include first-order phase transitions in the early Universe, which produce GWs through bubble collisions, sound waves, and magnetohydrodynamic turbulence; cosmic string networks that continuously emit gravitational radiation; parametric resonance and preheating after inflation; dynamics of scalar fields and moduli; and primordial magnetic fields or anisotropic stresses in the radiation era. Such mechanisms can lead to spectra spanning a wide range of frequencies and amplitudes, often with characteristic features that encode information about high-energy physics and thermal history well after inflation. Comprehensive treatments of these sources and their observational prospects can be found in Refs.~\cite{Maggiore:2000gv, Maggiore:2018sht, Caprini:2018mtu, Christensen:2018iqi, Mazumdar:2018dfl}, 
see also Refs.~\cite{Boyle:2005se, Watanabe:2006qe, Visinelli:2014qla, Saikawa:2018rcs, Bernal:2019lpc, Haque:2021dha, Addazi:2023jvg, Barman:2023ktz, Maity:2024cpq}. In particular, non-standard post-inflationary histories, such as an early matter-dominated or kination, modify the evolution of primordial tensor modes and imprint characteristic tilts in the GW spectrum, depending on the background equation of state~\cite{Cui:2018rwi, Ramberg:2019dgi, Ramberg:2020oct, Giovannini:2020wrx, Haque:2021dha,Maity:2024cpq,Harigaya:2023pmw, Eroncel:2025bcb}.

While the above constraints apply to the primordial GW background distorted by reheating, Primordial black holes (PBHs) formation introduces qualitatively new sources that shape the SGWB and are not captured by this framework. Large curvature perturbations responsible for PBH formation generate second-order tensor modes, while a temporary PBH-dominated era followed by evaporation (``PBH reheating'') produces additional GW sources. On small scales, the discrete and inhomogeneous spatial distribution of PBHs induces \textit{Poissonian density perturbations}, which act as secondary GW sources. On larger scales, PBHs behave as a pressureless fluid, and the sudden transition from PBH domination to radiation domination (RD), driven by PBH evaporation, amplifies adiabatic curvature perturbations, producing an additional GW signal~\cite{Inomata:2019ivs, Domenech:2020ssp,Papanikolaou:2020qtd,Domenech:2020ssp,Dalianis:2020gup,Domenech:2021wkk,Bhaumik:2022pil,Bhaumik:2022zdd,Domenech:2021and, Mazde:2022sdx, Harigaya:2023pmw, Inomata:2023drn,Papanikolaou:2022chm,Bhaumik:2024qzd,Domenech:2024wao,Gross:2024wkl,Gross:2025hia,Bhaumik:2024qzd,Domenech:2024wao}.

Finally, direct graviton emission during PBH evaporation can generate very high-frequency GWs~\cite{Dong:2015yjs, Inomata:2019ivs, Inomata:2020lmk, Kohri:2020qqd, Kohri:2024qpd}, which may become relevant for future cavity or resonator based experiments, offering a unique probe into the quantum-gravity regime of the early Universe~\cite{Aguiar:2010kn, Berlin:2021txa, Gatti:2024mde, Cai:2025fpe, Berlin:2026che}. In what follows, we first discuss the primary GWs generated during inflation, then move on to the scenario where the induced GW signals are associated with PBHs.

\subsection{Primary gravitational waves spectrum}

We begin with primary tensor perturbations generated during inflation and subsequently evolving freely. Possible additional sources such as anisotropic stress or particle production are neglected here, so that the tensor evolution is fully determined by the background expansion. The perturbed line element for a spatially flat FLRW spacetime is given by
\begin{equation}
    {\rm d}s^{2} = a^{2}(\eta)\,[-{\rm d}\eta^{2} + (\delta_{ij} + h_{ij})\,{\rm d}x^{i}{\rm d}x^{j}]\,,
\end{equation}
where $h_{ij}$ denotes the tensor perturbations, assumed to be transverse and traceless, i.e. $\partial_{i}h_{ij} = h^{i}_{i} = 0$. The Fourier components of these perturbations obey
\begin{equation}
    h_{k}'' + 2\frac{a'}{a}h_{k}' + k^{2}h_{k} = 0.
\end{equation}
Defining the Mukhanov--Sasaki variable $u_{k} = a h_{k}$, one obtains
\begin{equation}
    u_{k}'' + \left(k^{2} - \frac{a''}{a}\right)u_{k} = 0.
\end{equation}
This equation can be solved in terms of Bessel functions, with the form of the solution determined by the background evolution governed by the Hubble parameter. The tensor power spectrum is defined through the two-point correlation in Fourier space as
\begin{equation}
    \label{eq:PTk}
    \mathcal{P}_T(k) = \frac{4k^{3}}{2\pi^{2}}|h_{k}|^{2},
\end{equation}
where the factor 4 accounts for the two GW polarizations. The spectrum is evaluated on super-Hubble scales during inflation. Considering slow-roll inflation under the de Sitter approximation, the scale factor during inflation and reheating evolves as
\begin{equation}
    a_{I}(\eta) = \frac{a_{\rm inf}}{1 - H_{\rm inf}a_{\rm inf}(\eta - \eta_{\rm inf})},
\end{equation}
and
\begin{equation}
    a_{\rm RH}(\eta) = a_{\rm inf}\!\left[1 + \frac{H_{\rm inf}(1 + 3w_{\phi})}{2a_{\rm inf}}(\eta - \eta_{\rm inf})\right]^{\tfrac{2}{1 + 3w_{\phi}}},
\end{equation}
where $H_{\rm inf}$ denotes the Hubble parameter during inflation. As we assume de Sitter inflation, $H_{\rm inf}$ is constant throughout inflation. The scale factors are matched continuously at the end of inflation such that $a_{I}(\eta_{\rm inf}) = a_{\rm RH}(\eta_{\rm inf}) = a_{\rm inf}$. Introducing the conformal time variable
\begin{equation}
    \tau = \begin{cases}
    \eta - \eta_{\rm inf} - \frac{1}{a_{\rm inf}H_{\rm inf}} \quad \text{(during inflation)}, \\
    \eta - \eta_{\rm inf} - \frac{2}{a_{\rm inf}H_{\rm inf}(1 + 3w_{\phi})} \quad \text{(during reheating)}\,,
    \end{cases}
\end{equation}
the scale factors can be rewritten as
\begin{equation}
    a_{I}(\tau) = -\frac{1}{\tau H_{\rm inf}}, \qquad a_{\rm RH}(\tau) = a_{\rm inf}\left(\frac{\tau}{\tau_{\rm inf}}\right)^{\tfrac{2}{1 + 3w_{\phi}}},
\end{equation}
with $\tau_{\rm inf} = -1/(a_{\rm inf}H_{\rm inf})$. Using these expressions, the mode function at the end of inflation becomes
\begin{equation}\label{eq:hk}
    h_{k}(\tau_{\rm inf}) = \frac{\sqrt{2}}{M_{\rm Pl}}\frac{iH_{\rm inf}}{\sqrt{2k^{3}}}\left(1 - i\frac{k}{k_{\rm inf}}\right)e^{i k/k_{\rm inf}},
\end{equation}
where $k_{\rm inf}$ denotes the comoving mode leaving the Hubble radius at the end of inflation, $k_{\rm inf} = -1/\eta_{\rm inf}$, and $M_{\rm Pl}$ is the reduced Planck mass. Substituting Eq.~\eqref{eq:hk} into Eq.~\eqref{eq:PTk} gives the tensor power spectrum at the end of inflation,
\begin{equation}
    \mathcal{P}_T(k) = \frac{H_{\rm inf}^{2}}{2\pi^{2}M_{\rm Pl}^{2}}\left(1 + \frac{k^{2}}{k_{\rm inf}^{2}}\right).
\end{equation}
In the limit $k \ll k_{\rm inf}$, the de Sitter background yields a nearly scale-invariant tensor spectrum, while slow-roll effects induce only a small red tilt, which can be safely neglected. In the following, we assume sharp transitions between inflation, reheating, and RD. To describe the post-inflationary evolution of the tensor perturbations, we write
\begin{equation}
    h_{k}(\eta) = h_{k}(\tau_{\rm inf})\,\chi_{k}(\eta),
\end{equation}
where $h_{k}(\tau_{\rm inf})$ is given by Eq.~\eqref{eq:hk} and $\chi_{k}(\eta)$ denotes the transfer function satisfying
\begin{equation}
    \chi_{k}'' + 2\mathcal{H}\chi_{k}' + k^{2}\chi_{k} = 0\,,
\end{equation}
in terms of the Hubble parameter in conformal time, $\mathcal{H}\equiv a'/a$. During reheating, characterized by an effective equation of state $w_\phi$, the Hubble parameter evolves as $H = H_{\rm inf} (a/a_{\rm inf})^{-\frac{3(1+w_\phi)}{2}}$. Substituting this into the above expression gives
\begin{equation}
    \frac{{\rm d}^2\chi_k}{{\rm d}a^2} + \frac{5 - 3w_\phi}{2a} \frac{{\rm d}\chi_k}{{\rm d}a} + \frac{(k/(a_{\rm inf}\, k_{\rm inf}))^2}{(a/a_{\rm inf})^{1-3w_\phi}} \chi_k = 0.
\end{equation}
The solution during reheating is expressed in terms of Bessel functions as
\begin{equation}
    \chi_k^{\text{RH}}(a) = (a/a_{\rm inf})^{-\nu} \left[ C_{1k} J_{-\nu/\gamma}(x) + C_{2k} J_{\nu/\gamma}(x) \right],
\end{equation}
where $\nu = \tfrac{3}{4}(1 - w_\phi)$, $\gamma = \tfrac{1}{2}(1 + 3w_\phi)$, and $x = k (a/a_{\rm inf})^\gamma / (\gamma k_{\rm inf})$. The coefficients $C_{1k}$ and $C_{2k}$ are determined by continuity at the end of inflation:
\begin{align}
    C_{1k} &= \frac{\pi^2}{2\gamma k_{\rm inf}} \left[ \frac{k}{ik-k_{\rm inf}} J_{\nu/\gamma}\left(\frac{k}{\gamma k_{\rm inf}}\right) - J_{(\gamma+\nu)/\gamma}\left(\frac{k}{\gamma k_{\rm inf}}\right) \right]\csc\left(\frac{\pi\nu}{\gamma}\right), \\
    C_{2k} &= \frac{\pi^2}{2\gamma k_{\rm inf}} \left[ \frac{k}{ik-k_{\rm inf}} J_{-\nu/\gamma}\left(\frac{k}{\gamma k_{\rm inf}}\right) - J_{-(\gamma+\nu)/\gamma}\left(\frac{k}{\gamma k_{\rm inf}}\right) \right]\csc\left(\frac{\pi\nu}{\gamma}\right),
\end{align}
where $\csc(x) = 1/\sin(x)$. During RD, where $H \propto a^{-2}$, the transfer function takes the form
\begin{equation}
    \chi_k^{\text{RD}}(a) = \frac{1}{a/a_{\rm inf}} \left[D_{1k} e^{-i k (a/a_{\rm inf})/(k_{\text{RH}} a_{\rm RH}/a_{\rm inf})} + D_{2k} e^{i k (a/a_{\rm inf})/(k_{\text{RH}} a_{\rm RH}/a_{\rm inf})} \right],
\end{equation}
where the coefficients $D_{1k}$ and $D_{2k}$ are fixed by matching the mode functions and their derivatives at reheating:
\begin{align}
    D_{1k} &= \frac{a_{\rm RH}/a_{\rm inf}}{2} e^{i k / k_{\text{RH}}} \left[ \left(1 + \frac{i k_{\text{RH}}}{k}\right) \chi_k^{\text{RH}}(a_{\rm RH}) + i \frac{k_{\text{RH}}}{k} \frac{a_{\rm RH}}{a_{\rm inf}} \frac{\partial \chi_k^{\text{RH}}(a_{\rm RH})}{\partial (a/a_{\rm inf})} \right] = \nonumber \\ & = i \frac{(a_{\rm RH}/a_{\rm inf}) k_{\text{RH}}}{2 k} \mathcal{E}_{1k}, \\
    D_{2k} &= \frac{a_{\rm RH}/a_{\rm inf}}{2} e^{-i k / k_{\text{RH}}} \left[ \left(1 - \frac{i k_{\text{RH}}}{k}\right) \chi_k^{\text{RH}}(a_{\rm RH}) - i \frac{k_{\text{RH}}}{k} \frac{a_{\rm RH}}{a_{\rm inf}} \frac{\partial \chi_k^{\text{RH}}(a_{\rm RH})}{\partial(a/a_{\rm inf})} \right] = \nonumber \\
    & = -i \frac{(a_{\rm RH}/a_{\rm inf}) k_{\text{RH}}}{2 k} \mathcal{E}_{2k}.
\end{align}
Deep in the radiation era, the GW energy density per logarithmic interval is given by
\begin{equation}
    \rho_{\text{GW}}(k,\eta) = \frac{M_{\rm Pl}^2 k^2 a_{\rm inf}^2}{4 a^4} \mathcal{P}_T(k) (|D_{1k}|^2 + |D_{2k}|^2).
\end{equation}
Normalizing to the critical density, $\rho_c = 3M_{\rm Pl}^2 H^2$, yields the dimensionless GW energy density parameter,
\begin{equation}
    \Omega_{\rm GW}(k, \eta) = \frac{\mathcal{P}_T(k)}{48} (|\mathcal{E}_{1k}|^2 + |\mathcal{E}_{2k}|^2).
\end{equation}
At late times, after redshifting, this becomes
\begin{equation}
    \label{eq:GW_present}    
    \Omega_{\rm GW}(k) h^2 = c_g\, \Omega_{\text{rad,0}} h^2\, \Omega_{\rm GW}(k, \eta),
\end{equation}
where the factor accounting for changes in relativistic degrees of freedom is
\begin{equation}
    c_g=\left(\frac{g_{\ast k}}{g_{\ast 0}}\right) \left(\frac{g_{\ast s0}}{g_{\ast s k}}\right)^{4/3} \simeq 0.39,
\end{equation}
where we used $g_{\ast k}=g_{\ast s k}=106.75$ for the SM plasma, $g_{\ast 0}=3.36$, and $g_{\ast s0}=3.91$. We use the \textit{Planck} 2018 value for the present radiation density, $\Omega_{\rm rad,0}h^2 = 4.16\times10^{-5}$, including photons and three neutrino species with $N_{\rm eff}=3.046$~\cite{Planck:2018vyg}, so that the present-day GW energy density in Eq.~\eqref{eq:GW_present} is $\Omega_{\rm GW}^{(0)}(k)h^2\simeq1.61\times10^{-5}\,\Omega_{\rm GW}(k,\eta)$. In the long-wavelength regime ($k \ll k_{\text{RH}}$), the solutions satisfy $|\mathcal{E}_{1k}|^2 = |\mathcal{E}_{2k}|^2 \simeq 1$, indicating an undistorted spectrum. In contrast, for $k \gg k_{\text{RH}}$, the amplitude exhibits a scale-dependent enhancement,
\begin{equation}
    |\mathcal{E}_{1k}|^2 = |\mathcal{E}_{2k}|^2 = \frac{4\gamma^2}{\pi}\,\Gamma^2\!\left(1+\frac{\nu}{\gamma}\right)\left(\frac{k}{2\gamma k_{\text{RH}}}\right)^{n_{\text{GW}}},
\end{equation}
where $\Gamma(x)$ denotes the Euler gamma function, defined as
\begin{equation}
    \Gamma(x) \equiv \int_{0}^{\infty} t^{x-1} e^{-t}\, {\rm d}t \,, \qquad \Re(x) > 0,
\end{equation}
and the spectral index is
\begin{equation}
    n_{\text{GW}} = 1 - \frac{2\nu}{\gamma} = -\frac{2(1 - 3w_\phi)}{1 + 3w_\phi}.
\end{equation}
This spectral tilt arises from the redshifting of modes relative to the background during reheating and depends on the effective equation of state $w_\phi$. Consequently, the present-day GW energy density spectrum reads
\begin{equation}
    \Omega_{\rm GW}^{\text{rad}}(k)h^2 = \frac{\Omega_{\text{rad,0}} h^2}{12\pi^2} \frac{H_{\rm inf}^2}{M_{\rm Pl}^2}, \quad (k \ll k_{\text{RH}}),
\end{equation}
and in the high-frequency limit ($k \gg k_{\text{RH}}$),
\begin{equation} \label{eq: gw energy density final}
    \Omega_{\rm GW}(k)h^2 = \Omega_{\rm GW}^{\text{rad}}(k)h^2 \, \frac{4\gamma^2}{\pi} \Gamma^2\left(1+\frac{\nu}{\gamma}\right) \left(\frac{k}{2\gamma k_{\text{RH}}}\right)^{n_{\text{GW}}}.
\end{equation}
\begin{figure}
    \centering
    \includegraphics[scale=0.48]{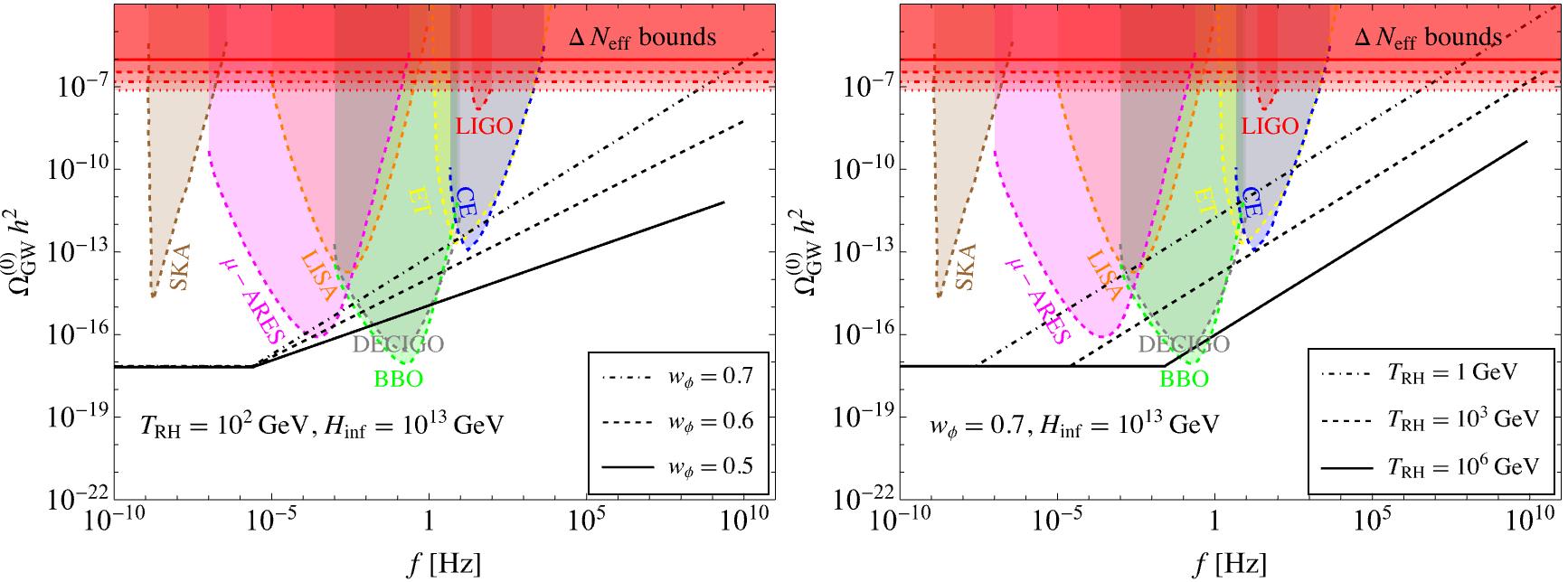}
    \caption{Primary GWs spectra at the present epoch as a function of frequency. Each plot represents the variation of a single parameter while the other two are kept fixed. The representation is referred to on the sensitivity curves of various future GW observatories. Additionally, the $\Delta N_{\rm eff}$ bound at BBN is shown (see, for instance Table-\ref{tab:DNeff}), as the GWs themselves behave as extra relativistic degrees of freedom.}
    \label{fig:PGW}
\end{figure}
This final expression encapsulates the influence of reheating dynamics on the high-frequency tail of the GW spectrum. The spectral shape depends on the background equation of state during reheating: it is red-tilted ($n_{\text{GW}} < 0$) for $w_\phi < 1/3$, scale-invariant ($n_{\text{GW}} = 0$) for $w_\phi = 1/3$, and blue-tilted ($n_{\text{GW}} > 0$) for $w_\phi > 1/3$. For instance, a matter-like reheating phase with $w_\phi = 0$ gives $n_{\text{GW}} = -2$, while a kination-dominated phase with $w_\phi = 1$ yields $n_{\text{GW}} = 1$ (see, for instance, Fig.~\ref{fig:PGW}). 
\begin{table}[t!]
    \begin{center}
        \begin{tabular}{|c||c|}
            \hline
            $\Delta N_{\rm eff}$ & Experiments  \\ 
            \hline\hline
            $0.17$ & Planck legacy data (combining BAO)~\cite{Planck:2018vyg}\\
            $0.14$ & BBN+CMB combined~\cite{Yeh:2022heq}\\
            $0.06$ & CMB-S4~\cite{Abazajian:2019eic} \\
            $0.027$ & CMB-HD~\cite{CMB-HD:2022bsz}\\
            $0.013$ &  COrE~\cite{COrE:2011bfs}, Euclid~\cite{EUCLID:2011zbd} \\
            $0.06$ & PICO~\cite{NASAPICO:2019thw} \\
            \hline
        \end{tabular}
    \end{center}
    \caption {Present and future constraints on $\Delta N_{\rm eff}$ from different experiments.}
    \label{tab:DNeff}
\end{table}.
From Fig. \ref{fig:PGW}, it is clear that a stiff post-inflationary phase
enhances the primordial GW background, which in turn
contributes to the total radiation energy density of the Universe. This
contribution can be quantified in terms of the excess effective number of
relativistic degrees of freedom, \(\Delta N_{\rm eff}\), constrained by Big
Bang Nucleosynthesis (BBN) and Cosmic Microwave Background (CMB)
observations. We therefore estimate the minimum reheating temperature,
\(T_{\rm RH}\), consistent with these bounds for fixed \(w_\phi\) and a given
inflationary energy scale.

The additional radiation energy from GWs is related to $\Delta N_{\rm eff}$ through~\cite{Jinno:2012xb}
\begin{equation}
    \Delta N_{\rm eff} = \frac{8}{7} \left( \frac{11}{4} \right)^{4/3}\frac{\rho_{\rm GW}}{\rho_\gamma}\,,
    \label{Eq: neff}
\end{equation}
where $\rho_{\rm GW}$ and $\rho_\gamma$ denote the energy densities of GWs and photons, respectively. This sets an upper limit on the total GW energy density today,
\begin{equation}
    \int_{k_{\rm RH}}^{k_{\rm inf}} \frac{{\rm d}k}{k}\,\Omega_{\rm GW}(k)h^2 \leq \frac{7}{8} \left( \frac{4}{11} \right)^{4/3} \Omega_\gamma h^2\,\Delta N_{\rm eff},
    \label{eq: deltaneff}
\end{equation}
with $\Omega_\gamma h^2 \simeq 2.47 \times 10^{-5}$ denoting photons only. For a blue-tilted spectrum, where the GW amplitude increases with $k$, the integral in Eq.~\eqref{eq: deltaneff} is dominated by high-frequency modes near $k_{\rm inf}$. Using the GW spectrum from Eq.~\eqref{eq: gw energy density final}, one finds
\begin{equation}
    \int_{k_{\rm RH}}^{k_{\rm inf}} \frac{{\rm d}k}{k}\,\Omega_{\rm GW}(k)h^2 \simeq\Omega_{\rm GW}^{\rm rad} h^2\,\zeta(w_\phi)\left( \frac{k_{\rm inf}}{k_{\rm RH}} \right)^{\frac{6w_\phi - 2}{1 + 3w_\phi}},
    \label{eq: BBNapprox}
\end{equation}
where
\begin{equation}
    \zeta(w_\phi) = (1 + 3w_\phi)^{\frac{4}{1 + 3w_\phi}} \Gamma^2\!\left(\frac{5 + 3w_\phi}{2 + 6w_\phi}\right)\frac{1 + 3w_\phi}{2\pi(3w_\phi - 1)}.
\end{equation}
The ratio of comoving scales $k_{\rm inf}/k_{\rm RH}$ follows from the reheating dynamics,
\begin{equation}
    \frac{k_{\rm inf}}{k_{\rm RH}} = \left( \frac{30\,\rho_{\rm inf}}{\pi^2 g_{\rm RH}} \right)^{\frac{1 + 3w_\phi}{6(1 + w_\phi)}} T_{\rm RH}^{-\frac{2}{3} \cdot \frac{1 + 3w_\phi}{1 + w_\phi}},
    \label{eq: kend kre}
\end{equation}
where $\rho_{\rm inf} = 3 M_{\rm Pl}^2 H_{\rm inf}^2$ denotes the total energy density at the end of inflation, and $T_{\rm RH}$ is the reheating temperature, defined at the onset of RD. Combining Eqs.~\eqref{eq: deltaneff}, \eqref{eq: BBNapprox}, and \eqref{eq: kend kre}, we obtain a lower limit on the reheating temperature:
\begin{equation}
    T_{\rm RH} \gtrsim \left[ \frac{\Omega_{\rm GW}^{\rm rad} h^2\,\zeta(w_\phi)}{5.61 \times 10^{-6}\,\Delta N_{\rm eff}} \right]^{\frac{3(1 + w_\phi)}{4(3w_\phi - 1)}}\left( \frac{30\,\rho_{\rm inf}}{\pi^2 g_{\rm RH}} \right)^{1/4}.
    \label{eq: BBNrestriction}
\end{equation}
Here, $g_{\rm RH}$ represents the effective relativistic and entropy degrees of freedom at the end of reheating. Equation~\eqref{eq: BBNrestriction} shows that a lower reheating temperature leads to a stronger GW contribution to $\Delta N_{\rm eff}$. Hence, for a given inflationary scale and $w_\phi$, $T_{\rm RH}$ must remain above this bound to satisfy BBN constraints.

\subsection{Induced Gravitational Waves and the Role of Primordial Black Holes}

\subsubsection{Overview of the PBH Reheating Scenario}

Up to this point we have discussed only primary GWs, generated during inflation and evolving freely under the background expansion. We now consider induced GWs arising at second order from scalar perturbations, with particular emphasis on scenarios where PBHs form and temporarily dominate the cosmic energy density prior to evaporation. Before proceeding to the details of GW production, we briefly outline the PBH reheating framework. After inflation, the inflaton oscillates around the minimum of its potential. Depending on the potential shape near the minimum, the effective equation of state (EoS), $w_\phi$, may differ from that of radiation ($w = 1/3$). 
The duration of reheating is determined by how efficiently the inflaton decays into SM particles. 
In the conventional picture, the Universe transitions to RD as the oscillating inflaton decays, with its energy initially dominating the total density~\cite{Garcia:2020eof,Haque:2020zco,Garcia:2020wiy, Haque:2022kez,Clery:2021bwz,Clery:2022wib}. 

We consider a non-standard background of EoS  $w_\phi$ in which PBHs form. The oscillations of the inflaton field can be expressed as
\begin{equation}
\phi(t) = \phi_0(t)\, \mathcal{P}(t)\,,
\end{equation}
where $\phi_0(t)$ represents the oscillation amplitude and $\mathcal{P}(t)$ its periodic behavior. 
Averaging over one oscillation gives $\langle\dot{\phi}^2\rangle \simeq \langle\phi\, dV_\phi/d\phi\rangle$, one gets an effective EoS~\cite{Garcia:2020eof, Bernal:2019mhf}
\begin{equation}
w_\phi \simeq \frac{n-2}{n+2}\,, 
\end{equation}
for a potential $V(\phi) \propto \phi^{2n}$. For example, $V(\phi)\sim \phi^6$ corresponds to $w_\phi=0.5$. After formation, PBHs behave as non-relativistic matter, and their energy density redshifts more slowly than that of the background with $0 < w_\phi \leq 1$.\footnote{Exact matter domination ($w=0$) is excluded, since PBH formation in such a phase would require a much longer collapse time~\cite{Escriva:2020tak}.} Consequently, PBHs can dominate the energy density before evaporating if their initial abundance is sufficiently large~\cite{Hidalgo:2011fj, Martin:2019nuw, Hooper:2019gtx, Hooper:2020evu, RiajulHaque:2023cqe}. 
We therefore envision a PBH-dominated epoch between the $w_\phi$-dominated and RD phases. 

The system is described by the following Boltzmann equations for the inflaton, radiation, and PBH energy densities:
\begin{align}
    \frac{{\rm d}\rho_\phi}{{\rm d}a} + 3(1+w_\phi)\frac{\rho_\phi}{a} &= -\frac{\Gamma_\phi(1+w_\phi)}{aH}\rho_\phi\,, \\
    \frac{{\rm d}\rho_r}{{\rm d}a} + 4\frac{\rho_r}{a} &= -\frac{\rho_{\rm BH}}{M_{\rm BH}}\frac{{\rm d}M_{\rm BH}}{{\rm d}a} + \frac{\Gamma_\phi\rho_\phi(1+w_\phi)}{aH} \simeq -\frac{\rho_{\rm BH}}{M_{\rm BH}}\frac{{\rm d}M_{\rm BH}}{{\rm d}a}\,, \\
    \frac{{\rm d}\rho_{\rm BH}}{{\rm d}a} + 3\frac{\rho_{\rm BH}}{a} &= \frac{\rho_{\rm BH}}{M_{\rm BH}}\frac{{\rm d}M_{\rm BH}}{{\rm d}a}\,, \\
    \frac{{\rm d}M_{\rm BH}}{{\rm d}a} &= -\epsilon\,\frac{M_{\rm Pl}^4}{M_{\rm BH}^2}\frac{1}{aH}\,,
\end{align}
where $\epsilon \equiv 3.8\,(\pi/480)\, g_\ast(T_{\rm BH})$ and the factor $3.8$ accounts for greybody effects~\cite{Arbey:2019mbc, Cheek:2021odj, Baldes:2020nuv}. Assuming $\Gamma_\phi \ll H$, the inflaton energy density primarily redshifts, while radiation is mainly produced through PBH evaporation. Solving for the PBH mass gives
\begin{equation}
    M_{\rm BH} = M_{\rm in}\left(1 - \Gamma_{\rm BH}(t - t_{\rm in})\right)^{1/3}, \qquad \Gamma_{\rm BH} = \frac{3\epsilon M_{\rm Pl}^4}{M_{\rm in}^3}\,,
\end{equation}
where, if PBHs form in a $w_\phi$-dominated phase, they have initial mass~\cite{Carr:1974nx}
\begin{equation}
    M_{\rm in} = \gamma\,\frac{4\pi M_{\rm Pl}^2}{H_{\rm in}}, \qquad \gamma \equiv w_\phi^{3/2}\,.
\end{equation}
A more precise calculation of $M_{\rm in}$ would depend on the fluctuation profile and $w_\phi$~\cite{Musco:2012au, Musco:2008hv, Hawke:2002rf,Niemeyer:1997mt,Escriva:2021pmf, Escriva:2019nsa,Escriva:2020tak, Escriva:2021aeh}, but for our purposes this expression suffices. The PBH lifetime is $t_{\rm ev} = 1/\Gamma_{\rm BH}$.
\begin{figure}
    \centering
    \includegraphics[scale=0.36]{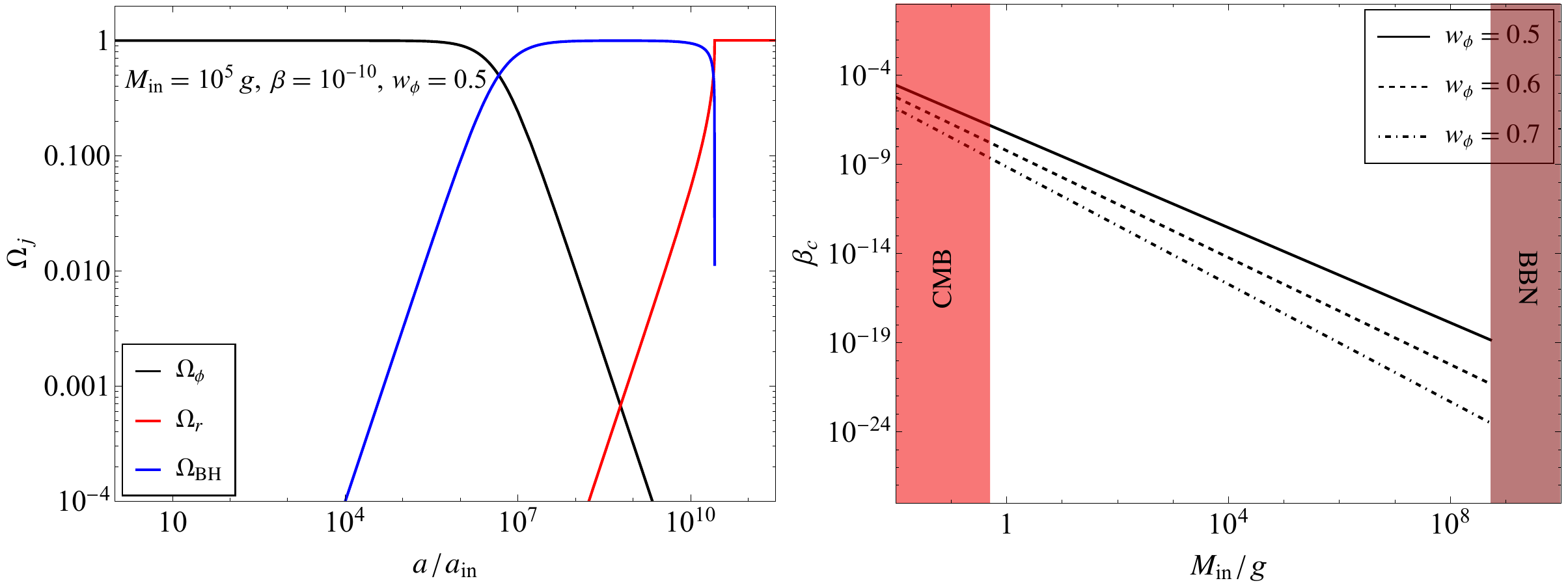}
    \caption{\it \textbf{Left panel:} Evolution of the normalized energy densities, $\Omega_j = \rho_j / (3 M_{\rm Pl}^2 H^2)$, as a function of the scale factor. The plot demonstrates a PBH-driven reheating scenario in which primordial black holes come to dominate the energy budget of the Universe before completely evaporating. \textbf{Right panel:} Critical abundance $\beta_c$ shown as a function of the initial PBH mass $M_{\rm in}$ for different values of $w_\phi$. The shaded regions denote excluded parameter space: the red band corresponds to CMB constraints on PBH formation, while the brown band indicates bounds arising from BBN.}
    \label{fig:betac}
\end{figure}

In Fig.~\ref{fig:betac}, we present the evolution of the different energy components as a function of the normalized scale factor, $a/a_{\rm in}$. The figure clearly shows the existence of a critical initial abundance, $\beta_c$, above which the energy density of PBHs surpasses that of the inflaton prior to their evaporation. In this regime, reheating is predominantly driven by PBH evaporation referred to as PBH reheating. The critical fraction is
\begin{equation}
    \beta_c = \left[\frac{\epsilon}{2\pi(1+w_\phi)\gamma}\frac{M_{\rm Pl}^2}{M_{\rm in}^2}\right]^{\frac{2w_\phi}{1+w_\phi}} \simeq \left(\frac{2.8\times10^{-6}}{w_\phi^{3/4}\sqrt{1+w_\phi}}\right)^{\frac{4w_\phi}{1+w_\phi}} \left(\frac{1\,{\rm g}}{M_{\rm in}}\right)^{\frac{4w_\phi}{1+w_\phi}}.
\end{equation}
For $w_\phi=1/3$, one finds $\beta_c \simeq 5.6\times10^{-6}(1\,{\rm g}/M_{\rm in})$, while for $w_\phi=1/2$, $\beta_c \simeq 6.1\times10^{-8}(1\,{\rm g}/M_{\rm in})^{4/3}$. A larger $w_\phi$ shifts $\beta_c$ to smaller values, making PBH domination easier to achieve. The end of PBH evaporation marks the reheating epoch, with the corresponding temperature given by
\begin{equation}
    T_{\rm ev} = \left(\frac{360\,\epsilon^2}{\pi^2g_\ast(T_{\rm ev})}\right)^{1/4} \left(\frac{M_{\rm Pl}}{M_{\rm in}}\right)^{3/2}M_{\rm Pl}\simeq 2.7\times10^{10}\left(\frac{1\,{\rm g}}{M_{\rm in}}\right)^{3/2},
\end{equation}
assuming $g_\ast(T_{\rm ev})=106.75$ and $g_\ast(T_{\rm BH})=108$. To ensure successful BBN, PBHs must evaporate before nucleosynthesis, requiring $T_{\rm ev} \geq 4\,{\rm MeV}$~\cite{Kawasaki:1999na,Kawasaki:2000en,Hasegawa:2019jsa}, which sets an upper bound on $M_{\rm in}$. The \textit{Planck} 2018 and BICEP2/Keck limits on the tensor-to-scalar ratio, $r_{0.05}<0.036$~\cite{BICEP:2021xfz, BICEP2:2015nss}, imply $H_{\rm inf}<4.8\times10^{13}\,{\rm GeV}$, providing a lower bound on $M_{\rm in}$. Together, these constraints give
\begin{equation}
    0.5\,{\rm g}\left(\frac{\gamma}{0.2}\right) \lesssim M_{\rm in} \lesssim 4.8\times10^8\,{\rm g}\left(\frac{g_\ast(T_{\rm ev})}{106.75}\right)^{-1/6}\left(\frac{g_\ast(T_{\rm BH})}{108}\right).
\end{equation}
Limits on the PBH mass for monochromatic distributions have also been derived based on the effects of PBH evaporation into the details of BBN~\cite{Carr:2020gox, Boccia:2024nly}.

Ultralight PBHs can act as efficient sources of GWs. 
Large curvature perturbations responsible for their formation induce GWs~\cite{Baumann:2007zm,Espinosa:2018eve,Domenech:2019quo,Ragavendra:2020sop,Inomata:2023zup,Franciolini:2023pbf,Firouzjahi:2023lzg,Maity:2024odg}, 
while their evaporation produces gravitons directly~\cite{Fujita:2014hha}. 
Moreover, fluctuations in the PBH number density generate isocurvature-induced GWs~\cite{Domenech:2020ssp,Papanikolaou:2020qtd,Domenech:2021wkk,Papanikolaou:2022chm,Bhaumik:2022pil,Bhaumik:2022zdd,Papanikolaou:2022hkg,Papanikolaou:2024kjb,Domenech:2024wao,Bhaumik:2024qzd,Gross:2024wkl}, 
and the sharp transition from PBH domination to RD amplifies adiabatic, second-order tensor modes~\cite{Inomata:2019ivs,Inomata:2020lmk,Bhaumik:2022pil,Bhaumik:2022zdd}. 
Since ultralight PBHs form near the end of inflation, the resulting GWs associated with PBH formation typically peak at high frequencies corresponding to the smallest scales and are not within the reach of most of the proposed future GW observatories. 

In the following, we focus on two principal GW sources: (i) GWs sourced by isocurvature perturbations arising from fluctuations in the PBH number density, and (ii) GWs induced by primordial adiabatic curvature perturbations originating during inflation, which are enhanced during the PBH-dominated era. Physically, the two GW channels differ in origin. Isocurvature-induced GWs directly trace the discrete nature of the PBH distribution, whereas adiabatic-induced GWs arise from curvature perturbations that are temporarily frozen during PBH domination and subsequently reactivated by the sudden transition to RD at evaporation. In both cases, the transition plays a central role in sourcing the GW signal. Together, these channels link the observable GW amplitude and spectral shape to the parameters, $(w_\phi, \beta, M_{\rm in})$, allowing future stochastic GW measurements to probe the post-inflationary epoch.

\subsubsection{Induced Gravitational Waves associated with PBH Reheating}
\label{sec:GW_PBH}

PBH reheating generically gives rise to secondary GWs sourced at second order by scalar perturbations. In this scenario, two physically distinct channels contribute to the induced GW background, namely isocurvature perturbations associated with fluctuations in the PBH number density, and primordial adiabatic curvature perturbations generated during inflation, whose evolution is modified during the PBH-dominated epoch. In this section, we analyze these two contributions separately and assess their detectability with future interferometric experiments such as LISA and the Einstein Telescope.

The spectral energy density of GWs is related to the oscillation-averaged tensor power spectrum $\overline{\mathcal{P}_h}$ through~\cite{Ando:2018qdb}
\begin{equation}
    \label{eq:omega_GW}
    \Omega_{\rm GW}(k,\eta)=\frac{k^2}{12\mathcal{H}^2}\,\overline{\mathcal{P}_h(k,\eta)}\,.
\end{equation}
The GW spectrum is evaluated during the RD epoch, after the relevant modes have re-entered the horizon and propagate freely. The oscillation-averaged tensor power spectrum can be expressed in terms of the curvature perturbation power spectrum $\mathcal{P}_\Phi$ as~\cite{Kohri:2018awv}
\begin{equation}
\label{eq:av_tensorPS}
    \overline{\mathcal{P}_h(k,\eta)}=8\!\int_0^\infty\!{\rm d}v\!\int_{|1-v|}^{1+v}\!{\rm d}u\left[\frac{(1+v^2-u^2)^2-4v^2}{4uv}\right]^2\mathcal{P}_\Phi(uk)\mathcal{P}_\Phi(vk)\,\overline{I^2}(x,u,v),
\end{equation}
where $x=k\eta$ and $I(x,u,v)$ represents the time evolution kernel of the tensor modes. In the presence of an early PBH-dominated phase followed by a sudden transition to RD, the dominant contribution to the induced tensor spectrum arises from modes sourced around the time of PBH evaporation~\cite{Inomata:2019ivs}. Consequently, the curvature power spectrum $\mathcal{P}_\Phi$ entering Eq.~\eqref{eq:av_tensorPS} can be evaluated at the end of PBH domination. In the following, we present the resulting GW spectra for the two source mechanisms and discuss their characteristic features and observational implications \footnote{ Note that the present-day spectrum of the induced GWs, including contributions from both isocurvature and adiabatic sources, can be written as
\begin{eqnarray}
\label{eq:GW_combined}
\Omega_{\rm GW,com}^{(0)}(f) h^2 
= \Omega_{\rm GW,iso}^{(0)}(f) h^2 
+ \Omega_{\rm GW,ad}^{(0)}(f) h^2 .
\end{eqnarray}}.

\subsubsection*{Gravitational Waves Induced by Isocurvature Fluctuations}
\label{subsec:density_fluctuation}

For $\beta>\beta_{\rm c}$, PBHs temporarily dominate the energy density of the Universe prior to their evaporation. In this regime, spatial inhomogeneities in the PBH number density act as isocurvature perturbations, which subsequently source curvature perturbations once PBHs become the dominant component. These scalar perturbations, in turn, generate GWs at second order. Assuming a Poisson-like initial PBH distribution, the isocurvature perturbation is
\begin{equation}
    S_i(\mathbf{x})\equiv \frac{\delta\rho_{\rm BH}(\mathbf{x})}{\rho_{\rm BH}}\,.
\end{equation} 
If $d$ denotes the mean comoving PBH separation, the real-space two-point correlation takes the form~\cite{Papanikolaou:2020qtd}
\begin{equation}
    \label{eq:density_contrast1}
    \left\langle S_i(\mathbf{x})S_i(\tilde{\mathbf{x}})\right\rangle=\frac{4\pi}{3}\left(\frac{d}{a}\right)^3\delta(\mathbf{x}-\tilde{\mathbf{x}}),
\end{equation}
which in Fourier space becomes~\cite{Papanikolaou:2020qtd,Domenech:2020ssp}
\begin{equation}
    \label{eq:density_contrast_fourier}
    \left\langle S_i(k)S_i(\tilde{k})\right\rangle=\frac{2\pi^2}{k^3}\,\mathcal{P}_{\rm BH,i}(k)\,\delta(k+\tilde{k}), \qquad \mathcal{P}_{\rm BH,i}(k)=\frac{2}{3\pi}\left(\frac{k}{k_{\rm UV}}\right)^3\,.
\end{equation}
The ultraviolet cutoff $k_{\rm UV}$ corresponds to the inverse mean PBH separation at formation and depends on the PBH mass scale~\cite{Domenech:2020ssp},
\begin{equation}
    \label{eq:kUV}
    k_{\rm UV}\simeq1.1\times10^4\,{\rm Hz}\left(\frac{g_{\ast s}(T_{\rm ev})}{106.75}\right)^{-1/3}\left(\frac{g_{\ast}(T_{\rm ev})}{106.75}\right)^{1/4}\left(\frac{g_{\ast}(T_{\rm BH})}{108}\right)^{1/6}\left(\frac{M_{\rm in}}{10^4\,{\rm g}}\right)^{-5/6}.
\end{equation}

The evolution of the Newtonian potential $\Phi$ in the presence of isocurvature perturbations is governed by
\begin{equation}
    \label{eq:PhiEq}
    \Phi''+3\mathcal{H}(1+c_s^2)\Phi'+\left[(1+3c_s^2)\mathcal{H}^2+2\mathcal{H}'+c_s^2k^2\right]\Phi=\tfrac{1}{2}a^2c_s^2\rho_{\rm BH}S_i\,,
\end{equation}
where $c_s$ is the sound speed and primes denote derivatives with respect to conformal time $\eta$. 
The source term on the right-hand side drives the growth of scalar perturbations during PBH domination, which subsequently induce tensor modes at second order. The resulting GW spectral energy density shows a resonant enhancement near the UV scale, with~\cite{Bhaumik:2024qzd,Domenech:2024wao}
\begin{equation}
    \label{eq:OGWres}
    \Omega_{\rm GW,res}(k)\simeq\Omega_{\rm GW,res}^{\rm peak}\left(\frac{k}{k_{\rm UV}}\right)^{11/3}\Theta_{\rm UV}^{\rm(iso)}(k)\,,
\end{equation}
where the peak amplitude is given by
\begin{eqnarray}
    \label{eq:OGWresPeak}
    \Omega_{\rm GW,res}^{\rm peak} &=& C^4(w_\phi)\frac{c_s^{7/3}(c_s^2-1)^2}{576\times6^{1/3}\pi}\left(\frac{k_{\rm BH}}{k_{\rm UV}}\right)^8\left(\frac{k_{\rm UV}}{k_{\rm ev}}\right)^{17/3}\,,\\
    C(w_\phi) &=& \frac{9}{20}\alpha_{\rm fit}^{-1/(3w_\phi)}\left(3+\frac{1-3w_\phi}{1+3w_\phi}\right)^{-1/(3w_\phi)}\,,
\end{eqnarray}
and where $\alpha_{\rm fit}\simeq 0.135$. Here, the PBH domination scale is
\begin{equation}
    \label{eq:kBH}
    k_{\rm BH}=\sqrt{2}\,\gamma^{1/3}\,\beta^{\frac{1+w_\phi}{6w_\phi}}\,k_{\rm UV}\,,
\end{equation}
and the cutoff function $\Theta_{\rm UV}^{\rm(iso)}(k)$ have the following form 
\begin{eqnarray}
    \Theta^{\rm (iso)}_{\rm UV}(k) &\equiv& \int_{-s_0(k)}^{s_0(k)} {\rm d}s
    \frac{\left(s^2-1\right)^2}{\left(1-c_s^2 s^2\right)^{5/3}} \label{eq:ThetaUV} \\
    &=&\frac{3 s_0 \left(5 c_s^4-2 c_s^2 \left(2 s_0^2+5\right)+9\right)-\left(5 c_s^2 \left(c_s^2+6\right)-27\right) s_0 \left(c_s^2 s_0^2-1\right) \, _2F_1\left(\frac{5}{6},1;\frac{3}{2};c_s^2 s_0^2\right)}{10 c_s^4 \left(1-c_s^2 s_0^2\right)^{2/3}}\,,\nonumber
\end{eqnarray}
and the $s_0(k)$ can be defined as 
\begin{eqnarray} \label{eq: s0}
 s_0(k)\equiv
 \begin{cases}
        1  \quad & \frac{k_{\rm UV}}{k}\geq \frac{1+c_s^{-1}}{2}\\
        2\frac{k_{\rm UV}}{k}-c_s^{-1} \quad & \frac{1+c_s^{-1}}{2}\geq \frac{k_{\rm UV}}{k}\geq\frac{c_s^{-1}}{2}\\
        0 \quad & \frac{c_s^{-1}}{2}\geq\frac{k_{\rm UV}}{k}
    \end{cases} \,.
\end{eqnarray} 
At frequencies well below the resonant regime, $k\ll k_{\rm UV}$, the spectrum is dominated by an infrared tail
\begin{align}
\label{eq:OGWIR}
\Omega_{\rm GW,IR}(k)
&=C^4(w_\phi)\frac{c_s^4}{120\pi^2}\left(\frac{2}{3}\right)^{1/3}
\left(\frac{k_{\rm BH}}{k_{\rm UV}}\right)^8
\left(\frac{k_{\rm UV}}{k_{\rm ev}}\right)^{14/3}
\left(\frac{k}{k_{\rm UV}}\right) \nonumber \\
&\simeq9.03\times10^{24}\,C^4(w_\phi)\,
\beta^{\frac{4(1+w_\phi)}{3w_\phi}}
\left(\frac{\gamma}{0.2}\right)^{8/3}
\left(\frac{M_{\rm in}}{10^4\,{\rm g}}\right)^{28/9}
\left(\frac{k}{k_{\rm UV}}\right).
\end{align}
The transition frequency separating the infrared and resonant regimes is defined implicitly by $\Omega_{\rm GW,IR}(f_{\rm T})=\Omega_{\rm GW,res}(f_{\rm T})$, yielding~\cite{Domenech:2024wao}
\begin{equation}
    f_{\rm T}=\frac{k_{\rm UV}}{2\pi}\left[\left(10^{-6}\frac{M_{\rm in}}{10^4\,{\rm g}}\right)^{-2/3}\right]^{3/8}\,.
\end{equation}
The complete isocurvature-induced GW spectrum is then
\begin{equation}
    \label{eq:GW_iso}
    \Omega_{\rm GW,iso}(f)=
    \begin{cases}
        \Omega_{\rm GW,res}(f), & f\ge f_{\rm T},\\[4pt]
        \Omega_{\rm GW,IR}(f), & f<f_{\rm T}.
    \end{cases}
\end{equation}
The present-day spectrum, $\Omega_{\rm GW,iso}^{(0)}(f)h^2$, is obtained using Eq.~\eqref{eq:GW_present}.
\begin{figure}
    \centering
    \includegraphics[scale=0.49]{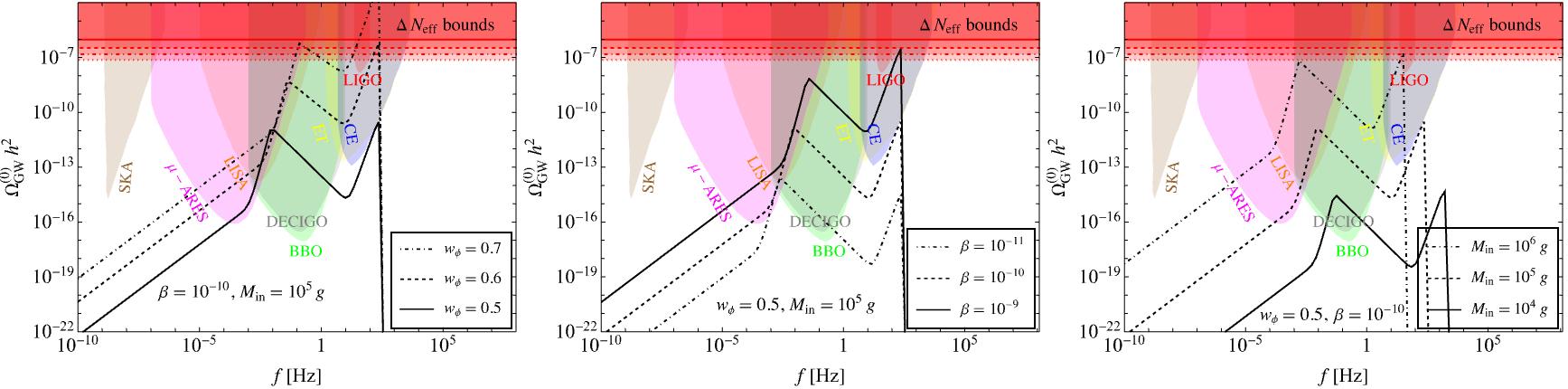}
    \caption{Induced GW spectra at the present epoch due to \textit{isocurvature fluctuations} (i.e., the peak placed on the right) and \textit{adiabatic fluctuations} (i.e., the peak on the left) are shown. Each plot represents the variation of a single parameter while the other two are kept fixed. The representation is referred to on the sensitivity curves of various future GW observatories. Additionally, the $\Delta N_{\rm eff}$ bound at BBN (see, for instance Table-\ref{tab:DNeff}) is shown, as the GWs themselves behave as extra relativistic degrees of freedom.}
    \label{fig:GWspectrum_mass}
\end{figure}
The peak on the right side of figure~\ref{fig:GWspectrum_mass} illustrates the resulting present-day spectrum for isocurvature fluctuations with representative parameter choices. The resonant slope $\Omega_{\rm GW,res} \propto k^{11/3}$ is a robust prediction of the isocurvature mechanism and is insensitive to $w_\phi$, while the peak frequency is set primarily by the PBH formation mass through $k_{\rm UV}$. Lighter PBHs evaporate earlier, shifting the GW peak to higher frequencies. The peak amplitude grows with both $w_\phi$ and $\beta$, reflecting the longer duration and enhanced efficiency of the PBH-dominated phase. This dependence follows from Eq.~\eqref{eq:OGWresPeak}: the amplitude scales as $C^4(w_\phi)\beta^{\frac{4(1+w_\phi)}{3w_\phi}}$, where $C(w_\phi)$ enhances the signal for smaller $w_\phi$ by reducing gravitational potential decay, while larger $w_\phi$ extends the PBH-dominated phase and strengthens the GW signal.  However, the amplitude's dependence on both $w_\phi$ and $\beta$ introduces a degeneracy that cannot be resolved by the isocurvature-induced signal alone. This motivates considering the adiabatic contribution discussed in the following subsection.

\subsubsection*{Gravitational Waves Induced by Adiabatic Fluctuations}
\label{subsec:adiabatic_induced_GW}

During the PBH-dominated phase, the gravitational potential remains approximately constant on subhorizon scales. Once PBHs evaporate, the Universe transitions rapidly into RD, where radiation pressure induces oscillations in the potential. Because PBH evaporation is nearly instantaneous, the transition is too brief for any significant decay of $\Phi$ before these oscillations begin. Consequently, strong GWs can be produced even for an almost scale-invariant primordial curvature spectrum. Following Ref.~\cite{Domenech:2024wao}, we model this class of adiabatic-induced GWs.

For adiabatic initial conditions, the scalar potential power spectrum is given by \footnote{At this point, we clarify that our study is carried out within the framework of linear cosmological perturbation theory, which provides a well-controlled and analytically tractable approach for computing the gravitational-wave spectrum, including the appearance of the characteristic double-peak structure. However, if the early matter-dominated era lasts long enough, density fluctuations on small scales grow proportionally to the scale factor and may eventually enter the non-linear regime. In that case, employing a fixed ultraviolet cutoff $k_{\rm UV}$ can potentially overestimate the signal. A proper treatment of the power spectrum beyond the linear regime therefore requires either dedicated numerical simulations~\cite{Fernandez:2023ddy} or more advanced analytical frameworks, such as kinetic field theory~\cite{Konrad:2022tdu, Bartelmann:2019unp}.}
\begin{equation}
\label{eq:ScalarPS}
\mathcal{P}_{\Phi_0}(k)=A_s\left(\frac{k}{k_*}\right)^{n_s-1}\Theta(k_{\rm UV}-k)\,\Theta(k-k_{\rm IR}),
\end{equation}
where $A_s$ and $n_s$ are the amplitude and spectral index of curvature perturbations, and $(k_{\rm IR},k_{\rm UV})$ are infrared and ultraviolet cutoffs. The cutoffs ensure linearity and consistency with CMB scales by excluding very large or highly nonlinear modes.

To compute the induced tensor spectrum, $\mathcal{P}_{\Phi_0}(k)$ and the adiabatic transfer function $\Phi(k)$ are substituted into the convolution integral of Eq.~\eqref{eq:av_tensorPS}. 
Unlike the isocurvature case, the resulting adiabatic spectrum depends explicitly on both $n_s$ and the post-inflationary EoS $w_\phi$, which shapes the transfer function. 
For modes re-entering during the pre-PBH epoch, the transfer function can be approximated as
\begin{equation}
\Phi_{\rm ad}(k\gg k_{\rm BH})=A_\Phi(w_\phi)\left(\frac{k}{k_{\rm BH}}\right)^{n(w_\phi)},
\end{equation}
where the effective exponent $n(w_\phi)$ depends on $w_\phi$ as~\cite{Domenech:2024wao}
\begin{equation}
\label{eq:nw}
n(w_\phi)\simeq
\begin{cases}
b/2-2, & w_\phi\lesssim0.2,\\
-1.83-0.285b+0.790b^2, & 0.2<w_\phi<0.67,\\
-(2+b), & w_\phi\gtrsim0.67,
\end{cases}
\quad b=\frac{1-3w_\phi}{1+3w_\phi},
\end{equation}
and $n(w_\phi)=0$ for $k\ll k_{\rm BH}$. 
The normalization $A_\Phi(w_\phi)$ follows the fit
\begin{equation}
\label{eq:a_phi}
A_\Phi(w_\phi)=
\begin{cases}
\frac{2}{3b}\left(7.76+18.3b+12.5b^2\right), & w_\phi<0.2~\text{or}~w_\phi>0.67,\\
15.6+39.2b+21.3b^2, & 0.2<w_\phi<0.67.
\end{cases}
\end{equation}
Substituting these expressions into Eq.~\eqref{eq:av_tensorPS} yields the schematic form
\begin{align}
\label{eq:TensorPS_Adi_Gen}
\overline{\mathcal{P}_{h,{\rm RD}}}(k,\eta,\bar{x}\gg1)
\propto
\int_0^\infty\!dv\!\int_{|1-v|}^{1+v}\!du\,
\left(4v^2-(1+v^2-u^2)^2\right)^2(uv)^{n_{\rm eff}(w_\phi)}\,\overline{\mathcal{I}^2_{\rm osc}}(\bar{x},u,v),
\end{align}
where $\bar{x}=k\bar{\eta}$ and $\overline{\mathcal{I}^2_{\rm osc}}$ is the oscillatory kernel evaluated at the end of PBH domination. 
The effective spectral index $n_{\rm eff}$, incorporating both $n_s$ and $w_\phi$, is given by~\cite{Domenech:2024wao}
\begin{equation}
\label{eq:neff}
n_{\rm eff}(w_\phi)=-\frac{5}{3}+n_s+2n(w_\phi).
\end{equation}
For $k\gg k_{\rm BH}$, the resonant contribution to the GW spectrum takes the form
\begin{equation}
\label{eq:OmegaGW_res_adi}
\Omega_{\rm GW,res}^{\rm ad}(k)
=\Omega_{\rm GW,res}^{\rm ad,peak}
\left(\frac{k}{k_{\rm UV}}\right)^{2n_{\rm eff}(w_\phi)+7}\Theta_{\rm UV}^{\rm(ad)}(k),
\end{equation}
where the amplitude at $k=k_{\rm UV}$ is
\begin{align}
\label{eq:OmegaGWres_Peak_adi}
\Omega_{\rm GW,res}^{\rm ad,peak}
\simeq9.7\times10^{32}\,
\frac{3^{2n(w_\phi)+n_s}}{4^{3n(w_\phi)+n_s}}
A_s^2A_\Phi(w_\phi)^4
\gamma^{-4n(w_\phi)/3}
\beta^{-\frac{2n(w_\phi)(1+w_\phi)}{3w_\phi}}
\left(\frac{g_H}{108}\right)^{-17/9}
\left(\frac{M_{\rm in}}{10^4\,{\rm g}}\right)^{34/9}.
\end{align}
The cutoff function $\Theta^{\rm (ad)}_{\rm UV}(k)$ for the resonant part for the adiabatic scenario, in Eq.~\eqref{eq:OmegaGW_res_adi}, can be expressed as
\begin{align}
    \Theta^{\rm (ad)}_{\rm UV}(k) \equiv&  
    \int_{-s_0(k)}^{s_0(k)} {\rm d}s\left(s^2-1\right)^2\left(1-c_s^2 s^2\right)^{n_{\rm eff}(w_\phi)} \label{eq:ThetaUV_ad} \\
    =&\frac{2}{5} s_0^5 \, _2F_1\left(\frac{5}{2},-n_{\rm eff}(w_\phi);\frac{7}{2};c_s^2 s_0^2\right)-\frac{4}{3} s_0^3 \, _2F_1\left(\frac{3}{2},-n_{\rm eff}(w_\phi);\frac{5}{2};c_s^2 s_0^2\right) \nonumber \\
    &+2 s_0 \, _2F_1\left(\frac{1}{2},-n_{\rm eff}(w_\phi);\frac{3}{2};c_s^2 s_0^2\right) \,. \nonumber
\end{align}
At $k\sim k_{\rm BH}$ and $k\ll k_{\rm BH}$, the intermediate and infrared (IR) regimes of the spectrum are described by
\begin{align}
\label{eq:OmegaGW_mid_adi}
\Omega_{\rm GW,mid}^{\rm ad}(k)
&=-\frac{A_s^2A_\Phi(w_\phi)^4c_s^4}{768(1+2n_{\rm eff})}
\left(\frac{3}{2}\right)^{2/3}\xi_1^{1+2n_{\rm eff}}
\left(\frac{k_{\rm BH}}{k_{\rm UV}}\right)^{2n_s-7/3}
\left(\frac{k_{\rm UV}}{k_{\rm ev}}\right)^{14/3}
\left(\frac{k}{k_{\rm UV}}\right)^5,\\
\label{eq:OmegaGW_IR_adi}
\Omega_{\rm GW,IR}^{\rm ad}(k)
&=\frac{A_s^2c_s^4(3w_\phi+5)^4}{10^4(6n_s+5)(1+w_\phi)^4}
\left(\frac{3}{2}\right)^{2/3}
\left(\frac{\xi_2k_{\rm BH}}{k_{\rm UV}}\right)^{5/3+2n_s}
\left(\frac{k_{\rm UV}}{k_{\rm ev}}\right)^{14/3}
\left(\frac{k}{k_{\rm UV}}\right),
\end{align}
where $\xi_1$ and $\xi_2$ are $\mathcal{O}(1)$ functions of $w_\phi$~\cite{Domenech:2024wao}, with
\begin{equation}
\label{eq:xi2}
\xi_2=
\begin{cases}
\left[\frac{7.76+18.3b+12.5b^2}{(3w_\phi+5)/(5w_\phi+5)}\right]^{(1+3w_\phi)/(3+3w_\phi)}, & w_\phi>1/3,\\[5pt]
\left[\frac{7.76+18.3b+12.5b^2}{(3w_\phi+5)/(5w_\phi+5)}\right]^{(2+6w_\phi)/(3+15w_\phi)}, & w_\phi<1/3.
\end{cases}
\end{equation}
The full adiabatic GW spectrum is the sum of the resonant, intermediate, and IR components, 
$\Omega_{\rm GW,ad}(k)=\Omega_{\rm GW,res}^{\rm ad}+\Omega_{\rm GW,mid}^{\rm ad}+\Omega_{\rm GW,IR}^{\rm ad}$. 
Its present-day value, $\Omega_{\rm GW,ad}^{(0)}(f)h^2$, is obtained using Eq.~\eqref{eq:GW_present}. 
Figure~\ref{fig:GWspectrum_mass} (i.e., the peak placed on the left for this doubly peaked spectrum, which is responsible for the adiabatic source) shows how varying the parameters $w_\phi$, $\beta$, and $M_{\rm in}$ modifies the spectrum. 
Compared with the isocurvature case, the adiabatic spectrum exhibits a broader resonant region, peaking near $k_{\rm BH}$ rather than $k_{\rm UV}$. 
Increasing either $w_\phi$ or $\beta$ shifts the peak to higher frequencies, as both lead to earlier PBH domination and hence smaller $k_{\rm BH}$. 
The spectral behavior follows Eqs.~\eqref{eq:OmegaGW_res_adi}–\eqref{eq:OmegaGW_IR_adi}: 
in the resonant regime $\Omega_{\rm GW}\propto k^{2n_{\rm eff}(w_\phi)+7}$, while in the intermediate and IR regions it scales as $k^5$ and $k$, respectively, as illustrated in Fig.~\ref{fig:GWspectrum_mass}.

\subsection{Detection Prospects}
The stochastic gravitational-wave background discussed above provides a powerful probe of the post-inflationary Universe and reheating dynamics. For the primary GW signal, Ref.~\cite{Maity:2024cpq} showed that future space-based interferometers such as LISA, ET, DECIGO, and BBO can significantly constrain the background equation of state and reheating temperature. In particular, LISA offers the strongest discriminatory power, restricting the reheating history to a narrow range characterized by 
$ w_\phi \simeq 1/2\text{-}3/5 $ 
with reheating temperatures below $\mathcal{O}(1)\,\mathrm{GeV}$ considering $\alpha-$ attractor model of inflation. 
In contrast, BBO allows comparatively broader regions of parameter space due to its different sensitivity window. For further details, we refer to Ref.~\cite{Maity:2024cpq}. The referred analysis is based on the signal-to-noise ratio (SNR) for each detector, modeling the sensitivity curves using instrumental noise only.

For the induced stochastic gravitational-wave background generated in PBH reheating scenarios, the detection prospects have been investigated in Ref.~\cite{Paul:2025kdd} using signal-to-noise ratio calculations, Fisher matrix forecasts, and Markov Chain Monte Carlo analyses based on mock LISA and ET data. The study finds that the ET is particularly sensitive to both isocurvature- and adiabatic-induced GW spectra, probing an initial PBH mass range $ M_{\rm in} \in (0.5\text{-}4\times10^7)\,\mathrm{g} $, although the reconstruction of isocurvature parameters is subject to relatively large uncertainties. In contrast, LISA is mainly sensitive to the adiabatic contribution and can probe the mass window $ M_{\rm in} \in (2\times10^4\text{-}5\times10^8)\,\mathrm{g} $. The complementarity between ET and LISA therefore opens a multi-frequency observational window onto the reheating epoch, enabling simultaneous constraints on the PBH formation mass, the initial abundance parameter $\beta$, and the effective equation-of-state parameter $w_\phi$ that characterizes the reheating history. A detailed discussion of these projected sensitivities can be found in Ref.~\cite{Paul:2025kdd}.

\newpage

%% file: 5_Radiation_particles.tex
\section{Radiation of new bosonic states}
Ultralight bosonic degrees of freedom with spin $0$, $1$, or $2$, which may constitute viable DM candidates, can be radiated from isolated stars or binary systems and thereby contribute to a variety of astrophysical observables, including pulsar spin-down luminosities and orbital period decay in compact binaries. In addition to radiation, such ultralight bosons can mediate long-range macroscopic forces between the components of a binary system, leading to observable modifications of the orbital dynamics. In the following, we consider the emission and dynamical effects of scalar, pseudoscalar, and vector bosons, as well as massive gravitons, in compact-star systems. By comparing their predicted signatures with current observational data, stringent constraints can be placed on the couplings of these ultralight degrees of freedom to SM particles.

\subsection{Radiation of scalars from quasi-stable compact binary systems}

In this section, we analyze the energy loss due to the radiation of a massive scalar field $\phi$ from a quasi-stable binary system. The scalar field may be sourced by the neutrons, electrons, or muons within the NS. The radiation of a massless scalar field from a compact binary system has been studied using a field-theoretic framework in~\cite{Mohanty:1994yi}. Contrary to the massless case, the radiation of a massive scalar field from a generic non-relativistic matter source has been investigated via a classical multipole expansion in~\cite{Krause:1994ar}.

In the following, we focus on computing the radiation of massive scalar particles originating from a non-relativistic nucleonic $(n)$, electronic $(e)$ or muonic $(\mu)$ source charge density, adopting a field-theoretic approach. To simplify the analysis, we model the stars as point-like sources. This assumption is justified because the Compton wavelength of the scalar field, $1/m_\phi \sim 1/\Omega_{\mathrm{orb}} \sim 10^9\,\mathrm{km}$, is much larger than the stellar radius, $R \sim 10\,\mathrm{km}$, where $\Omega_{\mathrm{orb}}$ is the orbital frequency of the binary system. The interaction between the scalar field and the fermionic number density is governed by the following Lagrangian
\begin{equation}
    \mathcal{L}_\mathrm{int}= g\phi \langle \bar{f}f\rangle,
    \label{ap1}
\end{equation}
where $g$ denotes the coupling of the scalar with the fermions $(f\in n,\mu,e )$ and for a point-like source, the non-relativistic fermionic current can be approximated as $\langle \bar{f}f \rangle = n(r)$, where the fermion number density is given by
\begin{equation}
    n(r) = \sum_{j=1,2} N_j \delta^3\left(\mathbf{r} - \mathbf{r}_j(t)\right),
\end{equation}
where, $N_j$ represents the total number of fermions in the $j$-th NS, and $\mathbf{r}_j(t)$ denotes its position vector in the center-of-mass frame of the binary system.

We consider the binary stars to follow Keplerian motion in the $x\text{-}y$ plane. The parametric form of their orbital trajectory is given by~\cite{Landau:1975pou}
\begin{equation}
    x=a(\cos\xi-e)\,,\quad y=a\sqrt{1-e^2}\sin\xi\,,\quad \Omega_{\mathrm{orb}} t=\xi-e\sin\xi\,,
    \label{ap2}
\end{equation}
where $e$, $a$ and $\xi$ represent the eccentricity, semi-major axis and the eccentric anomaly of the Keplerian orbit, respectively. For two stars of masses $M_1$ and $M_2$, the fundamental orbital frequency of the binary is
\begin{equation}
    \Omega_{\mathrm{orb}} = \sqrt{\frac{G(M_1 + M_2)}{a^3}}\,,
\end{equation}
and it is distinct from the spin frequency $\Omega$ of an individual star. Due to the elliptical nature of the orbit, the orbital angular velocity is not constant, and hence the radiation includes contributions from all harmonics of the fundamental frequency. To account for all the harmonics, we perform a decomposition of the orbital frequency into Fourier modes, and the corresponding coordinate representation of the Keplerian orbit in frequency space is given by~\cite{Landau:1975pou}
\begin{equation}
    x(\omega)=\frac{aJ_n^\prime(ne)}{n}\,,\qquad y(\omega)=\frac{ia\sqrt{1-e^2}J_n(ne)}{ne}\,,
    \label{ap3}
\end{equation} 
where $\omega = n \Omega_{\mathrm{orb}}$ denotes the $n$-th harmonic of the fundamental orbital frequency. The prime symbol (${}^\prime$) over the Bessel function represents the derivative of the Bessel function with respect to its argument. Therefore, the emission rate of scalar particles from the binary system can be expressed as
\begin{equation}
    {\rm d}\Gamma=g^2|n(\omega^\prime)|^2 2\pi\delta(\omega-\omega^\prime)\frac{{\rm d}^3k^\prime}{(2\pi)^3 2\omega^\prime},
    \label{ap4}
\end{equation}
where the source number density, when transformed into frequency space, takes the form
\begin{equation}
    n(\omega)=\frac{1}{2\pi}\int\int e^{i\mathbf{k}\cdot\mathbf{r}}e^{-i\omega t}\sum_{j=1,2}N_j\delta^3(\mathbf{r}-\mathbf{r}_j(t))\,{\rm d}^3{\bf r}\,{\rm d}t,
    \label{ap6}
\end{equation}
which can be simplified to
\begin{equation}
    n(\omega)=(N_1+N_2)\delta(\omega)+\Big(\frac{N_1}{M_1}-\frac{N_2}{M_2}\Big)\mu (ik_x x(\omega)+ik_yy(\omega))+\mathcal{O}(\mathbf{k}\cdot\mathbf{r})^2,
    \label{ap7}
\end{equation}
where $k$ denotes the scalar field momentum and $N_1$ and $N_2$ are the number of fermions in the binary. In the center-of-mass frame, the position vectors of the two stars are given by $\mathbf{r}_1 = \frac{\mu}{M_1} \mathbf{r}$ and $\mathbf{r}_2 = -\frac{\mu}{M_2} \mathbf{r}$, where $\mu = M_1 M_2/(M_1 + M_2)$ is the reduced mass of the binary system composed of stars with masses $M_1$ and $M_2$.

From Eq.~\eqref{ap4}, it follows that the rate of energy loss due to radiation of massive scalar particles from the binary system is obtained as
\begin{equation}
    \frac{{\rm d}E_\phi}{{\rm d}t}=\frac{g^2}{2\pi}\int|n(\omega^\prime)|^2\delta(\omega-\omega^\prime){\omega^\prime}^2\,\Big(1-\frac{m^2_\phi}{{\omega^\prime}^2}\Big)^\frac{1}{2}\,{\rm d}\omega^\prime,
    \label{ap5}
\end{equation}
where the scalar particle satisfies the dispersion relation $\omega^2=k^2+m^2_\phi$, with $m_\phi$ denoting the mass of the scalar field. 

Using Eqs.~\eqref{ap3} and~\eqref{ap6}, the leading non-vanishing contribution to $|n(\omega)|^2$ is given by
\begin{equation}
    |n(\omega)|^2=\frac{1}{3}\Big(\frac{N_1}{M_1}-\frac{N_2}{M_2}\Big)^2\mu^2a^2\Omega_{\mathrm{orb}}^2\Big[{J^\prime_n{(ne)}}^2+\frac{1-e^2}{e^2}J_n^2(ne)\Big]\Big(1-\frac{m^2_\phi}{n^2\Omega_{\mathrm{orb}}^2}\Big),
    \label{ap8}
\end{equation}
where, we have used the isotropic average $\langle k_x^2 \rangle = \langle k_y^2 \rangle = \frac{k^2}{3}$. Substituting Eq.~\eqref{ap8} into Eq.~\eqref{ap5}, we arrive at the expression for the rate of energy loss due to massive scalar radiation as~\cite{Poddar:2024thb}
\begin{equation}
    \frac{{\rm d}E_\phi}{{\rm d}t}=\frac{g^2}{6\pi}\Big(\frac{N_1}{M_1}-\frac{N_2}{M_2}\Big)^2\mu^2a^2\Omega_{\mathrm{orb}}^4\sum_{n>m_\phi/\Omega_{\mathrm{orb}}} n^2\Big[{J^\prime_n{(ne)}}^2+\frac{1-e^2}{e^2}J_n^2(ne)\Big]\Big(1-\frac{m^2_\phi}{n^2\Omega_{\mathrm{orb}}^2}\Big)^\frac{3}{2}.
    \label{ap9}
\end{equation}
In the massless scalar limit ($m_\phi\rightarrow 0$), Eq.~\eqref{ap9} becomes
\begin{equation}
    \frac{{\rm d}E_\phi}{{\rm d}t}=\frac{g^2}{12\pi}\Big(\frac{N_1}{M_1}-\frac{N_2}{M_2}\Big)^2\mu^2a^2\Omega_{\mathrm{orb}}^4 \frac{\Big(1+\frac{e^2}{2}\Big)}{(1-e^2)^\frac{5}{2}},
    \label{ap10}
\end{equation}
where we use the following relation
\begin{equation}
    \sum_{n} n^2\Big[{J^\prime_n{(ne)}}^2+\frac{1-e^2}{e^2}J_n^2(ne)\Big]=\sum_{n}f(n,e)=\frac{\Big(1+\frac{e^2}{2}\Big)}{2(1-e^2)^\frac{5}{2}}.
\end{equation}
Equation.~\eqref{ap9} describes the emission rate of massive scalar particles from a binary system, where the scalar field couples to the net non-relativistic fermion content of the binary. For scalar radiation to occur, there must be an asymmetry in the charge-to-mass ratios of the two stars. The resulting emission is dipolar in nature and scales with the fourth power of the orbital frequency, i.e., $\Omega_{\mathrm{orb}}^4$. This expression remains valid provided that $m_\phi < \Omega_{\mathrm{orb}}$, ensuring that the fundamental harmonic is kinematically accessible.

The expression for the scalar radiation energy loss, Eq.~\eqref{ap9}, is valid for arbitrary quasi-stable binary systems and for any non-relativistic fermions. However, the magnitude of the loss is strongly determined by the net fermionic content $(N_{1,2})$ of the compact stars, which in turn depends on their internal densities. For instance, in white dwarfs (WDs), the muon content is negligible, i.e., $N_\mu \simeq 0$.

The rate of energy loss due to the radiation of ultralight scalar particles, Eq.~\eqref{ap9}, provides an additional contribution to the orbital period decay of the binary system, which is an observable quantity, given by
\begin{equation}
    \frac{{\rm d}P_b}{{\rm d}t}=-6\pi G^{-3/2}(M_1 M_2)^{-1}(M_1+M_2)^{-1/2}a^{5/2}\Big(\frac{{\rm d}E_{\mathrm{GW}}}{{\rm d}t}+\frac{{\rm d}E_\phi}{{\rm d}t}\Big), 
    \label{op}
\end{equation}
where the quadrupole formula of the GW energy loss is given as
\begin{equation}
    \frac{{\rm d}E_{\mathrm{GW}}}{{\rm d}t}=\frac{32G}{5}\mu^2 a^4\Omega_{\mathrm{orb}}^6 (1-e^2)^{-7/2}\Big(1+\frac{73}{24}e^2+\frac{37}{96}e^4\Big).
    \label{opgw}
\end{equation}

Precision measurements of the orbital period loss of the binary put limits on the scalar-electron coupling as $g\lesssim 3\times 10^{-21}$ for $m_\phi\lesssim 10^{-19}~\mathrm{eV}$~\cite{Poddar:2024thb}.

\subsection{Radiation of vectors from quasi-stable compact binary systems}

Analogous to ultralight scalar fields, ultralight vector bosons can also be emitted from quasi-stable binary systems. In the following, we focus on the radiation of the $L_\mu - L_\tau$ vector gauge boson from compact object binaries. NSs contain an enormous number of muonic charges ($N_\mu \sim 10^{55}$), and as a result, massive $Z^\prime$ Proca vector bosons or gauge bosons of $L_\mu-L_\tau$ type coupled to muons can be radiated from a binary system in addition to the standard GWs. This additional radiation can contribute to the observed decay in the orbital period of binary systems. Given that a typical NS has a radius of about 10 km, it can be approximated as a point source, since the Compton wavelength of the radiated boson ($\lambda \sim 10^{12}\,\mathrm{m}$) is vastly larger than the star's size. In the following, we calculate the typical muon numbers in an NS and then derive the energy loss due to the muonphilic vector bosons from a binary.
\bigskip

\textit{Muon content inside an NS:} In addition to nucleons and electrons, NSs also contain a substantial population of muons. The chemical potential of relativistic, degenerate electrons in an NS is given by
\begin{equation}
\mu_e=(k^2_{fe}+m^2_e)^{1/2}=[(3\pi^2\rho Y_e)^{2/3}+m^2_e]^{1/2},
 \label{cons10} 
\end{equation}
where $m_e$ is the electron mass, $k_{fe}$ its Fermi momentum, $\rho$ the nucleon number density, and $Y_e$ the electron fraction.

Charge neutrality inside the star requires
\begin{equation}
Y_p=Y_e+Y_\mu,~~~~~~~~~Y_n+Y_p=1.
\label{newq1}
\end{equation}
where $Y_p$, $Y_n$, and $Y_\mu$ denote the fractions of protons, neutrons, and muons, respectively. At densities above nuclear saturation, once the electron chemical potential exceeds the muon rest mass ($m_\mu \simeq 105\,\mathrm{MeV}$), electron to muon conversion becomes energetically favorable through weak processes such as
\begin{equation}
e^{-}\to \mu^{-}+\nu_e+\bar{\nu}_\mu,~~~p+\mu^{-}\to n+\nu_\mu,~~~n\to p+\mu^{-}+\bar{\nu}_\mu.
\label{newq2}    
\end{equation}
As a result, both electrons and muons coexist in the NS and are stabilized through $\beta$ equilibrium. The corresponding equilibrium condition is~\cite{Zhang:2000my}
\begin{equation}
\mu_n-\mu_p=\mu_e=\mu_\mu=[m^2_{\mu}+(3\pi^2\rho Y_\mu)^{2/3}]^{1/2},
 \label{cons11} 
\end{equation}
which enforces equality of electron and muon chemical potentials.

The muon decay ($\mu^- \to e^- + \bar{\nu}_e + \nu_\mu$) is forbidden in the NS due to Fermi blocking. The electron Fermi energy is $\sim 100\,\mathrm{MeV}$ (relativistic), whereas the muon Fermi energy is only $\sim 30\,\mathrm{MeV}$ (non-relativistic). Since electron states are filled up to the Fermi surface, the muon decay cannot proceed. By contrast, in WDs, the muon Fermi energy is extremely small ($\sim 1\,\mathrm{eV}$), so Fermi suppression is negligible and muon decay is allowed.

From the $\beta$ equilibrium condition, one can relate the muon density to the electron fraction as
\begin{equation}
\rho_\mu=\frac{m^3_e}{3\pi^2}\Big(1+\frac{(3\pi^2\rho Y_e)^{2/3}}{m^2_e}-\frac{m^2_\mu}{m^2_e}\Big)^{3/2}.  
 \label{cons12} 
\end{equation}

The electron fraction itself can be parametrized as
\begin{equation}
Y_e=\frac{p_1+p_2\rho+p_6\rho^{3/2}+p_3\rho^{p_7}}{1+p_4\rho^{3/2}+P_5\rho^{p_7}},
 \label{cons13} 
\end{equation}
where the coefficients $p_i$ depend on the choice of nuclear equation of state (EoS).

For instance, adopting the BSk24 EoS~\cite{Pearson:2018tkr} with $\rho = 0.238\,\mathrm{fm}^{-3}$ and $Y_e = 0.052$ (corresponding to $\sim 10^{57}$ nucleons in the star), Eq.~\eqref{cons12} yields a muon number density of
\begin{equation}
    \rho_\mu\sim3.1\times 10^4\,\mathrm{MeV}^3.
\label{newq3}    
\end{equation}
Multiplying by the stellar volume, $(4/3)\pi R^3$ with $R = 10\,\mathrm{km}$, gives a total muon number of order
\begin{equation}
    N_\mu\approx 10^{55}.
    \label{newq4}    
\end{equation}
Since there is a significant number of muons inside an NS, ultralight $Z^\prime_\mu$ vector bosons can couple with the muon current inside an NS.

The vector boson $Z^\prime_\mu$ is radiated from a classical muonic current $J^\mu$ of the binary system which has a Keplerian motion. We assume an interaction vertex of the form 
\begin{equation}
    \mathcal{L}\supset g Z^\prime_\mu J^\mu, 
\end{equation}
where $g$ is the muonphilic coupling constant. Under these assumptions, the radiation rate of the massive $Z^\prime$ boson is given by
\begin{equation}
    {\rm d}\Gamma =g^2\sum^3_{\lambda=1}[J^\mu(k^\prime)J^{\nu*}(k^\prime)\epsilon^\lambda_\mu(k)\epsilon^{\lambda*}_\nu(k)]2\pi\delta(\omega-\omega^\prime)\frac{{\rm d}^3 k}{(2\pi)^3 2\omega},
\end{equation}
where, $J^\mu(k')$ denotes the Fourier transform of the spacetime current $J^\mu(x)$, and $\epsilon^\lambda_\mu(k)$ represents the polarization vector of the massive vector boson. The sum over polarization states of the massive vector boson is
\begin{equation}
    \sum^3_{\lambda=1}\epsilon^\lambda_\mu(k)\epsilon^{\lambda*}_\nu(k) = -g_{\mu\nu}+\frac{k_\mu k_\nu}{M^2_{Z^\prime}}\,.
    \label{eq:pol_sum}
\end{equation}
Therefore, the resulting emission rate becomes
\begin{equation}
    \begin{split}
    {\rm d}\Gamma = & \frac{g^2}{2(2\pi)^2}\int\Big[-|J^\mu(\omega^\prime)|^2+\frac{1}{M^2_{Z^\prime}}\Big(|J^0(\omega^\prime)|^2 \omega^2+J^i(\omega^\prime)J^{j*}(\omega^\prime)k_i k_j+2J^0(\omega^\prime) J^{i*}(\omega^\prime)k_0k_i\Big)\Big]\\ 
    & \times\delta(\omega-\omega^\prime)\,\omega \, \Big(1-\frac{M^2_{Z^\prime}}{\omega^2}\Big)^{\frac{1}{2}}{\rm d}\omega \, {\rm d} \Omega_k.
    \end{split}
    \label{edw}
\end{equation}
The four-momentum of the $Z^\prime$ boson is given by $k_\mu = (\omega, -\vec{k})$, with $k_i = |\vec{k}|\, \hat{n}_i$ and $k_j = |\vec{k}|\, \hat{n}_j$, where $\hat{n}_i$ and $\hat{n}_j$ are unit vectors indicating the direction of propagation. The third term in the first bracket of Eq.~\eqref{edw} does not contribute to the emission rate because
\begin{equation}
\int \hat{n_i}{\rm d}\Omega_k=0, \hspace{1cm}\int \hat{n_i}\hat{n_j}{\rm d}\Omega_k=\frac{4\pi}{3}\delta_{ij}.
\end{equation}
Thus, the rate of energy loss due to the radiation of massive $Z^\prime$ vector bosons is given by
\begin{equation}
\begin{split}
\frac{dE_V}{dt}=&\frac{g^2}{2\pi}\int\Big[-|J^0(\omega^\prime)|^2+|J^i(\omega^\prime)|^2+\frac{\omega^2}{M^2_{Z^\prime}}|J^0(\omega^\prime)|^2+\frac{\omega^2}{3M^2_{Z^\prime}}|J^i(\omega^\prime)|^2\Big(1-\frac{M^2_{Z^\prime}}{\omega^2}\Big)\Big]\\
&\times\delta(\omega-\omega^\prime)\,\omega^2 \, \Big(1-\frac{M^2_{Z^\prime}}{\omega^2}\Big)^{\frac{1}{2}}{\rm d}\omega.
\end{split}
\label{eq:first}
\end{equation}
The current density associated with the binary system can be expressed as
\begin{equation}
J^\mu(x)=\sum_{a=1,2}Q_a \delta^3(\textbf{x}-\textbf{x}_a(t))u^\mu_a,
\label{eq:current}
\end{equation}
where, the index $a = 1, 2$ labels the two stars in the binary system. $Q_a$ represents the total muonic charge of the $a$-th NS, and $\mathbf{x}_a(t)$ denotes its position as a function of time. The four-velocity of each star in the non-relativistic limit is given by $u^\mu_a = (1, \dot{x}_a, \dot{y}_a, 0)$, assuming motion confined to the $x\text{-}y$ plane. The Fourier transform of the spatial component of $J^\mu(\omega^\prime)$, evaluated at the harmonic frequency $\omega^\prime = n\Omega_{\mathrm{orb}}$, corresponding to Eq.~\eqref{eq:current}, is given by
\begin{equation}
    J^i(\omega^\prime)=\int  \frac{1}{T} \int^T_0 \, e^{in\Omega t}\,\dot{x}^i_a(t)\sum_{a=1,2} Q_a {\rm d}^3 \textbf{x}^\prime e^{-i\textbf{k}^\prime. \textbf{x}^\prime} \delta^3(\textbf{x}^\prime-\textbf{x}_a(t))\,{\rm d}t\,.
    \label{eq:Ji_omega}
\end{equation}
We expand the exponential as $e^{i\mathbf{k}^\prime \cdot \mathbf{x}^\prime} = 1 + i\, \mathbf{k}^\prime \cdot \mathbf{x}^\prime + \dots$, and retain only the leading-order term, since $\mathbf{k}' \cdot \mathbf{x}^\prime \sim \Omega_{\mathrm{orb}} a \ll 1$ for typical binary star orbits. Hence, Eq.~\eqref{eq:Ji_omega} simplifies to
\begin{equation}
    J^i(\omega^\prime)=\frac{Q_1}{T}\int^T_0 \, e^{in\Omega_{\mathrm{orb}} t}\dot{x}^i_1(t) \,{\rm d}t + \frac{Q_2}{T}\int^T_0 \, e^{in\Omega_{\mathrm{orb}} t}\dot{x}^i_2(t)\,{\rm d}t\,.
\end{equation}
Therefore, in the center-of-mass frame, the spatial part of the current density becomes
\begin{equation}
    J^i(\omega^\prime)=\frac{1}{T}\Big(\frac{Q_1}{M_1}-\frac{Q_2}{M_2}\Big)\mu\int^T_0 \, e^{in\Omega_{\mathrm{orb}} t}\dot{x^i}(t)\,{\rm d}t\,.
\end{equation}
The Fourier transform of the velocity along a Keplerian orbit can be computed using the following procedure.
\begin{eqnarray}
    \dot{x_n}&=&\frac{1}{T}\int^T_0 e^{i\Omega_{\mathrm{orb}} nt}\dot{x}{\rm d}t = \frac{\Omega_{\mathrm{orb}}}{2\pi}\int^{2\pi}_0 e^{in(\xi-e\sin\xi)}(-a\sin\xi){\rm d}\xi\,,
    \label{eq:xdot_omega}
\end{eqnarray}
where we have used the fact that $\dot{x}dt=-a\sin\xi {\rm d}\xi$ from Eq.~\eqref{ap2}. Similarly, 
\begin{equation}
    \dot{y_n}=\frac{1}{T}\int^T_0 e^{i\Omega_{\mathrm{orb}} nt}\dot{y}{\rm d}t\,.
\end{equation}
Using $\dot{y}{\rm d}t = a\sqrt{1-e^2}\cos\xi {\rm d}\xi$ from Eq.~\eqref{ap2} we can write
\begin{eqnarray}
    \dot{y_n}&=&\frac{\Omega_{\mathrm{orb}} a\sqrt{1-e^2}}{2\pi}\int^{2\pi}_0 e^{in(\xi-e\sin\xi)}\cos\xi {\rm d}\xi = \frac{\Omega_{\mathrm{orb}} a\sqrt{1-e^2}}{2\pi e}\int^{2\pi}_0 e^{in(\xi-e\sin\xi)} {\rm d}\xi.
    \label{eq:ydot_omega}
\end{eqnarray}
Using the Bessel function identity  
\begin{equation}
    J_n(z)=\frac{1}{2\pi}\int^{2\pi}_0 e^{i(n\xi-z\sin\xi)}\,{\rm d}\xi,
\label{eq:J_n_Z}
\end{equation}
in Eqs.~\eqref{eq:xdot_omega} and~\eqref{eq:ydot_omega} we obtain
\begin{equation}
    \dot{x}_n=-ia\Omega_{\mathrm{orb}} J^\prime_n(ne)\,,\quad\hbox{and}\quad\dot{y}_n=\frac{a\sqrt{1-e^2}\Omega_{\mathrm{orb}}}{e} J_n(ne),
\end{equation}
where the prime over the Bessel function denotes the derivative with respect to its argument as previously defined. Therefore, each component of the current density in the Fourier space becomes
 \begin{eqnarray}
    J^x(\omega^\prime)&=&\Omega_{\mathrm{orb}} \Big(\frac{Q_1}{M_1}-\frac{Q_2}{M_2}\Big)\mu \frac{1}{2\pi}\int^T_0 \, e^{in\Omega_{\mathrm{orb}} t}\dot{x}^i(t)\,{\rm d}t = -ia\Omega_{\mathrm{orb}}\Big(\frac{Q_1}{M_1}-\frac{Q_2}{M_2}\Big)\mu J^\prime_n(ne)\,.
\end{eqnarray}
and 
\begin{equation}
    J^y(\omega^\prime)=\Omega_{\mathrm{orb}}\Big(\frac{Q_1}{M_1}-\frac{Q_2}{M_2}\Big)\mu\frac{a\sqrt{1-e^2}}{e}J_n(ne)\,.
\end{equation}
Hence, the squared magnitude of the spatial component of $J^\mu(\omega^\prime)$ can be expressed as
\begin{eqnarray}
    |J^i(\omega^\prime)|^2&=&|J^x(\omega^\prime)|^2+|J^y(\omega^\prime)|^2 = a^2\Omega_{\mathrm{orb}}^2 \mu^2\Big(\frac{Q_1}{M_1}-\frac{Q_2}{M_2}\Big)^2\Big[J^{\prime^2}_n(ne)+\frac{(1-e^2)}{e^2}J^2_n(ne)\Big].
    \label{eq:ji_omega}
\end{eqnarray}
Similarly, from Eq.~\eqref{eq:current} we write the Fourier transform of the temporal component of $J^\mu(x)$ as 
\begin{equation}
    J^0(\omega)=\frac{1}{2\pi}\int e^{i\textbf{k}^\prime.\textbf{x}^\prime} e^{-i\omega t}\sum_{a=1,2} Q_a\,\delta^3(\textbf{x}^\prime-\textbf{x}_a(t)){\rm d}^3 \textbf{x}^\prime {\rm d}t.
    \label{kt}
\end{equation}
In the center-of-mass frame, Eq.~\eqref{kt} becomes
\begin{equation}
 J^0(\omega)= (Q_1+Q_2)\delta (\omega)+ i \mu \left(\frac{Q_1}{M_1} - \frac{Q_2}{M_2}\right)(k_x x(\omega) + k_y y(\omega))+ \mathcal{O}((\textbf{k}.\textbf{r})^2),
\label{eq:jzero}
\end{equation}
where $x(\omega)$ and $y(\omega)$ are the orbital coordinates, given in Eq.~\eqref{ap3}. The first term in Eq.~(\ref{eq:jzero}) vanishes due to the presence of the delta function $\delta(\omega)$. Thus, treating the second term as the leading-order contribution, we obtain
\begin{equation}
\vert J^0(\omega) \vert^2= \frac{1}{3} a^2 \mu^2 \Omega_{\mathrm{orb}}^2 \left(1- \frac{M_{Z^\prime}^2}{n^2\Omega^2}\right) \left(\frac{Q_1}{M_1}-\frac{Q_2}{M_2}\right)^2\left(J'^2_n(ne)+ \frac{1-e^2}{e^2} J^2_n(ne) \right),
\label{eq:j0_omega}
\end{equation}
where we use $\langle k_x^2 \rangle = \langle k_y^2 \rangle = k^2/3$ and $\omega = n\Omega_{\mathrm{orb}}$, and inserting Eqs.~\eqref{eq:ji_omega} and~\eqref{eq:j0_omega} into Eq.~\eqref{eq:first}, we find the energy loss rate as~\cite{KumarPoddar:2019ceq}
\begin{eqnarray}
\frac{dE_V}{dt}&=&\frac{g^2}{3\pi}a^2\mu^2\Big(\frac{Q_1}{M_1}-\frac{Q_2}{M_2}\Big)^2\left[\frac{\Omega_{\mathrm{orb}}^6}{M^2_{Z^\prime}}\sum_{n>n_0}n^4\Big[J^{\prime^2}_n(ne)+\frac{(1-e^2)}{e^2}J^2_n(ne)\Big]\Big(1-\frac{n^2_0}{n^2}\Big)^\frac{3}{2}\right. \nonumber\\
&&\left.+
\Omega_{\mathrm{orb}}^4\sum_{n>n_0}n^2\Big[J^{\prime^2}_n(ne)+\frac{(1-e^2)}{e^2}J^2_n(ne)\Big]\Big(1-\frac{n^2_0}{n^2}\Big)^\frac{1}{2}\Big(1+\frac{1}{2}\frac{n^2_0}{n^2}\Big)\right],
\label{eq:vecto}
\end{eqnarray}
with $n_0=M_{Z^\prime}/\Omega_{\mathrm{orb}}<1$.

We express Eq.~\eqref{eq:vecto} in a more simplified form as
\begin{equation}
\frac{dE_V}{dt}=\frac{g^2}{3\pi}a^2\mu^2\Big(\frac{Q_1}{M_1}-\frac{Q_2}{M_2}\Big)^2\Omega_{\mathrm{orb}}^4\Big(\frac{\Omega_{\mathrm{orb}}^2}{M^2_{Z^\prime}}K_1(n_0,e)+K_2(n_0,e)\Big),
\label{eq:vectory}
\end{equation}
where
\begin{eqnarray}
K_1(n_0,e)=\sum_{n>n_0}n^4\Big[J^{\prime^2}_n(ne)+\frac{(1-e^2)}{e^2}J^2_n(ne)\Big]\Big(1-\frac{n^2_0}{n^2}\Big)^\frac{3}{2},
\label{eq:k1}
\end{eqnarray}
and
\begin{eqnarray}
K_2(n_0,e)=\sum_{n>n_0}n^2\Big[J^{\prime^2}_n(ne)+\frac{(1-e^2)}{e^2}J^2_n(ne)\Big]\Big(1-\frac{n^2_0}{n^2}\Big)^\frac{1}{2}\Big(1+\frac{1}{2}\frac{n^2_0}{n^2}\Big).
\label{eq:k2}
\end{eqnarray}
Equation.~\eqref{eq:vectory} denotes the rate of energy loss due to the Proca massive vector boson radiation from compact binary systems.  

If the $Z^\prime$ boson is a gauge field such as in $L_\mu-L_\tau$ scenario, then gauge invariance implies $k_\mu J^\mu = 0$. Therefore, the second term in the polarization sum in Eq.~\eqref{eq:pol_sum} does not contribute to the energy loss. Following the same procedure, we obtain the rate of energy loss due to gauge boson radiation as 
\begin{equation}
\frac{dE_V}{dt}=\frac{g^2}{6\pi}a^2 \mu^2\Big(\frac{Q_1}{M_1}-\frac{Q_2}{M_2}\Big)^2 \Omega_{\mathrm{orb}}^4\sum_{n>n_0}2n^2\Big[J^{\prime^2}_n(ne)+\frac{(1-e^2)}{e^2}J^2_n(ne)\Big]\Big(1-\frac{n^2_0}{n^2}\Big)^\frac{1}{2}\Big(1+\frac{1}{2}\frac{n^2_0}{n^2}\Big),
\label{eq:vect}
\end{equation}
where for massless gauge boson radiation $(n_0=0)$, Eq.~\eqref{eq:vect} reduces to~\cite{KumarPoddar:2019ceq}
\begin{equation}
\frac{dE_V}{dt}=\frac{g^2}{6\pi}a^2 \mu^2\Big(\frac{Q_1}{M_1}-\frac{Q_2}{M_2}\Big)^2 \Omega_{\mathrm{orb}}^4\frac{(1+\frac{e^2}{2})}{(1-e^2)^\frac{5}{2}}.
\label{vectf}
\end{equation}
The above analysis applies to any ultralight massive spin-$1$ vector bosons emitted from binary systems, provided the stars contain a sufficiently large population of fermions to which the vector bosons can couple. Precision measurements of the orbital period loss from the binary system bounds the $L_\mu-L_\tau$ gauge coupling as $g\lesssim 10^{-20}$ for $M_{Z^\prime}\lesssim 10^{-19}\,\mathrm{eV}$~\cite{KumarPoddar:2019ceq}. Comparing Eqs.~\eqref{ap10} and~\eqref{vectf}, we obtain d$E_V/{\rm d}t=2 {\rm d}E_\phi/{\rm d}t$, which is due to the fact that a massless vector has two polarization modes and a massless scalar has one mode. Like scalar, the ultralight vector can also contribute to the orbital period loss of binary systems along with GW as given in Eq.~\eqref{op}.

Like scalar and vector fields, axions can also be radiated from compact binary systems if they couple to nucleons or other fermions. However, since the axion-fermion interaction is primarily spin-dependent, the corresponding axion radiation is usually suppressed. In contrast, if the underlying theory contains CP (charge conjugation-parity) violation, or if finite-density effects modify the axion potential such as when the standard QCD (quantum chromodynamics) sector interacts with a mirror QCD sector through axion exchange, the axion can acquire an effective scalar coupling. In such scenarios, axion emission can become significant and lead to a measurable enhancement in the orbital period loss of the binary system~\cite{Hook:2017psm,KumarPoddar:2019jxe}. In addition to the radiation of ultralight bosons, ultralight fermions such as neutrinos can also be emitted from compact binaries, thereby contributing to the overall orbital period decay of the system~\cite{Gavrilova:2023lxu}.

\subsection{Radiation of massless gravitons from quasi-stable compact binary systems}
\label{spin-2-massless}

In this section, we derive the GW quadrupole radiation formula from a field-theoretic perspective. This result was originally obtained using a multipole expansion approach in~\cite{Peters:1963ux}. The action for the graviton field $h_{\mu\nu}$ is derived by expanding the Einstein-Hilbert action, which describes gravity and its interaction with matter fields as
\begin{equation}
    S_{EH}= \int {\rm d}^4 x \sqrt{-g}\left[-\frac{1}{16 \pi G} R + {\cal L}_m \right].    
\end{equation}
The metric is expanded as $g_{\mu \nu} = \eta_{\mu \nu} + \kappa h_{\mu \nu}$, where $\kappa = \sqrt{32\pi G}$ denotes the gravitational coupling, and the expansion is carried out to linear order in the perturbation $h_{\mu \nu}$. For consistency, the inverse metric $g^{\mu \nu}$ and the square root of the determinant $\sqrt{-g}$ must be expanded up to quadratic order in $h_{\mu \nu}$. Therefore, 
\begin{eqnarray}
g_{\mu \nu}&=& \eta_{\mu \nu} + \kappa h_{\mu \nu}\,,\nonumber\\
g^{\mu \nu}&=& \eta^{\mu \nu} - \kappa h^{\mu \nu} + \kappa^2 h^{\mu \lambda} h^\nu_\lambda +\mathcal{O}(\kappa^3)\,,\nonumber\\
\sqrt{-g}&=& 1+ \frac{\kappa}{2} h +  \frac{\kappa^2}{8} h^2 -  \frac{\kappa^2}{4} h^{\mu \nu}h_{\mu \nu} + \mathcal{O}(\kappa^3),
\end{eqnarray}
where $\eta_{\mu \nu} = \text{diag}(1, -1, -1, -1)$ is the background Minkowski metric, and $h = h^\mu_{\ \mu}$ denotes the trace of the perturbation. Throughout, indices are raised and lowered using the Minkowski metric $\eta^{\mu \nu}$ and $\eta_{\mu \nu}$, respectively. 

The linearized Einstein-Hilbert action describing the dynamics of the graviton field $h_{\mu \nu}$ is then expressed as
\begin{eqnarray}
S_{EH}&=& \int {\rm d}^4 x \left[ -\frac{1}{2} (\partial_\mu h_{\nu \rho})^2 + \frac{1}{2} (\partial_\mu h)^2 - (\partial_\mu h)(\partial^\nu h^\mu_\nu)+(\partial_\mu h_{\nu \rho} )(\partial^\nu h^{\mu \rho})+ \frac{\kappa}{2} h_{\mu \nu}T^{\mu \nu} \right]\nonumber\\
&\equiv& \int {\rm d}^4 x \left[ \frac{1}{2} h_{\mu \nu} {\cal E}^{\mu \nu \alpha \beta}h_{\alpha \beta} + \frac{\kappa}{2} h_{\mu \nu }T^{\mu \nu} \right],
\label{EH}    
\end{eqnarray}
where ${\cal E}^{\mu \nu \alpha \beta}$ denotes the kinetic operator, given as
\begin{eqnarray}
{\cal E}^{\mu \nu \alpha \beta}= \left({\eta^{\mu(\alpha}} {\eta^{\beta)\nu} }-\eta^{\mu \nu}\eta^{\alpha \beta} \right) \Box - \eta^{\mu (\alpha }\partial^{ \beta) } \partial^\nu - \eta^{\nu (\alpha }\partial^{ \beta) } \partial^\mu+ \eta^{\alpha \beta} \partial^\mu \partial^\nu+ \eta^{\mu \nu} \partial^\alpha \partial^\beta ,
\label{kineticOp}    
\end{eqnarray}
where indices enclosed in parentheses indicate symmetrization, i.e., $A^{(\mu} B^{\nu)} = \frac{1}{2} (A^\mu B^\nu + A^\nu B^\mu)$. The propagator for a massless graviton, denoted $D^{(0)}_{\mu \nu \alpha \beta}$, is defined as the inverse of the kinetic operator $\mathcal{E}^{\mu \nu \alpha \beta}$
\begin{equation}
    \label{prop1}    
    {\cal E}^{\mu \nu \alpha \beta}D^{(0)}_{\alpha \beta \rho \sigma}(x-y) = \delta^\mu_{(\rho} \delta^\nu_{\sigma)} \delta^4(x-y)\,.
\end{equation}
The massless graviton action in Eq.~\eqref{EH} exhibits a gauge symmetry under the transformation $h_{\mu \nu} \rightarrow h_{\mu \nu} - \partial_\mu \xi_\nu - \partial_\nu \xi_\mu$. As a result of this gauge freedom, the kinetic operator $\mathcal{E}^{\mu \nu \alpha \beta}$ is not directly invertible, and the propagator cannot be obtained from Eq.~\eqref{prop1} without fixing a gauge. To facilitate the inversion of the kinetic operator, a specific gauge must be chosen. The de Donder gauge is particularly convenient, as it yields the simplest form of the graviton propagator. In this gauge
\begin{equation}
 \partial^\mu h_{\mu \nu} - \frac{1}{2} \partial_\nu h=0,    
\end{equation}
where, $h = {h^\alpha}_{\alpha}$ denotes the trace of the graviton field. This gauge condition can be implemented by adding the following gauge-fixing term to the Lagrangian in Eq.~\eqref{EH} as
\begin{equation}
    S_{gf} = -\int {\rm d}^4x \left( \partial^\mu h_{\mu \nu} - \frac{1}{2} \partial_\nu h\right)^2\,.
\end{equation}
The complete action, including the gauge-fixing term, takes the following form
\begin{eqnarray}
    S_{EH} + S_{gf} &=& \int {\rm d}^4 x  \left(\frac{1}{2} h_{\mu \nu} \Box h^{\mu \nu} -\frac{1}{4} h \Box h + \frac{\kappa}{2} h_{\mu \nu} T^{\mu \nu} \right) \nonumber\\
    &=& \int {\rm d}^4 x  \left( \frac{1}{2} h_{\mu \nu} {\cal K}^{\mu \nu \alpha \beta} h_{\alpha \beta}  + \frac{\kappa}{2} h_{\mu \nu} T^{\mu \nu} \right)\,,
    \label{Smassless}   
\end{eqnarray}
where $ {\cal K}^{\mu \nu \alpha \beta}$ denotes the kinetic operator in the de Donder gauge, which is given by
\begin{equation}
{\cal K}^{\mu \nu \alpha \beta}= \left(\frac{1}{2} (\eta^{\mu \alpha} \eta^{\nu \beta} +\eta^{\mu \beta}\eta^{\nu \alpha} ) - \frac{1}{2} \eta^{\mu \nu} \eta^{\alpha \beta} \right) \Box\,. 
 \label{Kop}     
\end{equation}
The graviton propagator in the de Donder gauge is obtained by inverting the kinetic operator given in Eq.~\eqref{Kop}, and is expressed as
\begin{equation}
    \label{eq:Prop1}
    {\cal K}^{\mu \nu \alpha \beta}D^{(0)}_{\alpha \beta \rho \sigma}(x-y)=  \delta^\mu_{(\rho} \delta^\nu_{\sigma)}\delta^4(x-y)\,.
\end{equation}
Using this relation, the propagator $D^{(0)}_{\alpha \beta \rho \sigma}$ can be explicitly computed in momentum space, where $\partial_\mu \rightarrow i k_\mu$, and is given by
\begin{equation}
    \label{eq:prop1D}
    D^{(0)}_{\mu \nu \alpha \beta}(k)=  \frac{1}{-k^2} \left( \frac{1}{2}(\eta_{\mu\alpha}\eta_{\nu\beta}+\eta_{\mu\beta}\eta_{\nu\alpha})- \frac{1}{2}\eta_{\mu\nu}\eta_{\alpha\beta} \right)\,.
\end{equation}
The graviton is treated as a quantum field by expressing it as a mode expansion involving creation and annihilation operators
\begin{equation}
\hat {h}_{\mu \nu}(x)= \sum_\lambda \int \frac{{\rm d}^3 k}{(2 \pi)^3} \frac{1}{\sqrt{2 \omega_k}} \left[ \epsilon^\lambda_{\mu \nu}(k) a_\lambda(k) e^{-i k \cdot x} +  \epsilon^{*\lambda}_{\mu \nu}(k) a^\dagger_\lambda(k) e^{i k \cdot x}  \right]\,.
\label{hhat}    
\end{equation}
The polarization tensors $\epsilon^\lambda_{\mu \nu}(k)$ satisfy the following orthogonality condition
\begin{equation}
\epsilon^\lambda_{\mu \nu}(k) 
\epsilon^{*\lambda^\prime \,\mu \nu}(k)= \delta_{\lambda \lambda^\prime},    
\end{equation}
where, $a_\lambda(k)$ and $a^\dagger_\lambda(k)$ are the annihilation and creation operators associated with the graviton modes, obeying the standard canonical commutation relations
\begin{equation}
\left[a_\lambda(k), a^\dagger_{\lambda^\prime}(k^\prime)\right]= \delta^4(k-k^\prime) \delta_{\lambda \lambda^\prime}\,.    
\end{equation}
The Feynman propagator for gravitons is defined as the time-ordered two-point correlation function
\begin{equation}
D^{(0)}_{\mu \nu \alpha \beta}(x-y) \equiv \langle 0 | T (\hat {h}_{\mu \nu}(x) \hat {h}_{\alpha \beta}(y))|0\rangle. 
\end{equation}
By applying Eq.~\eqref{hhat}, the Feynman propagator can be computed as
\begin{equation}
D^{(0)}_{\mu \nu \alpha \beta}(x-y)=\int \frac{{\rm d}^4 k}{(2 \pi)^4}\frac{1}{-k^2 +i \epsilon} e^{i k(x-y)} 
\sum_{\lambda} \epsilon_{\mu\nu}^\lambda(k)\epsilon_{ \alpha\beta}^{*\lambda}(k).
\label{prop2}    
\end{equation}
By comparing Eq.~\eqref{eq:prop1D} and Eq.~\eqref{prop2}, we obtain the expression for the polarization sum of massless spin-2 gravitons
\begin{equation}
\sum_{\lambda=1}^2 \epsilon_{\mu\nu}^\lambda(k)\epsilon_{ \alpha\beta}^{*\lambda}(k)= \frac{1}{2}(\eta_{\mu\alpha}\eta_{\nu\beta}+\eta_{\mu\beta}\eta_{\nu\alpha})- \frac{1}{2}\eta_{\mu\nu}\eta_{\alpha\beta}  .
\label{app3}
\end{equation}
This result is used in calculating the radiation of massless gravitons emitted by classical sources.

We now compute the energy loss due to the emission of massless gravitons from compact binary systems~\cite{Mohanty:1994yi}. In this analysis, the energy-momentum tensor $T_{\mu \nu}$ of the binary stars is treated as a classical source, while the gravitons are quantized fields. From the interaction Lagrangian in Eq.~\eqref{Smassless}, we identify the interaction vertex as $\frac{1}{2} \kappa h^{\mu \nu} T_{\mu \nu}$. Using this, the emission rate of massless gravitons with polarization tensor $\epsilon_\lambda^{\mu \nu}(k')$ from the classical source $T_{\mu \nu}(k)$ can be expressed as
\begin{equation}
    {\rm d}\Gamma= \frac{\kappa^2}{4}\sum_{\lambda=1}^2|T_{\mu\nu}(k^\prime)\epsilon^{\mu\nu}_\lambda (k)|^2 2\pi \delta(\omega-\omega^\prime)\frac{{\rm d}^3k}{(2\pi)^3}\frac{1}{2\omega}.
\label{app1}
\end{equation}   
Expanding the squared amplitude in Eq.~\eqref{app1}, we can write it as
\begin{equation}
    {\rm d}\Gamma = \frac{\kappa^2}{8(2\pi)^2}\sum_{\lambda=1}^2\Big(T_{\mu\nu}(k^\prime)T^*_{\alpha\beta}(k^\prime)\epsilon^{\mu\nu}_\lambda(k)\epsilon^{*\alpha\beta}_\lambda(k)\Big)\frac{{\rm d}^3k}{\omega}\delta(\omega-\omega^\prime)\,.
    \label{app2}
\end{equation}
Using the polarization sum of massless spin-2 gravitons given in Eq.~\eqref{app3}, the expression for the emission rate becomes
\begin{eqnarray}
    {\rm d}\Gamma&=&\frac{\kappa^2}{8(2\pi)^2}\int \Big[T_{\mu\nu}(k^\prime)T^*_{\alpha\beta}(k^\prime)\Big]\Big[\frac{1}{2}(\eta^{\mu\alpha}\eta^{\nu\beta}+\eta^{\mu\beta}\eta^{\nu\alpha}-\eta^{\mu\nu}\eta^{\alpha\beta})\Big]\frac{{\rm d}^3k}{\omega}\delta(\omega-\omega^\prime)\,\nonumber\\
    &=&\frac{\kappa^2}{8(2\pi)^2}\int \Big[|T_{\mu\nu}(k^\prime)|^2-\frac{1}{2}|T^{\mu}{}_{\mu}(k^\prime)|^2\Big]\delta(\omega-\omega^\prime)\,\omega \,{\rm d}\omega\, {\rm d}\Omega_k\,.
    \label{app4}
\end{eqnarray}
Using the relation ${\rm d}^3k = k^2\, {\rm d}k\, {\rm d}\Omega$, the energy loss rate due to massless graviton radiation can be expressed as
\begin{equation}
    \frac{dE_g}{dt}=\frac{\kappa^2}{8(2\pi)^2}\int \Big[|T_{\mu\nu}(k^\prime)|^2-\frac{1}{2}|T^{\mu}{}_{\mu}(k^\prime)|^2\Big]\delta(\omega-\omega^\prime)\, \omega^2\,{\rm d}\omega\, {\rm d}\Omega_k.
\label{app5}
\end{equation}
Applying the current conservation condition $k_\mu T^{\mu\nu} = 0$, the components $T^{00}$ and $T^{i0}$ of the stress-energy tensor can be expressed in terms of the spatial components $T^{ij}$ as
\begin{equation}
T_{0j}=-\hat{k^i}T_{ij},\hspace{0.5cm} T_{00}=\hat{k^i}\hat{k^j}T_{ij}.
\label{app6}
\end{equation}
We also write
\begin{equation}
\Big[|T_{\mu\nu}(k^\prime)|^2-\frac{1}{2}|T^{\mu}{}_{\mu}(k^\prime)|^2\Big]={\Lambda^0_{ij,lm}}T^{ij*}T^{lm},
\label{app7}
\end{equation}
where 
\begin{equation}
{\Lambda^0_{ij,lm}}=\Big[\delta_{il}\delta_{jm}-2\hat{k_j}\hat{k_m}\delta_{il}+\frac{1}{2}\hat{k_i}\hat{k_j}\hat{k_l}\hat{k_m}-\frac{1}{2}\delta_{ij}\delta_{lm}+\frac{1}{2}\Big(\delta_{ij}\hat{k_l}\hat{k_m}+\delta_{lm}\hat{k_i}\hat{k_j}\Big)\Big].
\label{app8}
\end{equation}
We perform the angular integrals
\begin{equation}
\int {\rm d}\Omega_k \Lambda^0_{ij,lm}T^{ij*}(\omega^\prime)T^{lm}({\omega^\prime})=\frac{8\pi}{5}\Big(T_{ij}(\omega^\prime)T^*_{ji}(\omega^\prime)-\frac{1}{3}|T^{i}{}_{i}(\omega^\prime)|^2\Big),
\label{app9}
\end{equation}
where we use the relations
\begin{equation}
\int {\rm d}\Omega \hat{k^i}\hat{k^j}=\frac{4\pi}{3}\delta_{ij}, \hspace{0.5cm} \int {\rm d}\Omega \hat{k^i}\hat{k^j}\hat{k^l}\hat{k^m}=\frac{4\pi}{15}(\delta_{ij}\delta_{lm}+\delta_{il}\delta_{jm}+\delta_{im}\delta_{jl}).
\label{dOmegak}
\end{equation}
The stress-energy tensor for the compact binary system is given as
\begin{equation}
T_{\mu\nu}(x^\prime)=\mu \delta^3(\textbf{x}^\prime-\textbf{x}(t))U_\mu U_\nu,
\label{eq:7}
\end{equation}
where, $\mu$ denotes the reduced mass of the binary system. The four-velocity of the reduced mass in the non-relativistic limit is given by $U_\mu = (1, \dot{x}, \dot{y}, 0)$, assuming motion confined to the $x\text{-}y$ plane in a Keplerian orbit.

The stress-energy tensor described above pertains solely to the matter contribution. In general, the total stress-energy tensor includes both the matter and GW components and is written as
\begin{equation}
T_{\mu \nu} = T^{\text{matter}}_{\mu \nu} + T^{\text{GW}}_{\mu \nu},
\end{equation}
where $T^{\text{matter}}_{\mu \nu}$ represents the conventional stress-energy tensor for matter fields, and $T^{\text{GW}}_{\mu \nu}$ accounts for the energy-momentum carried by GWs. The explicit form of $T^{\text{GW}}_{\mu \nu}$ is given by
\begin{equation}
T_{\mu\nu}^{GW}= \langle h_{\alpha\beta,\mu}h^{\alpha\beta}{}_{,\nu} -\frac{1}{2}h_{,\mu}h_{,\nu} \rangle.
\end{equation}
Using the equation of motion for $h_{\alpha\beta}$ at tree level, we can write the following expression
\begin{equation}
h_{\alpha\beta}\sim \frac{1}{M_{\rm Pl}}(\Box -m^2_g)^{-1}T^{\text{matter}}_{\alpha\beta} .
\end{equation}
Hence,
\begin{equation}
T_{\mu\nu}^{GW}(k_{\alpha})\sim \frac{1}{M_{\rm Pl}^2}\left((T^{\text{matter}}_{\alpha\beta})^2-\frac{(T^{\text{matter}})^2}{2}\right)\left(\frac{k_{\mu}k_{\nu}}{(k^{\alpha}k_{\alpha}-m^2_g)^2}\right) .
\end{equation}
In the radiation zone, i.e., far from the source, the GW contribution $T^{\text{GW}}_{\mu\nu}$ is suppressed by a factor of $1/M_{\text{Pl}}^2$ relative to the matter component $T^{\text{matter}}_{\mu\nu}$. Consequently, for gravitational radiation from compact binary systems, it is a good approximation to take $T_{\mu\nu} \simeq T^{\text{matter}}_{\mu\nu}$.

We now proceed to compute the Fourier transforms of the various components of the stress-energy tensor, evaluated at frequencies $\omega' = n\Omega_\mathrm{orb}$, as detailed below. Therefore,
\begin{eqnarray}
    T_{ij}(\textbf{k}^\prime,\omega^\prime)&=&\frac{1}{T}\int_0^T\int T_{ij}(\textbf{x},t) e^{-i(\textbf{k}^\prime\cdot \textbf{x}-\omega^\prime t)}{\rm d}^3x {\rm d}t\nonumber\\
    &=& \int T_{ij}(\textbf{x},\omega^{\prime}) e^{-i\textbf{k}^\prime\cdot \textbf{x}}{\rm d}^3x\,.
    \label{eq:12} 
\end{eqnarray}
Expanding $e^{-i\mathbf{k}' \cdot \mathbf{x}} \approx 1 - i\mathbf{k}' \cdot \mathbf{x} + \cdots$ and keeping only the leading-order term under the assumption $\mathbf{k}' \cdot \mathbf{x} \sim \Omega_{\mathrm{orb}} a \ll 1$, appropriate for a binary orbit, Eq.~\eqref{eq:12} simplifies to
\begin{equation}
T_{ij}(\textbf{k}^\prime,\omega^\prime)\simeq T_{ij}(\omega^\prime)= \int T_{ij}(\textbf{x},\omega^{\prime}) {\rm d}^3x.
\end{equation}
Applying the conservation condition $\partial_\mu T^{\mu\nu}(\mathbf{x}, t) = 0$, we obtain
\begin{equation}
\partial^i\partial^j T_{ij}(\textbf{x},\omega^{\prime})= -\omega^{\prime}{}^{2}T_{00}(\textbf{x},\omega^{\prime}).
\label{eq:cons_Tmunu}
\end{equation}
Multiplying both sides of Eq.~\eqref{eq:cons_Tmunu} by $x_k x_l$ and integrating over all space, we obtain
\begin{eqnarray}
T_{kl}(\omega^{\prime})&=& -\frac{\omega^{\prime}{}^2}{2}\int T_{00}(\textbf{x},\omega^{\prime})x_kx_l {\rm d}^3x \label{eq:step1}\\
&=& -\frac{\mu\omega^{\prime}{}^2}{2T}\int_0^T\int \delta^3(\textbf{x}^\prime-\textbf{x}(t))e^{i\omega^{\prime}t}x_kx_l {\rm d}^3x {\rm d}t \label{eq:step2}\\
&=& -\frac{\mu\omega^{\prime}{}^2}{2T} \int_0^T x^{\prime}_k(t)x^{\prime}_l(t)e^{i\omega^{\prime}t}{\rm d}t\label{eq:ak2},
\end{eqnarray}
where in Eq.~\eqref{eq:step2}, we have made use of Eq.~\eqref{eq:7}. By performing integration by parts on Eq.~\eqref{eq:ak2} and applying standard Bessel function identities\footnote{$J_{n-1}(z) - J_{n+1}(z) = 2J_n'(z), \hspace{0.5cm} J_{n-1}(z) + J_{n+1}(z) = \frac{2n}{z}J_n(z)$}, we can express the various components of the stress-energy tensor in the $x\text{-}y$ plane. The $xx$-component of the stress tensor in Fourier space is given by
\begin{eqnarray}
T_{xx}(\omega^{\prime})&=&-\frac{\mu\omega^{\prime}{}^2}{2T}\int_0^Tx^2(t)e^{i\omega^{\prime}t}dt \nonumber\\
&=& -\frac{\mu\omega^{\prime}{}^2}{4\pi}\int_0^{2\pi} a^2(\cos \xi -e)^2e^{in\beta}{\rm d}\beta,
\label{eq:ak2_xx}    
\end{eqnarray}
where, we have used Eq.~\eqref{ap2} along with the relations $\omega' = n\Omega_{\mathrm{orb}}$, $\beta = \Omega_{\mathrm{orb}} t$, and $T = 2\pi/\Omega_{\mathrm{orb}}$. Performing integration by parts on Eq.~\eqref{eq:ak2_xx}, we obtain
\begin{eqnarray}
T_{xx}(\omega^{\prime})&=& \frac{\mu\omega^{\prime}{}^2}{4\pi i n}\int^{2\pi}_0 e^{i n\beta} \frac{{\rm d}}{{\rm d}\beta} (\cos \xi -e)^2\, {\rm d}\beta \nonumber\\
&=& -\frac{\mu\omega^{\prime}{}^2}{4 n}\left[J_{n-2}(ne)-2eJ_{n-1}(ne)+2eJ_{n+1}(ne)-J_{n+2}(ne)\right],
\label{eq:Txx}
\end{eqnarray}
where in the final step, we have used the definition of the Bessel function along with the relation $\beta = \Omega t = \xi - e \sin \xi$.

Similarly, the $yy$ component of the stress tensor is 
\begin{eqnarray}
    T_{yy}(\omega^{\prime})&=&-\frac{\mu\omega^{\prime}{}^2}{2T}\int_0^Ty^2(t)e^{i\omega^{\prime}t}{\rm d}t = -\frac{\mu\omega^{\prime}{}^2 (1-e^2)}{4\pi}\int_0^{2\pi} a^2\sin^2\xi \, e^{in\beta}{\rm d}\beta.
\label{eq:ak2_yy}    
\end{eqnarray}
By summing Eqs.~\eqref{eq:ak2_xx} and~\eqref{eq:ak2_yy}, we obtain
\begin{eqnarray}
    T_{yy}(\omega^{\prime})+ T_{xx}(\omega^{\prime})&=& -\frac{\mu\omega^{\prime}{}^2 a^2}{4\pi}\int^{2\pi}_0 (1-e\cos \xi)^2 e^{in\beta} {\rm d}\beta = \frac{\mu \omega^{\prime}{}^{2} a^2 }{n^2}J^2_n(ne)\,.
\end{eqnarray}
Hence,
\begin{eqnarray}
    T_{yy}(\omega^\prime)&=&-T_{xx}(\omega^{\prime})+\frac{\mu \omega^{\prime}{}^{2} a^2 }{n^2}J^2_n(ne)\nonumber\\
    &=& \frac{\mu\omega^{\prime2}a^2}{4n}\Big[J_{n-2}(ne)-2eJ_{n-1}(ne)+\frac{4}{n}J_n(ne)+2eJ_{n+1}(ne)-J_{n+2}(ne)\Big].
    \label{werw}
\end{eqnarray}
Lastly, the $xy$-component of the stress tensor in the Fourier space is 
\begin{eqnarray}
    T_{xy}(\omega^{\prime})&=&-\frac{\mu\omega^{\prime}{}^2}{2T}\int_0^Tx(t)y(t)e^{i\omega^{\prime}t}{\rm d}t  \nonumber \\
    & = & -i\frac{\mu\omega^{\prime}{}^2 a^2 \sqrt{1-e^2}}{4n}\left[J_{n+2}(ne)-2J_n(ne)+J_{n-2}(ne)\right].
    \label{eq:ak2_xy}
\end{eqnarray}
Therefore, using Eqs.~\eqref{eq:Txx},~\eqref{werw} and~\eqref{eq:ak2_xy} we obtain two useful results
\begin{eqnarray}
T_{ij}(\omega^{\prime})T^{ij*}(\omega^{\prime})&=&\frac{\mu^2\omega^{\prime}{}^4a^4}{8n^2}\Big\{[J_{n-2}(ne)-2eJ_{n-1}(ne)+2eJ_{n+1}(ne)+\frac{2}{n}J_n(ne)-J_{n+2}(ne)]^2 \nonumber\\
&& +(1-e^2)[J_{n-2}(ne)-2J_n(ne)+J_{n+2}(ne)]^2+\frac{4}{n^2}J^2_{n}(ne)\Big\}\nonumber\\
&=& 4\mu^2\omega'^4a^4\left(f(n,e)+\frac{J^2_n(ne)}{12n^4}\right),
\end{eqnarray}
where $f(n,e)$ is given as
\begin{equation}
\begin{split}
f(n,e)=\frac{1}{32n^2}\Big\{[J_{n-2}(ne)-2eJ_{n-1}(ne)+2eJ_{n+1}(ne)+\frac{2}{n}J_n(ne)-J_{n+2}(ne)]^2+\\
(1-e^2)[J_{n-2}(ne)-2J_n(ne)+J_{n+2}(ne)]^2+\frac{4}{3n^2}J^2_{n}(ne)\Big\},
\end{split}
\end{equation}
and
\begin{equation}
    \vert T^i{}_i\vert^2= \frac{\mu^2\omega^{\prime}{}^4a^4}{n^4}J^2_{n}(ne).
\end{equation}
Hence, using Eq.~\eqref{app5} we obtain the rate of energy loss due to massless graviton radiation as
\begin{eqnarray}
    \frac{d E_g}{dt}&=&\frac{\kappa^2}{8(2\pi)^2}\int\frac{8\pi}{5}\Big[T_{ij}(\omega^\prime)T^*_{ji}(\omega^\prime)-\frac{1}{3}|T^{i}{}_{i}(\omega^\prime)|^2\Big]\delta(\omega-\omega^\prime)\omega^2 {\rm d}\omega,\nonumber\\
    &=& \frac{32G}{5}\sum^\infty_{n=1}(n\Omega_{\mathrm{orb}})^2\mu^2a^4(n\Omega_{\mathrm{orb}})^4f(n,e)\nonumber\\
    &=&\frac{32G}{5}\Omega_{\mathrm{orb}}^6\Big(\frac{M_1M_2}{M_1+M_2}\Big)^2a^4(1-e^2)^{-7/2}\Big(1+\frac{73}{24}e^2+\frac{37}{96}e^4\Big).
\label{eq:app10}
\end{eqnarray}
This expression corresponds to Einstein's quadrupole formula for gravitational radiation and agrees with the result obtained by Peters and Mathews~\cite{Peters:1963ux}. From the energy loss rate, one can determine the change in the orbital period, $P_b = 2\pi/\Omega_{\mathrm{orb}}$. According to Kepler's third law, $\Omega_{\mathrm{orb}}^2 a^3 = G(M_1 + M_2)$, which leads to the relation $\dot{a}/a = \frac{2}{3}(\dot{P}_b/P_b)$. The gravitational binding energy of the binary system is given by $E = -\frac{G M_1 M_2}{2a}$, implying $\dot{a}/a = -(\dot{E}/E)$. Combining these two expressions, we arrive at the relation
\begin{equation}
    \frac{\dot{P}_b}{P_b} = -\frac{3}{2} \frac{\dot{E}}{E}.
\end{equation}

\subsection{Radiation of gravitons in massive gravity theories from quasi-stable compact binary systems}

Like standard massless gravitons, massive gravitons in different massive gravity theories can also radiate from binary systems. Massive gravity is a theoretical modification of Einstein's GR, where the graviton, the hypothetical spin-$2$ particle that mediates gravity, is assumed to have a non-zero mass.

In standard GR, the gravitational interaction is governed by two fundamental constants, Newton's gravitational constant $G$, and the cosmological constant $\Lambda$, which accounts for the accelerated expansion of the universe and is often associated with dark energy~\cite{deRham:2014zqa}. A major unresolved issue in modern cosmology, known as the cosmological constant problem or vacuum catastrophe, arises from a severe mismatch between theory and observation. Quantum field theory predicts an enormous vacuum energy density due to the sum of zero-point energies at every point in spacetime. However, the observed value of $\Lambda$ is smaller by $60$ to $120$ orders of magnitude than the theoretical prediction-a discrepancy that remains one of the greatest puzzles in physics~\cite{Weinberg:1988cp}.

One of the key motivations for studying massive gravity theories, such as bigravity, is that they offer a natural resolution to the cosmological constant problem. These theories can account for the accelerated expansion of the universe without requiring dark energy. Additionally, massive gravity provides an alternative perspective on DM. Since DM is inferred solely through its gravitational effects and no particle candidate has yet been experimentally confirmed, modified gravity theories present a compelling framework to explain phenomena typically attributed to DM~\cite{Volkov:2011an,Lust:2021jps}.

The graviton, the hypothetical quantum of the gravitational field, can be understood as small perturbations $\kappa h_{\mu\nu}$ around a flat Minkowski spacetime background. In the regime where $\kappa h_{\mu\nu} \ll 1$, the linearized version of Einstein's GR becomes applicable, allowing gravitational interactions to be treated as perturbations at different length scales. This approximation holds true for distances $r > R_s$, where $R_s$ is the Schwarzschild radius of a massive object. For instance, the Sun has a Schwarzschild radius of approximately 3 km, so linearized gravity works well throughout the solar system.

At scales between the Planck length $R_{\text{pl}} \sim 10^{-35} \, \text{m}$ and $R_s$, classical non-linear gravitational effects dominate, particularly in the vicinity of BHs. However, quantum corrections are still negligible in this regime. When $r < R_{\text{pl}}$, quantum gravitational effects become significant, corresponding to the region near a gravitational singularity.

There is also a critical scale known as the Vainshtein radius $r_V$, beyond which linear perturbation theory remains valid~\cite{Babichev:2013usa}. The concept of massive gravity, introducing a nonzero mass for the graviton-emerged in the early 20th century. A foundational attempt was made by Fierz and Pauli~\cite{Fierz:1939ix}, who formulated a theory of a massive graviton in flat spacetime. However, their model encountered issues, most notably the presence of unphysical ghost states.

Another fundamental challenge is the van Dam-Veltman-Zakharov (vDVZ) discontinuity~\cite{vanDam:1970vg,Zakharov:1970cc}. Although the massive gravity theory formally reduces to GR in the limit of zero graviton mass at the action level, the same is not true at the level of the graviton propagator. This discrepancy arises because a massless graviton has two polarization states, while a massive one has five. The additional scalar degree of freedom in the massive theory leads to the vDVZ discontinuity.

Various attempts have been made to resolve these issues. For example, the Dvali-Gabadadze-Porrati (DGP) model~\cite{Dvali:2000rv} based on a five-dimensional framework and modifications of the Fierz-Pauli (FP) theory attempted to eliminate ghosts or cancel problematic scalar modes~\cite{Finn:2001qi}. However, these models often failed to resolve both the ghost problem and the vDVZ discontinuity simultaneously or relied on restrictive assumptions.

A major breakthrough came in 2010 with the work of Claudia de Rham, Gregory Gabadadze, and Andrew Tolley, who constructed a nonlinear, ghost-free theory of massive gravity in four dimensions, now known as the dRGT (de Rham-Gabadadze-Tolley) model. This theory successfully addressed the aforementioned issues~\cite{deRham:2010kj}. Additionally, other models like bimetric gravity have shown promise, particularly in offering solutions to the cosmological constant problem~\cite{Schmidt-May:2015vnx,Akrami:2015qga}.

An important prediction of massive gravity theories is that the graviton must propagate at a speed less than that of light. Observational data have placed stringent limits on the graviton mass. For example, the GW event GW170104 set an upper bound of $m_g < 7.7 \times 10^{-23} \, \text{eV}$ on the graviton mass~\cite{LIGOScientific:2017bnn}.

\subsubsection{Fierz-Pauli theory}

The action of the FP theory is given as 
\begin{eqnarray}
S&=&\int {\rm d}^4x\Big[ -\frac{1}{2} (\partial_\mu h_{\nu \rho})^2 + \frac{1}{2} (\partial_\mu h)^2 
- (\partial_\mu h)(\partial^\nu h^\mu_\nu)+(\partial_\mu h_{\nu \rho} )(\partial^\nu h^{\mu \rho})\nonumber\\
&&\quad \quad \quad\quad+\frac{1}{2}m^2_g\Big(h_{\mu\nu}h^{\mu\nu}-h^2\Big)+\frac{\kappa}{2}h_{\mu\nu}T^{\mu\nu} \Big] \nonumber\\
&=&\int {\rm d}^4x \left[ \frac{1}{2} h_{\mu \nu} {\cal E}^{\mu \nu \alpha \beta}h_{\alpha \beta} 
+\frac{1}{2}m^2_g h_{\mu \nu} (\eta^{\mu (\alpha } \eta^{\beta) \nu}- \eta^{\mu \nu}\eta^{\alpha \beta})h_{\alpha \beta}+ \frac{\kappa}{2} h_{\mu \nu }T^{\mu \nu} \right],
\label{eq:FP_action}
\end{eqnarray}
where ${\cal E}^{\mu \nu \alpha \beta}$ is the kinetic operator, given in Eq.~\eqref{kineticOp} and the mass term breaks the gauge symmetry $h_{\mu \nu} \rightarrow h_{\mu \nu} - \partial_\mu \xi_\nu -\partial_\nu \xi_\mu$. Also, the conservation of the energy-momentum implies $\partial_\mu T^{\mu \nu}=0$. Therefore, we can write the equation of motion of the graviton from Eq.~\eqref{eq:FP_action} as
\begin{eqnarray}
\left( \Box +m_g^2 \right) h_{\mu \nu} -\eta_{\mu \nu} \left(\Box +m_g^2\right) h - \partial_\mu \partial^\alpha h_{\alpha \nu} - \partial_\nu \partial^\alpha h_{\alpha \mu} + \eta_{\mu \nu} \partial^\alpha \partial^\beta h_{\alpha \beta}+ \partial_\mu \partial_\nu h = -\kappa T_{\mu \nu}. \nonumber\\
\label{eom1}
\end{eqnarray}
The divergence of Eq.~\eqref{eom1} yields
\begin{equation}
    m_g^2 \left( \partial^\mu h_{\mu \nu} -\partial_\nu h \right)=0, 
    \label{eom2}
\end{equation}
which yields four constraints equations for graviton and reduces its physical degrees of freedom from ten to six. 
Inserting Eq.~\eqref{eom2} into Eq.~\eqref{eom1} yields
\begin{equation}
    \Box h_{\mu \nu} -\partial_\mu \partial_\nu h +m_g^2\left(h_{\mu \nu} -\eta_{\mu \nu}h\right)= -\kappa T_{\mu \nu}.
\label{vc}
\end{equation}
The trace of Eq.~\eqref{vc} results 
\begin{equation}
    h = \frac{\kappa }{3 m_g^2} T.
\end{equation}

Consequently, the trace component $h$ is not a dynamical field but is instead fixed algebraically by the trace of the energy-momentum tensor. This mode is identified as a ghost, since its kinetic term in Eq.~\eqref{eom1} carries the wrong sign. However, in the FP theory, this ghost mode is rendered non-propagating. As a result, the theory contains five independent propagating degrees of freedom, two tensor modes, two vector modes (which do not couple to the energy-momentum tensor), and one scalar mode that couples to its trace.

The propagator in the FP theory is given as
\begin{equation}
\left[ {\cal E}^{\mu \nu \alpha \beta}+m_g^2 \left(\eta^{\mu (\alpha } \eta^{\beta) \nu}- \eta^{\mu \nu}\eta^{\alpha \beta} \right)\right] D^{(m)}_{\alpha \beta \rho \sigma}(x-y)=\delta^\mu_{(\rho}\delta^\nu_{\sigma)}\delta^4(x-y).
\label{Dm1}
\end{equation}
In the momentum space, Eq.~\eqref{Dm1} can be written as
\begin{equation}
D^{(m)}_{\alpha \beta \rho \sigma}(k)=  \frac{1}{-k^2 +m_g^2} \left (\frac{1}{2} (P_{\alpha \rho}  P_{\beta \sigma} + P_{\alpha \sigma }P_{\beta \rho}) - \frac{1}{3} P_{\alpha  \beta}P_{\rho \sigma}\right),
\label{Dm2}    
\end{equation}
where 
\begin{equation}
P_{\alpha \beta}\equiv\eta_{\alpha \beta}- \frac{k_\alpha k_\beta}{m_g^2}.
\end{equation}

At tree level, processes involving graviton exchange between conserved currents yield an amplitude of the form
\begin{equation}
{\cal A}_{FP} =\frac{\kappa^2}{4}  T^{\alpha \beta} D^{(m)}_{\alpha \beta \mu \nu}T^{\prime \mu \nu}.
\label{ampFP}
\end{equation}

Therefore, we can write Eq.~\eqref{Dm2} as
\begin{equation}
    D^{(m)}_{\mu \nu \alpha \beta }(k)=  \frac{1}{-k^2 +m_g^2} \left (\frac{1}{2} (\eta_{\alpha \mu}  \eta_{\beta \nu} + \eta_{\alpha \nu }\eta_{\beta \mu}) - \frac{1}{3} \eta_{\alpha  \beta}\eta_{\mu \nu}  + (k{\rm -dependent\, terms}) \right).
    \label{Dm3}
\end{equation}

In the quantum field treatment of the graviton, the Feynman propagator is defined analogously to the massless case, as presented in Eq.~\eqref{prop2},
\begin{eqnarray}
    D^{(m)}_{\mu \nu \alpha \beta}(x-y)&=& \langle 0 | T (\hat {h}_{\mu \nu}(x) \hat {h}_{\alpha \beta}(y))|0\rangle\nonumber\\
    &=&\int \frac{{\rm d}^4 k}{(2 \pi)^4}\frac{1}{-k^2 + m_g^2 + i \epsilon} e^{i k(x-y)} \sum_{\lambda} \epsilon_{\mu\nu}^\lambda(k)\epsilon_{ \alpha\beta}^{*\lambda}(k).
    \label{Dm4}
\end{eqnarray}
Comparing Eqs.~\eqref{Dm3} and~\eqref{Dm4} we obtain the polarization sum of the massive graviton in FP theory as
\begin{equation}
\sum_{\lambda} \epsilon_{\mu\nu}^\lambda(k)\epsilon_{ \alpha\beta}^{*\lambda}(k)= \frac{1}{2} (\eta_{\mu \alpha}  \eta_{\nu \beta} + \eta_{\nu \alpha }\eta_{\mu \beta }) - \frac{1}{3} \eta_{\alpha  \beta}\eta_{\mu \nu}  + (k{\rm-dependent\, terms}).
\label{polsumFP}
\end{equation}

The conservation of $T^{\mu \nu}$ ensures that the momentum-dependent parts of the polarization sum yield no contribution, allowing us to discard them from Eq.~\eqref{polsumFP} when computing diagrams with graviton emission from external lines.

When we compare the propagator in Eq.~\eqref{eq:prop1D} and the polarization sum in Eq.~\eqref{app3} in the massless graviton theory with their counterparts in the massive FP theory in Eq.~\eqref{Dm3} and Eq.~\eqref{polsumFP}, we find a notable difference that persists even in the limit $m_g \rightarrow 0$. Specifically, the amplitude in the FP theory Eq.~\eqref{ampFP} includes an additional term proportional to $(1/6) T^* T'$, which is absent in the massless case. This extra contribution arises from the scalar degree of freedom in $g_{\mu \nu}$, which fails to decouple as $m_g \rightarrow 0$.

Using Eq.~\eqref{polsumFP} we write 
\begin{equation}
{\cal A}_{FP} = \frac{\kappa^2}{4}\frac{1}{-k^2+m_g^2} \left(T_{\mu \nu}-\frac{1}{3} \eta_{\mu \nu} T^{\alpha}_{\alpha}\right) T^{\prime \mu \nu}.
\end{equation}
For massive bodies at rest in a given reference frame, the stress-energy tensors take the form $T^{\mu \nu} = (M_1, 0, 0, 0)$ and $T^{\prime \alpha \beta} = (M_2, 0, 0, 0)$, where $M_1$ and $M_2$ are the respective masses. Therefore, the gravitational potential between two massive bodies in the FP theory becomes
\begin{eqnarray}
V_{FP}= &= &\frac{\kappa^2}{4}\int\frac{{\rm d}^3k}{(2\pi)^3} e^{i k\cdot r}\frac{1}{-k^2+m_g^2} \left(T_{\mu \nu}-\frac{1}{3} \eta_{\mu \nu} T^{\alpha}_{\alpha}\right) T^{\prime \mu \nu}\nonumber\\
&=& \left( \frac{4}{3}\right) \frac{G M_1 M_2}{r}e^{-m_g r}.
\label{VFP}    
\end{eqnarray}
The FP theory yields a Yukawa-type gravitational potential. However, in the limit $m_g \to 0$, the potential between massive bodies remains larger by a factor of $4/3$ compared to the Newtonian potential of GR, which is ruled out by solar system tests~\cite{Talmadge:1988qz}. In contrast, light bending remains consistent with GR in this limit since the photon's stress-energy tensor $T^\mu_\nu=(\omega,0,0,-\omega)$ is traceless, and the scattering amplitudes in FP and GR match: $\mathcal{A}_{\text{FP}}(m_g \to 0) = \mathcal{A}_{\text{GR}}$~\cite{Fomalont:2009zg}. Together, these imply that the $4/3$ enhancement cannot be removed by redefining $G$, revealing a mismatch known as the vDVZ discontinuity between the FP action and its propagator in the $m_g \to 0$ limit~\cite{vanDam:1970vg, Zakharov:1970cc}.

Vainshtein~\cite{Vainshtein:1972sx, Babichev:2013usa} pointed out that the linearized FP theory breaks down well outside the Schwarzschild radius $R_s = 2GM$, below which even linearized GR becomes invalid due to strong gravitational fields ($\kappa h_{\mu \nu} \sim 1$). As the graviton mass $m_g$ decreases, the scalar mode in FP theory becomes increasingly strongly coupled. This limits the regime where the linear approximation holds, introducing a characteristic scale known as the Vainshtein radius, beyond which non-linear effects must be considered. This radius is given by
\begin{equation}
r_V = \left( \frac{R_s}{m_g^4} \right)^{1/5}.
\end{equation}

When gravitons are radiated from external legs, typical of GW emission from classical sources, the corresponding amplitude squared takes the following form
\begin{equation}
|{\cal M}|^2 = \left(\frac{\kappa^2}{4} \right )\sum_\lambda  | \epsilon_{\mu\nu}^\lambda(k) T^{\mu\nu}(k^\prime)|^2=  \left(\frac{\kappa^2}{4} \right ) \sum_\lambda \epsilon_{\mu\nu}^\lambda(k)\epsilon_{ \alpha\beta}^{*\lambda}(k) T^{\mu\nu}(k^\prime) T^{*\alpha \beta}(k^\prime).
\end{equation}
In the following, we calculate the massive graviton radiation from a compact binary system in FP theory.

\textbf{Graviton radiation in Fierz-Pauli theory:}
We analyze graviton radiation from compact binary systems within a classical framework. The classical energy-momentum tensor $T^{\mu\nu}$, which acts as the graviton source, is derived from the Keplerian orbital motion of the binary constituents. The interaction between the gravitational field and the source is described by the vertex
\begin{equation}
\frac{1}{2}\kappa\, h_{\mu\nu} T^{\mu\nu}.
\end{equation}

We work in the linearized gravity limit and extend it by including a nonzero graviton mass term as in Eq.~\eqref{eq:FP_action}, in order to compute the energy loss due to graviton emission from the binary system.

The graviton emission rate, arising from the above interaction Lagrangian, is given by
\begin{equation}
{\rm d}\Gamma= \frac{\kappa^2}{4}\sum_{\lambda} |T_{\mu\nu}(k^\prime)\epsilon^{\mu\nu}_\lambda(k)|^2 2\pi \delta(\omega-\omega^\prime)\frac{{\rm d}^3k}{(2\pi)^3}\frac{1}{2\omega},
\label{eq:1}
\end{equation}
where $T_{\mu\nu}(k')$ denotes the classical graviton current in momentum space. By expanding the squared modulus in Eq.~\eqref{eq:1}, we obtain
\begin{equation}
{\rm d}\Gamma=\frac{\kappa^2}{8(2\pi)^2}\sum_{\lambda}\Big(T_{\mu\nu}(k^\prime)T^*_{\alpha\beta}(k^\prime)\epsilon^{\mu\nu}_\lambda(k)\epsilon^{*\alpha\beta}(k)\Big)\frac{{\rm d}^3k}{\omega}\delta(\omega-\omega^\prime).
\label{eq:2}
\end{equation}
Applying the polarization sum from FP theory Eq.~\eqref{polsumFP}, the expression simplifies to
\begin{eqnarray}
    {\rm d}\Gamma&=&\frac{\kappa^2}{8(2\pi)^2}\int \Big[T_{\mu\nu}(k^\prime)T^*_{\alpha\beta}(k^\prime)\Big]\Big[\frac{1}{2}(\eta^{\mu\alpha}\eta^{\nu\beta}+\eta^{\mu\beta}\eta^{\nu\alpha}-\eta^{\mu\nu}\eta^{\alpha\beta})+\frac{1}{6}\eta^{\mu\nu}\eta^{\alpha\beta}\Big]\frac{{\rm d}^3k}{\omega}\delta(\omega-\omega^\prime)\,.\nonumber\\
\end{eqnarray}
The additional term $(1/6)\eta^{\mu\nu}\eta^{\alpha\beta}$, which is absent in the massless graviton case, arises due to the contribution of the scalar mode in FP theory. Upon simplification, we obtain
\begin{eqnarray}
    {\rm d}\Gamma&=&\frac{\kappa^2}{8(2\pi)^2}\int \Big[|T_{\mu\nu}(k^\prime)|^2-\frac{1}{3}|T^{\mu}{}_{\mu}(k^\prime)|^2\Big]\delta(\omega-\omega^\prime)\omega \Big(1-\frac{m^2_{g}}{\omega^2}\Big)^\frac{1}{2}{\rm d}\omega {\rm d}\Omega_k,
    \label{eq:4}
\end{eqnarray}
where we have used ${\rm d}^3k = k^2\,{\rm d}k\,{\rm d}\Omega$ along with the dispersion relation $k^2 = \omega^2 - m_g^2$. From the emission rate, we then compute the corresponding energy loss rate due to massive graviton emission, given by
\begin{equation}
\frac{{\rm d}E}{{\rm d}t}=\frac{\kappa^2}{8(2\pi)^2}\int \Big[|T_{\mu\nu}(k^\prime)|^2-\frac{1}{3}|T^{\mu}{}_{\mu}(k^\prime)|^2\Big]\delta(\omega-\omega^\prime)\omega^2 \Big(1-\frac{m^2_{g}}{\omega^2}\Big)^\frac{1}{2}{\rm d}\omega {\rm d}\Omega_k.
\label{eq:5}
\end{equation}
The dispersion relation for the massive graviton is
\begin{equation}
|\textbf{k}|^2=\omega^2\Big(1-\frac{m^2_g}{\omega^2}\Big).
\label{eq:a1}
\end{equation}
Therefore, the unit vector in the direction of the graviton's momentum is given by $\hat{k}^i = k^i/(\omega\sqrt{1 - \frac{m_g^2}{\omega^2}})$. Using the transversality condition $k_\mu T^{\mu\nu} = 0$ along with Eq.~\eqref{eq:a1}, the components $T_{00}$ and $T_{i0}$ of the stress-energy tensor can be expressed in terms of $T_{ij}$ as
\begin{equation}
T_{0j}=-\sqrt{1-\frac{m^2_g}{\omega^2}}\hat{k^i}T_{ij},\hspace{0.5cm} T_{00}=\Big(1-\frac{m^2_g}{\omega^2}\Big)\hat{k^i}\hat{k^j}T_{ij}.
\end{equation}
Therefore, we write
\begin{equation}
\Big[|T_{\mu\nu}(k^\prime)|^2-\frac{1}{3}|T^{\mu}{}_{\mu}(k^\prime)|^2\Big]\equiv {\Lambda_{ij,lm}}T^{ij*}T^{lm},
\end{equation}
where
\begin{equation}
{\Lambda_{ij,lm}}=\delta_{il}\delta_{jm}-2\Big(1-\frac{m^2_g}{\omega^2}\Big)\hat{k_j}\hat{k_m}\delta_{il}+\frac{2}{3}\Big(1-\frac{m^2_g}{\omega^2}\Big)^2\hat{k_i}\hat{k_j}\hat{k_l}\hat{k_m}-\frac{1}{3}\delta_{ij}\delta_{lm}+\frac{1}{3}\Big(1-\frac{m^2_g}{\omega^2}\Big)\Big(\delta_{ij}\hat{k_l}\hat{k_m}+\delta_{lm}\hat{k_i}\hat{k_j}\Big).
\label{rat1}
\end{equation}
Therefore, we write Eq.~\eqref{eq:5} as
\begin{equation}
\frac{{\rm d}E}{{\rm d}t}=\frac{\kappa^2}{8(2\pi)^2}\int {\Lambda_{ij,lm}}T^{ij*}T^{lm} \delta(\omega-\omega^\prime)\omega^2 \Big(1-\frac{m^2_{g}}{\omega^2}\Big)^\frac{1}{2}{\rm d}\omega {\rm d}\Omega_k.
\label{wer}
\end{equation}
Performing the angular integrals (see Eq.~\eqref{dOmegak})), 
one gets that Eq.~\eqref{wer} for the rate of energy loss becomes
\begin{eqnarray}
\frac{{\rm d}E}{{\rm d}t}&=& \frac{8G}{5}\int \left[\left\lbrace\frac{5}{2}-\frac{5}{3}\left(1-\frac{m_g^2}{\omega'^2}\right)+\frac{2}{9}\left(1-\frac{m_g^2}{\omega'^2}\right)^2\right\rbrace T^{ij}T^*_{ij}\right.\nonumber\\
&& \left. +\left\lbrace -\frac{5}{6}+\frac{5}{9}\left(1-\frac{m_g^2}{\omega'^2}\right)+\frac{1}{9}\left(1-\frac{m_g^2}{\omega'^2}\right)^2\right\rbrace\vert T^i{}_{i}\vert^2\right]\delta(\omega-\omega^\prime)\omega^2 \Big(1-\frac{m^2_{g}}{\omega^2}\Big)^\frac{1}{2}{\rm d}\omega. \nonumber\\
\label{eq:a4}
\end{eqnarray}
The prefactors of $T^{ij}T^*_{ij}$ and $\vert T^i{}_i\vert^2$ correspond to massless gravity theory Eq.~\eqref{eq:app10} are $1$ and $-1/3$ respectively. However, in the $m_g\rightarrow 0$ limit of Eq.~\eqref{eq:a4}, the prefactors are different which is due to the contribution of five polarization components of graviton instead of two as in the massless limit. Therefore, we will not obtain Eq.~\eqref{eq:app10} from Eq.~\eqref{eq:a4} in the $m_g\rightarrow 0$ limit. This is the consequence of the vDVZ discontinuity. 

Using Eqs.~\eqref{eq:Txx},~\eqref{werw} and~\eqref{eq:ak2_xy}, we obtain the rate of energy loss from Eq.~\eqref{eq:a4} as~\cite{Poddar:2021yjd}
\begin{equation}
\begin{split}
\frac{{\rm d}E}{{\rm d}t}= \frac{32G}{5} \mu^2 a^4\Omega^6_\mathrm{orb} \sum_{n=1}^{\infty} n^6\sqrt{1-\frac{n_0^2}{n^2}}\left[f(n,e)\left( \frac{19}{18}+\frac{11}{9}\frac{n_0^2}{n^2}+\frac{2}{9}\frac{n_0^4}{n^4}\right) + \frac{5J^2_n(ne)}{108n^4}\left(1-\frac{n_0^2}{n^2}\right)^2\right].
\end{split}
\label{eq:dedt_FP}
\end{equation}
We rewrite Eq.~\eqref{eq:dedt_FP} as
\begin{equation}
\begin{split}
\frac{{\rm d}E}{{\rm d}t}= \frac{32G}{5} \mu^2 a^4\Omega^6_\mathrm{orb} \sum_{n=1}^{\infty} n^6\sqrt{1-\frac{n_0^2}{n^2}}\left[f(n,e)\left( 1+\frac{4}{3}\frac{n_0^2}{n^2}+\frac{1}{6}\frac{n_0^4}{n^4}\right) -\frac{5J^2_n(ne)}{36n^4}\frac{n_0^2}{n^2}\left(1-\frac{n_0^2}{4n^2}\right)\right]+\\
\frac{32G}{5} \mu^2 a^4\Omega^6_\mathrm{orb} \sum_{n=1}^{\infty} n^6\sqrt{1-\frac{n_0^2}{n^2}}\Big[\frac{1}{18}f(n,e)\Big(1-\frac{n^2_0}{n^2}\Big)^2+\frac{5J^2_n(ne)}{108n^4}\Big(1+\frac{n^2_0}{2n^2}\Big)^2\Big],
\end{split}
\label{eq:wa}
\end{equation}
where the first term represents the energy loss predicted by the massive gravity theory in the absence of the vDVZ discontinuity, as given in Eq.~\eqref{eq:dedt_FPd}, while the second term captures the additional contribution from the scalar mode, associated with the $\frac{1}{6}\eta_{\mu\nu}\eta_{\alpha\beta}$ structure.

We write Eq.~\eqref{eq:wa} in the leading order of $n^2_0$ as 
\begin{equation}
\frac{{\rm d}E}{{\rm d}t}=\frac{32G}{5}\mu^2 a^4\Omega^6_\mathrm{orb}\Big[\sum_{n=1}^\infty\Big(\frac{19}{18}n^6f(n,e)+\frac{5}{108}n^2J^2_n(ne)\Big)+n^2_0\sum _{n=1}^\infty\Big(\frac{25}{36}n^4f(n,e)-\frac{25}{216}J^2_n(ne)\Big) \Big]+\mathcal{O}(n^4_0).
\end{equation}
In massive gravity theories, the gravitational potential deviates from that predicted by GR, leading to modifications in the Keplerian orbits of binary systems. Specifically, in FP massive gravity, the potential takes a Yukawa-like form with an additional prefactor of $4/3$ in the absence of screening effects. However, for GW emission to occur, the conditions $n_0 < 1$ or $m_g < \Omega_{\mathrm{orb}}$, and $a < R_V$ must be satisfied. Under these conditions, the Newtonian potential governing the orbital dynamics is subject to Vainshtein screening. As a result, the potential receives corrections from the screened scalar mode. 

\subsubsection{Massive graviton theory without vDVZ discontinuity}

In the FP theory Eq.~\eqref{eq:FP_action}, the absence of ghosts is ensured by the specific choice of the relative coefficient between the $h^2$ and $h_{\mu \nu} h^{\mu \nu}$ terms, set to $-1$. Deviating from this precise relation generally introduces ghost-like instabilities. However, there exists a particular modification in which the ghost contribution cancels the unwanted scalar mode in the propagator. This special tuning results in a theory that is both ghost-free and free from the vDVZ discontinuity~\cite{Gambuti:2020onb, Gambuti:2021meo}. From a phenomenological perspective, this leads to a clean generalization of the spin-2 graviton with only two physical polarizations, satisfying the dispersion relation
\begin{equation}
k_0^2 = |\vec{k}|^2 + m_g^2.
\end{equation}
The one parameter generalization of the FP theory is given by the action 
\begin{eqnarray}
    S=\int {\rm d}^4x \left[ \frac{1}{2} h_{\mu \nu} {\cal E}^{\mu \nu \alpha \beta}h_{\alpha \beta} +\frac{1}{2}m^2_g h_{\mu \nu} \left(\eta^{\mu (\alpha } \eta^{\beta) \nu}-(1-a) \eta^{\mu \nu}\eta^{\alpha \beta})\right) h_{\alpha \beta}+ \frac{\kappa}{2} h_{\mu \nu }T^{\mu \nu} \right]\,.
    \label{eq:FP_deformed}    
\end{eqnarray}
The case $a = 0$ corresponds to the original FP theory Eq.~\eqref{eq:FP_action}. In the following, we will derive the equations for a general $a \neq 0$ and investigate which values of $a$ can address the vDVZ discontinuity, a generic feature of massive gravity theories.

Therefore, the equation of motion from Eq.~\eqref{eq:FP_deformed} becomes
\begin{eqnarray}
\left( \Box +m_g^2 \right) h_{\mu \nu} -\eta_{\mu \nu} \left(\Box +m_g^2(1-a)\right) h - \partial_\mu \partial^\alpha h_{\alpha \nu} - \partial_\nu \partial^\alpha h_{\alpha \mu} + \eta_{\mu \nu} \partial^\alpha \partial^\beta h_{\alpha \beta}+ \partial_\mu \partial_\nu h = -\kappa  T_{\mu \nu}. \nonumber\\
\label{deom1}
\end{eqnarray}
The divergence of Eq.~\eqref{deom1} yields
\begin{equation}
    m_g^2 \left( \partial^\mu h_{\mu \nu} -(1-a) \partial_\nu h \right)=0.
\label{deom2}
\end{equation}
These four constraint equations reduce the number of independent components of the graviton field from $10$ to $6$.
Inserting Eq.~\eqref{deom2} into Eq.~\eqref{deom1}, one gets
\begin{equation}
(\Box +m_g^2) h_{\mu \nu}- a \eta_{\mu \nu} \Box h -(1-2 a)\partial_\mu\partial_\nu h - m_g^2 \eta_{\mu \nu}(1-a)h  = -\kappa T_{\mu \nu}. 
\label{ry}
\end{equation}
The trace of Eq.~\eqref{ry} gives
\begin{equation}
-2 a \Box h-(3m^2_g -4 m^2_g a)h=- \kappa T.   
\label{ry1}
\end{equation}
The trace equation \eqref{ry1} implies that $h$ is a propagating field if $a\neq 0$. Also, $h$ is a ghost field as its kinetic term appears with a negative sign. Therefore, the homogeneous equation of motion for $h$ becomes
\begin{equation}
\Box h-m_h^2 h =0,    
\end{equation}
where $m_h$ is the ghost mass given by
\begin{equation}
 m_h^2 = \frac{m^2_g}{2} \left( 1+3 \left(1- \frac{1}{a}\right) \right).
\label{mh}   
\end{equation}
The propagator of this theory is
\begin{equation}
\left[ {\cal E}^{\mu \nu \alpha \beta}+m_g^2 \left(\eta^{\mu (\alpha } \eta^{\beta) \nu}- \eta^{\mu \nu}\eta^{\alpha \beta} (1-a) \right)\right] D^{(a)}_{\alpha \beta \rho \sigma}(x-y)=\delta^\mu_{(\rho}\delta^\nu_{\sigma)}\delta^4(x-y).
\label{Dalpha1}    
\end{equation}
In momentum space, $D^{(a)}_{\alpha \beta \rho \sigma}$ becomes
\begin{eqnarray}
D^{(a)}_{\alpha \beta \mu \nu}(k) &=&  \frac{1}{-k^2 +m_g^2} \left (\frac{1}{2} (\eta_{\alpha \mu}  \eta_{\beta \nu} + \eta_{\alpha \nu }\eta_{\beta \mu}) - \frac{1}{3} \eta_{\alpha  \beta}\eta_{\mu \nu} \right) + \frac{i}{k^2 +m_h^2} \left( \frac{1}{6} \eta_{\alpha  \beta}\eta_{\mu \nu} \right ) \nonumber\\ &+& (k{\rm -dependent\, terms}). 
\label{Dalpha2}    
\end{eqnarray}
This analysis reveals that the propagator consists of two main contributions-- one from the helicity-2 modes of the massive spin-2 graviton (along with helicity-1 and helicity-0 components), and another from a massive scalar field with mass $m_h$. This part of the propagator matches that of the original Fierz-Pauli theory. In Eq.~\eqref{Dalpha2}, an additional contribution arises from a ghost mode with mass $m_h$ given in Eq.~\eqref{mh} and a kinetic term with the wrong sign, characteristic of a ghost. The three helicity-1 (vector) modes do not couple to the energy-momentum tensor and are therefore neglected in this discussion.

If we choose the parameter $a = 1/2$, the ghost mass becomes $m_h^2 = -m_g^2$, rendering the ghost tachyonic. Substituting this into Eq.~\eqref{Dalpha2}, the propagator simplifies to

\begin{equation}
D^{(1/2)}_{\alpha \beta \mu \nu}(k) =  \frac{1}{-k^2 + m_g^2} \left( \frac{1}{2} (\eta_{\alpha \mu} \eta_{\beta \nu} + \eta_{\alpha \nu} \eta_{\beta \mu}) - \frac{1}{2} \eta_{\alpha \beta} \eta_{\mu \nu} \right) + \text{($k$-dependent terms)}.
\label{Dalpha3}
\end{equation}
The tachyonic ghost cancels the extra scalar contribution to the propagator, leaving only the tensor structure, which matches that of the massless graviton propagator Eq.~\eqref{eq:prop1D}, but with the dispersion relation for a massive graviton
\begin{equation}
    k_0^2 = |\vec{k}|^2 + m_g^2\,.
\end{equation}
As $m_g \rightarrow 0$, the propagator smoothly approaches the massless form of Eq.~\eqref{eq:prop1D}, eliminating the vDVZ discontinuity. The tensor structure in Eq.~\eqref{Dalpha3} also confirms that in this limit, the polarization sum reduces to that of the massless theory Eq.~\eqref{app3}.

In this theory, the gravitational potential takes the Yukawa form
\begin{equation}
    V^{(1/2)}(r) = \frac{G M_1 M_2}{r}\, e^{-m_g r}.
\end{equation}
Unlike the FP theory Eq.~\eqref{VFP}, the characteristic $4/3$ enhancement is absent, as the scalar graviton contribution is exactly canceled by the ghost mode in the propagator. The exponential correction to the $1/r$ potential leads to perihelion precession in planetary orbits~\cite{KumarPoddar:2020kdz}. Constraints on such Yukawa-type deviations from standard gravity, derived from solar system dynamics, place limits on the mass of the exchanged particle. Additionally, long-range Yukawa interactions such as those potentially mediated by ultralight axions can also influence gravitational lensing and Shapiro time delay~\cite{Poddar:2021sbc}.

Because this theory successfully eliminates the problematic scalar mode, it stands out as one of the most phenomenologically viable models of massive gravity. A classical calculation of gravitational energy loss from binary systems in this framework was carried out by Finn and Sutton~\cite{Finn:2001qi}. The QFT-based computation provides a tree-level derivation of the same quantity, and we find that our leading-order result is consistent with their classical analysis.

\vspace{0.2in}

\textbf{Massive graviton radiation in theories without vDVZ discontinuity:}
Following Section~\ref{spin-2-massless}, we obtain the rate of energy loss due to the massive graviton radiation as
\begin{eqnarray}
\frac{{\rm d}E}{{\rm d}t}&=&\frac{\kappa^2}{8(2\pi)^2}\int \Big[|T_{\mu\nu}(k^\prime)|^2-\frac{1}{2}|T^{\mu}{}_{\mu}(k^\prime)|^2\Big]\delta(\omega-\omega^\prime)\omega^2 \Big(1-\frac{m^2_g}{\omega^2}\Big)^\frac{1}{2}{\rm d}\omega {\rm d}\Omega_k\nonumber\\
&=& \frac{\kappa^2}{8(2\pi)^2}\int \tilde{\Lambda}_{ij,lm}T^{ij*}T^{lm} \delta(\omega-\omega^\prime)\omega^2 \Big(1-\frac{m^2_g}{\omega^2}\Big)^\frac{1}{2}{\rm d}\omega {\rm d}\Omega_k,
\label{eq:dedt_deformedFP}
\end{eqnarray}
where the projection operator
\begin{equation}
\begin{split}
{\tilde{\Lambda}_{ij,lm}}=\Big[\delta_{il}\delta_{jm}-2\left(1-\frac{m_g^2}{\omega^2}\right)\hat{k_j}\hat{k_m}\delta_{il}+\frac{1}{2}\left(1-\frac{m_g^2}{\omega^2}\right)^2\hat{k_i}\hat{k_j}\hat{k_l}\hat{k_m}-\frac{1}{2}\delta_{ij}\delta_{lm}
\\
+\frac{1}{2}\left(1-\frac{m_g^2}{\omega^2}\right)\Big(\delta_{ij}\hat{k_l}\hat{k_m}+\delta_{lm}\hat{k_i}\hat{k_j}\Big)\Big].
\end{split}
\label{eq;Lambda_deformedFP}
\end{equation}
After performing the angular integration, Eq.~\eqref{eq:dedt_deformedFP} becomes
\begin{eqnarray}
\frac{{\rm d}E}{{\rm d}t}&=& \frac{8G}{5}\int \left[\left\lbrace\frac{5}{2}-\frac{5}{3}\left(1-\frac{m_g^2}{\omega'^2}\right)+\frac{1}{6}\left(1-\frac{m_g^2}{\omega'^2}\right)^2\right\rbrace T^{ij}T^*_{ij}\right. \label{eq:a4_deformedFP}
\\
&& \left. +\left\lbrace -\frac{5}{4}+\frac{5}{6}\left(1-\frac{m_g^2}{\omega'^2}\right)+\frac{1}{12}\left(1-\frac{m_g^2}{\omega'^2}\right)^2\right\rbrace\vert T^i{}_{i}\vert^2\right]\delta(\omega-\omega^\prime)\omega^2 \Big(1-\frac{m^2_{g}}{\omega^2}\Big)^\frac{1}{2}{\rm d}\omega. \nonumber
\end{eqnarray}
Equations ~\eqref{eq:Txx},~\eqref{werw} and~\eqref{eq:ak2_xy}, yield the rate of energy loss expression~\cite{Poddar:2021yjd}
\begin{equation}
\begin{split}
\frac{{\rm d}E}{{\rm d}t}= \frac{32G}{5} \mu^2 a^4\Omega^6_\mathrm{orb} \sum_{n=1}^{\infty} n^6\sqrt{1-\frac{n_0^2}{n^2}}\left[f(n,e)\left( 1+\frac{4}{3}\frac{n_0^2}{n^2}+\frac{1}{6}\frac{n_0^4}{n^4}\right) -\frac{5J^2_n(ne)}{36n^4}\frac{n_0^2}{n^2}\left(1-\frac{n_0^2}{4n^2}\right)\right].
\end{split}
\label{eq:dedt_FPd}
\end{equation}
To the leading order in $n^2_0$, Eq.~\eqref{eq:dedt_FPd} reads
\begin{equation}
\frac{{\rm d}E}{{\rm d}t}=\frac{32G}{5}\mu^2 a^4 \Omega^6_\mathrm{orb} \Big[\sum_{n=1}^\infty n^6 f(n,e)+n^2_0\sum_{n=1}^\infty\Big(\frac{5}{6}n^4f(n,e)-\frac{5}{36}J^2_n(ne)\Big)\Big]+\mathcal{O}(n^4_0).   
\end{equation}
We observe that the expression smoothly reduces to the result of GR, Eq.~\eqref{eq:app10} in the limit $n_0 = 0$, indicating the absence of the vDVZ discontinuity. Moreover, to leading order in $n_0^2$, this result is consistent with the classical analysis by Finn and Sutton~\cite{Finn:2001qi}. Precision measurements of the orbital period loss of binary system yields upper bound on the graviton mass in this model as $m_g\lesssim 10^{-20}\,\mathrm{eV}$~\cite{Poddar:2021yjd}.

\subsubsection{Dvali-Gabadadze-Porrati (DGP) theory}

GR is a nonlinear theory invariant under diffeomorphisms. This symmetry is broken in massive gravity theories, such as FP, where expanding around curved spacetime introduces a ghost instability, the Boulware-Deser (BD) ghost~\cite{Boulware:1972yco}. To construct a ghost-free massive gravity theory, one can move to higher dimensions. The Dvali-Gabadadze-Porrati (DGP) model~\cite{Dvali:2000hr, Dvali:2000rv, Dvali:2000xg, Dvali:2006su} achieves this via a 5D braneworld framework.

In 5D, the massless graviton has five polarization states and respects general covariance. When the extra dimension is compactified, these degrees of freedom appear as a massive graviton in 4D, free of BD ghosts. The DGP model can explain late-time cosmic acceleration without a cosmological constant~\cite{Deffayet:2001pu}, as the graviton acquires a momentum-dependent mass-modifying gravity at cosmological scales while recovering Newtonian gravity at small scales.

However, the scalar graviton mode introduces the vDVZ discontinuity, posing challenges for phenomenology~\cite{Dvali:2006su}. In this setup, matter is confined to a 4D brane within a 5D bulk, leading to an induced curvature term on the brane. The five and four dimensional Planck masses are denoted by $M_5$ and $M_{\rm Pl}$, respectively.

 The action for the five-dimensional DGP model~\cite{Dvali:2000hr,Dvali:2000rv,Dvali:2000xg}, where matter fields are confined to a four-dimensional brane located at $y = 0$, is given as
\begin{equation}
\mathcal{S}\supset\int {\rm d}^4x {\rm d}y\Big(\frac{M^3_5}{4}\sqrt{-^{(5)}g} {}^{(5)}R+\delta(y)\Big[\sqrt{-g}\frac{M_{\rm Pl}^2}{2}R[g]+\mathcal{L}_m(g,\psi_i)\Big]\Big),
\label{j1}
\end{equation}
where $\psi_i$ represents the matter fields on the brane, with an associated energy-momentum tensor $T_{\mu\nu}$.
Therefore, the resulting modified linearized Einstein equation on the brane located at $y = 0$ becomes~\cite{deRham:2014zqa} 
\begin{equation}
\left(\Box h_{\mu\nu} - \partial_{\mu}\partial_{\nu} h\right) -m_0\sqrt{-\Box}\left(h_{\mu\nu}-h\eta_{\mu\nu}\right)=- \frac{\kappa}{2}T_{\mu\nu}(x),
\label{xsw}
\end{equation}
where $m_0=M^3_5/M_{\rm Pl}^2$ and $M_{\rm Pl}^2=1/{8\pi G}=4/\kappa^2$. Therefore, $(h_{\mu\nu}-h\eta_{\mu\nu})$, which is the FP mass term naturally appears in the higher-dimensional DGP model. The expression Eq.~\eqref{xsw} corresponds to the linearized massive gravity with a scale-dependent graviton mass $m^2_g(\Box)=m_0\sqrt{-\Box}$. The equivalent graviton propagator in DGP model becomes
\begin{eqnarray}
D^{(5)}_{\alpha \beta \mu \nu}(k) &=&  \frac{i}{(-\omega^2 +|\textbf{k}|^2) +m_0 (\omega^2 -|\textbf{k}|^2)^{1/2}} \left (\frac{1}{2} (\eta_{\alpha \mu}  \eta_{\beta \nu} + \eta_{\alpha \nu }\eta_{\beta \mu}) - \frac{1}{3} \eta_{\alpha  \beta}\eta_{\mu \nu} \right).
\label{propDGP}
\end{eqnarray}
The term $ (\frac{1}{2} (\eta_{\alpha \mu}  \eta_{\beta \nu} + \eta_{\alpha \nu }\eta_{\beta \mu}) - \frac{1}{3} \eta_{\alpha  \beta}\eta_{\mu \nu}$ represents the polarization sum of the graviton mode, which matches that of the FP theory. However, in the limit $m_0 \rightarrow 0$, the propagator does not recover its massless form, indicating that the vDVZ discontinuity persists in this theory.

\vspace{0.1in}

\textbf{Massive graviton radiation in the Dvali-Gabadadze-Porrati theory:}

The dispersion relation of the real graviton in the DGP model is governed by the pole of the propagator and is obtained as
\begin{equation}
\omega^2= |\textbf{k}|^2 - m_0^2,
\end{equation}
where $|\mathbf{k}|$ denotes the magnitude of the propagation vector. It is important to note that, in the DGP model, the graviton acquires a tachyonic mass.

Therefore, in the DGP framework, all pertinent expressions in the calculation of the rate of energy loss differ from those in the FP theory by the replacement $m_g^2 \rightarrow -m_0^2$, such that $\tilde{n}_0^2 = m_0^2 / \Omega^2_{\mathrm{orb}} = -n_0^2$. Hence, 
\begin{equation}
\frac{{\rm d}E}{{\rm d}t}=\frac{\kappa^2}{8(2\pi)^2}\int \Big[|T_{\mu\nu}(k^\prime)|^2-\frac{1}{3}|T^{\mu}{}_{\mu}(k^\prime)|^2\Big]\delta(\omega-\omega^\prime)\omega^2 \Big(1+\frac{m^2_{0}}{\omega^2}\Big)^\frac{1}{2}{\rm d}\omega {\rm d}\Omega_k.
\label{j2}
\end{equation}
From the dispersion relation one gerts $\hat{k}^i=k^i/(\omega\sqrt{1+\frac{m^2_0}{\omega^2}})$. Further, the current conservation relation $k_\mu T^{\mu\nu}=0$ yields
\begin{equation}
T_{0j}=-\sqrt{1+\frac{m^2_0}{\omega^2}}\hat{k^i}T_{ij},\hspace{0.5cm} T_{00}=\Big(1+\frac{m^2_0}{\omega^2}\Big)\hat{k^i}\hat{k^j}T_{ij}.
\label{j3}
\end{equation}
Therefore, the term within the third bracket of Eq.~\eqref{j2} can be expressed using the projection operator $\tilde{\Lambda}_{ij,lm}$ as 
\begin{equation}
    |T_{\mu\nu}(k^\prime)|^2-\frac{1}{3}|T^{\mu}{}_{\mu}(k^\prime)|^2 = \tilde{\Lambda}_{ij,lm}T^{ij*}T^{lm},
    \label{j4}
\end{equation}
where the projection operator
\begin{equation}
    {\tilde{\Lambda}_{ij,lm}} = \delta_{il}\delta_{jm}-2\Big(1+\frac{m^2_0}{\omega^2}\Big)\hat{k_j}\hat{k_m}\delta_{il}+\frac{2}{3}\Big(1+\frac{m^2_0}{\omega^2}\Big)^2\hat{k_i}\hat{k_j}\hat{k_l}\hat{k_m}-\frac{1}{3}\delta_{ij}\delta_{lm}+\frac{1}{3}\Big(1+\frac{m^2_0}{\omega^2}\Big)\Big(\delta_{ij}\hat{k_l}\hat{k_m}+\delta_{lm}\hat{k_i}\hat{k_j}\Big)\,.
\label{j5}
\end{equation}
Therefore, in the massive DGP framework, the final form of $dE/dt$ can be expressed as~\cite{Poddar:2021yjd}
\begin{equation}
\begin{split}
\frac{{\rm d}E}{{\rm d}t}= \frac{32G}{5} \mu^2 a^4\Omega^6_{\mathrm{orb}} \sum_{n=1}^{\infty} n^6\sqrt{1+\frac{\tilde{n}_0^2}{n^2}}\left[f(n,e)\left( \frac{19}{18}-\frac{11}{9}\frac{\tilde{n}_0^2}{n^2}+\frac{2}{9}\frac{\tilde{n}_0^4}{n^4}\right) + \frac{5J^2_n(ne)}{108n^4}\left(1+\frac{\tilde{n}_0^2}{n^2}\right)^2\right].
\end{split}
\label{eq:dedt_DGP}
\end{equation}
To the leading order of $\tilde{n}^2_0$, Eq.~\eqref{eq:dedt_DGP} reads
\begin{equation}
\frac{{\rm d}E}{{\rm d}t}\simeq \frac{32G}{5}\mu^2 a^4\Omega^6_{\mathrm{orb}}\Big[\sum_{n=1}^\infty\Big(\frac{19}{18}n^6f(n,e)+\frac{5}{108}n^2 J^2_n(ne)\Big)-\tilde{n}^2_0\sum_{n=1}^\infty\Big(\frac{25}{36}n^4f(n,e)-\frac{25}{216}J^2_n(ne)\Big)\Big]+\mathcal{O}(\tilde{n}^4_0). 
\end{equation}

In addition to the standard emission of massless gravitons, quasi-stable binary systems may also lose energy through the radiation of massive gravitons. This additional energy loss contributes to the secular decay of the orbital period. By comparing the predicted contribution from massive graviton radiation with high-precision binary pulsar timing measurements of orbital period decay, one can place robust constraints on the graviton mass~\cite{Poddar:2021yjd,Lambiase:2025qyl}.

\subsection{Effects of ultralight bosons in coalescing binaries}

In earlier discussions, we considered the radiation of ultralight bosons from quasi-stable binary systems, where both the orbital angular frequency and the orbital separation remain approximately constant over time. In contrast, for a coalescing binary, the situation is different as the two stars spiral inward, the orbital separation decreases and the GW frequency increases with time. The orbital eccentricity also evolves while the system may start with a finite eccentricity, it gradually circularizes during the inspiral phase and ultimately enters the sensitivity band of GW detectors in an almost circular orbit. Therefore, the following analysis has been discussed in the limit $e\to 0$. The masses of ultralight bosons emitted from coalescing binaries are typically much larger than those associated with quasi-stable binary systems~\cite{Kopp:2018jom,Poddar:2023pfj}.

In the presence of a long-range Yukawa-type potential,
\begin{equation}
V(r) = \frac{\alpha Q_1 Q_2}{r} e^{-m r},
\label{coal1}
\end{equation}
originating from a possible BSM interaction, the net force between two stars takes the form
\begin{equation}
F(r) = \frac{GM_1 M_2}{r^2} \left[ 1 + \frac{\alpha Q_1 Q_2}{GM_1 M_2} (1+m r)e^{-m r} \right].
\label{coal2}
\end{equation}

The corresponding orbital angular frequency, obtained from the relation
\begin{equation}
\Omega^2_{\mathrm{orb}} = \frac{(M_1+M_2)}{M_1 M_2 r}\frac{{\rm d}V}{{\rm d}r},
\label{coal3}
\end{equation}
becomes
\begin{equation}
\Omega^2_{\mathrm{orb}} = \frac{GM_1 M_2}{r^3} \left[ 1 + \frac{\alpha Q_1 Q_2}{GM_1 M_2}(1+mr)e^{-m r} \right].
\label{coal4}
\end{equation}

The total energy of the binary system is likewise modified as
\begin{equation}
E = -\frac{GM_1 M_2}{2r} \left( 1 + \frac{\alpha Q_1 Q_2}{GM_1 M_2} e^{-m r} \right).
\label{coal5}
\end{equation}

Here, $Q_1$ and $Q_2$ denote the charges of the stars under the new interaction, while $\alpha$ is the analogue of a fine-structure constant. The force may be mediated by a particle of mass $m$, with $\alpha = g^2/4\pi$, where $g$ characterizes the mediator's coupling to baryonic charge. The interaction range is $\lambda = 1/m$. Unlike gravity, this new force cannot be infinite in range, since in that case its effect could simply be absorbed into the definition of the stellar mass. Thus, the GW emission from the binary remains consistent with GR as long as the orbital separation is much larger than the force range $\lambda$. Deviations from GR predictions may only appear once the separation becomes comparable to or smaller than $\lambda$, when the Yukawa suppression is lifted.

The orbital frequency $\Omega_{\mathrm{orb}}$ in Eq.~\eqref{coal4} is not constant for a coalescing binary, since the orbital separation $r$ evolves with time. Using Eq.~\eqref{coal4}, we can express $r$ in terms of $\Omega_{\mathrm{orb}}$ as
\begin{equation}
    r(\Omega_{\mathrm{orb}})=\Big[\frac{G(M_1+M_2)}{\Omega^2_{\mathrm{orb}}}\Big]^{1/3}\Bigg[1+\frac{\alpha_m}{3}\Big\{1+(G(M_1+M_2))^{1/3}\frac{m}{\Omega^{2/3}_{\mathrm{orb}}}\Big\}\exp{\Big(-\frac{(G(M_1+M_2))^{1/3}m}{\Omega^{2/3}_{\mathrm{orb}}}\Big)}\Bigg],
\label{coal6}
\end{equation}
where $\alpha_m$ is defined as
\begin{equation}
    \alpha_m=\frac{\alpha Q_1 Q_2}{GM_1M_2}\,.
    \label{coal7}
\end{equation}
The rate of total energy change is obtained from Eq.~\eqref{coal5} as
\begin{equation}
    \frac{{\rm d}E}{{\rm d}t}=\frac{GM_1M_2}{2r^2}\frac{{\rm d}r}{{\rm d}t}\Big[1+\alpha_m e^{-m r}(1+m r+m^2 r^2)\Big].
    \label{coal8}
\end{equation}
Equating Eq.~\eqref{coal8} with the rate of energy loss due to radiation of ultralight particles, one obtains
\begin{equation}
\frac{{\rm d}r}{{\rm d}t}=-\frac{64}{5}\frac{G^3 M_1 M_2(M_1+M_2)}{r^3}\Big[1+2\alpha_m e^{-m r}(1+mr)-\alpha_m  m^2r^2 e^{-m r}\Big] \Big[1+\Big(\frac{{\rm d}E_\mathrm{rad}}{{\rm d}t}\Big)_{\mathrm{red}}\Big],  
\label{coal9}
\end{equation}
where 
\begin{equation}
    \Big(\frac{{\rm d}E_\mathrm{rad}}{{\rm d}t}\Big)_\mathrm{red}=\frac{1}{\frac{32G}{5}\mu^2 r^4\Omega_{\mathrm{orb}}^6}\,\frac{{\rm d}E_\mathrm{rad}}{{\rm d}t}\,,
    \label{coal10}
\end{equation}
where d$E_\mathrm{rad}/{\rm d}t$ is the rate of energy loss due to the radiation of ultralight particles. If there are no other new particles involved in the theory, then Eq.~\eqref{coal9} reduces to 
\begin{equation}
\frac{{\rm d}r}{{\rm d}t}=-\frac{64}{5}\frac{G^3 M_1 M_2(M_1+M_2)}{r^3}.    
\label{coal11}
\end{equation}
Differentiating Eq.~\eqref{coal4} with respect to time as $r$ is a function of $t$ and using Eqs.~\eqref{coal9} and~\eqref{coal6}, one obtains
\begin{equation}
    \begin{split}
    \dot{\Omega}_{\mathrm{orb}}=\frac{96}{5}(G \mathcal{M}_{\rm ch})^{5/3}\Omega^{11/3}_{\mathrm{orb}}\Big[1+\frac{2\alpha_m}{3}e^{-m (\mathcal{M} \Omega^{-2}_\mathrm{orb})^{1/3}}\{1+m(\mathcal{M}\Omega^{-2}_\mathrm{orb})^{1/3}-m^2(\mathcal{M}\Omega^{-2}_{\mathrm{orb}})^{2/3}\}\Big]\times\\
    \Big[1+\Big(\frac{{\rm d}E_\mathrm{rad}}{{\rm d}t}\Big)_\mathrm{red}\Big],
    \end{split}
    \label{coal13}
\end{equation}
where 
\begin{equation}
    \mathcal{M}=G(M_1+M_2)\,, \quad \mathcal{M}_{\rm ch}=\frac{(M_1M_2)^{3/5}}{(M_1+M_2)^{1/5}}\,.    
\end{equation}
$\mathcal{M}_{\rm ch}$ is called the chirp mass of a coalescing binary, a quantity that can be measured directly from observations. In standard GR, where there is no other new particle, Eq.~\eqref{coal13} reduces to
\begin{equation}
\dot{\Omega}_\mathrm{orb}=\frac{96}{5}(G\mathcal{M}_{\rm ch})^{5/3}\Omega^{11/3}_\mathrm{orb}.  
\label{coal14}
\end{equation}
Any deviation from standard GR may arise either from the emission of additional light particles or from a modification of the underlying force law, both of which alter the effective expression of the chirp mass. Therefore, precise measurements of the chirp mass can serve as a powerful probe to constrain the presence of such light particles. Similarly, constraints can also be derived for scalar-tensor~\cite{Alsing:2011er} and higher-curvature extensions of gravity~\cite{Lambiase:2025qyl}.

Apart from the above frequency chirp measurements, proper waveform modeling of the GW phase and strain can also constrain these extra modes and tests gravity.

\subsection{Radiation of ultralight bosons from isolated stars}
In addition to compact binary systems, where the emission of ultralight particles can contribute to the orbital period decay and the frequency chirp, the same particles may also be radiated from isolated pulsars if they couple to electromagnetic (EM) fields, or to SM constituents such as nucleons or electrons within the star. In what follows, we discuss how the emission of such ultralight particles from a compact, magnetized star can affect several key observables such as the pulsar spin-down rate, photon redshift, photon propagation speed, birefringence, and EM luminosity. Such ultralight fields also give rise to a long-range field configuration (or ``hair") extending beyond the surface of the star. These ultralight degrees of freedom may correspond to scalar, pseudoscalar (axion-like), or other light bosonic fields~\cite{Poddar:2024thb,Poddar:2025oew}.

\subsubsection{Scalar coupling with electromagnetic fields of a magnetized star}

Assuming the magnetized star (or pulsar) behaves as a perfect conductor and is surrounded by vacuum, we consider the aligned rotator model, in which the star's magnetic axis is parallel to its rotation axis. In such a configuration, no EM radiation is produced. The radiation arises only when there is a finite inclination angle between the magnetic and rotational axes and hence the EM fields are time-dependent. Under this assumption of time-independent configuration of the EM fields, the dipolar magnetic field of the star can be expressed as~\cite{Goldreich:1969sb,Shapiro:1983du,Mohanty:1993nh}
\begin{equation}
\mathbf{B}(r,\theta)= \frac{B_0 R^3}{r^3}\Big(\cos\theta \hat{r}+ \frac{\sin\theta}{2}\hat{\theta}\Big), ~~~~r>R  
\label{isolated1}
\end{equation}
where $B_0$ denotes the surface magnetic field strength, $R$ is the stellar radius, and $\theta$ represents the angle between the rotation axis and the line of sight (i.e. the polar angle). Using the boundary condition that the tangential component of the electric field is continuous at the stellar surface $r=R$ while the normal component may exhibit a discontinuity, the electric field outside the star can be obtained as
\begin{equation}
\mathbf{E}(r,\theta)=-\frac{B_0\Omega R^5}{r^4}\Big[\Big(1-\frac{3}{2}\sin^2\theta\Big)\hat{r}+\sin\theta\cos\theta \hat{\theta}\Big],~~~r>R
\label{isolated2}
\end{equation}
where $\Omega$ denotes the angular velocity of the rotating star.

A CP-even ultralight scalar field can couple to the EM current density through the interaction term $(1/2)F_{\mu\nu}F^{\mu\nu}$ which, in terms of the EM fields, reads
\begin{equation}
\frac{1}{2}F_{\mu\nu}F^{\mu\nu}=\mathbf{B}^2-\mathbf{E}^2=\frac{B^2_0R^6}{4r^6}(3\cos^2\theta+1)-\frac{B^2_0\Omega^2R^{10}}{4r^8}(5\cos^4\theta-2\cos^2\theta+1),
\label{isolated3}
\end{equation}
where Eqs.~\eqref{isolated1} and~\eqref{isolated2} have been used, and $F_{\mu\nu}$ denotes the EM field strength tensor.

The dynamics of the scalar field $\phi$ coupled to the stellar EM field can be described by the Lagrangian
\begin{equation}
\mathcal{L}=\frac{1}{2}\partial_\mu\phi\partial^\mu\phi-\frac{1}{4}F_{\mu\nu}F^{\mu\nu}-\frac{1}{2}g_{\phi\gamma\gamma}\phi F_{\mu\nu}F^{\mu\nu},
\label{isolated4}
\end{equation}
where $g_{\phi\gamma\gamma}$ represents the effective scalar-photon coupling constant. The sign of $g_{\phi\gamma\gamma}$ depends on the dominant contribution in the loop: it is positive if the loop is dominated by fermions (e.g., leptons or heavy quarks), and negative if the dominant contribution arises from $W$ bosons loop.

The corresponding equation of motion for the scalar field is then given by
\begin{equation}
\Box\phi=-g_{\phi\gamma\gamma}(\mathbf{B}^2-\mathbf{E}^2),
\label{isolated5}
\end{equation}
where $\Box=\frac{\partial^2}{\partial t^2}-\nabla^2$ denotes the d’Alembertian operator in background spacetime. To have a non-trivial scalar field profile outside the star, one needs a non-trivial scalar source charge density $\rho_\phi=g_{\phi\gamma\gamma}(\mathbf{B}^2-\mathbf{E}^2)$ outside the star. In a magnetically dominated region, where $|\mathbf{B}|\gg |\mathbf{E}|$, and if the star is slowly rotating $\Omega R\ll 1$, then one can neglect the $\mathbf{E}^2$ term and, in that case, Eq.~\eqref{isolated5} reduces to 
\begin{equation}
\Box\phi\approx -g_{\phi\gamma\gamma}\frac{B^2_0R^6}{4r^6}(3\cos^2\theta+1),
\label{isolated6}
\end{equation}
whose solution  outside of the star is (using the Green's function method)
\begin{equation}
\phi(r)\approx \frac{Q_\phi^{\mathrm{eff}}}{r}+\mathcal{O}\Big(\frac{1}{r^2}\Big),
\label{isolated7}
\end{equation}
Here, $Q_\phi^{\mathrm{eff}}$ denotes the effective scalar-induced charge, which is given by~\cite{Poddar:2025oew}
\begin{equation}
Q_\phi^{\mathrm{eff}} =\frac{g_{\phi\gamma\gamma}B^2_0R^3}{6}.
\label{isolated8}
\end{equation}
 Note that $Q_\phi^{\mathrm{eff}}$ can have either sign depending on the sign of $g_{\phi\gamma\gamma}$. Thus, the rotating magnetized star has a long-range ``hair" outside the star $(\phi\propto 1/r)$ associated with the chrge $Q_\phi^{\mathrm{eff}}$. The above derivation assumes a massless scalar field, but the results remain valid as long as the scalar's Compton wavelength exceeds the stellar radius, i.e., $m_\phi\lesssim 1/R$.

Taking GRB 080905A as a representative example, the quantitative estimate of the induced scalar charge is
\begin{equation}
Q_\phi^{\mathrm{eff}} \simeq 10^{37}\Big(\frac{g_{\phi\gamma\gamma}}{10^{-15}\,\mathrm{GeV^{-1}}}\Big)\Big(\frac{B_0}{3.93\times 10^{16}\,\mathrm{G}}\Big)^2\Big(\frac{R}{10\,\mathrm{km}}\Big)^3.   
\label{isolated9}
\end{equation}

The interaction between a CP-even scalar field and the EM fields of a star modifies the vacuum Maxwell equations, leading to corrections in both the electric and magnetic field configurations of the star.
To capture this effect, the EM field strength tensor can be expanded perturbatively in powers of the scalar-photon coupling $g_{\phi\gamma\gamma}$ as
\begin{equation}
F^{\mu\nu}=F_{(0)}^{\mu\nu}+F^{\mu\nu}_\phi +\mathcal{O}(g^2_{\phi\gamma\gamma}),
\label{isolated12}
\end{equation}
where $F_{(0)}^{\mu\nu}$ denotes the background field in the absence of scalar interaction $(g_{\phi\gamma\gamma}=0)$.
The corresponding equation of motion for the scalar-induced correction of the EM field strength tensor $F^{\mu\nu}_\phi$ is then 
\begin{equation}
\partial_\mu F^{\mu\nu}_\phi=-g_{\phi\gamma\gamma}(\partial_\mu \phi)F^{\mu\nu}_{(0)},  \label{isolated13}
\end{equation}
which yields the inhomogeneous Maxwell equations for the scalar-induced electric and magnetic fields as
\begin{eqnarray}
 \nabla\cdot \mathbf{E}_\phi & = & -g_{\phi\gamma\gamma} \, \mathbf{E}_{(0)} \cdot \nabla\phi \, , \nonumber \\
\nabla\times \mathbf{B}_\phi & = & \frac{\partial \mathbf{E}_\phi}{\partial t}-g_{\phi\gamma\gamma} \, \nabla\phi\times\mathbf{B}_{(0)} 
+ g_{\phi\gamma\gamma} \left(\frac{\partial \phi}{\partial t} \right) \mathbf{E}_{(0)}, 
\label{isolated14}
\end{eqnarray}
where $\mathbf{E}_\phi$ and $\mathbf{B}_\phi$ present the scalar-induced electric and magnetic fields, respectively. The Bianchi identity
\begin{equation}
    \partial_\mu \tilde F^{\mu\nu}_\phi=0,
    \label{isolated15}
\end{equation}
implies the homogeneous Maxwell equations
\begin{eqnarray}
    \nabla\cdot \mathbf{B}_\phi & = & 0 \, , \nonumber \\
    \nabla\times \mathbf{E}_\phi & = & -\frac{\partial \mathbf{B}_\phi}{\partial t}.
    \label{isolated16}
\end{eqnarray}
In the aligned rotator model, the background fields $\mathbf{E}_{(0)}$ and $\mathbf{B}_{(0)}$ are stationary; hence, any time dependence of $\phi, \mathbf{E}_\phi, \mathbf{B}_\phi$ can be neglected. Combining Eqs.~\eqref{isolated14} and~\eqref{isolated16} gives the wave equations for the scalar-induced fields

\begin{equation}
\begin{split}
\nabla^2 \mathbf{B}_\phi=g_{\phi\gamma\gamma} \,(\nabla\phi\cdot\nabla)\, \mathbf{B}_{(0)}, \\
\nabla^2 \mathbf{E}_\phi=g_{\phi\gamma\gamma} \, (\nabla\phi\cdot\nabla)\, \mathbf{E}_{(0)},
 \label{isolated17}
\end{split}    
\end{equation}
where terms involving second derivatives of $\phi$ are neglected since $\phi\propto 1/r$ and $\mathbf{B}_\phi$, $\mathbf{E}_\phi$ are time-independent as the scalar field and the background EM fields are time-independent. 

In the slow-rotation limit $(\Omega R\ll1)$, the scalar-induced magnetic field takes the form~\cite{Poddar:2025oew}
\begin{equation}
    \mathbf{B}_\phi(r,\theta)\simeq\frac{3}{10}g_{\phi\gamma\gamma}B_0R^3Q^{\mathrm{eff}}_\phi\Big(\frac{\cos\theta}{r^4}\Big)\hat{\mathbf{r}}+\frac{3}{20}g_{\phi\gamma\gamma}B_0R^3Q^{\mathrm{eff}}_\phi\Big(\frac{\sin\theta}{r^4}\Big)\hat{\theta}\,,
    \label{isolated18}   
\end{equation}
showing that $\mathbf{B}_\phi$ falls as $1/r^4$, in contrast to the background field $\mathbf{B}_{(0)}$ which falls as $1/r^3$. The deviation of the stellar surface magnetic field from its background value is small, scaling as $B_\phi\propto g^2_{\phi\gamma\gamma}$ since $Q^\mathrm{eff}_{\phi}\propto g_{\phi\gamma\gamma}$. 

The propagation of photons (or EM waves) through a background scalar field modifies the effective Maxwell equations, and hence the photon dispersion relation and group velocity.
Consider a photon emitted from a compact star, propagating through a region endowed with static EM fields and a background scalar field $\phi$.
In the absence of external plasma charge and current densities, the modified Maxwell equations are
\begin{eqnarray}
\nabla\cdot \mathbf{E} & = & -g_{\phi\gamma\gamma}\mathbf{E}\cdot \nabla\phi \, , \nonumber \\
\nabla\times \mathbf{B} & = & \frac{\partial \mathbf{E}}{\partial t}-g_{\phi\gamma\gamma}\nabla\phi\times \mathbf{B} \, , \nonumber \\
\nabla\cdot \mathbf{B} & = & 0 \, , \nonumber \\
\nabla\times \mathbf{E} & = & -\frac{\partial \mathbf{B}}{\partial t} \, .
 \label{isolated20}   
\end{eqnarray}

where $\mathbf{E}$ and $\mathbf{B}$ are the electric and magnetic fields of the propagating wave.
Since the scalar field profile decays as $\phi\propto 1/r$, second derivatives of
$\phi$ are subdominant and have been neglected.
Using Eq.~\eqref{isolated20}, the corresponding wave equations for the EM fields become
\begin{equation}
\begin{split}
\Box\mathbf{B}=g_{\phi\gamma\gamma}(\nabla\phi\cdot\nabla)\mathbf{B}\, ,\\
\Box\mathbf{E}=g_{\phi\gamma\gamma}(\nabla\phi\cdot\nabla)\mathbf{E}\, .
 \label{isolated21}   
\end{split}
\end{equation}

Employing the eikonal ansatz $\mathbf{B}(x,t)=\mathcal{B} \, e^{iS(x,t)}$, where the phase $S$ defines the photon frequency and wave vector through $\omega=-\partial S/\partial t$ and $\mathbf{k}=\nabla S$ respectively, the dispersion relation in an asymptotically flat spacetime reads
\begin{equation}
\omega^2=k^2-ig_{\phi\gamma\gamma} (\nabla\phi\cdot \mathbf{k}), 
 \label{isolated22}   
\end{equation}

The imaginary term indicates that the wave number is complex in the scalar-field background. Assuming radial propagation, one can define an effective 
scalar-induced photon mass
\begin{equation}
m_\gamma=|g_{\phi\gamma\gamma}\nabla \phi|.    
\end{equation} 

The photon's group velocity is then given by
\begin{equation}
v_g=\frac{d\omega}{dk}=\frac{2k-i m_\gamma}{2\omega(k)},  
\label{isolated23}
\end{equation}
assuming the photon propagation is radial. 
Solving Eq.~\eqref{isolated22} for the complex wave number $k=k_R+ik_I$ yields

\begin{equation}
k=\frac{\sqrt{4\omega^2-m_\gamma^2}}{2}+\frac{im_\gamma}{2},
\label{isolated24}   
\end{equation}
with the real and imaginary components
\begin{equation}
k_R=\frac{\sqrt{4\omega^2-m_\gamma^2}}{2},~~~   k_I=\frac{m_\gamma}{2}.
\label{ngwq}
\end{equation}
The real part is well defined for $\omega> m_\gamma/2$, corresponding to propagating modes. The group velocity follows from Eqs.~\eqref{isolated23} and \eqref{isolated24}, and is given by~\cite{Poddar:2025oew}
\begin{equation}
v_g=\Big(1-\frac{m_\gamma^2}{4\omega^2}\Big)^\frac{1}{2},    
\label{isolated25}
\end{equation}
showing that photon propagation is subluminal in the presence of a scalar background. The correction to $v_g$ is of order $\mathcal{O}(g^4_{\phi\gamma\gamma})$. In the limit $g_{\phi\gamma\gamma}\to 0$, one recovers the standard luminal propagation, $v_g\to 1$.

The wavelength of the EM wave is governed by $k_R$; for $\omega>m_\gamma/2$,
\begin{equation}
k_R\simeq\omega-\frac{m^2_\gamma}{8\omega}.    
\end{equation}
Hence, the apparent redshift between emission near the stellar surface $(r_1)$ and observation at infinity $(r_2)$ can be expressed as
\begin{equation}
\delta z
= \frac{\lambda(r_2) - \lambda(r_1)}{\lambda(r_1)}
\simeq \frac{k_R(r_1) - k_R(r_2)}{k_R(r_2)}
\simeq \frac{m_\gamma^2}{8\omega^2}.
\label{eq:isolated26}
\end{equation}
Since $\phi\propto 1/r$, we set $m_\gamma(r_2)=0$ at large distances. Expressing $m_\gamma$ in terms of the stellar parameters, one obtains
\begin{equation}
\delta z
\simeq \frac{g^4_{\phi\gamma\gamma}B^4_0R^2}{288\omega^2}.
\label{eq:redshift_scaling}
\end{equation}

For the benchmark GRB080905A system, the estimated fractional wavelength shift is~\cite{Poddar:2025oew}
\begin{equation}
\delta z = \frac{\Delta k}{k}\simeq 10^{-4} \Big(\frac{g_{\phi\gamma\gamma}}{ 2.8\times 10^{-15}\,\mathrm{GeV}^{-1}}\Big)^4 \Big(\frac{2.1\,\mathrm{GHz}}{\omega}\Big)^2\Big(\frac{B_0}{3.93\times 10^{16}\,\mathrm{G}}\Big)^4 \Big(\frac{R}{10\,\mathrm{km}}\Big)^2,
\label{isolated27}
\end{equation}
Thus, a redshift measurement with precision at the level of  $10^{-4}$, comparable to that currently achieved for GRB~080905A host galaxy, can probe scalar-photon couplings down to $g_{\phi\gamma\gamma}\sim 2.8\times 10^{-15}\,\mathrm{GeV}^{-1}$. The effect is enhanced for objects with stronger magnetic fields, larger radii, and for lower photon frequencies. The equivalent change in the redshift prediction for the time-oscillating scalar field has been discussed in~\cite{Poddar:2026myk}.

The imaginary part of the wave number $k_I$ describes photon absorption in the scalar-field medium.
The intensity of the wave decays as
\begin{equation}
I\propto \mathbf{E}^2\propto E^2_0 e^{-2k_I x}.    
\end{equation}
Defining the absorption coefficient $\alpha$ through $I=I_0e^{-\alpha x}$, one finds
\begin{equation}
\alpha=2k_I=m_\gamma=\frac{1}{x}\ln \Big(\frac{I_0}{I}\Big),  
\label{isolated28}
\end{equation}
which scales as $g^2_{\phi\gamma\gamma}$.
Hence, the scalar field not only induces a frequency-dependent redshift but also leads to attenuation of photon intensity during propagation. There can also be exponential enhancement of the field amplitude and hence intensity if $\nabla\phi\cdot \mathbf{k}$ picks a minus sign.

To induce time-dependent scalar emission, the scalar field must couple to time-varying EM fields, which arise naturally if the magnetic and spin axes of the star are misaligned by an angle $\alpha$. In this case, the magnetic field rotates with angular frequency $\Omega$, and the EM fields take the form  
\begin{equation}
\begin{split}
\textbf{B}=\frac{B_0R^3}{2r^3}\Big[(3\cos\theta_m\sin\theta\cos\varphi-\sin\alpha\cos\Omega t)(\sin\theta\cos\varphi\,\hat{r} \, +\cos\theta\cos\varphi\,\hat{\theta}-\sin\varphi\,\hat{\varphi}) \, +\\
(3\cos\theta_m\sin\theta\sin\varphi-\sin\alpha\sin\Omega t)(\sin\theta\sin\varphi\,\hat{r}+\cos\theta\sin\varphi\,\hat{\theta}+\cos\varphi\,\hat{\varphi})+ \\(3\cos\theta_m\cos\theta-
\cos\alpha)\times
(\cos\theta\,\hat{r}-sin\theta\,\hat{\theta})\Big],
\end{split}
\label{isolated29}
\end{equation}
and
\begin{equation}
\begin{split}
\textbf{E}=-\frac{B_0\Omega R^5}{2r^4}\Big[\{\cos\alpha(3\cos^2\theta-1)+3\sin\alpha\sin\theta\cos\theta\cos(\Omega t-\varphi)\}\hat{r} \, + \\ 2\{\cos\alpha\sin\theta\cos\theta
+\sin\alpha\sin^2\theta\cos(\Omega t-\varphi)\}\hat{\theta}\Big],
\label{isolated30}
\end{split}
\end{equation}
where the magnetic colatitude $\theta_m$ is defined through $\cos\theta_m=\cos\alpha\cos\theta+\sin\alpha\sin\theta\cos(\Omega t-\varphi)$, which denotes the angle between the magnetic axis and the line of sight. In the limit $\alpha\to 0$, Eqs.~\eqref{isolated29} and~\eqref{isolated30} reduce to the static expressions given in Eqs.~\eqref{isolated1} and~\eqref{isolated2}, respectively. 

The time-dependent scalar source density is $\rho_\phi(\mathbf{r},t)=g_{\phi\gamma\gamma}
(\mathbf{B}^2-\mathbf{E}^2)$, where Eqs.~\eqref{isolated29} and~\eqref{isolated30}, give
\begin{equation}
\mathbf{B}^2-\mathbf{E}^2\approx -\frac{3}{2}\frac{B^2_0R^6}{r^6}\sin\alpha\cos\alpha\sin\theta\cos\theta\cos(\Omega t-\varphi).
\label{isolated31}
\end{equation}
In deriving Eq.~\eqref{isolated31}, higher-order terms in $\Omega R\ll 1$ and all time-independent contributions have been neglected, as they do not contribute to radiation. Similarly, higher harmonics of $\Omega$ are suppressed by powers of the stellar rotational velocity and are omitted. 
Therefore, the rate of energy loss due to the scalar radiation is obtained as~\cite{Poddar:2025oew}
\begin{equation}
\frac{dE}{dt}\approx \frac{1}{80} g^2_{\phi\gamma\gamma}B^4_0R^{10}\Omega^6\sin^2(2\alpha)\Big(1-\frac{m^2_\phi}{\Omega^2}\Big)^{5/2}.
\label{isolated34}
\end{equation}
This radiative energy loss contributes to the pulsar spin-down whenever the scalar field is light enough to be radiated, i.e., for $m_\phi<\Omega$. This photophilic scalar radiation can contribute to the pulsar spin-down which can be compared with the observational results to obtain constraints on the scalar-photon coupling~\cite{Poddar:2025oew}. The pulsar spin-down luminosity can be measured with great precision which results bound on the scalar-phoron coupling to be $g_{\phi\gamma\gamma}\lesssim 10^{-14}\,\mathrm{GeV}^{-1}$ for $m_\phi\lesssim 10^{-13}\,\mathrm{eV}$~\cite{Poddar:2025oew}.

\subsubsection{Pseudoscalar coupling with electromagnetic fields of a magnetized star}
The pseudoscalar, CP-odd axion can couple to the EM fields of a pulsar through the interaction term
\begin{equation}
\mathcal{L}_\mathrm{int}\supset \frac{1}{4}g_{a\gamma\gamma} a F_{\mu\nu}\tilde{F}^{\mu\nu},
\label{bireaxion1}
\end{equation}
which leads to the equation of motion for a massless axion field,
\begin{equation}
\Box a = g_{a\gamma\gamma}\mathbf{E}\cdot\mathbf{B}.
\label{bireaxion2}
\end{equation}
Thus, the axion field is sourced by an effective charge density
$(\rho_a = g_{a\gamma\gamma}\mathbf{E}\cdot\mathbf{B})$.
Following the same method used for the scalar case, the axion field profile outside the star is obtained as
\begin{equation}
    a(r,\theta) = -\Big(\frac{2}{575}\Big)\frac{g_{a\gamma\gamma}B_0^2R^8\Omega\cos\alpha}{M^3} \frac{\cos\theta}{r^2} + \mathcal{O}\left(\frac{1}{r^3}\right).
    \label{bireaxion3}
\end{equation}
The interaction in Eq.~\eqref{bireaxion1} modifies the Maxwell equations in vacuum to~\cite{Sikivie:1983ip, Wilczek:1987mv, Visinelli:2013mzg}
\begin{equation}
\begin{split}
\nabla\cdot\mathbf{E} &= -g_{a\gamma\gamma}\mathbf{B}\cdot\nabla a,\\
\nabla\times\mathbf{B} - \dot{\mathbf{E}} &= g_{a\gamma\gamma}(\dot{a}\mathbf{B} + \nabla a \times \mathbf{E}),\\
\nabla\cdot\mathbf{B} &= 0,\\
\dot{\mathbf{B}} + \nabla\times\mathbf{E} &= 0,
\end{split}
\label{bireaxion4}
\end{equation}
where the coupling induces parity-violating terms through the pseudoscalar nature of axion.

Because of its CP-odd structure, photon propagation through an axion background becomes polarization-dependent, leading to axion-induced birefringence. Consequently, when pulsar light traverses this axion field, the left- and right-circularly polarized components experience distinct phase velocities. The resulting differential time delay between the two polarization modes is given by~\cite{Mohanty:1993nh}
\begin{equation}
\delta t = \frac{1}{4}\Big(\frac{2}{575}\Big)^3
\frac{g_{a\gamma\gamma}^6 B_0^6 R^{16}\Omega^3}{M^9 \omega^3}.
\label{bireaxion5}
\end{equation}
Equations~\eqref{bireaxion3} and \eqref{bireaxion5} are derived using the pulsar magnetic and electric field configurations given in Eqs.~\eqref{isolated1} and \eqref{isolated2}.

Pseudoscalar axions can also be emitted from pulsars when sourced by the time-varying EM fields. In the Vacuum Dipole Model (VDM), the exterior of the pulsar is assumed to be vacuum, and the magnetic field is described by a rotating dipole
$(\boldsymbol{\mu}(t) = B_0 R^3 \hat{\boldsymbol{\mu}}(t))$.
Since the stellar interior behaves as a perfect conductor, one has $(\mathbf{E}\cdot\mathbf{B} = 0)$ inside the star. The corresponding exterior field is obtained by solving Laplace's equation with conducting boundary conditions at the surface, yielding
\begin{equation}
    \mathbf{E}\cdot\mathbf{B} = \frac{|\boldsymbol{\mu}|^2 \Omega}{4r^5}\left[\cos\theta\sin^2\theta_m + \frac{1}{2}\sin\theta\sin(2\theta_m)\big(\Omega r \cos(\phi-\Omega t) - \sin(\phi-\Omega t)\big)\right].
    \label{bireaxion6}
\end{equation}
Following the same procedure as in the scalar case, the rate of energy loss due to axion emission is estimated as~\cite{Khelashvili:2024sup}
\begin{equation}
    \frac{{\rm d}E_a}{{\rm d}t} \approx \Big(\frac{\pi}{432}\Big)g_{a\gamma\gamma}^2B_0^4\Omega^6R^{10}\sin^2(2\theta_m).
    \label{bireaxion7}
\end{equation}
If the magnetosphere is screened, axion emission mainly originates from the polar cap (PC) region. In the PC model, the corresponding energy loss rate becomes~\cite{Khelashvili:2024sup}
\begin{equation}
    \frac{{\rm d}E_a}{{\rm d}t} \approx \frac{1}{3\pi}R^2\Omega^4Q_a^2\sin^2\theta_m,
    \label{bireaxion8}
\end{equation}
where $Q_a$ is the effective axion charge in the gap region, defined by
\begin{equation}
    Q_a = g_{a\gamma\gamma}\int_{\mathrm{gap}} {\rm d}^3x\langle \mathbf{E}\cdot\mathbf{B}(\mathbf{x},t) \rangle_t = \pi g_{a\gamma\gamma} B_0^2 \Omega r_{\mathrm{pc}}^2 h^2.
    \label{bireaxion9}
\end{equation}
Here $r_{\mathrm{pc}} = R\sqrt{\Omega R}$ denotes the radius of the PC, and h represents the height of the gap (or equivalently, the mean free path of pair-producing photons), approximately given by
\begin{equation}
h \sim 7\,\mathrm{m}
\left(\frac{\Omega}{2\pi\times30\,\mathrm{Hz}}
\frac{B_0}{8.5\times10^{12}\,\mathrm{G}}\right)^{-4/7}.
\label{bireaxion10}
\end{equation}
The radiation of these ultralight axions can contribute to the pulsar spin-down luminosity and comparing with the observational data, constraint on the axion-photon coupling could be obtained. Precision measurement of the energy loss yields bound on the axion-photon coupling as $g_{a\gamma\gamma}\lesssim 10^{-11}\,\mathrm{GeV}^{-1}$ for $m_a\lesssim 10^{-13}\,\mathrm{eV}$~\cite{Khelashvili:2024sup}. 

In addition to radiation processes, ultralight particles can mix with photons, leading to oscillation phenomena analogous to neutrino oscillations. Such mixing can imprint characteristic features in observed photon spectra, allowing these particles to be constrained through spectroscopic observations. Related studies can be found in~\cite{Smarra:2024vzg,Poddar:2026wyd,McDonald:2024nxj}.

To conclude this section, we briefly summarize the possible radiation of new light bosonic degrees of freedom including spin-$0$ (scalar and pseudoscalar), spin--$1$ and spin--$2$ particles from astrophysical systems, which can provide an additional channel of energy loss. The orbital decay of binary systems, the spin-down luminosity of pulsars, and related spectroscopic observables such as frequency shifts and redshifts are measured with high precision. Consequently observations from GW detectors and complementary EM measurements can place strong constraints on the masses and coupling of such particles. The analyses in this section shows that efficient radiation occurs only if the emitted particles are ultralight, with masses set by the characteristic length and frequency scales of the system, such as the orbital separation, orbital frequency, or stellar spin frequency. Importantly the radiated particles need not constitute DM, therefore, many existing DM bounds are not directly applicable. Instead the constraints discussed here arise purely from energy loss considerations in astrophysical dynamics.

\newpage

%% file: 6_Superradiance_and_search_for_bosonic_DM.tex
\section{Superradiance and search for Bosonic dark matter}
\label{Super}

Dissipative systems in relativistic and astrophysical settings can exhibit superradiance (SR), where incident radiation is amplified through energy extraction from the background system. SR is a kinetic process associated with rotating or charged media and appears in a variety of physical contexts beyond GR. Here, we focus on BH superradiance (BH-SR) and its relevance for fundamental physics and astrophysics, in particular as a probe of light bosonic fields. BH-SR arises from the interaction between bosonic waves and rotating BHs, and is closely connected to nonlinear gravitational dynamics near the event horizon. It is also related to fundamental concepts such as the BH area theorem, BH thermodynamics, the no-hair theorem, and Hawking radiation~\cite{Brito:2015oca}. The scattering processes associated with SR may also help understanding various high-energy astrophysical phenomena.\footnote{Key examples include the Penrose process~\cite{Penrose:1969pc}, the Blandford-Znajek mechanism~\cite{Blandford:1977ds}, and magnetic reconnection~\cite{Comisso:2020ykg}, together with extensions in modified gravity theories~\cite{Khodadi:2022dff,Khodadi:2023juk,Zhang:2024ptp,Long:2024tws,Shaymatov:2023dtt}.} In particular, the extraction of rotational energy\footnote{Thermodynamically, the efficiency of SR is bounded by $\lesssim 0.29$, though nonlinear effects may reduce this to $\lesssim 0.1$~\cite{Herdeiro:2021znw}, highlighting the role of backreaction and self-interactions in realistic BH systems.} from rotating BHs provides a plausible mechanism for powering relativistic jets in active galactic nuclei (AGN)~\cite{Bardeen:1972fi}. Amplification becomes especially significant in the presence of confinement mechanisms such as massive bosonic fields, anti-de Sitter (AdS) boundaries, or nonlinear interactions, which can trigger spacetime instabilities commonly referred to as BH bombs~\cite{Press:1972zz}, see Ref.~\cite{Cardoso:2004nk} for a modern treatment. This possibility has attracted considerable interest because it offers a way to probe DM and physics beyond the SM, particularly through ultralight bosonic fields (e.g.,~\cite{Brito:2020lup,Ng:2020ruv,Yuan:2022bem}). By contrast, fermionic fields do not appear to undergo SR around BHs.\footnote{Although not rigorously proven, existing analyses indicate that massless and massive Dirac fields scattered by Kerr BHs do not exhibit SR; see Ref.~\cite{Kim:1997hy} for a systematic discussion.}

Broadly, two approaches address the gravitational anomalies attributed to DM. One modifies gravity on large scales (e.g.,~\cite{Milgrom:1983ca,Moffat:2005si,Chamseddine:2013kea}), while the other posits new particle species beyond the SM. In the particle-DM paradigm, the universe is assumed to contain both baryonic matter and a non-baryonic component that accounts for observations such as flat galactic rotation curves, strong lensing in clusters, and the large-scale evolution of the universe~\cite{Boehm:2024ana}.\footnote{Despite these successes, the absence of direct detection in laboratory experiments means that the particle nature of DM remains unconfirmed, motivating complementary probes that combine particle physics, astrophysics, and cosmology.} Ultralight bosonic fields are particularly difficult to study with conventional laboratory techniques because of their tiny masses and weak couplings to SM particles. In this regime, gravitational interactions provide one of the few viable observational windows. Rotational SR offers a powerful method to search for axion-like and other ultralight fields through their indirect effects on BH dynamics and astrophysical observables~\cite{Bekenstein:1998nt}.

\begin{figure}[ht!]
	\centering
	\includegraphics[width=0.85\columnwidth]{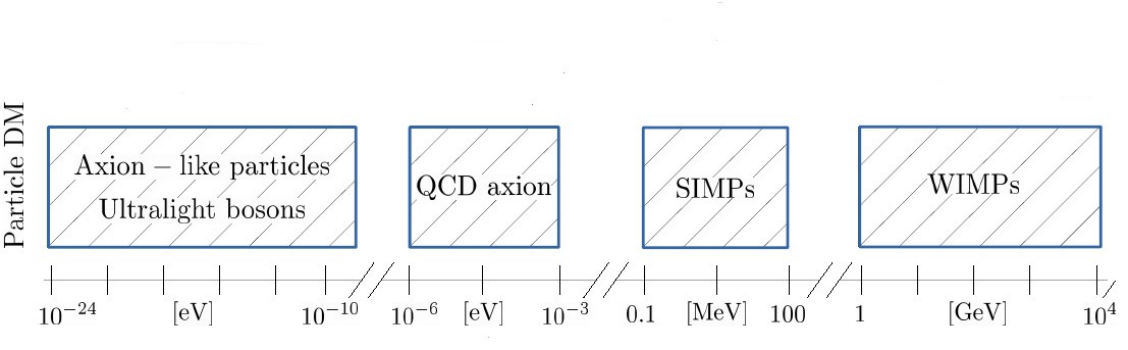}~~~
	\caption{The mass range of theoretically well-motivated particle DM candidates spans a broad spectrum. On the higher end, two prominent examples are SIMPs~\cite{Hochberg:2014dra} and WIMPs~\cite{Overduin:2004sz}, both of which are considered more massive candidates for DM. These particles are distinguished by their interactions and mass scales, with SIMPs characterized by strong interactions and WIMPs by weak interactions, making them key targets in the search for DM. } 
	\label{Mass}
\end{figure}

In Fig.~\ref{Mass}, we show the particle mass spectrum of DM where the axions or generic ultralight bosons could cause SR.

\subsection{Superradiant scattering in rotating Black holes}

Rotating BHs possess two key geometric features: an event horizon and an ergoregion. While the horizon enforces one-way propagation, the ergoregion allows negative-energy states for timelike observers. Although SR is often framed in terms of ingoing boundary conditions at the horizon~\cite{Richartz:2009mi,Cardoso:2012zn}, this interpretation is incomplete: SR can occur even without such conditions~\cite{Vicente:2018mxl}. Energy extraction is instead fundamentally tied to the presence of an ergoregion, which provides the effective dissipation required for amplification~\cite{Eskin:2015ssa}. Spacetimes with ergoregions but no horizons can therefore exhibit SR; however, the horizon plays a stabilizing role, and its absence generically leads to superradiant instabilities (SRIs). We consider test scalar, EM, and GW perturbations of an asymptotically flat Kerr BH. In Boyer-Lindquist coordinates $(t,r,\theta,\phi)$, the Kerr metric is~\cite{Teukolsky:2014vca}
\begin{align}
	{\rm d}s^2 = -\left(1-\frac{2Mr}{\varrho^2}\right){\rm d}t^2 + \frac{\varrho^2}{\Delta}{\rm d}r^2 + \varrho^2 {\rm d}\theta^2 + \frac{A\sin^2\theta}{\varrho^2}{\rm d}\phi^2 - \frac{4Mra\sin^2\theta}{\varrho^2}{\rm d}t {\rm d}\phi,
\end{align}
where $\varrho^2 = r^2 + a^2\cos^2\theta$, $\Delta = r^2 - 2Mr + a^2$, $A = (r^2 + a^2)^2 - a^2\Delta\sin^2\theta$, and the horizons lie at $r_{\pm} = M \pm \sqrt{M^2 - a^2}$.

For a scalar field $\Psi$ of mass $\mu_S$ satisfying $(\nabla_\alpha\nabla^\alpha + \mu_S^2)\Psi = 0$, we adopt
\begin{equation}
	\Psi = R_{\omega lm}(r)S_{\omega lm}(\theta)e^{-i\omega t}e^{im\phi},
\end{equation}
with frequency $\omega > 0$ and angular quantum numbers $(l,m)$. The separated equations are
\begin{align}
	&\frac{\rm d}{{\rm d}r}\left(\Delta\frac{{\rm d}R}{{\rm d}r}\right) + \left[\frac{[(r^2 + a^2)\omega - am]^2}{\Delta} - \left(\mu_S^2 r^2 + l(l+1) + a^2\omega^2 - 2am\omega\right)\right]R = 0,\\
	&\frac{1}{\sin\theta}\frac{{\rm d}}{{\rm d}\theta}\left(\sin\theta\frac{dS}{d\theta}\right) + \left[l(l+1) - a^2\omega^2\sin^2\theta - \frac{(a\omega\sin^2\theta - m)^2}{\sin^2\theta} - a^2\mu_S^2\cos^2\theta\right]S = 0,
\end{align}
where the angular $\mu_S^2$ term is sometimes included for approximate treatments of the massive scalar. Introducing the tortoise coordinate via ${\rm d}r_*/{\rm d}r = (r^2 + a^2)/\Delta$, the asymptotic behavior is
\begin{equation}
	R \to 
	\begin{cases}
		\mathcal{A}^+_{\mathrm{in}} \Delta^{-s}e^{-ik_+ r_*}, & r\to r_+,\\
		\mathcal{A}^\infty_{\mathrm{in}}\dfrac{e^{-ik_\infty r_*}}{r} + \mathcal{A}^\infty_{\mathrm{ref}}\dfrac{e^{ik_\infty r_*}}{r}, & r\to \infty,
	\end{cases}
\end{equation}
where $k_+ = \omega - m\Omega_+$, $k_\infty = \sqrt{\omega^2 - \mu_S^2}$, and $\Omega_+ = a/(2Mr_+)$. The Wronskian identity gives
\begin{equation}
	|\mathcal{A}^\infty_{\mathrm{ref}}|^2 = |\mathcal{A}^\infty_{\mathrm{in}}|^2 - \frac{k_+}{k_\infty}|\mathcal{A}^+_{\mathrm{in}}|^2\,,
\end{equation}
so that for $\omega < m\Omega_+$ (i.e., $k_+<0$) the reflected flux exceeds the incident one, signaling SR.

More generally, linear perturbations of spin $s$ (scalar $s=0$, EM $s=\pm1$, GW $s=\pm2$) obey the Teukolsky equation~\cite{Teukolsky:1972my}. Writing $\Psi = e^{-i\omega t}e^{im\varphi}S_{slm}(\theta)R_{slm}(r)$, one finds
\begin{align}\label{teu_radial}
	&\Delta^{-s}\frac{{\rm d}}{{\rm d}r}\left(\Delta^{s+1}\frac{{\rm d}R}{{\rm d}r}\right) + \left[\frac{K^2 - 2is(r-M)K}{\Delta} + 4is\omega r - \lambda\right]R = 0, \\ 
	&\frac{1}{\sin\theta}\frac{{\rm d}}{{\rm d}\theta}\left(\sin\theta\frac{{\rm d}S}{{\rm d}\theta}\right) + \left[a^2\omega^2\cos^2\theta - \frac{m^2}{\sin^2\theta} - 2a\omega s\cos\theta - \frac{2ms\cos\theta}{\sin^2\theta} - s^2\cot^2\theta + s + \lambda_{slm}\right]S = 0,
\end{align}
where $K = (r^2 + a^2)\omega - am$ and $\lambda = \lambda_{slm} + a^2\omega^2 - 2am\omega$. The angular functions $S_{slm}$ are spin-weighted spheroidal harmonics, reducing to $Y_{slm}$ when $a\omega = 0$. Regularity at the horizon enforces purely ingoing modes, and asymptotically
\begin{equation}
    \label{bc_kerr}
	R_{slm} \sim 
	\begin{cases}
		\mathcal{A}^+_{\mathrm{in}} \Delta^{-s} e^{-ik_+ r_*}, & r\to r_+,\\
		\mathcal{A}^\infty_{\mathrm{in}}\dfrac{e^{-i\omega r_*}}{r} + \mathcal{A}^\infty_{\mathrm{ref}}\dfrac{e^{i\omega r_*}}{r^{2s+1}}, & r\to \infty.
	\end{cases}
\end{equation}
The energy fluxes at infinity and the horizon can be written in terms of the asymptotic amplitudes~\cite{Teukolsky:1974yv,Brito:2015oca}. Energy conservation implies
\begin{equation}
	\frac{{\rm d}E_{\mathrm{in}}}{{\rm d}t} - \frac{{\rm d}E_{\mathrm{out}}}{{\rm d}t} = \frac{{\rm d}E_{\mathrm{hole}}}{{\rm d}t}.
\end{equation}
When $\omega < m\Omega_+$, $k_+ < 0$ and the horizon flux becomes negative, indicating energy extraction from the BH. The amplification factor $Z_{slm}$ (positive for SR, negative for absorption) is
\begin{equation}
	Z_{slm} \equiv \frac{{\rm d}E_{\mathrm{out}}}{{\rm d}E_{\mathrm{in}}} - 1.
\end{equation}
Explicitly, for scalar ($s=0$), electromagnetic ($s=\pm1$), and gravitational ($s=\pm2$) perturbations:
\begin{align}
	&Z_{0lm} = \frac{|\mathcal{A}^\infty_{\mathrm{ref}}|^2}{|\mathcal{A}^\infty_{\mathrm{in}}|^2} - 1,\\
	&Z_{\pm1lm} = \frac{|\mathcal{A}^\infty_{\mathrm{ref}}|^2}{|\mathcal{A}^\infty_{\mathrm{in}}|^2}\left(\frac{16\omega^4}{B^2}\right)^{\pm1} - 1,\\
	&Z_{\pm2lm} = \frac{|\mathcal{A}^\infty_{\mathrm{ref}}|^2}{|\mathcal{A}^\infty_{\mathrm{in}}|^2}\left(\frac{256\omega^8}{|C|^2}\right)^{\pm1} - 1,
\end{align}
with $B^2 = [\lambda + s(s+1)]^2 + 4ma\omega - 4a^2\omega^2$ and $|C|^2$ defined in Ref.~\cite{Brito:2015oca}. Stars with dissipation mechanisms can also exhibit superradiance. For a conducting, spinning compact star with conductivity $\sigma$, the axial perturbation equation for vector fields is~\cite{Cardoso:2017kgn}
\begin{equation}
	\frac{{\rm d}^2 a}{{\rm d}r_*^2}+\left[\omega^2-2m\omega\zeta(r)-V\right]a=0,
\end{equation}
with $V=F\left(l(l+1)/r^2+\mu_V^2-4i\pi\sigma(\omega-m\Omega)/\sqrt{F}\right)$. The amplification factor $Z>0$ when $\omega<m\Omega$, increasing with $\sigma$ and compactness.

A nonzero cosmological constant modifies this picture. In de Sitter backgrounds the SR window narrows, while in AdS spacetime confinement can trigger BH-bomb instabilities. For Kerr-dS, imposing ingoing conditions at the event horizon yields~\cite{Tachizawa:1992ue}
\begin{equation}
	|\mathcal{A}^{\infty}_{\mathrm{ref}}|^2 = |\mathcal{A}^{\infty}_{\mathrm{in}}|^2 - \frac{\omega - m\Omega_{+}}{\omega - m\Omega_{\mathrm{c}}}\, |\mathcal{A}^{+}_{\mathrm{in}}|^2\,,
\end{equation}
so that SR occurs for
\begin{equation}
	m\Omega_{\mathrm{c}} < \omega < m\Omega_{+}.
\end{equation}
Compared to asymptotically flat space, the SR band is narrower, though peak amplification is slightly higher. Few studies extend SR to dS backgrounds; examples include charged BHs~\cite{Zhu:2014sya, Konoplya:2014lha}. In AdS backgrounds, SR depends sensitively on boundary conditions: reflective conditions suppress scalar SR, whereas transmissive ones permit it in any dimension~\cite{Winstanley:2001nx,Jorge:2014kra}. In the extremal limit where the Hawking temperature goes to zero, $T_{\rm H}\to0$, Hawking emission occurs only in the superradiant regime $\omega<m\Omega_{\rm H}$, with rate $\mp Z_{slm}$ (minus for fermions, plus for bosons)~\cite{Hawking:1974sw}. Finally, string theory and AdS/CFT provide microscopic descriptions where superradiant emission arises from interactions in the dual CFT~\cite{Dias:2007nj,Bredberg:2009pv}. In the fuzzball proposal, microstate geometries without horizons reproduce Hawking radiation via the ergoregion instability~\cite{Chowdhury:2007jx,Chowdhury:2008bd}.\footnote{Horizonless objects with ergoregions suffer from the ergoregion instability~\cite{Sato:1978ue,Friedman:1978ygc}. For ultracompact stars, the instability timescale scales as $\tau_{\rm ergo}\sim4\alpha e^{2\beta m}$ in the eikonal limit~\cite{1978RSPSA.364..211C}. A consistent treatment requires including second-order rotational terms in the metric.}

Recent GW and horizon-scale imaging observations~\cite{LIGOScientific:2016aoc, EventHorizonTelescope:2019dse, EventHorizonTelescope:2022wkp} remain consistent with GR but leave room for deviations from the Kerr metric. SR is not exclusive to GR and can occur in any gravitational theory admitting BH solutions. Modified gravity theories, which typically introduce new degrees of freedom or alter field equations, can significantly enhance or suppress SR amplification. Many beyond-GR frameworks yield rotating BH metrics that reduce to Kerr in the GR limit. These modified geometries, along with altered wave dynamics, can lead to stronger SRIs or, conversely, weaken or eliminate SR altogether~\cite{Bini:2003sy,Bini:2014kga, Wondrak:2018fza, Fukuda:2019ewf, Khodadi:2020cht,Franzin:2021kvj,Khodadi:2021owg, Khodadi:2021mct, Chesler:2021ehz, Jha:2022tdl,Yang:2022uze,Jha:2022ewi, Lingetti:2022psy, Wang:2022hra,Cannizzaro:2023ltu, Dolan:2024qqr,Dorlis:2025amf, Tanaka:2025bfl}.

\begin{center}
	\begin{figure}[ht]
		\begin{center}
			\begin{tabular}{cc}
				\epsfig{file=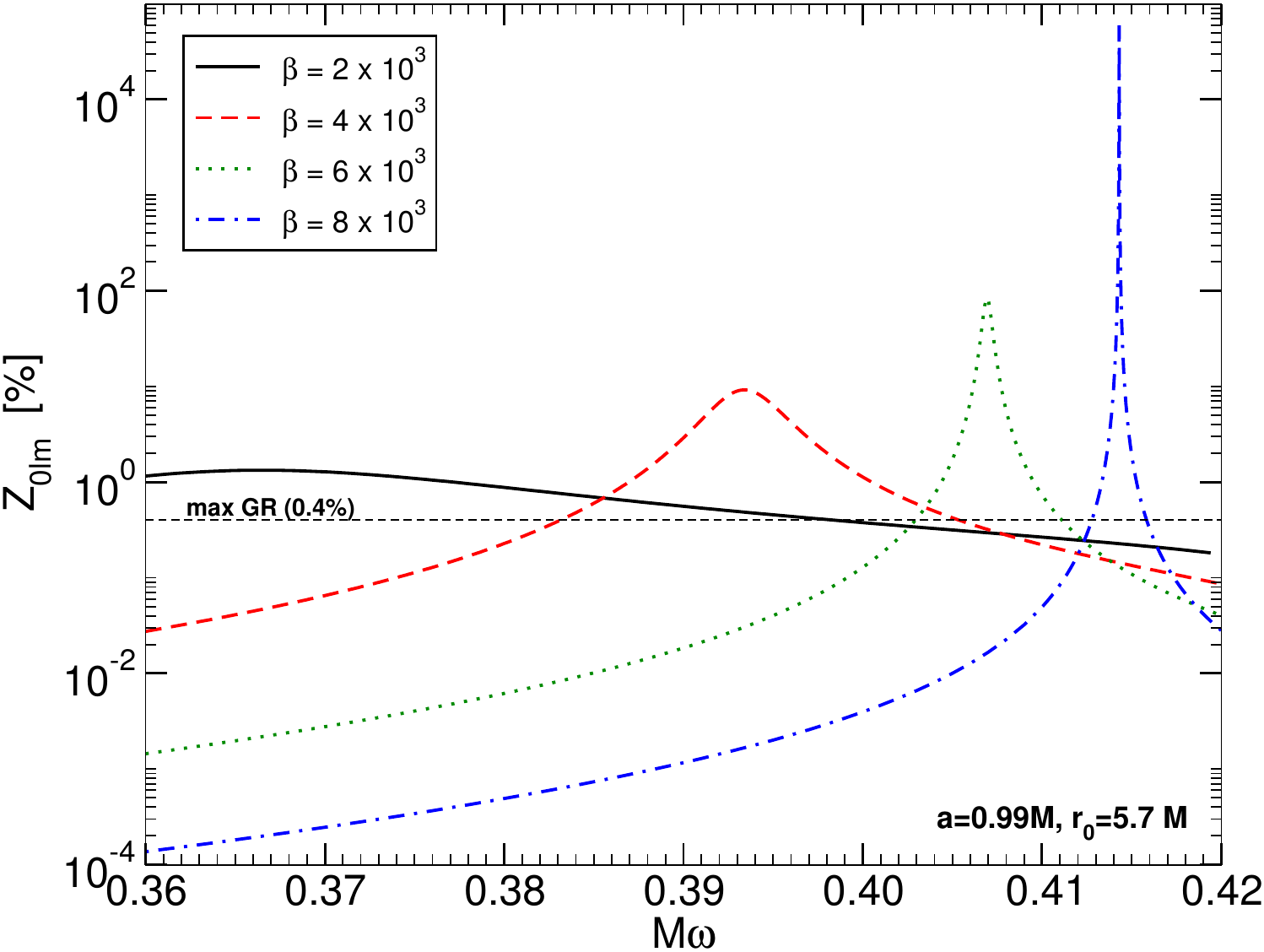,width=0.6\textwidth,angle=0,clip=true}
			\end{tabular}
		\end{center}
		\caption{\label{fig:STamplification}
			The SR amplification factor, $ Z_{0lm} $, expressed as a percentage, plotted against the wave frequency $ \omega $ for a scalar wave interacting with a Kerr BH in a scalar-tensor theory. The matter profile is defined by Eq.~\eqref{eq:eff_mass_profile}, with the model parameter $ \beta $. The horizontal line indicates the maximum amplification in GR ($ \beta = 0 $). Adapted from Ref.~\cite{Cardoso:2013opa}.}
	\end{figure}
\end{center}

Exact spinning BH solutions in modified gravity are often difficult to obtain. In the slow-rotation regime ($a \ll M$), analytical results exist; for example, in quadratic gravity, where SR amplification is reduced compared to Kerr~\cite{Pani:2011gy}. Some theories yield Kerr-like metrics but with modified field equations, leading to distinct SR phenomenology~\cite{Psaltis:2007cw,Berti:2015itd}. Certain configurations are particularly susceptible to BH-bomb instabilities, where superradiant modes become trapped by potential barriers, leading to exponential growth~\cite{Myung:2011we,Myung:2013oca}. Matter distributions around BHs also influence SR scattering. Accretion disks, DM halos, or other environmental factors can modify amplification factors and mode stability~\cite{Cardoso:2013fwa,Cardoso:2013opa,Khodadi:2021mct,Cuadros-Melgar:2021sjy}. In scalar-tensor theories, for instance, a matter profile of the form
\begin{equation}
	\mu_{\rm eff}^2(r,\theta) = \frac{2{\cal G}(r)}{a^2 + 2r^2 + a^2\cos 2\theta}, \quad
	{\cal G}(r) = \beta \,\Theta(r - r_0)\frac{r - r_0}{r^3},
	\label{eq:eff_mass_profile}
\end{equation}
can enhance SR by several orders of magnitude, as illustrated in Fig.~\ref{fig:STamplification}. This strong dependence makes SR a promising observational probe for testing deviations from GR.

\subsection{Superradiant instabilities and bosonic clouds}

\begin{figure}[ht]
	\centering
	\includegraphics[width=0.9\textwidth]{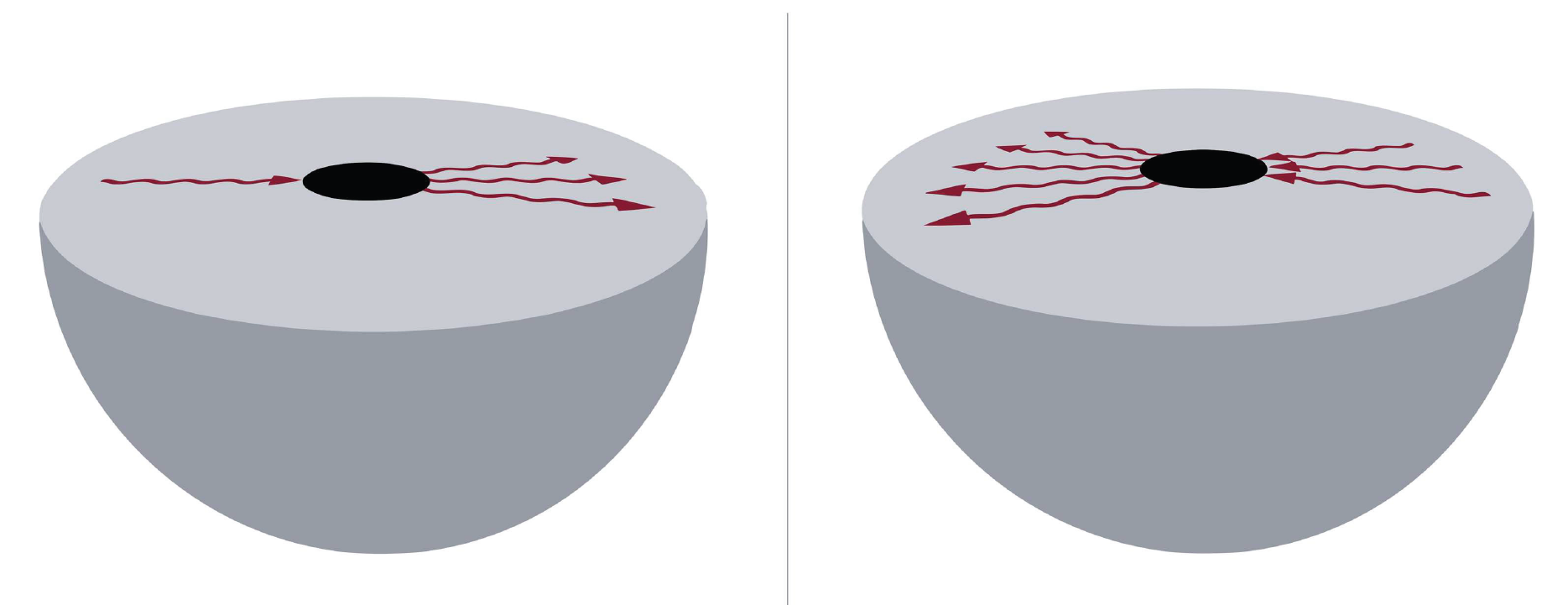}
	\caption{Schematic of the BH bomb: a wave (red arrow) undergoes repeated reflection and SR amplification inside a confining boundary. Adapted from Ref.~\cite{Brito:2015oca}.}
	\label{fig:bhbomb}
\end{figure}

SR amplification plays a dual role in rotating BH spacetimes: it allows for rotational energy extraction and, in the presence of confinement, can trigger spacetime instabilities. These SRIs spin down the BH and can seed long-lived bosonic condensates, or in some cases lead to stationary ``hairy'' BH configurations. The classic analysis by Press and Teukolsky~\cite{Press:1972zz} established that repeated scattering alone cannot destabilize a Kerr BH, but showed that confinement converts superradiant amplification into exponential growth, giving rise to the so-called BH bomb. The mechanism is illustrated in Fig.~\ref{fig:bhbomb}: an incoming wave is amplified in the ergoregion and, if reflected back by a confining boundary such as a mirror, an AdS boundary, or the effective potential of a massive field, it undergoes repeated interactions with the BH. Each cycle increases the amplitude, leading to exponential growth. Two important realizations are provided by BHs in asymptotically AdS spacetimes, where the boundary naturally confines radiation, and by systems in which a massive bosonic field creates an effective potential barrier that traps modes outside the horizon.

When confinement is present, the perturbation spectrum consists of quasi-normal modes with complex frequencies $\omega=\omega_R+i\omega_I$. Stability is controlled by the sign of $\omega_I$, with positive values corresponding to exponential growth and negative values describing damping. For a small BH of mass $M$ inside a cavity of size $r_0\gg M$, and in the low-frequency regime $M\omega\ll1$, a wave of initial amplitude $A_i$ evolves after $N=t/r_0$ reflections as
\begin{equation}
    A(t) \simeq A_i \left(1 - |\mathcal{A}_a|^2\right)^{t/r_0} \approx A_i \left(1 - \frac{t}{r_0}|\mathcal{A}_a|^2\right),
\end{equation}
where $|\mathcal{A}_a|^2$ is the absorption (or amplification) probability at the horizon. Matching to the exponential form $A(t) \sim A_i e^{\omega_I t}$ yields the growth rate
\begin{equation}
	\omega_I \simeq \frac{|\mathcal{A}_a|^2}{r_0}.
    \label{eq:growth_rate}
\end{equation}
Rotation modifies $|\mathcal{A}_a|^2$ through the superradiant factor $(1-m\Omega_+/\omega)$. As shown by Starobinsky~\cite{1973ZhETF..64...48S,1973ZhETF..65....3S}, for moderate spins one finds
\begin{equation}
    |\mathcal{A}_a|^2 \propto (\omega - m\Omega_+) \, \omega^{2l+1},
\end{equation}
instead of the non-rotating scaling $\propto \omega^{2l+2}$. Consequently, when $\omega < m\Omega_+$, $|\mathcal{A}_a|^2 < 0$ and Eq.~\eqref{eq:growth_rate} gives $\omega_I > 0$ and instability. The corresponding timescale $\tau\sim1/\omega_I$ decreases with decreasing confinement radius $r_0$. This behavior is generic for scalar, EM, and gravitational perturbations~\cite{Cardoso:2004nk,Cardoso:2013pza,Brito:2014nja}.

Three main realizations of the BH-bomb paradigm have been explored. The most idealized consists of a spinning BH surrounded by a perfectly reflecting mirror~\cite{1972JETP...35.1085Z, Press:1972zz, Cardoso:2004nk}. Superradiant modes grow exponentially until nonlinear effects or field pressure overwhelm the confinement. For scalar perturbations governed by Eq.~\eqref{teu_radial}, boundary conditions impose purely ingoing waves at the horizon and vanishing flux at the mirror radius $r_m$, implemented through Dirichlet, Neumann, or more general Robin conditions~\cite{Ferreira:2017tnc}. Numerous extensions and numerical studies can be found in~\cite{Hod:2013fvl, Witek:2010qc,Herdeiro:2013pia, Astrakhantsev:2018dbb,Borges:2018fso, Li:2015bfa,Luo:2024gqo}.

AdS spacetimes provide a natural confining environment because their timelike boundary prevents geodesics from escaping to infinity. For a massless particle in pure AdS ($M=0$, $\Lambda<0$), radial null geodesics satisfy
\begin{equation}
	\frac{{\rm d}r}{{\rm d}t} = 1 + \frac{r^2}{L^2}, \quad L^2 \equiv -\frac{3}{\Lambda},
\end{equation}
where $L$ is the AdS curvature radius. A light ray from the origin reaches spatial infinity in finite coordinate time $t = \pi L/2$, demonstrating the necessity of boundary conditions at infinity. Kerr-AdS BHs behave similarly to mirror-confined systems. SRI occur for small BHs ($r_+\ll L$), while large BHs ($r_+\gtrsim L$) are typically stable because their characteristic frequencies exceed the superradiant threshold. Slowly rotating configurations with $\Omega_+L<1$ are stable, whereas rapidly spinning BHs become unstable~\cite{Hawking:1999dp,Rahmani:2020wlq, Ganchev:2016zag,Aliev:2015wla, Rahmani:2018hgp,Chesler:2018txn}.

\begin{figure}[t]
	\centering
	\includegraphics[width=0.9\textwidth]{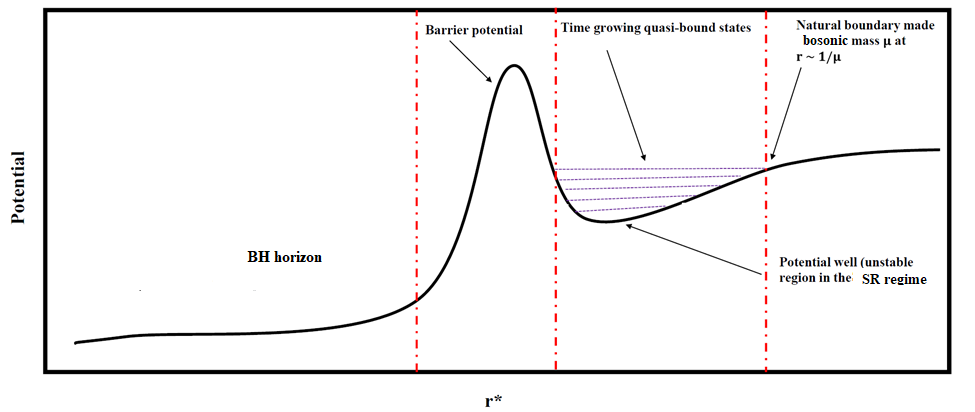}
	\caption{Effective gravitational potential showing a trapping region outside the BH. A bosonic field of mass $\mu_i$ confined in this region grows exponentially, leading to instability. Adapted from Ref.~\cite{Khodadi:2022dyi}.}
	\label{Gr}
\end{figure}

In asymptotically flat spacetimes, a third and particularly compelling confinement mechanism arises from massive bosonic fields. Their Yukawa falloff $\sim e^{-\mu r}/r$ acts as an effective mirror, producing quasi-bound states localized in a potential well outside the BH, as illustrated in Fig.~\ref{Gr} and discussed in~\cite{Press:1972zz,Cardoso:2004nk,Hod:2015goa,Hod:2017gvn}. These instabilities become astrophysically relevant when $M\mu_i\lesssim1$, where $\mu_i$ denotes the boson mass ($i=S,V,T$). For a solar-mass BH, a boson of mass $\mu_i\sim10^{-10}$\,eV corresponds to $\mu_i M \sim 0.75$, motivating searches for ultralight fields as discussed in Sec.~\ref{sec:ultralight_motivations}. For massive scalars on Kerr backgrounds, the Teukolsky equation remains separable, with the radial potential acquiring a term proportional to $\mu_S^2$. The dynamics are controlled by $M\mu_S$, and analytic results exist in both small coupling ($M\mu_S\ll1$) and large coupling limits ($M\mu_S\gg1$). In the small-coupling regime, the spectrum resembles that of hydrogen~\cite{Detweiler:1980uk,Rosa:2009ei,Pani:2012bp}, with
\begin{align}
	\omega\sim M\mu_S\Bigl[1-\frac12\Bigl(\frac{M\mu_S}{l+n+1}\Bigr)^2\Bigr]
	+ i (\mu_S M)^{4l+5}\bigl(m\chi-2\mu_S r_+\bigr)
	\frac{2^{4l+2}(2l+1+n)!}{(l+1+n)^{2l+4}n!}\nonumber\\
	\times\biggl[\frac{l!}{(2l)!(2l+1)!}\biggr]^2
	\prod_{k=1}^{l}\Bigl[k^2\bigl(1-\chi^2\bigr)+\bigl(m\chi-2\mu_S r_+\bigr)^2\Bigr],
	\label{eq:omegaDetweiler}
\end{align}
where $\chi=a/M$. The instability sets in for
\begin{equation}
    a_{\text{crit}}\approx\frac{2\mu_S M r_+}{m}\,,
\end{equation}
with maximal growth for $l=m=1$, $n=0$, and near-extremal spin. In contrast, for $M\mu_S\gg1$ the instability is exponentially suppressed~\cite{Zouros:1979iw},
\begin{equation}
    \tau_S\sim10^7 M\,e^{1.84 M\mu_S}\qquad (M\mu_S\gg1),
\end{equation}
making this regime phenomenologically irrelevant for astrophysical BHs. The SRI is bounded by $\mu_S<\sqrt{2}\,m\Omega$, which can be approached arbitrarily closely in the eikonal limit $M\mu_S\gg1$~\cite{Hod:2012zza}. The most interesting regime is $M\mu_S\lesssim1$, where timescales can be short.

No-hair theorems imply that, in asymptotically flat vacuum GR, axisymmetric stationary BHs are uniquely Kerr~\cite{Hawking:1973uf, Heusler:1995qj, Sotiriou:2011dz,Graham:2014ina}. Thus, the end-state of the SRI is typically a Kerr BH with reduced mass and spin. Exceptions exist in the form of \emph{hairy} BH solutions~\cite{Herdeiro:2014goa,Herdeiro:2017phl}, where the metric remains stationary while the scalar field oscillates in time, evading the no-hair assumptions and suppressing GW emission. Charged scalars around Kerr-Newman BHs also exhibit SRIs, with growth rates controlled by the product $qQ$~\cite{Furuhashi:2004jk}. While static charged BHs in asymptotically flat spacetime are stable~\cite{Hod:2013eea,Hod:2013nn,Hod:2016bas}, rotation enables bound superradiant modes. Additional couplings can further modify this picture by generating effective trapping potentials~\cite{Kolyvaris:2018zxl}.

\subsection{Astrophysical implications of superradiance}

\begin{figure}[ht]
	\centering
	\includegraphics[width=0.7\textwidth]{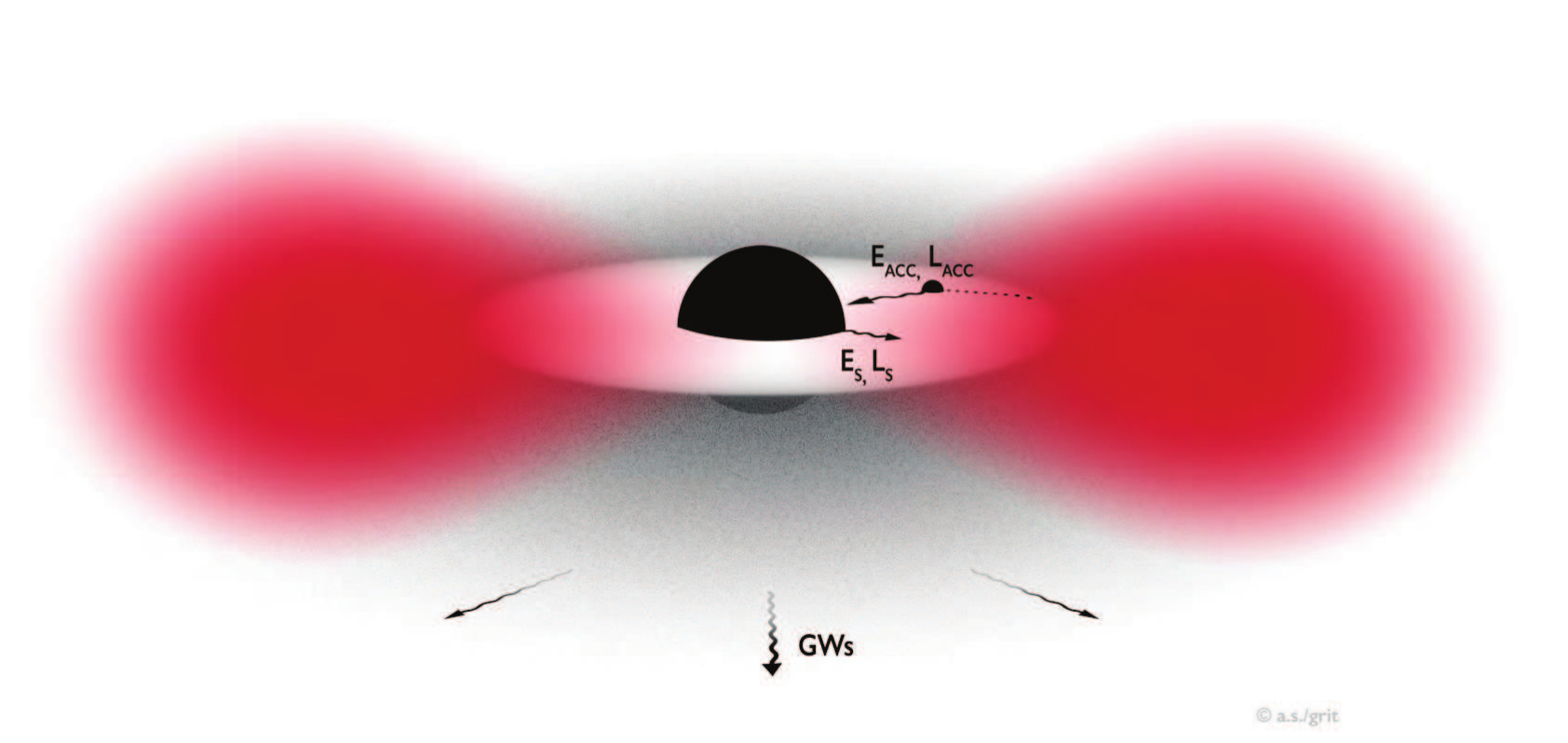}
	\caption{Schematic of a bosonic cloud around a rotating BH. Superradiance extracts energy $E_S$ and angular momentum $L_S$, while accretion from a disk supplies $E_{\rm ACC}$ and $L_{\rm ACC}$. Material inside the innermost stable circular orbit (ISCO) plunges into the BH. Adapted from~\cite{Brito:2015oca}.}
	\label{fig:draw}
\end{figure}

Astrophysical BHs span a wide mass range, from stellar-mass systems ($\sim5$-$30\,M_\odot$) observed in X-ray binaries to supermassive objects ($\sim10^6$-$10^9\,M_\odot$) residing in galactic centers. Direct GW detections have revealed binary BHs up to $\sim80\,M_\odot$~\cite{LIGOScientific:2018mvr}. Within GR, isolated vacuum BHs are uniquely described by the Kerr metric, making deviations from this paradigm a probe of new physics. The universal coupling of gravity ensures that BHs interact with any bosonic field, enabling SRIs in realistic astrophysical settings. Massive bosonic fields can trigger SRIs around Kerr BHs on timescales competitive with astrophysical processes. For an ultralight scalar of mass $\mu_S$, the shortest instability timescale scales approximately as $\tau_S\sim(M/10^6M_\odot)$\,yr~\cite{Cardoso:2005vk,Dolan:2007mj}. Vector and tensor fields with masses $\mu_V$ and $\mu_T$ can trigger even faster instabilities, with characteristic timescales
\begin{align}
	\tau_V &\sim \frac{M(M\mu_V)^{-7}}{4(a/M - 2 \mu_V r_+)}, \label{eq:tau_vector} \\
	\tau_T &\sim \frac{M(M\mu_T)^{-3}}{\gamma_{\rm polar}(a/M - 2 r_+ \omega_R)}\,,\label{eq:tau_tensor}
\end{align}
where $\gamma_{\rm polar}\sim\mathcal{O}(1)$. The nonlinear evolution of the instability produces a bosonic cloud that extracts energy and angular momentum from the BH until the superradiant condition saturates. The cloud then dissipates through GW emission on longer timescales~\cite{Arvanitaki:2009fg,Brito:2014wla,East:2017ovw,Roy:2021uye,Chen:2022nbb}, giving the BH transient hair. The cloud typically extends to radii $\sim1/(M\mu_S^2)$, see Fig.~\ref{fig:draw}. These systems provide a natural detector for ultralight bosons: rapidly spinning BHs would be spun down if a suitable field exists, producing characteristic GW emission and leaving gaps in the distribution of BH spins.

The SRI evolution can be modeled in a quasi-adiabatic framework including accretion and GW losses. For a massive scalar $\Psi$ minimally coupled to gravity,
\begin{equation}
	S = \int {\rm d}^4x \sqrt{-g} \left( \frac{R-2\Lambda}{\kappa} - \frac{1}{2}g^{\mu\nu}\Psi^{\ast}_{,\mu}\Psi_{,\nu} - \frac{\mu_S^2}{2}\Psi^{\ast}\Psi \right) + S_M,
\end{equation}
the hierarchy of timescales, $M\ll\mu_S^{-1}\ll\tau_{\rm SRI}$, allows the scalar to evolve on a fixed Kerr background when the mass of the bosonic cloud $M_s\ll M$~\cite{Brito:2014wla}. For $M\mu_S\lesssim0.2$, the dominant $l=m=1$ mode has a hydrogenic profile with characteristic size
\begin{equation}
	\langle r \rangle \approx \frac{2l^2+3n^2+l(6n+5)+3(2n+1)}{2M\mu_S^2},
\end{equation}
The cloud emits quasi-monochromatic GWs at frequency $\sim2\mu_S$ with energy flux~\cite{Brito:2014wla,Brito:2017zvb}
\begin{align}
    \dot{E}_{\rm GW} &= \frac{484+9\pi^2}{23040}\left(\frac{M_s^2}{M^2}\right)(M\mu_S)^{14},\\
    \dot{J}_{\rm GW} &= \frac{1}{\omega_R}\dot{E}_{\rm GW}.
\end{align}
Gas accretion introduces competing effects. For sub-Eddington accretion,
\begin{equation}
	\dot{M}_{\rm ACC} \equiv f_{\rm Edd}\dot{M}_{\rm Edd} \sim 0.02 \, f_{\rm Edd} \frac{M(t)}{10^6 M_\odot} M_\odot {\rm yr}^{-1},
\end{equation}
with $f_{\rm Edd}$ the Eddington ratio. For angular momentum accretion, assuming equatorial thin disk extending to ISCO:
\begin{equation}
	\dot{J}_{\rm ACC} \equiv \frac{L(M,J)}{E(M,J)}\dot{M}_{\rm ACC},
\end{equation}
where $L$ and $E$ are the specific angular momentum and energy at the ISCO. Energy–angular-momentum conservation then yields
\begin{align}
    \dot M +\dot M_s &= -\dot E_{\rm GW}+\dot M_{\rm ACC},\\
    \dot J +\dot J_s &= -\frac{1}{\mu_S}\dot E_{\rm GW}+\dot J_{\rm ACC}.
\end{align}
The superradiant extraction rate is
\begin{equation}
    \dot E_S=2M_s\omega_I,\qquad 
    M\omega_I=\frac{1}{48}\left(\frac{a}{M}-2\mu_S r_+\right)(M\mu_S)^9,
\end{equation}
so accretion can both suppress and trigger the instability by spinning the BH into the superradiant regime. Typical evolutionary tracks show that BHs either enter the instability after sufficient mass growth or begin within it and rapidly spin down. In both cases, the system approaches an attractor trajectory in which the dimensionless spin follows the critical superradiant threshold.

\begin{figure}[ht]
	\centering
	\includegraphics[width=0.48\textwidth]{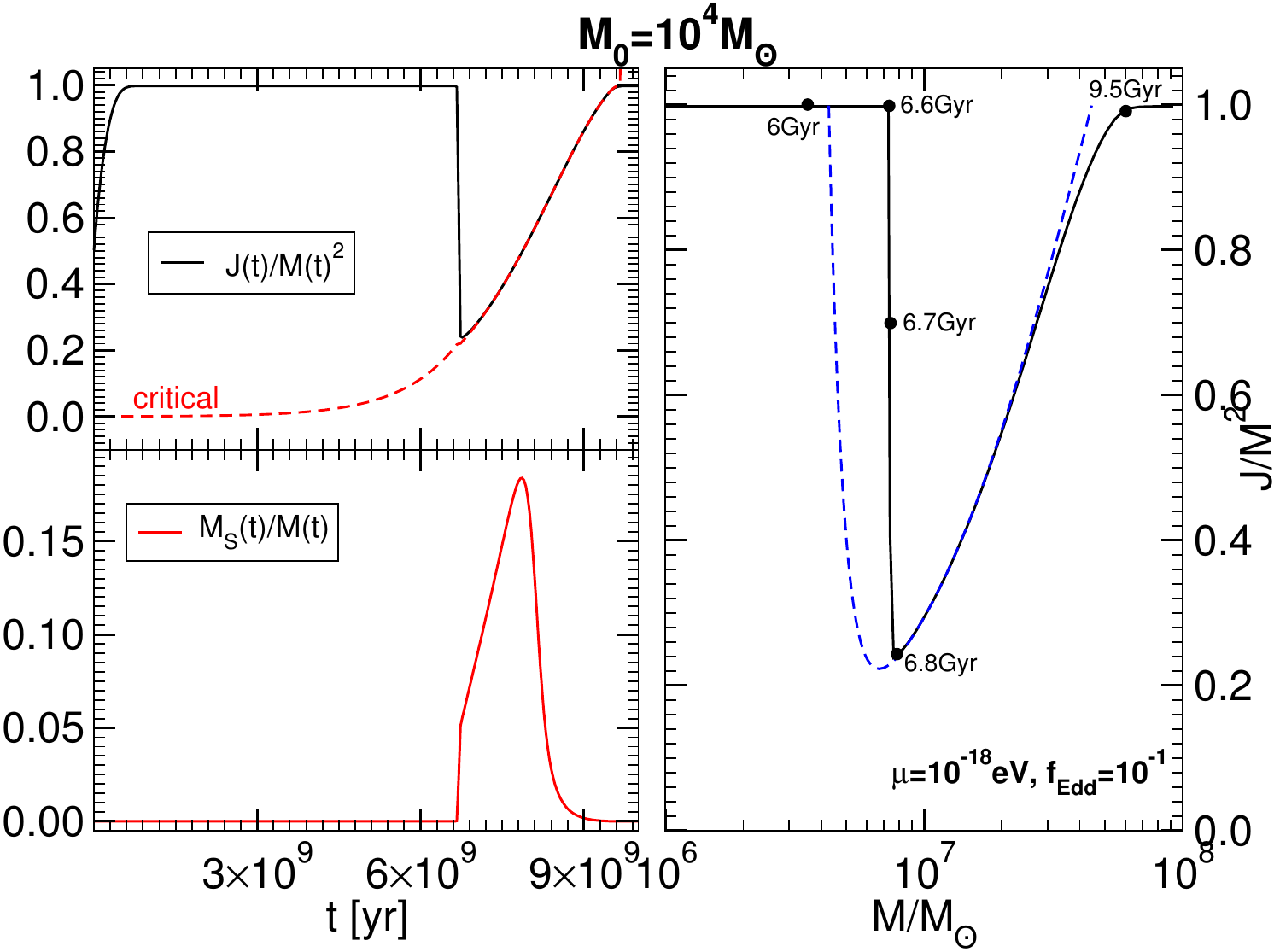}
	\includegraphics[width=0.48\textwidth]{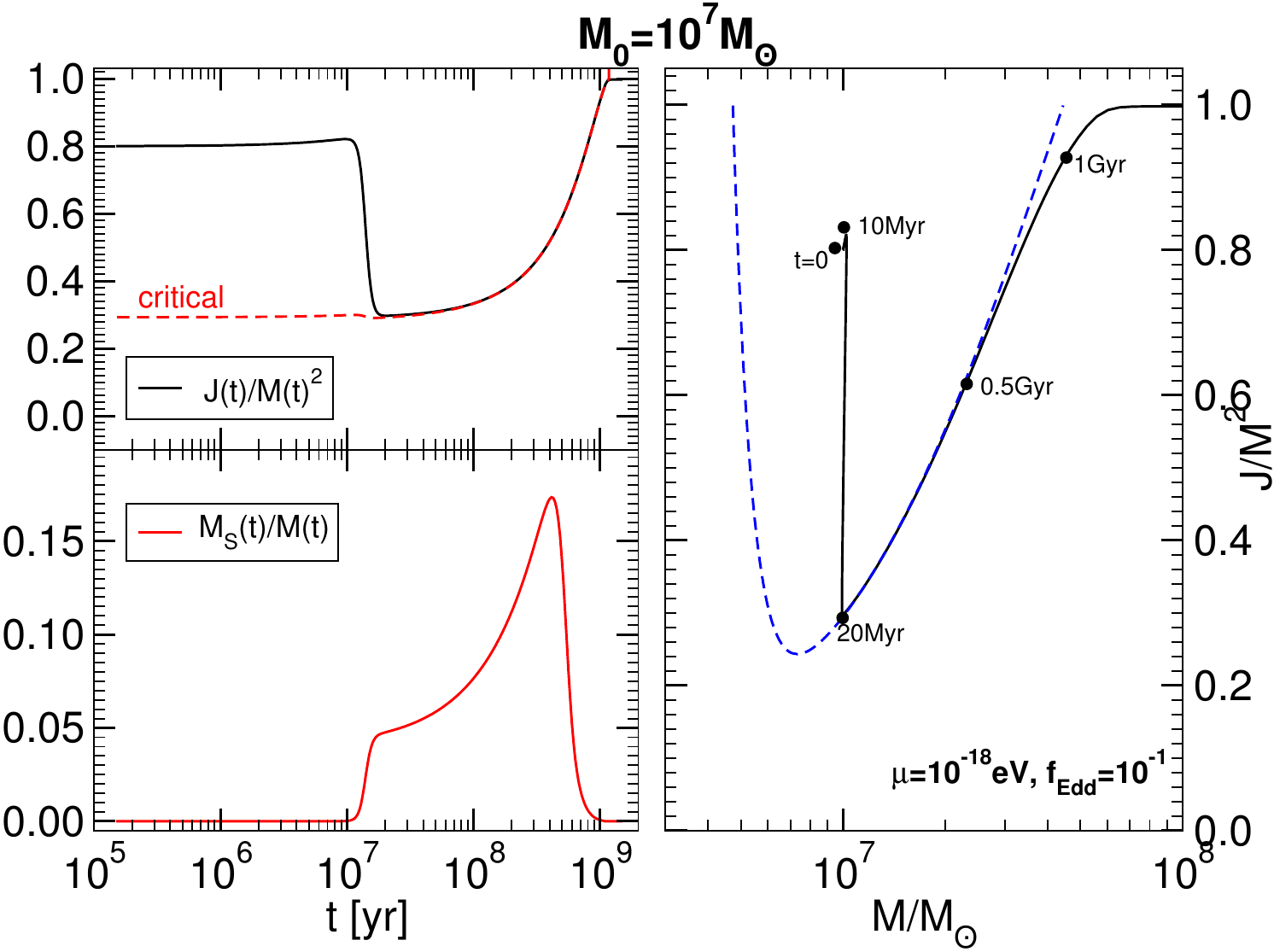}
	\caption{\label{fig:evolution}
		Evolution of BH mass, spin, and scalar cloud mass for $\mu_S=10^{-18}$\,eV, $f_{\rm Edd}=0.1$. 
		\textbf{Left:} Case 1: $M_0=10^4M_\odot$, $a_0/M_0=0.5$. BH enters SRI region at $t\sim6$\,Gyr when $M\sim10^7M_\odot$. 
		\textbf{Right:} Case 2: $M_0=10^7M_\odot$, $a_0/M_0=0.8$. BH begins in SRI region. 
		Top panels show spin $J/M^2$ and critical threshold $a_{\rm crit}/M$; bottom panels show $M_s/M$. Dashed blue lines indicate depleted region where $\tau_{\rm SRI}\sim\tau_{\rm ACC}$. From Ref.~\cite{Brito:2014wla}.
	}
\end{figure}

Representative results in Fig.~\ref{fig:evolution} illustrate two regimes: \textbf{Case 1:} Initially negligible SRI ($\mu_S M_0\sim10^{-4}$). Accretion drives spin to extremality ($J/M^2\sim0.998$) within $\tau_{\rm ACC}\sim10\tau_{\rm Salpeter}$, where $\tau_{\rm Salpeter}\sim4.5\times10^7$\,yr is the Salpeter timescale~\cite{Salpeter:1964kb}. At $t\sim6$\,Gyr, $\mu_S M$ becomes significant; cloud grows exponentially, extracting BH mass and spin until $a\sim a_{\rm crit}$. Subsequent evolution follows $J/M^2\sim a_{\rm crit}/M$ as an attractor. \textbf{Case 2:} BH initially within SRI regime. Spin increases slightly before superradiant extraction dominates ($\sim10$\,Myr). Post-extraction, evolution tracks $a_{\rm crit}/M$ during continued accretion. In both cases, GW emission dissipates cloud energy while accretion replenishes BH mass/spin. The Regge trajectory $J(t)/M(t)^2\sim a_{\rm crit}/M$ serves as an evolutionary attractor until accretion saturates near extremal spin.

\begin{figure}[ht]
	\centering
	\includegraphics[width=0.65\textwidth]{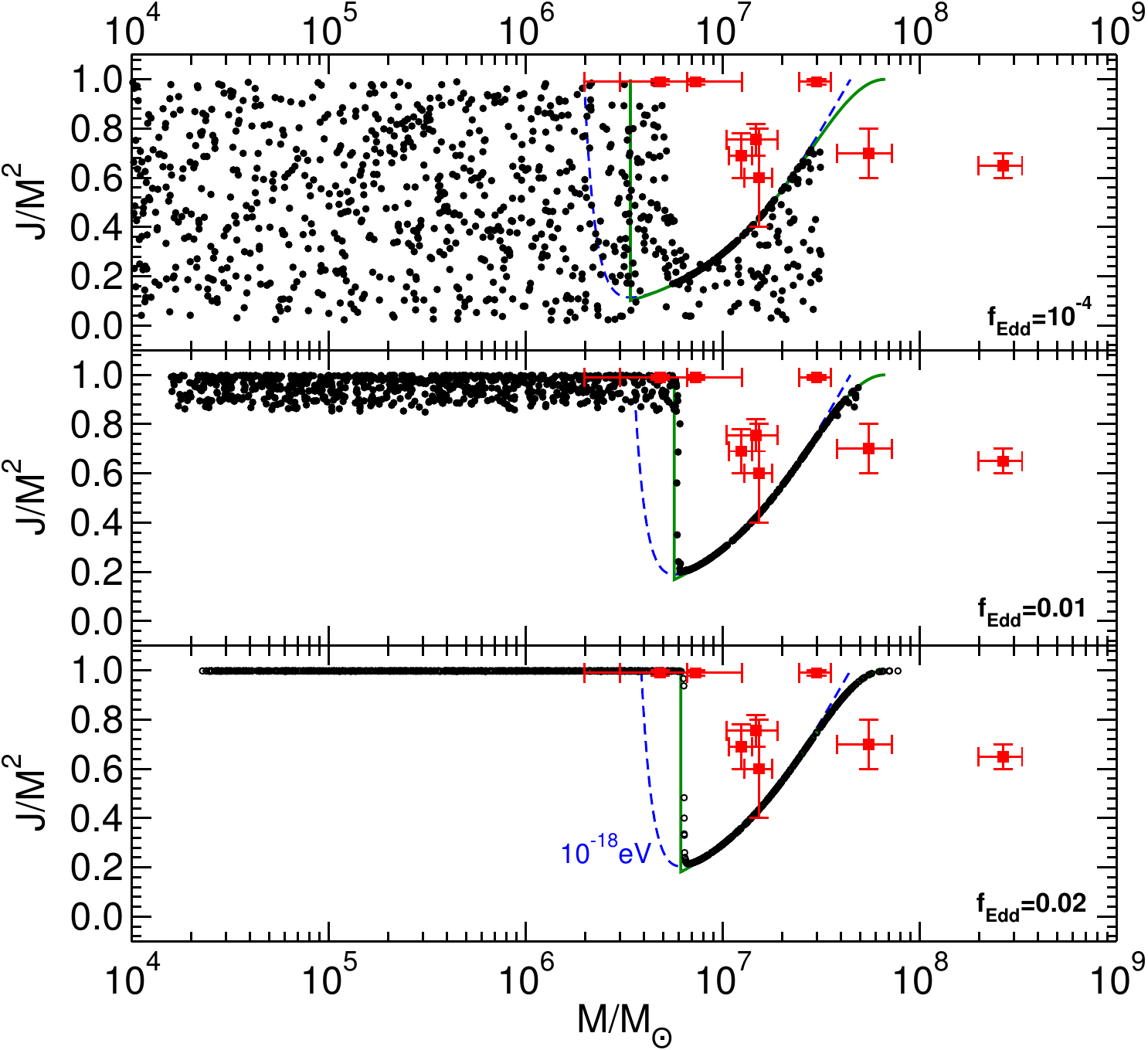}
	\caption{Final BH mass and spin in the Regge plane for $N=10^3$ BHs with random initial $\log_{10} M_0 \in [4,7.5]$ and $J_0/M_0^2 \in [0.001,0.99]$, extracted at $t_F \sim 2\times 10^9$ years (Gaussian width $\sigma=0.1\bar t_F$). Scalar mass $\mu_S=10^{-18}$\,eV. Dashed blue: linearized threshold $\tau \approx \tau_{\mathrm{Salpeter}}/f_{\mathrm{Edd}}$. Solid green: depleted region from Eq.~\eqref{region}. Points with error bars are supermassive BH data from Ref.~\cite{Brenneman:2011wz}. Adapted from Ref.~\cite{Brito:2014wla}.}
	\label{fig:ReggeMC}
\end{figure}

The combined effect of accretion and superradiance is conveniently visualized in the Regge plane, where BHs evolve toward curves defined by the superradiant condition. Regions above these curves become depleted because BHs cannot remain there without spinning down. For scalar fields, the approximate boundary of the depleted region can be written as
\begin{equation}
	\frac{J}{M^2} \gtrsim \frac{a_{\mathrm{crit}}}{M} \sim 4\mu M \quad \cup \quad M \gtrsim \left(\frac{96}{\mu^{10} \tau_{\mathrm{ACC}}}\right)^{1/9},
	\label{region}
\end{equation}
as shown by the solid green line in Fig.~\ref{fig:ReggeMC}. The absence of BHs within such regions would be inconsistent with the existence of a light boson of mass $\mu_S=10^{-18}$\,eV~\cite{Brenneman:2011wz}. Monte Carlo simulations indicate that these depleted regions are robust against variations in accretion rate and can be compared directly with observed BH spin distributions~\cite{Brenneman:2011wz}. Current observations already constrain ultralight bosons over a range of masses~\cite{Arvanitaki:2009fg,Arvanitaki:2010sy,Pani:2012vp,Brito:2013wya,Baryakhtar:2017ngi,Baryakhtar:2020gao,Witte:2024drg}. Higher multipoles with $l=m>1$ can also contribute to these depleted regions when their superradiant condition is satisfied, leading to a union of multiple exclusion zones~\cite{Ficarra:2018rfu}. During the SRI evolution, the bosonic cloud can retain a significant fraction of the BH mass over cosmological timescales while remaining sufficiently dilute that the spacetime geometry is well approximated by Kerr. Accretion therefore competes with superradiant extraction but can also catalyze it by spinning the BH into the instability regime~\cite{Guo:2025dkx,Sarmah:2024nst}. The presence of depletion regions in the Regge plane is thus a generic prediction of superradiance, and for near-Eddington accretion they are well described by Eq.~\eqref{region}. Similar qualitative behavior holds for vector fields, although with shorter instability and GW-emission timescales.

Since superradiant instabilities can significantly reduce the spin of astrophysical BHs, they may indirectly affect processes that depend on rotation, such as the launching of relativistic jets in active galactic nuclei. Relativistic jets from active galactic nuclei are widely believed to be powered either by accretion energy or by the rotational energy of the central BH. The Blandford-Znajek mechanism~\cite{Blandford:1977ds} provides a canonical example of the latter, extracting spin energy through magnetic fields anchored in an accretion disk. Frame dragging twists the magnetic field lines, generating an EM outflow that accelerates plasma into collimated jets along the rotation axis. Recent general-relativistic magnetohydrodynamic simulations support a significant role for this mechanism in jet production~\cite{Tchekhovskoy:2012up,McKinney:2012vh,Penna:2013rga}. The efficiency of the process correlates with BH spin and magnetic flux, offering a natural explanation for observed correlations between jet power and spin. Distinguishing between Blandford-Znajek and purely accretion-powered contributions remains challenging, but future observations and improved simulations are expected to clarify the relative roles of these mechanisms.

\subsection{Implications for ultralight bosons and DM}
\label{sec:ultralight_motivations}

SR occurs only for integer-spin bosonic fields. While all known elementary bosons are either massless (photon, gluons) or heavy ($W/Z$, Higgs $\sim 100\,\text{GeV}$), astrophysical BHs can efficiently amplify \textit{ultralight} bosons via SR when $\mu M \lesssim 1$. For stellar mass BHs, this requires $\mu \lesssim 10^{-10}\,\text{eV}$, while for supermassive BHs ($M \sim 10^9 M_\odot$) the relevant window is $\mu \lesssim 10^{-17}\,\text{eV}$. Such ultralight bosons arise naturally in SM extensions and modified gravity. Light axion with masses $\mu_a \sim 10^{-11}\,\text{eV}$ appear in string compactifications within the ``axiverse'' framework, which predicts multiple axion-like fields spanning a wide mass range down to $\sim 10^{-33}\,\text{eV}$~\cite{Arvanitaki:2009fg, Mehta:2020kwu}. Familions~\cite{Wilczek:1982rv} and Majorons~\cite{Chikashige:1980qk,Chikashige:1980ui} emerge from broken family or lepton-number symmetries and couple weakly due to the high symmetry-breaking scale. Dark photons as massive $U(1)$ vectors arise in hidden sectors~\cite{Goodsell:2009xc, Jaeckel:2010ni}, while massive spin-2 fields appear in nonlinear massive gravity and bimetric theories~\cite{Hinterbichler:2011tt, deRham:2014zqa}. Modified gravity models often introduce effective scalar degrees of freedom: in scalar--tensor theories a massive scalar can trigger SRIs even when no-hair theorems hold for massless fields~\cite{Sotiriou:2011dz}, and in $f(R)$ gravity the effective scalar is tied to the Ricci curvature and may drive deviations from the Kerr solution~\cite{Hersh:1985hz}. Astrophysical BHs therefore provide natural laboratories to probe ultralight bosons across mass scales inaccessible to terrestrial experiments.

\begin{figure}[t]
	\centering
	\includegraphics[width=0.75\textwidth]{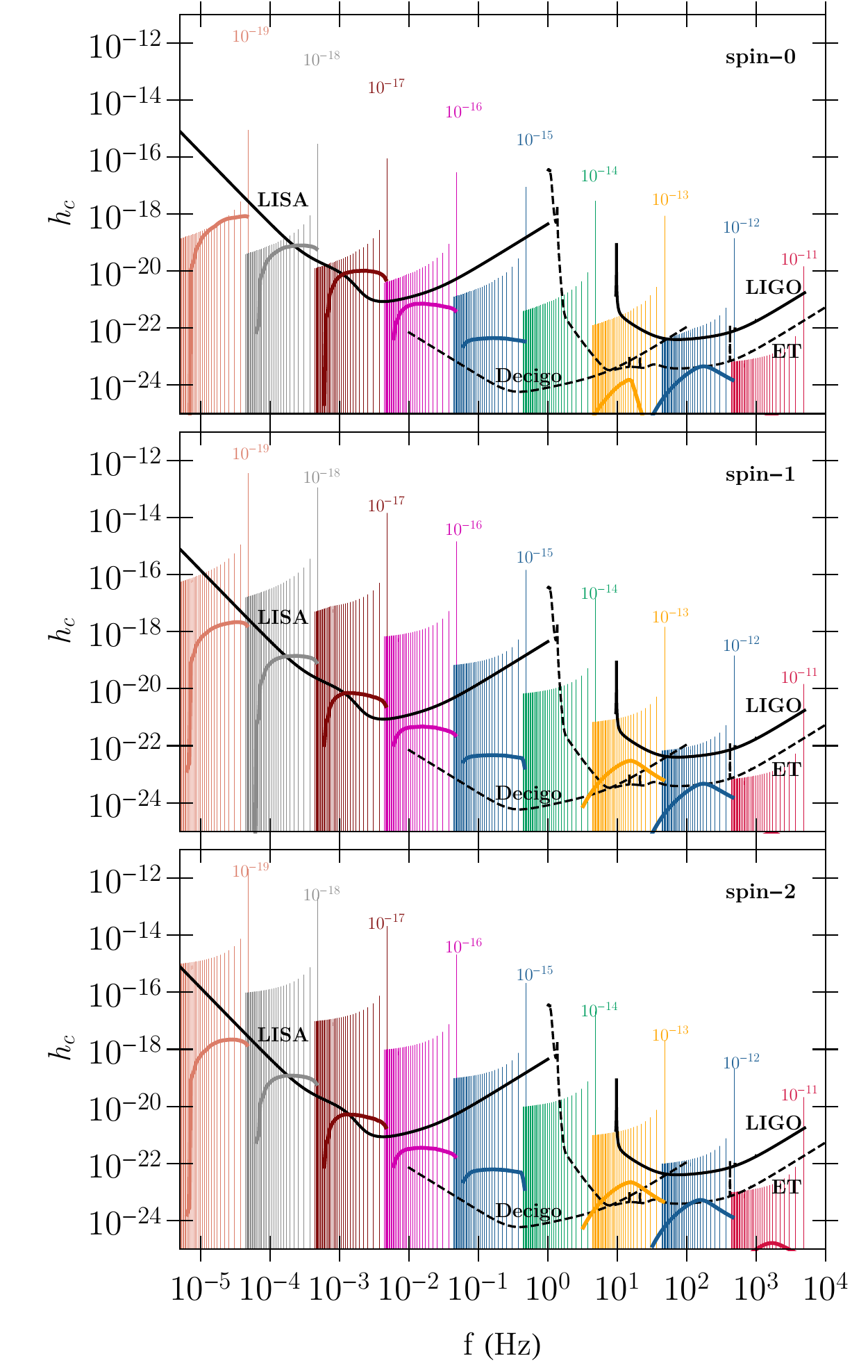}
	\caption{Characteristic strain for BH-boson condensates with scalar ($S=0$), vector ($S=1$), and tensor ($S=2$) fields. Lines show strains for specific boson masses at redshifts $z\in(0.001,10)$ (right to left). Noise curves for Advanced LIGO, LISA, DECIGO, and ET are shown. Adapted from Refs.~\cite{Brito:2017wnc,Brito:2020lup}.
		\label{fig:sensitivity}}
\end{figure}

SRIs generically extract angular momentum from spinning BHs, imposing upper bounds on their spins below the Kerr limit and producing distinctive GW signatures. These effects have been extensively studied for ultralight scalars~\cite{Arvanitaki:2009fg,Arvanitaki:2010sy,Brito:2014wla,Brito:2017zvb,Cardoso:2018tly}, vectors~\cite{Pani:2012vp, Baryakhtar:2017ngi}, and tensors~\cite{Brito:2013wya, Brito:2020lup}. The key parameter controlling the instability is $\mu M$, with maximal growth typically near $\mu M\sim0.5$ and faster rates for higher-spin fields. Once the boson cloud saturates, GW emission from the cloud-BH system provides observable signatures. A monochromatic boson cloud emits GWs via annihilation, with energy flux scaling as~\cite{Arvanitaki:2009fg,Arvanitaki:2010sy,Yoshino:2013ofa,Brito:2014wla,Arvanitaki:2014wva,Arvanitaki:2016qwi,Baryakhtar:2017ngi,East:2017mrj,Brito:2017zvb,Brito:2017wnc,Isi:2018pzk,East:2018glu}
\begin{equation}
	\dot{E}_{\rm GW}\propto \left(\frac{M_S}{M}\right)^2 (\mu M)^{4l+4S+10},
\end{equation}
where $M_S$ is the cloud mass and $S$ the boson spin. The radiation is monochromatic with frequency
\begin{equation}
	f_{\rm GW}\sim \frac{\omega_R}{\pi} \sim 5\,{\rm kHz}\left(\frac{\mu}{10^{-11}{\rm\,eV}}\right).
\end{equation}
The corresponding emission timescale is~\cite{Yoshino:2013ofa,Brito:2017zvb,Baryakhtar:2017ngi,Siemonsen:2019ebd,Brito:2020lup}
\begin{align}
	\tau_{\rm GW}^{S} &\approx 1.3 \times 10^{5}\, {\rm yr} \left(\frac{M}{10\, M_{\odot}}\right) \left(\frac{0.1}{M\mu_S}\right)^{15}\left(\frac{0.5}{\chi_i-\chi_f}\right),\\
	\tau_{\rm GW}^{V,T}&\approx 2\,\, {\rm days} \left(\frac{M}{10\, M_{\odot}}\right) \left(\frac{0.1}{M\mu_{V,T}}\right)^{11}\left(\frac{0.5}{\chi_i-\chi_f}\right),
\end{align}
where $\chi_{i,f}$ are initial and final dimensionless spins. The characteristic strain for dominant hydrogen modes ($M\mu \ll 1$) is~\cite{Yoshino:2013ofa,Arvanitaki:2014wva,Brito:2014wla,Brito:2017zvb,Baryakhtar:2017ngi,Siemonsen:2019ebd,Brito:2020lup}
\begin{align}
	h^{S} &\approx 5\times 10^{-27} \left(\frac{M}{10 M_{\odot}}\right)\left(\frac{M\mu_S}{0.1}\right)^7 \left(\frac{\rm Mpc}{d}\right) \left(\frac{\chi_i - \chi_f}{0.5}\right),\\
	h^{V,T} &\approx 10^{-23} \left(\frac{M}{10 M_{\odot}}\right)\left(\frac{M\mu_{V,T}}{0.1}\right)^5 \left(\frac{\rm Mpc}{d}\right) \left(\frac{\chi_i - \chi_f}{0.5}\right).
\end{align}
Figure.~\ref{fig:sensitivity} shows the characteristic strain for various boson and BH masses compared to detector sensitivities. LISA and third-generation ground-based detectors could detect sources at high redshifts, particularly for vector and tensor fields~\cite{Brito:2020lup}. 

\begin{figure}[thb]
	\begin{center}
	\begin{tabular}{ccc}
	   \includegraphics[width=0.25\textwidth]{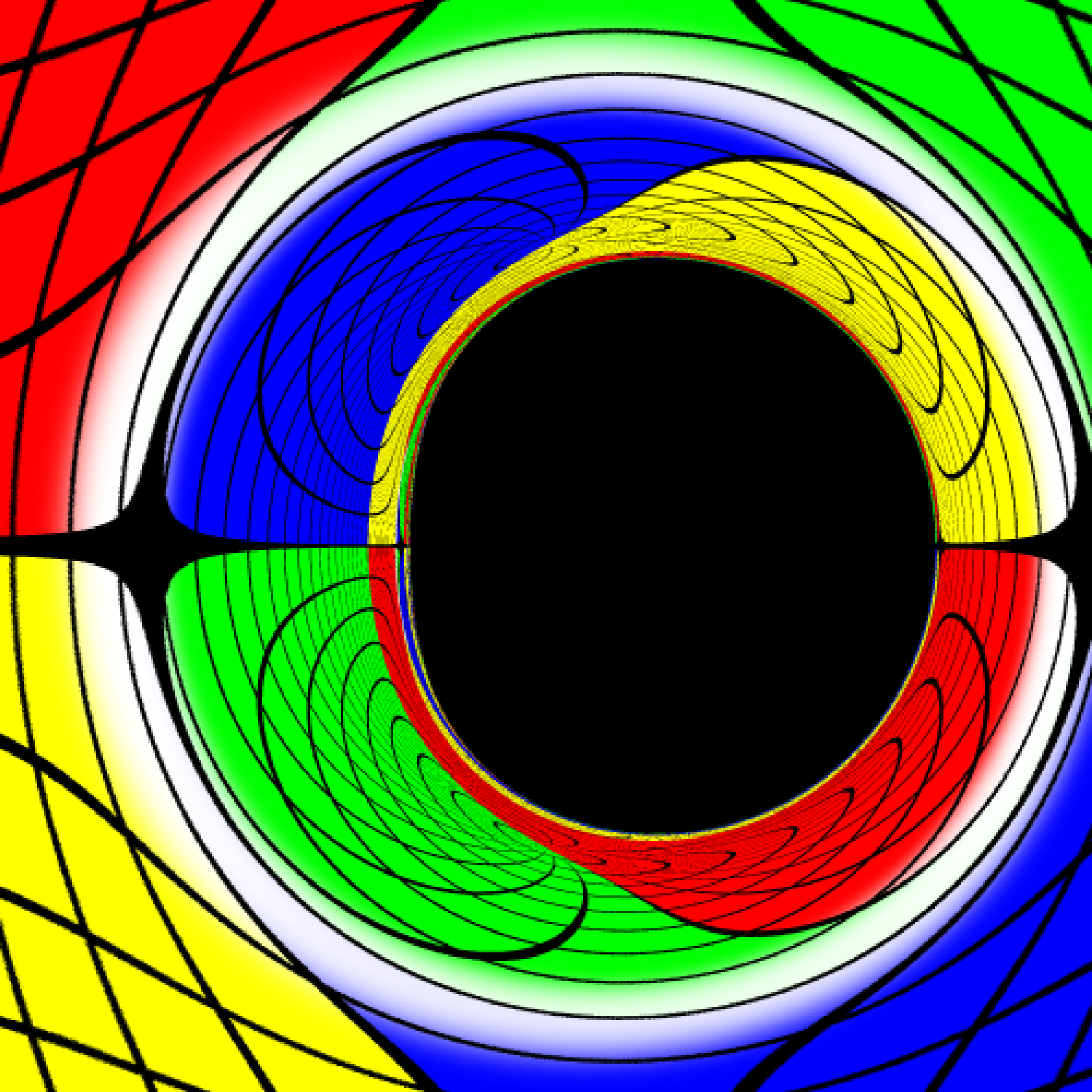} & \includegraphics[width=0.25\textwidth]{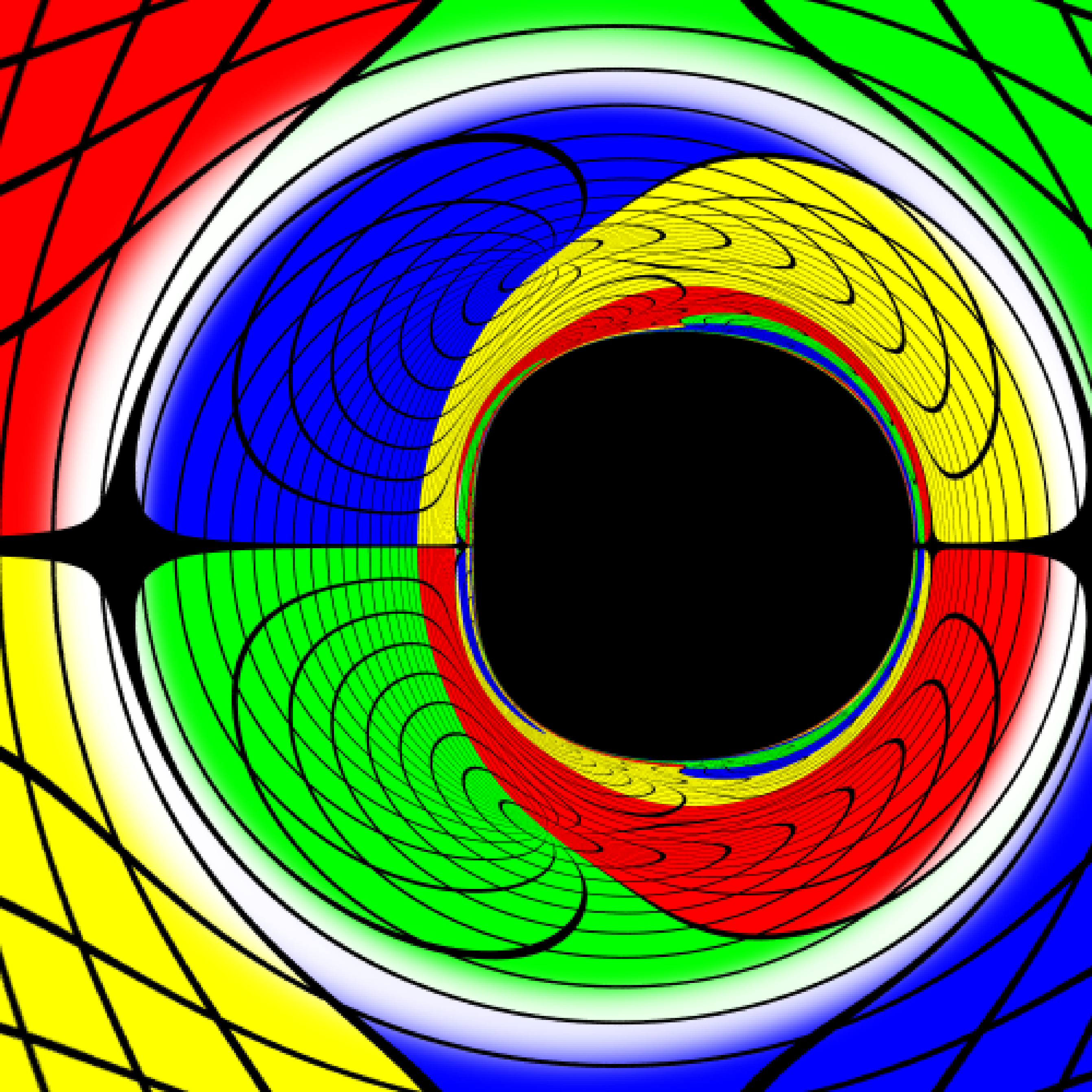}& \includegraphics[width=0.25\textwidth]{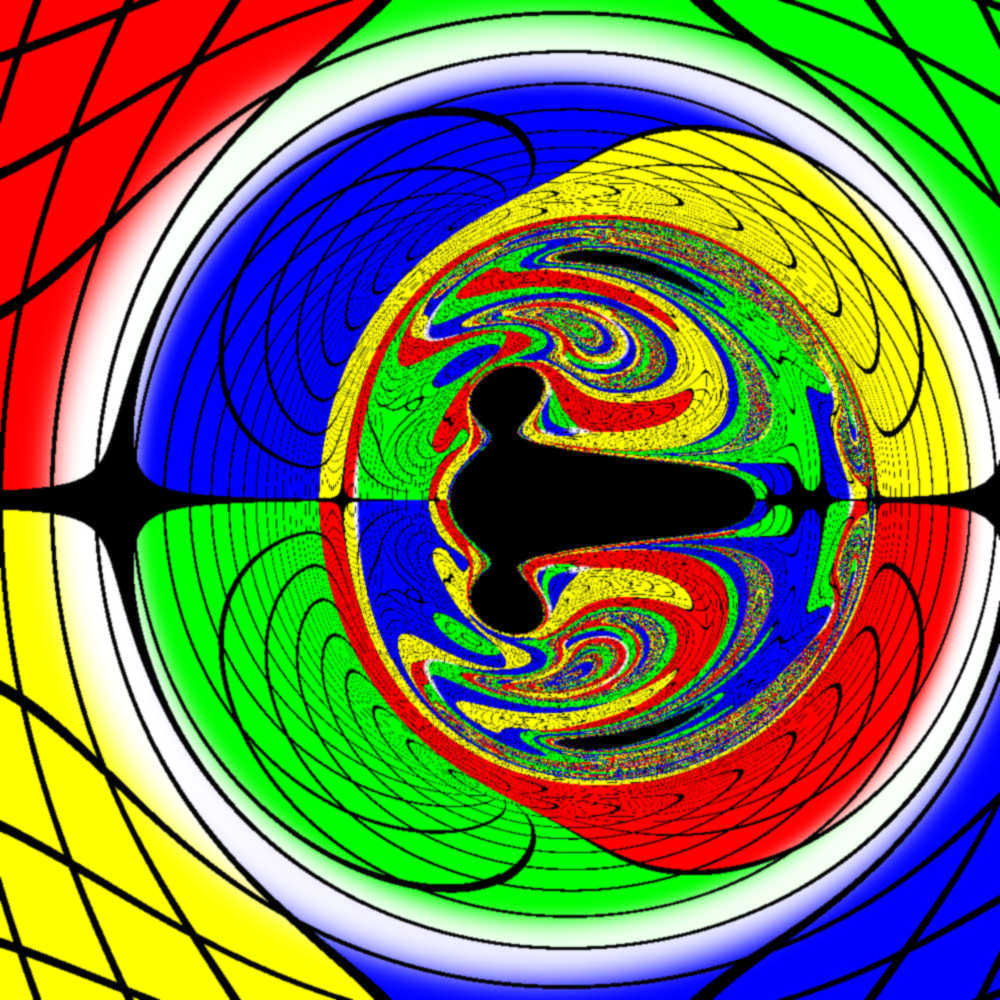}\\
	   \includegraphics[width=0.25\textwidth]{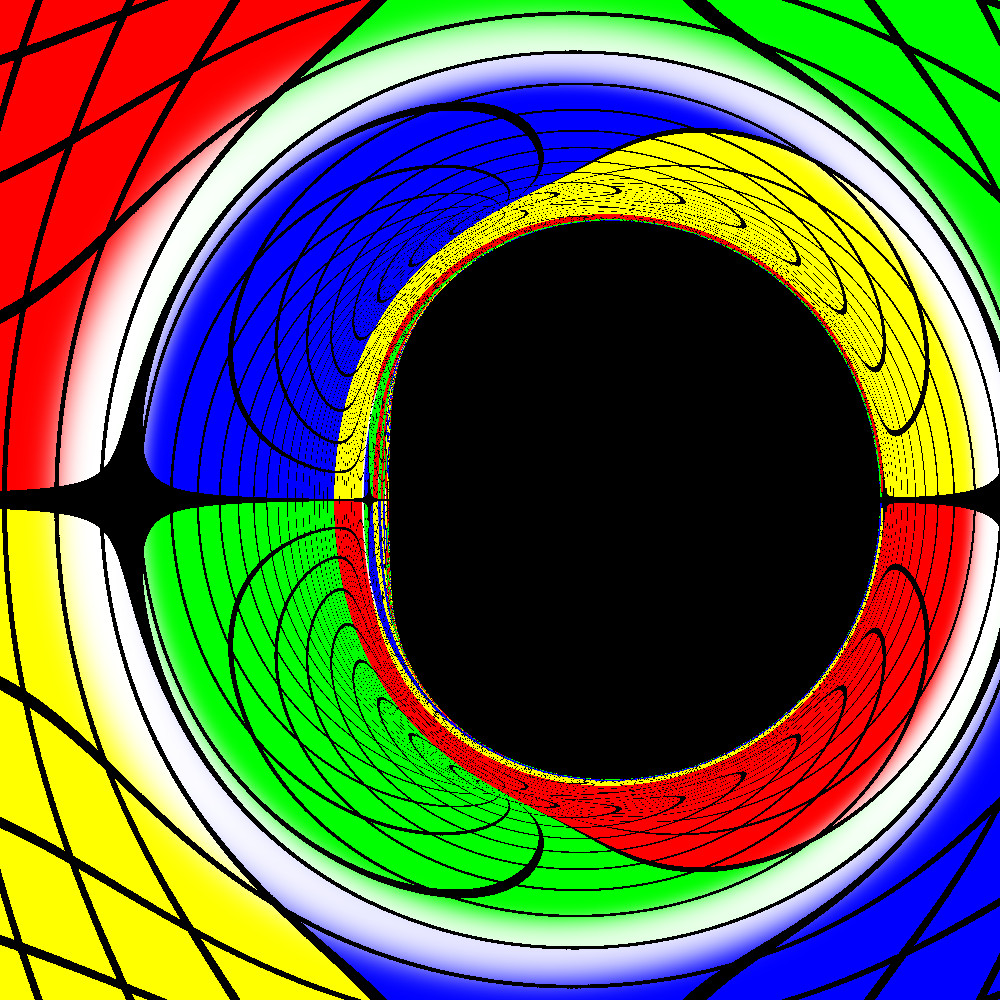} & \includegraphics[width=0.25\textwidth]{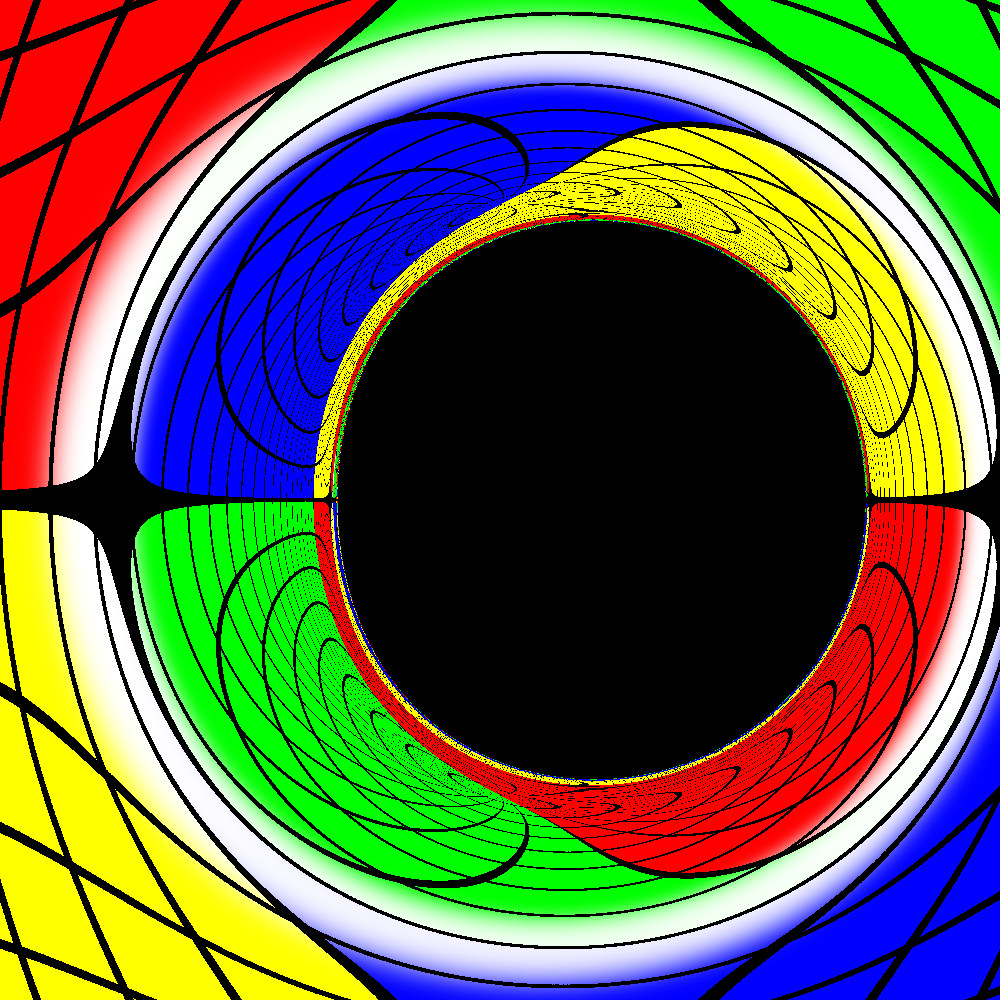}& \includegraphics[width=0.25\textwidth]{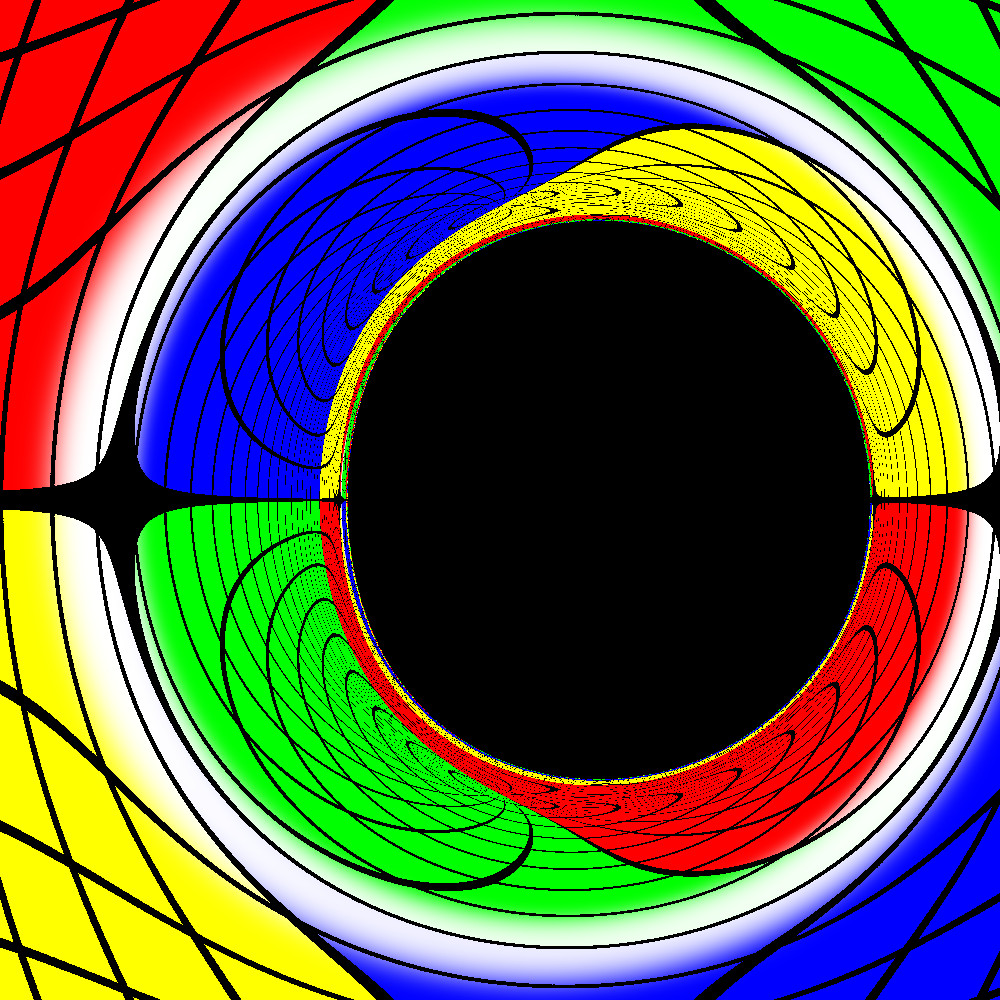}
	\end{tabular}
    \caption{Shadows of rotating BHs surrounded by a dipolar scalar cloud. The top row shows three scenarios in which the cloud contributes approximately $5\%$, $75\%$, and $98.2\%$ of the total ADM mass, moving from left to right. For each case, the corresponding Kerr BH shadow with the same ADM mass and spin is shown for comparison. The bottom row indicates the dimensionless spin values, approximately $a \approx 0.999M$, $0.85M$, and $0.894M$, from left to right. The strong gravitational lensing produced by the BH is illustrated by dividing the celestial sphere into four colored regions. This figure highlights how dense bosonic clouds can significantly alter the appearance of BH shadows. Figure adapted from Ref.~\cite{Cunha:2015yba}.}
    \label{fig:shadows}
	\end{center}
\end{figure}

Besides GW probes, EM counterparts provide complementary ways to detect superradiant clouds. One prominent example is BH shadows: dense bosonic clouds modify the spacetime geometry, altering photon geodesics and hence the observed shadow morphology~\cite{Cunha:2015yba, Cunha:2016bpi, Vincent:2016sjq, Bambi:2019tjh, Vagnozzi:2019apd, Vagnozzi:2022moj}. Fig.~\ref{fig:shadows} illustrates how scalar clouds of varying mass fractions affect the shadows of rotating BHs. While dense clouds produce pronounced deviations from the Kerr shadow, current Event Horizon Telescope (EHT) measurements do not yet rule out the presence of superradiantly-generated clouds around the supermassive BH M87$^\star$~\cite{Cunha:2019ikd}. In addition to shadow distortions, changes in the underlying geodesic structure can induce quasi-periodic oscillations in X-ray emissions from the accretion flow~\cite{Franchini:2016yvq}, offering another potential observable signature. Light or moderately dense clouds remain challenging to detect with current instruments, but future high-resolution imaging may enhance sensitivity. Moreover, if scalar fields couple to photons, boson clouds can produce bursts of EM radiation with characteristic frequency
\begin{equation}
	f_{\rm EM}\sim \frac{\omega_R}{4\pi} \sim 1200\,{\rm Hz}\left(\frac{\mu_S}{10^{-11}{\rm\,eV}}\right),
\end{equation}
potentially relevant for fast radio bursts in scenarios involving primordial BHs (PBHs) and heavier axions~\cite{Rosa:2017ury,Sen:2018cjt,Ikeda:2019fvj,Boskovic:2018lkj}.

PBHs extend the reach of superradiance to earlier cosmic epochs and to much smaller mass scales than those typically associated with astrophysical BHs. If PBHs exist across a broad mass spectrum~\cite{Carr:2020gox}, superradiant instabilities can operate efficiently in the early Universe, modifying PBH spin distributions and generating GW or EM signals. The instability develops whenever its characteristic timescale is shorter than the relevant competing processes, namely the age of the Universe $t_U$, Hawking evaporation $\tau_{\text{evap}}$, or accretion $\tau_{\text{acc}}$, as
\begin{equation}
	\tau_{\text{SRI}} < \min\left(t_U, \tau_{\text{evap}}, \tau_{\text{acc}}\right).
\end{equation}
When this inequality is satisfied, superradiant growth can significantly alter the PBH evolution. Conversely, the absence of such effects in observational data constrains the properties of ultralight bosons by requiring consistency with observed BH spins or with the allowed PBH population~\cite{Arvanitaki:2010sy,Stott:2020gjj,Bernal:2022oha}. In this way, PBHs provide a complementary laboratory for probing BSM particles across a wide range of masses and couplings.

The mass-spin evolution of PBHs is particularly important for determining the efficiency of superradiance. Although PBHs may form with negligible initial spin, several mechanisms can subsequently endow them with substantial rotation, including mergers, accretion, or torques from surrounding matter and radiation fields. The evolution of the dimensionless spin parameter can be written schematically as
\begin{equation}
	\frac{{\rm d}a}{{\rm d}t} = \left.\frac{{\rm d}a}{{\rm d}t}\right|_{\text{SRI}} + \left.\frac{{\rm d}a}{{\rm d}t}\right|_{\text{acc}} + \left.\frac{{\rm d}a}{{\rm d}t}\right|_{\text{merger}},
\end{equation}
where the SRI contribution extracts angular momentum when $\omega < m\Omega_+$. For ultralight scalar fields, the instability saturates through bosenova collapse or GW emission, leading to spin-down to the threshold $\omega \approx m\Omega_+$~\cite{Brito:2015oca}. The resulting evolution imprints characteristic features on the PBH mass-spin distribution that can be searched for observationally. Through this mechanism, PBH superradiance provides a means of probing particle physics scenarios that are inaccessible to terrestrial experiments. PBHs in the mass range $M\sim10^{-12}$–$10^{-8}M_\odot$ are sensitive to axion-like particles in the QCD axion window with masses $m_a\sim10^{-12}$–$10^{-8}\,$eV~\cite{Arvanitaki:2014wva}. Stellar-mass PBHs can instead probe vector bosons such as dark photons with masses in the range $m_V\sim10^{-18}$–$10^{-10}\,$eV~\cite{Pani:2012vp}. In all of these cases, the formation of gravitationally bound boson clouds around spinning PBHs produces configurations sometimes referred to as ``gravitational atoms,'' whose dynamics lead to continuous GW emission that may be detectable by space-based observatories such as LISA or by future ground-based detectors~\cite{Baryakhtar:2017ngi}. The presence or absence of these signals can therefore be used to constrain new light degrees of freedom.

A variety of observable signatures may arise from superradiant dynamics around PBHs. The boson clouds that form around spinning PBHs emit quasi-monochromatic GWs with characteristic frequency $f_{\text{GW}}\approx\mu/\pi$, potentially within the sensitivity range of LISA, the Einstein Telescope, or Cosmic Explorer. In scenarios involving axions coupled to photons, nonlinear collapse events such as bosenovae may also produce transient EM signals, including gamma-ray bursts arising from axion decays $a\to\gamma\gamma$~\cite{Yoshino:2012kn}. On a population level, PBHs with masses in the range $\sim1$–$100M_\odot$ are expected to exhibit suppressed spins, typically $a_\ast\lesssim0.1$, if ultralight bosons exist within the corresponding superradiant mass windows~\cite{Bernal:2022oha}. Furthermore, photon emission from decaying boson clouds could contribute to early-Universe reionization, providing an additional cosmological probe of these scenarios~\cite{Stott:2020gjj}. Current constraints on PBH abundances already exclude portions of parameter space for ultralight bosons. For example, scalar fields with masses $m_s\sim10^{-13}\,$eV are incompatible with scenarios in which PBHs with masses $M_{\text{PBH}}\sim0.1$-$1M_\odot$ constitute the entirety of DM~\cite{Chen:2020cef}. Future GW observatories will significantly enhance sensitivity to these effects, allowing either the detection of individual boson clouds or statistical studies of PBH populations through their collective GW backgrounds. The combined study of PBH formation, cosmic evolution, and superradiant instabilities therefore provides a powerful multi-messenger framework for exploring both BH physics and particle phenomenology.

In addition to constraining ultralight bosons, superradiance from PBHs offers a compelling mechanism for the non-thermal production of DM in the early Universe. PBHs lose mass through Hawking evaporation, but their angular momentum can simultaneously drive superradiant instabilities, establishing a competitive channel for generating heavy DM~\cite{Lennon:2017tqq, Allahverdi:2017sks, Hooper:2019gtx, Gondolo:2020uqv, Masina:2020xhk, Bernal:2022oha}. This interplay introduces a resonant production mechanism that complements conventional gravitational production channels. Standard gravitational production of DM from light PBHs, whether through Hawking evaporation or ultraviolet freeze-in, is typically efficient only for particle masses below the PBH horizon temperature. Superradiance instead provides a resonant amplification mechanism capable of producing heavier bosonic particles, including those with masses in the TeV range. The amplification is maximized when the Compton wavelength of the bosonic DM candidate matches the gravitational scale of the PBH, corresponding to the dimensionless parameter $M_{\rm BH} m_{\rm dm} \simeq \mathcal{O}(1)$, where $M_{\rm BH}$ is the PBH mass and $m_{\rm dm}$ the DM particle mass. For PBHs with masses $M_{\rm BH}\lesssim10^9$ g, which evaporate prior to BBN, this resonance can efficiently populate bound gravitational clouds of heavy bosonic states that would otherwise be underproduced~\cite{Bernal:2022oha}. The dynamics of this process involve the coupled evolution of the PBH mass $M_{\rm BH}$, the dimensionless spin parameter $a_\ast$, and the occupation number $N_{\rm dm}$ of particles in the gravitational cloud. Superradiance extracts rotational energy and angular momentum from the PBH, reducing its spin, while Hawking radiation continuously decreases its mass. At the same time, the bosonic occupation number grows exponentially during the superradiant phase when the resonance condition $\alpha\sim1$ is satisfied. As shown in~\cite{Bernal:2022oha}, efficient superradiance can substantially enhance the final DM abundance relative to scenarios that include Hawking evaporation alone. Consequently, achieving the observed DM relic density may require a smaller initial PBH energy fraction $\beta\equiv\rho_{\rm BH}(T_{\rm in})/\rho_R(T_{\rm in})$, thereby relaxing constraints on PBH formation models.

The cosmic neutrino background (C$\nu$B) can suppress SRI via Pauli blocking if the boson couples to neutrinos ($a\to\nu\bar{\nu}$). The decay timescale $\tau_{\rm decay}$ competes with the superradiant timescale $\tau_{\rm SR}$. When $\tau_{\rm decay}\lesssim\tau_{\rm SR}$, the instability is suppressed, modifying exclusion bounds on ultralight bosons~\cite{Lambiase:2025twn}. This effect was stronger at higher redshifts when the C$\nu$B was denser.

\subsection{BH Lasers: Stimulated axion decay in SR clouds}

The dense environment of a superradiant cloud can profoundly alter the decay dynamics of the bosonic particles within it. While the spontaneous decay of axions via $\phi \to \gamma \gamma$ is negligible over cosmological timescales, a dense axion cloud can undergo stimulated decay, leading to a coherent, explosive emission of photons. This process results in the formation of an extremely bright, transient astrophysical source: a BH Laser powered by Axion SuperradianT instabilities (BLAST)~\cite{Rosa:2017ury}.

Within a dense axion cloud, the phase space density of photons can become high enough that the probability of an axion decaying is significantly enhanced by the presence of other photons (stimulated emission). The evolution of the photon number density $n_\gamma$ is governed by a Boltzmann equation that includes terms for spontaneous decay, stimulated decay, and inverse decays (photon annihilation). The key condition for lasing is that the stimulated decay rate overwhelms both the spontaneous decay and the photon escape rate from the cloud. Analysis shows this occurs when the number of axions in the cloud exceeds a critical threshold:
\begin{equation}
	N_\phi^c \simeq \frac{\Gamma_e}{2A\Gamma_\phi},
\end{equation}
where $\Gamma_e \sim c/r_0$ is the photon escape rate ($r_0$ is the cloud's effective Bohr radius), $\Gamma_\phi = \tau_\phi^{-1}$ is the spontaneous decay width, and $A$ is a geometric factor depending on the dimensionless coupling $\alpha_\mu$. This threshold can be attained in superradiant clouds around stellar-mass and PBHs for axion masses $\mu \gtrsim 10^{-8}$\,eV. Once the lasing threshold is crossed, the system dynamics are described by coupled equations for the axion number $N_\phi$ and photon number $N_\gamma$:
\begin{align}
	\frac{{\rm d}N_\phi}{{\rm d}t} &= \Gamma_s N_\phi - \Gamma_\phi \left[ N_\phi (1 + A N_\gamma) - B_1 N_\gamma^2 \right], \\
	\frac{{\rm d}N_\gamma}{{\rm d}t} &= -\Gamma_e N_\gamma + 2\Gamma_\phi \left[ N_\phi (1 + A N_\gamma) - (B_1+B_2) N_\gamma^2 \right],
\end{align}
where $\Gamma_s$ is the SR growth rate, and $A$, $B_1$, $B_2$ are constants. The numerical solution reveals a rapid, non-linear feedback: a sharp rise in $N_\gamma$ due to stimulated decay depletes the axion cloud ($N_\phi$), which then quenches the lasing. Photons escape, the axion cloud regrows via SR, and the cycle can repeat, leading to a series of damped oscillations manifesting as discrete, intense laser bursts.

The peak luminosity, total energy, and duration of these bursts can be estimated. For a representative axion mass of $\mu \sim 10^{-5}$\,eV around a BH of mass $M_{\text{BH}} \sim 10^{24}$ kg (parameters where PBHs could constitute all DM), the bursts occur in the millisecond duration range, emit in the GHz radio band, and have peak luminosities of
\begin{equation}
	L_B \sim 10^{42}\,\text{erg/s}.
\end{equation}
This places their phenomenal brightness and temporal signature in the same realm as observed Fast Radio Bursts (FRBs), suggesting a potential physical link.

An intriguing additional physical effect can modulate the BLAST phenomenon: Schwinger pair production. The intense EM field during a laser burst can be strong enough to spontaneously produce electron-positron pairs, creating a critical plasma within the axion cloud. This plasma can absorb photons and quench the lasing. The laser can only restart once this plasma annihilates and its density drops sufficiently, introducing a delay between bursts and contributing to the episodic nature of the emission. The BLAST mechanism establishes a direct astrophysical probe connecting fundamental particle physics (axion mass $\mu$, coupling constant) with BH physics (mass $M_{\text{BH}}$, spin $a_*$). The predicted transient, coherent radio emission offers a multi-messenger signature that could be identified in radio surveys. Furthermore, it provides a specific formation channel for highly spinning PBHs (required for SR) through subsequent mergers of initially non-rotating PBHs.

\subsection{Bounds on ultralight particles from pulsar observations}
\label{sec:bounds_pulsar}

Pulsars with measured spindown timescales $\tau_{\text{spindown}}$ provide stringent probes of SRIs involving ultralight bosons. Systems where the predicted SRI timescale $\tau$ satisfies $\tau < \tau_{\text{spindown}}$ are excluded, yielding constraints on boson masses and couplings. For isolated BHs, current spin measurements (e.g., LMC X-3, Cygnus X-1) yield only weak bounds; pulsar timing offers far greater precision. Many pulsars in the ATNF catalogue~\cite{Manchester:2004bp} have spin frequencies $f_{\text{spin}} \simeq 500\text{--}700$ Hz and spindown timescales $\tau_{\text{spindown}} \gtrsim 10^9$–$10^{10}$ yr. Three representative systems are:
\begin{itemize}
	\item PSR J1938+2012: $f_{\text{spin}} \simeq 380$ Hz, $\tau_{\text{spindown}} \simeq 1.1 \times 10^{11}$ yr
	\item PSR J1748-2446ad: $f_{\text{spin}} \simeq 716$ Hz, $\tau_{\text{spindown}} > 7.6 \times 10^7$ yr
	\item PSR B1957+20: $f_{\text{spin}} \simeq 622$ Hz, $\tau_{\text{spindown}} \simeq 3 \times 10^9$ yr
\end{itemize}

\begin{figure}[t]
	\centering
	\includegraphics[width=0.7\textwidth]{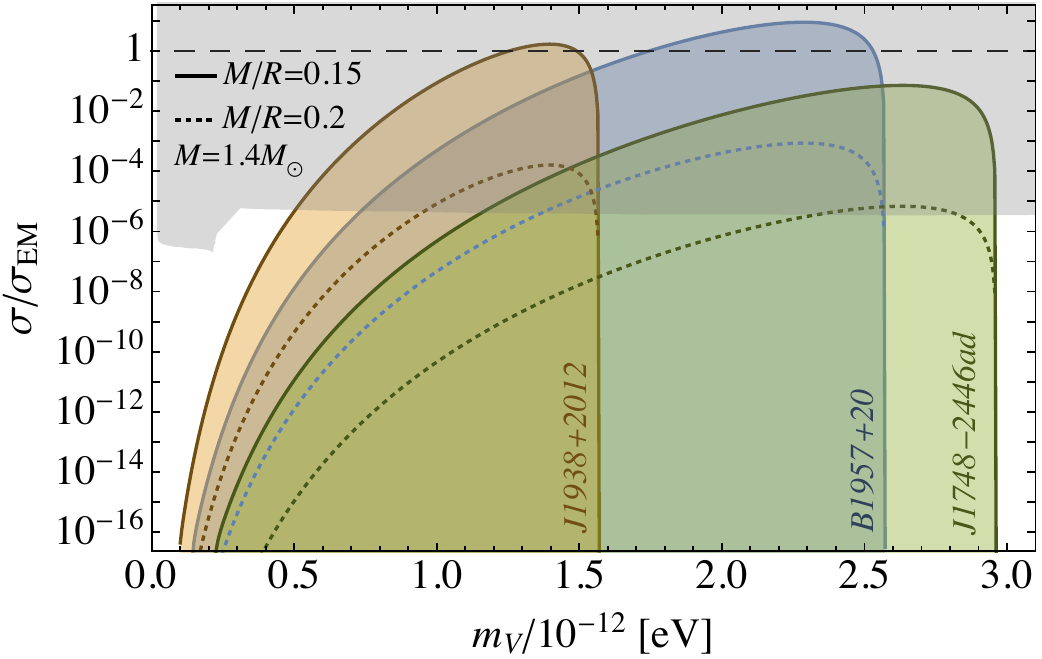}
	\caption{
		Exclusion regions in the $\sigma/\sigma_{\text{EM}}$–$m_V$ plane, derived from pulsar spindown measurements. 
		Colored shaded regions correspond to $\tau < \tau_{\text{spindown}}$ for each pulsar (solid: $M/R = 0.15$; dotted: $M/R = 0.2$). 
		Gray shading shows the CMB distortion constraint~\cite{Mirizzi:2009iz}. 
		Horizontal dashed line indicates standard electric conductivity $\sigma = \sigma_{\text{EM}}$.
		Figure adapted from Ref.~\cite{Cardoso:2017kgn}.
	}
	\label{fig:pulsar}
\end{figure}

For a conducting star, the axial Proca mode instability (with $l=m=1$) in the small $\mu_V M$, small $\Omega/\Omega_K$ limit is approximated by~\cite{Cardoso:2017kgn}
\begin{align}
	\omega_R^2 &\sim \mu_V^2 \left(1 - \frac{\mu_V^2 M^2}{8}\right), \\
	\omega_I &\sim -\left[\frac{\alpha_1 \sigma M}{\alpha_2 + (\sigma M)^{3/2}}\right] (\mu_V M)^8 (\mu_V - m\Omega), \label{wIaxial}
\end{align}
where $\sigma$ is conductivity and $\alpha_i$ are compactness-dependent constants. Constraints are derived by requiring $\tau \equiv 1/|\omega_I| < \tau_{\text{spindown}}$, with Fig.~\ref{fig:pulsar} showing the resulting excluded regions. Similar analyses apply to other boson-star coupling scenarios~\cite{Kaplan:2019ako}. A comprehensive summary of current bounds on ultralight scalar, vector, and tensor fields from SRI considerations is provided in Table 5.1 of Ref.~\cite{Brito:2015oca}. For spin-2 fields, only hydrogenic modes have been fully analyzed; the special dipolar mode~\cite{Brito:2013wya} could yield stronger constraints but remains limited by perturbative treatments.

\newpage

%% file: 7_Compact_stars_properties.tex
\section{Impact of dark matter on the characterization of compact star properties and possible spectral signatures}
\label{sec:NS}

In this section, we review NSs admixed with bosonic and fermionic DM and examine how the presence of DM can modify NS properties. We discuss the resulting imprints on GW signals, as well as the possible formation of exotic compact objects. We also summarize potential spectral signatures of specific DM candidates, with particular emphasis on axions and Majorons.

\subsection{Probing Dark Matter through its impact on neutron star properties}

“Another proposed way to probe DM properties is to study its effects on compact stars, such as NSs, compact remnants of massive stars and exist under conditions of extreme gravity and high matter densities on the order of $\gtrsim 10^{14}\,$g/cm$^3$, or GeV/fm$^3$. Such extreme conditions make NSs natural laboratories for probing new physics, including DM. The basic idea is that DM could accumulate in or around NSs, for example, through accretion or via interactions of a DM particle with the dense nuclear matter. The presence of DM could then modify the observable NS properties, such as radius, gravitational mass, or tidal deformability. The dynamical phenomenology during NS mergers might also be affected. All of these quantities have been measured through EM observations (e.g. NICER, XMM-Newton) and/or GWs. The magnitude and shape of DM effects on the NS observables depend on the DM mass, interaction strength, and DM model (e.g., if DM is a fermion or boson). It is therefore possible to use compact stars to probe the properties of DM through them. We give here a first overview of the relevant concepts. Other, more detailed, review papers on this topic were written by, e.g., Grippa et al.~\cite{Grippa:2024ach} and Bramante et al.~\cite{Bramante:2023djs}.

\subsubsection{DM accumulation in and around compact objects}

There are multiple channels through which DM could accumulate in or around NSs. The viability of each channel depends on the specific DM model. We discuss some of them in the following.

\paragraph{Accretion of DM particles onto NSs.}
First, we discuss the accretion of DM particles onto NSs. This scenario typically assumes that DM consists of massive particles in the GeV range, which have a finite interaction cross-section with neutrons and/or protons but do not self-interact significantly. A NS in the galactic disk will be permeated by DM particles depending on the average DM velocity and density. Some of these DM particles will interact with nucleons in the NS. If the DM particle mass is comparable to or larger than the nucleon mass, it can be captured in the NS's gravitational well and eventually thermalize. This capture mechanism can lead to the accumulation of a significant amount of DM, which we estimate below.

The first study on this topic was carried out by Goldmann and Nussinov~\cite{Goldman:1989nd}. They considered massive DM particles that can interact weakly with nucleons. If the DM particles lose enough energy by scattering with nucleons in their first passage through the NS, they can become gravitationally bound and captured. Goldmann and Nussinov estimated that this occurs for DM particles below $1.3\times 10^6\,$GeV, with nucleon-DM cross-sections greater than $2.5\times 10^{-45}\,$cm$^2$. With this, they found that the amount of DM accumulated inside a NS as a function of time, $t$, is
\begin{align}
    M_\mathrm{DM} = 5.4\times 10^{-16}\,M_\odot \, R_{12\mathrm{km}} \, M_{1.4M_\odot}  \left( \frac{\rho_\mathrm{DM}}{0.008\,M_\odot /\mathrm{pc}^{3}} \right) \left( \frac{220\,\mathrm{km/s}}{v_\mathrm{DM}} \right) \, \frac{ \left(t /  10^9\,\mathrm{yrs} \right)}{1 - 0.345 \, R_{12\mathrm{km}} \, M_{1.4M_\odot}} \: .
\end{align}
Here, $R_{12\mathrm{km}}$ is the NS radius in units of $12\,$km, $M_{1.4M_\odot}$ is the NS mass in units of $1.4M_\odot$, $\rho_\mathrm{DM}$ is the ambient DM density around the NS, $v_\mathrm{DM}$ is the average ambient streaming velocity of DM around the NS, and $t$ is the elapsed time since the NS formation. The DM background velocity and density have been normalized to typical values expected in the galactic disk close to the Sun (see~\cite{Goldman:1989nd}).

The DM particles captured in this way would eventually coalesce and thermalize in the center of the NS. Thermalization occurs relatively quickly -- on the order of years -- for the parameter range considered here (see~\cite{Goldman:1989nd}). If the DM thermal de Broglie wavelength remains smaller than the average particle separation, the DM distribution follows a Maxwell-Boltzmann profile. The average radius of the thermalized DM distribution in the NS center is on the order of cm to m for particles in the GeV/c$^2$ and above mass range,
\begin{align}
    r_{\rm DM} \sim 6.4 \mathrm{cm} \left( \frac{T}{10^5\,\mathrm{K}} \right)^{1/2} \left( \frac{\rho_\mathrm{c}}{10^{15}\,\mathrm{g/cm}^{3}} \right)^{-1/2} \left( \frac{m_\mathrm{DM}}{1000\,\mathrm{GeV/c}^{2}} \right)^{-1/2} \: .
\end{align}
Here, $T$ is the DM temperature in the NS core, $\rho_\mathrm{c}$ is the central density of the NS, and $m_\mathrm{DM}$ is the DM particle mass. As DM continues to accumulate, the density of the central DM clump increases. When it becomes similar to the surrounding NS density, the DM core becomes self-gravitating. At this point, it is no longer supported by the gravity of the NS and becomes gravitationally unstable. If no repulsive forces (e.g., from degeneracy pressure or self-interactions) prevent the collapse, the DM core will undergo gravitational collapse, forming a mini BH.

The mini BH will subsequently accrete the NS on timescales of a few thousand years. Consequently, the time of BH formation, $t_\mathrm{BH}$, sets an upper bound on the NS lifetime. This formation time depends on the DM particle mass and, e.g., the DM self-interaction cross-section. Accordingly, the age of NSs can be used to constrain these DM parameters.

The stability -- and thus the time of collapse -- of the DM clump depends on whether DM is fermionic or bosonic. For fermions, the degeneracy pressure can stabilize the DM clump.~\cite{Goldman:1989nd} found that in this way only DM fermions with masses $\gtrsim 10^7\,$GeV could form a mini BH, even when assuming a NS lifetime of the age of the universe. However, such DM masses are likely excluded by constraints from DM self-annihilation in the early universe (see section 2 in~\cite{Goldman:1989nd}). In the case of bosons, the collapse is facilitated if the bosons can form a Bose-Einstein condensate. The characteristic radius of the condensate is on the order of nanometers, and the critical density for self-gravitation can be reached much sooner. Without DM self-interactions, this can easily lead to mini BH formation by bosonic DM within a typical NS lifetime. DM self-interaction could, however, stabilize the DM clump and delay the collapse. Based on the age of observed NSs, it is therefore possible to constrain the product of the DM mass and the self-interaction cross-section.

Several studies have constrained the DM parameters in this way. Bertone and Fairbairn~\cite{Bertone:2007ae} derived constraints for DM using NSs and white dwarfs, while De Lavallaz and Fairbairn~\cite{deLavallaz:2010wp} subsequently investigated the possible heating of NSs by self-annihilating DM. Kouvaris and Tinyakov~\cite{Kouvaris:2011fi} argued that NSs exclude non-self-interacting DM over a wide range of masses $m_\mathrm{DM} \sim 2\,\mathrm{keV} -16\,\mathrm{GeV}$, and provide strong bounds on repulsive self-interacting DM that are complementary to the Bullet cluster constraint. Kouvaris~\cite{Kouvaris:2011gb} then did a similar analysis for self-interacting DM with an attractive Yukawa-like coupling, which excludes DM particles with $m_\mathrm{DM}>10\,$GeV with this type of coupling. A short review of the field up to 2013 was written by Kouvaris~\cite{Kouvaris:2013awa}. 

McDermott et al.~\cite{McDermott:2011jp} constrained the DM-nucleon cross-section for $m_\mathrm{DM} \sim 5\,\mathrm{MeV} -13\,\mathrm{GeV}$ to be $\sigma_\mathrm{DM,n}\sim 10^{-45}-10^{-47} \,\mathrm{cm}^2$. In contrast to earlier work mostly focusing on non-self-interacting DM, Bell et al.~\cite{Bell:2013xk} found that a DM-DM self-interaction is essentially unavoidable, with a self-interaction cross-section being of a similar order as the DM-nucleon cross-section. This is because the DM-DM interaction can be mediated by nucleons. Since even a small DM self-interaction can significantly delay the formation of a mini-BH, this weakens many of the exclusion bounds previously established, see~\cite{Bell:2013xk}. Bramante et al.\ derived constraints on self-interacting and self-annihilating boson~\cite{Bramante:2013hn} and fermion~\cite{Bramante:2013nma} DM. They also found that a sufficiently large self-annihilation cross-section can essentially prevent the accumulation of the critical DM mass needed for collapse into a mini-BH. G\"uver et al.~\cite{Guver:2012ba} investigated the possibility of DM self-capturing via DM-DM interactions, which could lead to an exponential growth of DM particles accumulated inside the NS -- instead of the linear growth expected from DM-nucleon interactions. Liu et al.~\cite{Liu:2025qco} constrained the cross-section in the case where the DM-nucleon cross section is different for the DM particle and its anti-particle. Jamison~\cite{Jamison:2013yya} improved bounds on the condensation of bosonic DM by noticing that Bose-Einstein condensation is facilitated when the external gravitational potential is fully accounted for. Finally, isospin-violating DM, which couples differently to protons and neutrons, was explored by Zheng et al.~\cite{Zheng:2014fya,Zheng:2016ygg}.

\paragraph{Decay of Standard Model particles into DM.}

Another possible channel for DM accumulation inside NSs is via the decay of SM particles into DM. In contrast to the accretion of DM, DM production through particle decay depends solely on intrinsic NS properties and is therefore independent of the ambient DM properties, such as DM density and velocity distribution, in the vicinity around the NS. One channel of DM creation is motivated by the neutron decay anomaly. It consists of the different decay times of neutrons trapped in a potential bottle $\tau_n^\mathrm{Bottle} = 879.6 \pm 0.6\,$s~\cite{Serebrov:2004zf,Pichlmaier:2010zz,Arzumanov:2015tea} and neutrons in a beam $\tau_n^\mathrm{Beam} = 888.0 \pm 2.0\,$s~\cite{Byrne:1990ig,Yue:2013qrc}, which persists to a $4\sigma$ confidence. The different decay rates could be explained by an additional decay channel of the neutron to an undetected dark particle, as argued by Fornal and Grinstein~\cite{Fornal:2018eol}. This has led many authors to contemplate the existence of DM accumulation inside NSs as a result of neutron decay. We refer to Berryman et al.~\cite{Berryman:2022zic} for a review of a wide selection of SM particle decay to DM in NSs.

One of the first studies came from Motta~\cite{Motta:2018rxp}, who found that the inclusion of a dark neutron decay channel significantly softens the nuclear matter EoS, such that only NSs of masses $<0.7\,M_\odot$ are possible. This is excluded by the existence of NSs of over $2\,M_\odot$. In a follow-up work, Motta~\cite{Motta:2018bil} showed that introducing a strong repulsive DM self-interaction can create more massive NSs, consistent with the maximum NS mass. But in the process, one obtains interaction-cross sections which are far too large to be consistent with observations of cosmological DM. Thus, the hypothetical DM particle from neutron decay can only be a sub-dominant part of the total DM and only be at most a few percent of the total NS mass. Baym et al.~\cite{Baym:2018ljz} and MyKeen~\cite{McKeen:2018xwc} independently found similar results to Motta's works. Husain~\cite{Husain:2022bxl} additionally found that such high cross-sections would lead to young NSs increasing their temperature and rotation rate on timescales of days, creating a possible observable signature for young NSs. Cline and Cornell~\cite{Cline:2018ami} considered the neutron decay into a dark fermion and dark photon, and found that NSs, together with direct-detection experiments, strongly constrain the admissible mass range of the fermion to $937.9-939.6\,$MeV and the DM mass fraction of the NS to $\sim1\,\%$.

Oscillations between neutrons and sterile mirror-neutrons (similar to neutrino oscillation) have been proposed as as an alternative explanation of the neutron decay anomaly~\cite{Berezhiani:2005hv,Serebrov:2007gw,Berezhiani:2018eds}. Such oscillations have been searched for in the laboratory~\cite{Berezhiani:2017jkn,Ayres:2021zbh}. Some astrophysical consequences regarding the stability of the hydrogen atom have been considered by Berezhiani~\cite{Berezhiani:2018udo}.

\paragraph{Accumulation of baryonic matter in DM potential wells.}

Instead of accreting DM onto existing stars, this scenario follows essentially the opposite idea. In the presence of a DM accumulation, or gravitational potential well, baryonic matter could instead accumulate around the DM and eventually lead to mixed systems of DM and baryons.

The basic idea that DM clouds could contract and form self-gravitating compact objects was first explored by Kaup~\cite{Kaup:1968zz} and Ruffini and Bonazzola~\cite{Ruffini:1969qy}, and subsequently Colpi~\cite{Colpi:1986ye} for the case of bosonic DM. These systems can be stable if they are either stabilized by the Heisenberg uncertainty principle or by a repulsive self-interaction between DM particles. Kouvaris~\cite{Kouvaris:2015rea} later studied self-gravitating systems of DM fermions. These systems are called boson stars or DM stars, see Refs.~\cite{Schunck:2003kk, Liebling:2012fv, Torres:2000dw,Visinelli:2017ooc, Visinelli:2021uve} for recent reviews. We will discuss them in more detail in the next section.

If such dark stars exist, their gravitational potential wells could enable ordinary matter to accumulate within or around them~\cite{Meliani:2016rfe}. They could also interact with binary star systems~\cite{Monroy-Rodriguez:2014ula} and potentially get captured in a bound orbit via three-body interactions. Eventually, the star-dark star system could merge after GW emission~\cite{Maselli:2017vfi,Hippert:2022snq,Diedrichs:2023trk}. Another way how dark stars could accumulate baryonic matter is if there is some interaction between DM and regular matter. Kamenetskaia et al.~\cite{Kamenetskaia:2022lbf} studied the scenario where the dark star captures hydrogen from the interstellar medium, which then accumulates in the dark star. If the capture process is efficient enough, a star could potentially form inside the dark star and later form a compact object after the stellar evolution~\cite{Kouvaris:2010vv}. Through these mechanisms, compact hybrid objects consisting of significant amounts of DM and baryonic matter may form. If the baryonic component is a NS, these are commonly called DM-admixed neutron stars (DANS).

Recent work has also focused on mergers involving DANS analyzed the GW signatures of fermion-boson star (FBS) mergers~\cite{Giangrandi:2025rko,Srikanth:2025lic}. These studies suggest that FBSs can generate distinctive inspiral and post-merger signals, offering a potential observational pathway to identify DANS or DANS-NS mergers in future GW detections.

\subsubsection{Models for DM admixed NS}

Several theoretical frameworks have been developed to describe DANS. The appropriate description depends on the assumed microphysical nature of DM -- fermionic, bosonic, or field-like -- as well as on the assumed DM-baryon interactions. Below, we outline the most commonly used modeling approaches.\\

\paragraph{Modified nuclear equation of state}

One way to model the influence that DM has on the structure of NSs is by modifying the nuclear EoS. This approach is relevant when the amount of DM is not large enough to significantly change the system through its gravitational mass, but the DM-nucleon interactions are sufficiently strong to alter the properties of dense nuclear matter. This can be relevant, for example, when neutrons can decay into a DM particle (e.g., motivated by the lifetime discrepancy between neutrons in a beam vs. bottle, see above). In this case, the conversion of neutrons into DM particles will reduce the neutron number density, thereby lowering the effective pressure contribution from neutrons at high densities and softening the EOS. The modified EOS is obtained by adding the energy and pressure contributions of the DM component and by imposing a chemical-equilibrium condition that couples the neutron and DM populations. Therefore, due to the equilibrium of the DM-neutron reaction and beta-equilibrium, the chemical potentials will obey the relations (see Motta et al.~\cite{Motta:2018rxp}):
\begin{align}
    \mu_\mathrm{n} = \mu_\mathrm{p} + \mu_\mathrm{e} \:\:\:\: , \:\:\:\: \mu_\mathrm{n} = \mu_\mathrm{DM} \: ,
\end{align}
where $\mu_\mathrm{n}$, $\mu_\mathrm{p}$, $\mu_\mathrm{e}$ and $\mu_\mathrm{DM}$ are the chemical potentials of neutrons, protons, electrons and DM, respectively. The pressure, which stabilizes the NS, however, comes only from baryonic particles:
\begin{align}
    P = \sum_\mathrm{n,p,e} \mu_i n_i - \epsilon \: ,
\end{align}
where $n_i$ are the respective number densities and $\epsilon$ is the total energy density. Motta et al.~\cite{Motta:2018rxp} investigated this scenario and found that even a small abundance of dark-sector particles can significantly soften the EOS, limiting the maximum mass of neutron stars to $\lesssim 0.7~M_\odot$. 

Other authors have considered variations of this idea. For example, works on Higgs-portal fermionic DM (e.g., Das et al.~\cite{Das:2018frc}) show that DM-nucleon scalar interactions can reduce the effective nucleon mass at high density, leading to additional softening of the EoS. Studies involving mirror DM (e.g., Sandin et al.~\cite{Sandin:2008db}) have similarly demonstrated that DM-baryon interactions, even when only gravitational or very weakly coupled, can modify the stiffness of the EoS and thereby affect the maximum mass, radius, and tidal deformability of NSs.\\

\paragraph{Two-fluid approaches}
A widely used framework for modeling DANS is the relativistic two-fluid formalism, in which baryonic matter and DM are treated as distinct, interpenetrating fluids. This approach is applicable when the DM constitutes a significant fraction of the total mass such that its gravitational influence cannot be neglected, but where the DM does not need to be modeled as a coherent scalar or vector field. In its simplest formulation, the fluids interact only through gravity, each obeying their own EoS. Mathematically, the two-fluid approach is described using the TOV equations of two gravitating fluids. Following Sandin et al.~\cite{Sandin:2008db} (also see Leung et al.~\cite{Leung:2022wcf}), the total energy-momentum tensor is written as the sum of the energy-momentum tensors of the two fluids:
\begin{align}
    T^{\mu\nu}_\mathrm{tot} = T^{\mu\nu}_\mathrm{baryons} + T^{\mu\nu}_\mathrm{DM} = (e+P) u^\mu u^\nu - P g^\mathrm{\mu\nu} \: .
\end{align}
Here, $e = e_{\rm baryons} + e_{\rm DM}$ and $P = P_{\rm baryons} + P_{\rm DM}$ are the total energy density and pressure. For this case, the resulting TOV equations separate into two equations that describe the pressure gradients of both fluids independently of the pressure contribution from the other fluid. For a spherically symmetric metric with line element
\begin{align}
    ds^2 = - e^{\nu(r)} {\rm d}t^2 + e^{\lambda(r)} {\rm d}r^2 + r^2 ({\rm d}\theta^2 + \sin(\theta)^2 {\rm d}\phi^2) \: , \label{sec5:Compact-star-properties:spherical-metric-ansatz}
\end{align}
and $e^{-\lambda(r)} = 1 - 2M(r)/r$, where $M(r)$ is the enclosed gravitational mass within $r$, we get:
\begin{subequations}
\begin{align}
    \frac{{\rm d}P_\mathrm{baryons}}{{\rm d}r} &= (e_\mathrm{baryons}+P_\mathrm{baryons}) \frac{{\rm d}\nu}{{\rm d}r} \: , \\
    \frac{{\rm d}P_\mathrm{DM}}{{\rm d}r} &= (e_\mathrm{DM}+P_\mathrm{DM}) \frac{{\rm d}\nu}{{\rm d}r} \: , \\
    \frac{{\rm d}\nu}{{\rm d}r} &= \frac{(M(r) + 4\pi r^2 P)}{r ( r - 2M(r))} \: , \\
    \frac{{\rm d}M}{{\rm d}r} &= 4\pi r^2 (e_\mathrm{baryons} + e_\mathrm{DM}) \: .
\end{align}
\end{subequations}
To solve this system, one must specify an EoS for both the baryonic and DM components. There is a wide variety of realistic EoS for baryonic matter. In the case where the DM does not interact with the baryonic matter except for gravity, they can be used directly. Otherwise, custom/modified nuclear EoS can be used, see above. For the DM component, a variety of microphysical models have been implemented, including relativistic or non-relativistic Fermi gases~\cite{Ivanytskyi:2019wxd, Sagun:2021oml, Kain:2021hpk, Wu:2022wzw, Ruter:2023uzc, Giangrandi:2025rko}, asymmetric DM~\cite{Rutherford:2022xeb,Sagun:2022ezx,Rutherford:2024uix,Routaray:2024fcq,Kumar:2025yei}, bosonic DM modeled through effective polytropic EoSs~\cite{Colpi:1986ye,Leung:2022wcf,Karkevandi:2021ygv,Giangrandi:2022wht,Shakeri:2022dwg}, and mirror-DM, which has the same properties as baryonic matter but only couples gravitationally to it~\cite{Kain:2021hpk,Yang:2024ycl,Issifu:2025gsq,Kumar:2025oyx,Emma:2022xjs,Ciarcelluti:2010ji}.

The two-fluid framework has been applied and extended in several recent studies. Cipriani et al.~\cite{Cipriani:2025tga} and Issifu~\cite{Issifu:2025gsq} investigated rotating two-fluid configurations, allowing the baryonic and DM components to have different rotation rates. Cipriani showed how the presence of a DM component modifies the mass-radius relation, moment of inertia, and overall stability of the rotating star. Giangrandi et al.~\cite{Giangrandi:2022wht} used the two-fluid formalism for non-rotating NSs to place constraints on DM properties based on astrophysical observables, while Sagun et al.~\cite{Sagun:2022ezx} explored how repulsive, vector-mediated DM interactions affect global NS properties, such as maximum mass and tidal deformability. Earlier work by Sagun et al.\cite{Sagun:2021oml} modeled DM as a relativistic Fermi gas within the two-fluid framework, demonstrating how even simple microphysical assumptions can yield significant changes in the NS structure. More recently, Giangrandi et al.~\cite{Giangrandi:2024qdb} extended the two-fluid model to study the thermal evolution of NSs containing a DM component, highlighting the impact of DM annihilation or accumulation on the cooling behavior of the star. 
Two-fluid models have also been incorporated into numerical relativity simulations of mergers, providing physically consistent initial data and fully dynamical evolution. Numerical simulations of the merger of two-fluid NSs with DM have been performed for fermionic DM cores (Giangrandi et al.~\cite{Giangrandi:2025rko}), for mirror-DM-admixed stars (Emma et al.~\cite{Emma:2022xjs}), and for two-fluid Fermi gas DM (Ruter et al.~\cite{Ruter:2023uzc}). A two-fluid DANS scenario has also been investigated in \cite{Grippa:2024sfu}, where BM and DM interact only gravitationally. The study examines linear and quadratic couplings of scalar dark mediators and their impact on the mass-radius relation, tidal deformability, and sound speed, using X-ray and GW observations to constrain the DM parameters.\\

\paragraph{Field-theoretic (scalar/vector) approaches}

In some regimes, it is more appropriate to treat DM not as a collection of individual particles but as a macroscopic, coherent matter field. This situation arises when the Compton wavelength of the DM particle becomes similar in size to the NS itself. For instance, ultralight DM with masses $\sim 10^{-11}$\,eV/c$^2$ has a Compton wavelength comparable to the NS radius, on the order of kilometers. In this limit, the DM inside NSs no longer behaves as a dilute gas but rather as a self-gravitating matter wave, which contributes to the gravitational field through a classical scalar (or vector) field configuration. The system of self-gravitating macroscopic matter waves has been studied since the late 1960s, when the concept of a boson star was first introduced by Kaup~\cite{Kaup:1968zz} and by Ruffini and Bonazzola~\cite{Ruffini:1969qy}. These works established that a complex scalar field minimally coupled to gravity can form stable, self-gravitating configurations even without self-interactions. Later, Colpi, Shapiro, and Wasserman~\cite{Colpi:1986ye} demonstrated that repulsive quartic self-interactions can significantly increase the maximum mass of such objects. Reviews of boson-star physics were provided by Jetzer~\cite{Jetzer:1991jr}, and Liebling and Palenzuela~\cite{Liebling:2012fv}. The extension of these ideas to systems containing both a scalar-field component and baryonic matter naturally leads to field-theoretic models of DANS, which were first explored in an astrophysical context by Henriques et al.~\cite{Henriques:1989ar}. The DM component was then modeled as a boson star, which coincides in location with a baryonic NS. Through the gravitational back-reaction, the DM component influences the properties of the baryonic NS component, while modifying its observables such as mass, radius and tidal deformability. These fermion-boson stars have been studied by Jockel~\cite{Jockel:2023qnq}. We refer the reader to this work for a more comprehensive overview and detailed derivations, which also form the basis of the following discussion. 

In the simplest case, the DM wave is modeled as a complex scalar field, and the DANS can then be described by a matter field that is minimally coupled to this scalar field. The corresponding action can be written as
\begin{align}
    S_\mathrm{DANS} = \int {\rm d}^4x \sqrt{-g} \left( \frac{R}{2\kappa} - \mathcal{L}_\mathrm{matter} - \nabla_\alpha \phi \nabla^\alpha \bar{\phi} - V(\phi\bar{\phi}) \right) \: ,
\end{align}
where $\kappa=8\pi$ in units with $c=G=1$. The first part is the Einstein-Hilbert action, the second term describes the nucleonic matter, and the other terms describe a complex scalar field $\phi$ with a potential $V$. The equations of motion follow from the variation of $S_\mathrm{DANS}$, yielding the Einstein-Hilbert-Klein-Gordon system:
\begin{align}
    G_\mathrm{\mu\nu} &= \kappa \left( T_{\mu\nu}^\mathrm{baryons} + T_{\mu\nu}^\mathrm{DM}\right) \: ,\\
    \nabla_\alpha \nabla^\alpha \phi &= \phi\,\frac{{\rm d} V(\phi\bar{\phi})}{{\rm d} |\phi|^2} \: .
\end{align}
These are the equations of motion for gravity (Einstein equations) with an additional matter term corresponding to the scalar field, coupled to the Klein-Gordon equations in curved spacetime. For the baryonic part, $T_{\mu\nu}^\mathrm{baryons}$, one usually assumes a perfect fluid. The energy-momentum tensor of the scalar field can be derived from the action and is given by
\begin{align}
    T_{\mu\nu}^\mathrm{DM} = \partial_\mu \phi \partial_\nu \bar{\phi} + \partial_\nu \phi \partial_\mu \bar{\phi} - g_{\mu\nu} \left( \partial_\alpha \phi \partial^\alpha \bar{\phi} + V(\phi\bar{\phi}) \right) \: .
\end{align}
In the case of spherical symmetry -- with the line element from Eq. \eqref{sec5:Compact-star-properties:spherical-metric-ansatz} -- these equations reduce to a system of coupled ordinary differential equations. For the scalar field, one usually makes a harmonic wave ansatz $\phi(x^\mu) = \phi(r) e^{i\omega t}$, which introduces different solutions in the form of modes in $\omega$. The resulting field equations are:
\begin{subequations}
\begin{align}
    \frac{{\rm d}\nu}{{\rm d}r} &= \frac{e^{\lambda(r)} -1}{r} + 8\pi r e^{\lambda(r)} \left( P + \omega^2 \phi(r)^2 e^{-\nu(r)} + \Psi^2 e^{-\lambda(r)} - V(\phi\bar{\phi})\right) \: , \\
    \frac{{\rm d}\lambda}{{\rm d}r} &= \frac{1 - e^{\lambda(r)}}{r} + 8\pi r e^{\lambda(r)} \left( e + \omega^2 \phi(r)^2 e^{-\nu(r)} + \Psi^2 e^{-\lambda(r)} + V(\phi\bar{\phi})\right) \: , \\
    \frac{{\rm d}\phi(r)}{{\rm d}r} &= \Psi \: , \\
    \frac{{\rm d}\Psi}{{\rm d}r} &= \left( e^{\lambda(r)} \frac{{\rm d}V(\phi\bar{\phi})}{{\rm d} |\phi|^2} - \omega^2 e^{\lambda(r)-\nu(r)} \right) \phi(r) + \left( \frac{\lambda'(r)}{2} - \frac{\nu'(r)}{2} - \frac{2}{r}\right) \Psi \: , \\
    \frac{{\rm d}P}{{\rm d}r} &= - \frac{1}{2} (e + P) \frac{{\rm d}\nu}{{\rm d}r} \: .
\end{align}
\end{subequations}
The scalar-field frequency, $\omega$, is not directly fixed by the equations of motion but needs to be tuned to specific discrete values, the field modes. Each mode frequency can be found by imposing that the solution of the equations is regular at the center and at infinity, and that the final solution converges to the Schwarzschild solution at large radii. The modes are characterized by the number of zero-crossings of the field $\phi(r)$ before asymptotically approaching zero. The lowest mode has no zero crossings.
The final solution is characterized by a set of global quantities. First, the total gravitational mass is obtained in the limit of large radii as
\begin{align}
    M_\mathrm{tot} = \lim_{r\rightarrow\infty} \frac{r}{2}\left( 1 - e^{-\lambda(r)}\right) \: .
\end{align}
Different characteristic radii can also be defined. The fermionic radius, or NS radius, $R_\mathrm{NS}$, is the radius of the baryonic NS component. The bosonic radius, or DM radius $R_\mathrm{DM}$, is usually defined as the radius that contains $99\%$ of the DM rest mass. The rest masses of the NS component and the DM component are defined as
\begin{align}
    M_\mathrm{rest,NS} = 4\pi \int_0^{R_\mathrm{NS}} e^{\lambda(r)/2} \rho r^2 {\rm d}r\,,\qquad  M_\mathrm{rest,DM} = 8\pi \int_0^\infty e^{(\lambda(r)-\nu(r))/2}\; \omega \phi(r)^2 r^2 {\rm d}r \; ,
\end{align}
where $\rho$ is the baryonic rest mass density. It is common to then define the DM fraction as $f_\mathrm{DM} = M_\mathrm{rest,DM} / (M_\mathrm{rest,NS} + M_\mathrm{rest,DM})$.

The first studies of DANS with scalar-field DM came in 1989 from Henriques et al.~\cite{Henriques:1989ar}, who first used the field-theoretic approach to study isolated systems and investigated their basic principles. Shortly thereafter, Henriques et al.~\cite{Henriques:1990xg} followed with an investigation into the stability of scalar-field DANS with respect to radial perturbations, finding stable parameter regions for a range of DM masses (in the range $m_\mathrm{DM}\sim 10^{-16}–10^{-11}\,$eV/c$^2$) with respect to the central nuclear matter densities and scalar-field amplitudes.

One of the first works that numerically studied scalar DANS and their stability was by Valdez-Alvarado et al.~\cite{Valdez-Alvarado:2012rct}, who studied the dynamical stability numerically, including the scalar normal modes. Later, in a series of papers, Di Giovanni et al.~\cite{DiGiovanni:2021ejn,DiGiovanni:2022mkn} investigated the cases where the scalar-field potential is given by a mass term $V = \frac{1}{2}|\phi|^2$. In~\cite{DiGiovanni:2021ejn}, they investigated the impact of different DM particle masses and DM fractions, $f_\mathrm{DM}$, on stable DANS configurations. They especially focused on the mass-radius curves of the baryonic NS part of the DANS, which would be visible to EM observations. Di Giovanni et al. 
\cite{DiGiovanni:2022mkn} then studied the impact of scalar DM on the formation and evolution of the bar mode instability in unstable, differentially rotation NS, and found that scalar DM can have an impact on the GW signal in the post-merger phase of a binary DANS merger.
Different scalar potentials with quartic self-interaction terms ($V = \frac{1}{2}|\phi|^2 + \frac{1}{4}|\phi|^4$) were then adopted by Di Giovanni et al.~\cite{DiGiovanni:2020frc} and Valdez-Alvarado et al.~\cite{Valdez-Alvarado:2020vqa}. These works performed numerical simulations of NSs with a coalescing scalar cloud, and then did a stability analysis with different central scalar-field amplitudes and baryonic matter densities. Di Giovanni et al.~\cite{DiGiovanni:2021vlu} later investigated the dynamical stability of DANS where the scalar field is in the excited state and found that higher scalar-field modes can be stable due to the gravitational effect from the NS component (the higher modes are unstable for pure boson stars).

Diedrichs et al.~\cite{Diedrichs:2023trk} later expanded the study of DANS with scalar quartic self-interactions to systematically probe DANS properties such as gravitational mass, radius, and tidal deformability for a wide range of DM fractions. They also analyzed the stability of scalar DANS for various DM properties/parameters, matching previous numerical predictions. DANS in which the DM component is composed of axions were also explicitly studied by Di Giovanni et al.~\cite{DiGiovanni:2022xcc}, who applied a special axion potential, $V_\mathrm{axion}$, for the scalar field. In this work, the stability of these axion DM DANS was investigated and multiple stable and unstable branches were found based on the axion particle mass and decay constant.

In parallel to fully field-theoretical treatments, a two-fluid model with an effective bosonic EoS has been developed to mimic the field-theoretic approach and approximate the scalar-field DM in a hydrodynamic framework. This approach was used by Karkevandi et al.~\cite{Karkevandi:2021ygv} and Leung et al.~\cite{Leung:2022wcf}, who used the effective bosonic EoS derived by Colpi~\cite{Colpi:1986ye}. This approach is useful for scalar fields with strong self-interactions, because it reduces computational cost by solving the two-fluid model instead of the full scalar-field equations. In the paper by Diedrichs et al.~\cite{Diedrichs:2023trk}, the authors also compared the DANS solutions for the effective bosonic EoS with the full scalar theory. They found that both approaches agree well when the self-interaction parameter is $\Lambda_\mathrm{int}=\lambda/(8\pi m^2) \gtrsim 400$, while for low self-interaction strengths, it remains necessary to solve the full field-theoretic system.

Field-theoretic DANS have also been investigated in numerical-relativity simulations. Using the field-theory approach, Bezares et al.~\cite{Bezares:2019jcb}, and also Srikanth et al.~\cite{Srikanth:2025lic}, performed fully general-relativistic simulations of DANS with a scalar-field DM, following the inspiral and merger of two DANS and studying the impact that DM has on the merger evolution and final GW signal.
Different types of DANS in which the DM component is modeled as a complex vector field were also studied by Jockel and Sagunski~\cite{Jockel:2023rrm}. Such systems correspond to a NS admixed with a Proca star, see Brito et al.~\cite{Brito:2015pxa}, and can be used to constrain spin-1 (vector) boson DM candidates.

\subsubsection{Observational consequences of DM admixed NSs}

Independent of the specific mechanism through which DM accumulates inside of NSs, many models predict that DM could accumulate in sufficient amounts to modify the internal structure and observable properties of NSs. This allows us to indirectly probe the DM properties -- such as the particle mass, DM-DM and DM-nucleon interaction strengths, and whether DM is bosonic or fermionic -- through astrophysical observations of NSs. In the following we give an overview of the most important observational consequences of DM in NSs for different DM and DANS models. We start with isolated NSs and subsequently discuss binary systems.

\paragraph{Isolated neutron stars}

\begin{figure}
    \centering
    \includegraphics[width=0.45\textwidth]{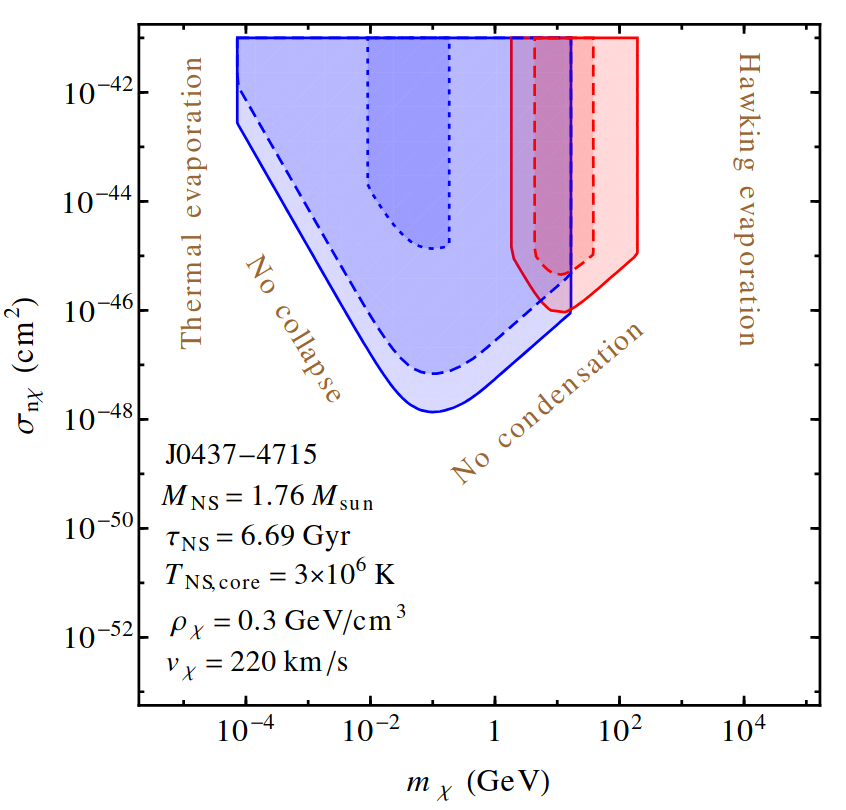}
    \includegraphics[width=0.45\textwidth]{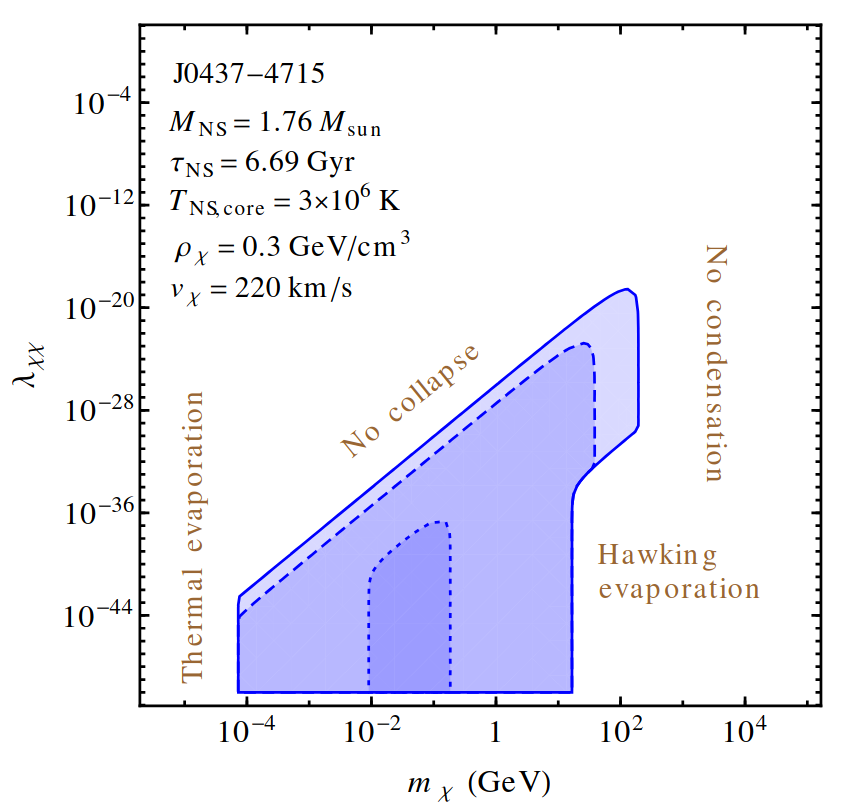}
    \caption{Exclusion bounds for the DM-nucleon cross-section, $\sigma_{n\chi}$ (left), and the DM self-interaction parameter, $\lambda_{\chi\chi}$ (right), related to the self-interaction cross-section through $\sigma_{\chi\chi}=\lambda^2_{\chi\chi}/64\pi m_\mathrm{DM}^2$, obtained by Bell et al.~\cite{Bell:2013xk}. The shaded regions mark the mass and interaction cross-sections, which are excluded by the pulsar J0437-4715. The blue region assumes no DM self-interaction, and the red region is the bound if a small DM self-interaction is present. Adapted from Ref.~\cite{Bell:2013xk}.}
    \label{fig:DANS-Exclusion-bounds-single-NS-Bell}
\end{figure}

First, we discuss scenarios where DM consists of particles that interact with nucleons and accumulate in small amounts near the NS center. As discussed above, the most significant constraints on DM in this case comes from the lifetime of NSs. If the DM particle mass and self-interaction are such that the accumulated DM becomes self-gravitating, it may collapse into a mini BH. The existence of old NSs implies that this collapse has not occurred, placing lower bounds on the DM parameters. For example, McDermott et al.~\cite{McDermott:2011jp} constrained the DM-nucleon cross-section for DM with masses of $m_\mathrm{DM} \sim 5\,\mathrm{MeV} -13\,\mathrm{GeV}$ to be $\sigma_\mathrm{DM,n}\sim 10^{-45}-10^{-47} \,\mathrm{cm}^2$. Bell et al.~\cite{Bell:2013xk} then derived bounds on the DM-nucleon and DM-DM cross section using the age of old NSs. Specifically, they used the NS J0437-4715 and derived bounds on the DM-nucleon cross-section, $\sigma_{{\rm DM},n}$, and found that this observation can exclude the range $\sigma_{{\rm DM},n}\geq 10^{-48}$cm$^2$ or $\sigma_{{\rm DM},n}\geq 10^{-52}$cm$^2$, for DM masses in the range $10^{-4}\leq m_\mathrm{DM}\:c^2/\mathrm{GeV} \leq 10^3$, depending on the local DM density around the pulsar. But they also stressed that very small amounts of DM self-interactions can provide sufficient stability to the DM core such that all constraints are effectively avoided. This is the case for $\sigma_{\rm DM,DM}\geq 10^{-70}\sim 10^{-76}$cm$^2$, which is, according to Bell et al., not unlikely to happen because weak self-interactions could be mediated by nucleons directly (DM-nucleon-DM) and evade the bounds altogether. In Fig. \ref{fig:DANS-Exclusion-bounds-single-NS-Bell} we show two plots from Bell et al.~\cite{Bell:2013xk} (see their figures 2 and 3) that show the exclusion bounds obtained from NS lifetimes. In the left panel, the DM-nucleon cross-section ($\sigma_{n\chi}$ in the notation of Bell et al.) is shown. The blue region marks the case without any DM self-interaction and the red region marks the case where weak DM self-interactions are present. As discussed above, the nucleon mediated DM self-interaction can significantly weaken the bounds because it increases the accumulated DM mass necessary for a collapse to a mini BH to happen. The right panel shows the exclusion zone for the DM-DM self-interaction strength.
For a more complete overview of the bounds obtained from NS lifetimes, we refer to the short review by Kouvaris~\cite{Kouvaris:2013awa}.

In scenarios where DM accumulates inside NSs but remains stable, its presence can still leave measurable imprints on the global properties of isolated stars, if the total mass of DM is a significant fraction of the baryonic mass. These effects have been studied using both two-fluid formalisms and field-theoretic models, particularly for non-rotating DANSs.

The typical qualitative behaviors of the DANS properties are as follows. Irrespective of whether the DM particle is a boson or a fermion, if the DM particle mass is large and if the self-interaction is weak, the DM tends to cluster in the center of the DANS as a DM core with characteristic radii which are smaller than the baryonic NS radius of $\sim12-15\,$km. The exact size of the DM core depends on the DM fraction and the details of the model. Conversely, when the DM particle mass is small and the self-interaction is strong and repulsive, the DM tends to form an extended cloud/halo around the NS, with characteristic radii that are significantly larger than the baryonic component. In this case, a significant amount of mass can be contained outside of the baryonic component.

The observable quantities in the case of DANS are the gravitational mass, the radius of the baryonic component -- even in the case of a DM halo, only the baryonic matter is visibly by EM radiation -- and the tidal properties such as the tidal deformability. As an example, we discuss the results by Karkevandi et al.~\cite{Karkevandi:2021ygv} and Giangrandi et al.~\cite{Giangrandi:2022wht}, who both studied DANS admixed with self-interacting bosonic DM. Karkevandi et al. studied the case where the DM self-interactions are strong and the effective bosonic EoS from Colpi~\cite{Colpi:1986ye} can be used to model the DM component, while Giangrandi et al. considered a DM particle with a vector-boson mediated self-interaction. Karkevandi et al.~\cite{Karkevandi:2021ygv} studied DANS with DM particles in the range of $100-500\,$MeV and high DM fractions up to $50\,\%$. Regarding the radial distribution of DM and baryonic matter, they found that a DM core will increase the compactness of the system and decrease the radius of the baryonic component. The transition between a DM core and halo was found in the range of $m_\mathrm{DM}=100-400\,$MeV for a self-interaction parameter of $\lambda_{\rm DM,DM}=\pi$ and for different DM fractions. A higher DM fraction leads to a larger DM core/halo. Also, Karkevandi et al.~\cite{Karkevandi:2021ygv} shows that a DM halo is essentially always formed for $f_\mathrm{DM}=10\,\%$ if $m_\mathrm{DM}<140\,$MeV and $\lambda_{\rm DM,DM}=0.5\pi-2\pi$. Regarding the maximal possible NS mass, they found that it decreases when increasing the DM fraction. But once the DM fraction is large enough and the systems gravitational potential becomes dominated by the DM, then the maximal possible mass increases and can be even larger than an isolated NS without DM. Thus, observing an overmassive NS could be a signature of a DANS.

The mass-radius (MR) relation of DANS was studied by Giangrandi et al.~\cite{Giangrandi:2022wht}, who studied in detail the cases where DM forms a core and the particle mass is $100-1000\,$MeV range. They focused on the cases of DANS with DM fractions of a few percent and computed the MR relation of DANS with different EoS for the baryonic component. They found that a DM core reduces the radius of the baryonic component, thus reducing the visible NS radius. For DM fractions of $5\,\%$, the reduction in baryonic radius can be up to $500\,$m. They also find that the additional DM mass reduces the  DANS mass compared to a pure NS for a given central baryonic density. The resulting shift in the MR relation could be used to test the DM content of NSs using current measurements of NS properties through EM observations and GW inferred quantities. 
Karkevandi et al.~\cite{Karkevandi:2021ygv} and Giangrandi et al.~\cite{Giangrandi:2022wht}, also studied the tidal deformability of DANS with different DM fractions. Because the tidal deformability depends on the matter distribution within the object, it is also sensitive to the DM content and can thus be used to probe the DM properties. For example, in the case of a DM core, the tidal deformability is found to decrease because the characteristic radius of the DANS decreases. This effectively mimics a softer EOS and makes the object less susceptible to outside tidal forces. In contrast, in the case of a DM cloud/halo, the characteristic radius of the DANS extends to radii significantly larger than the baryonic component, which makes the DANS easier to tidally disrupt. The tidal deformability is thus increased. This behavior was also found by other authors, Ellis et al.~\cite{Ellis:2018bkr}, Nelson et al.~\cite{Nelson:2018xtr}, Dengler et al.~\cite{Dengler:2021qcq}, and Rutherford et al.~\cite{Rutherford:2022xeb}, for two-fluid DANS, and also by Diedrichs et al.~\cite{Diedrichs:2023trk} for DANS with a scalar field DM model.

\begin{figure}
    \centering
    \includegraphics[width=0.44\textwidth]{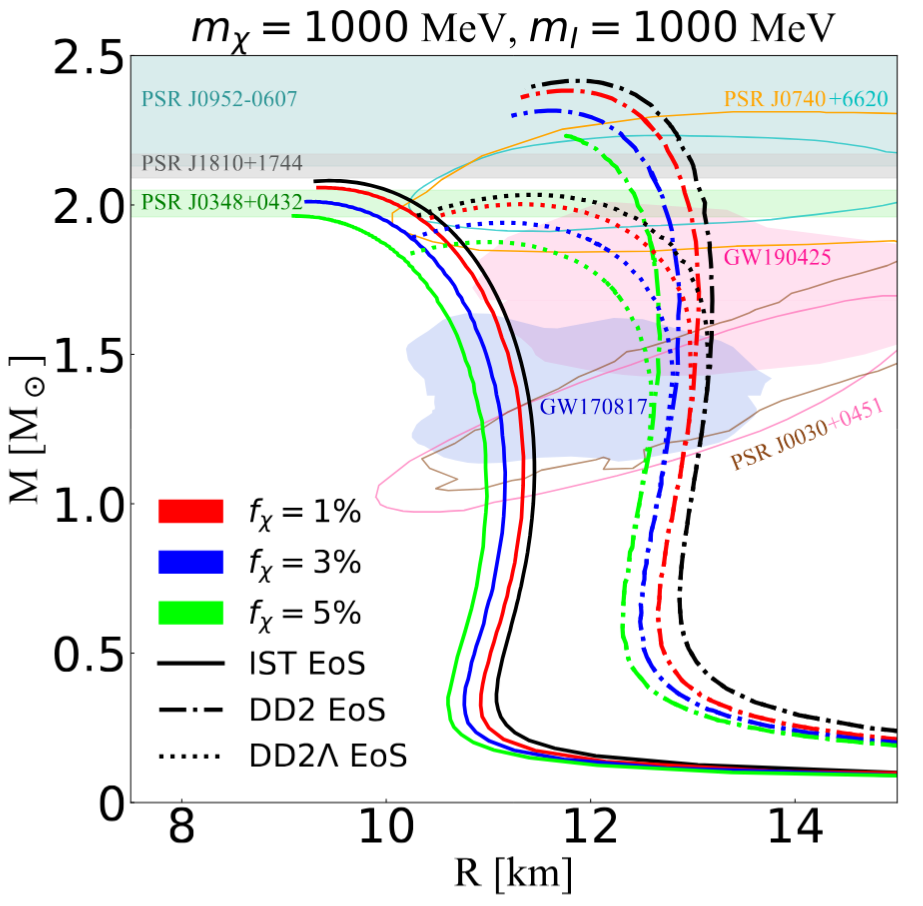}
    \includegraphics[width=0.461\textwidth]{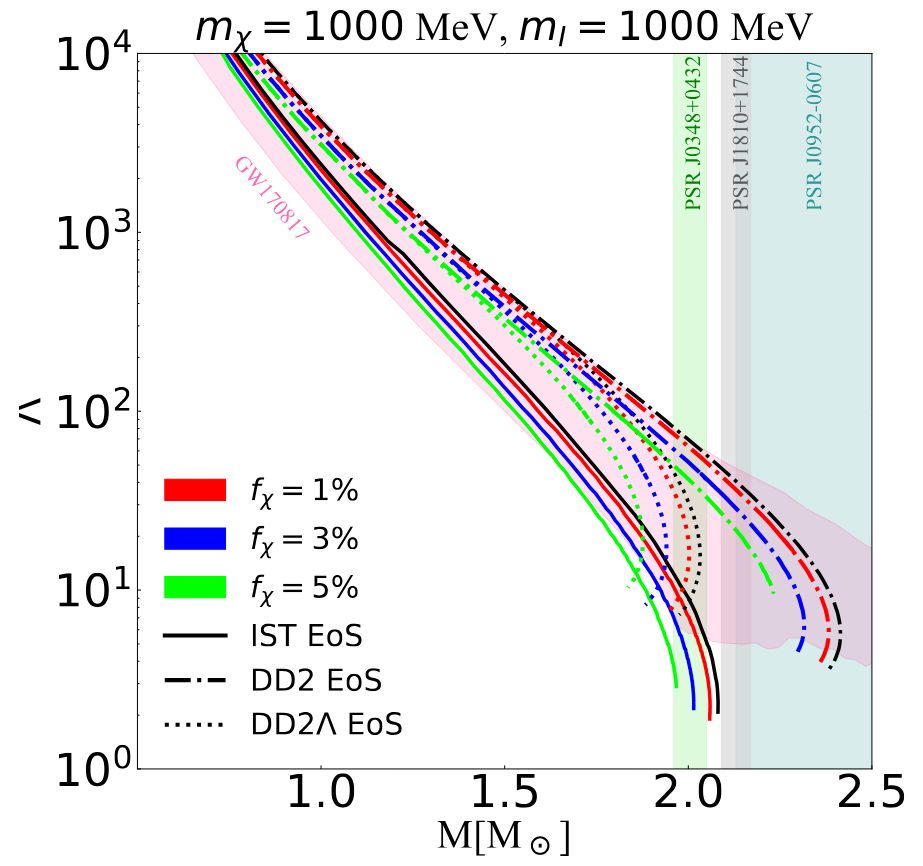}
    \caption{Mass-radius relation (left) and tidal deformability (right) of DANS with different DM fractions, obtained by Giangrandi et al.~\cite{Giangrandi:2022wht}. The DM particle mass is $m_\chi=1000\,$MeV and $m_I$ is a parameter that sets the strength of the DM self-interaction.  The shaded regions show observational bounds on the NS mass, radius and tidal deformability from various experiments. For the above figures, the DANS have a DM core, which is why the radius of the baryonic component decreases with increasing DM fraction. The tidal deformability decreases as well because the NS becomes more compact for fixed mass. Adapted from Ref.~\cite{Giangrandi:2022wht}.}
    \label{fig:DANS-MR-relation-and_tidal-deformability-Giangrandi}
\end{figure}

We show some exemplary results for the DANS MR relation and tidal deformability from Giangrandi et al., in Fig. \ref{fig:DANS-MR-relation-and_tidal-deformability-Giangrandi}. In the left panel, we show MR relations for DANS with different nucleonic EoS and different DM fractions ($f_\chi$ in their notation) for a DM particle with mass $m_\mathrm{DM}=1000\,$MeV ($m_\chi$ in their notation). $m_I$ is a parameter which sets the DM self-interaction strength. The shaded regions mark current observational bounds from NS EM and GW observations. For an increasing DM fraction, the MR relation can be seen to shift to lower masses and radii. Through this shift in the MR curve one can constrain the DM fraction using the observational bounds. In the right panel, we show the tidal deformability as a function of the DANS gravitational mass for the same DM parameters as above. Giangrandi et al.\ studied DM core configurations and accordingly, the presence of a DM core can be seen to decrease the tidal deformability. The effect becomes stronger with increasing DM fraction.

\begin{figure}
    \centering
    \includegraphics[width=0.95\textwidth]{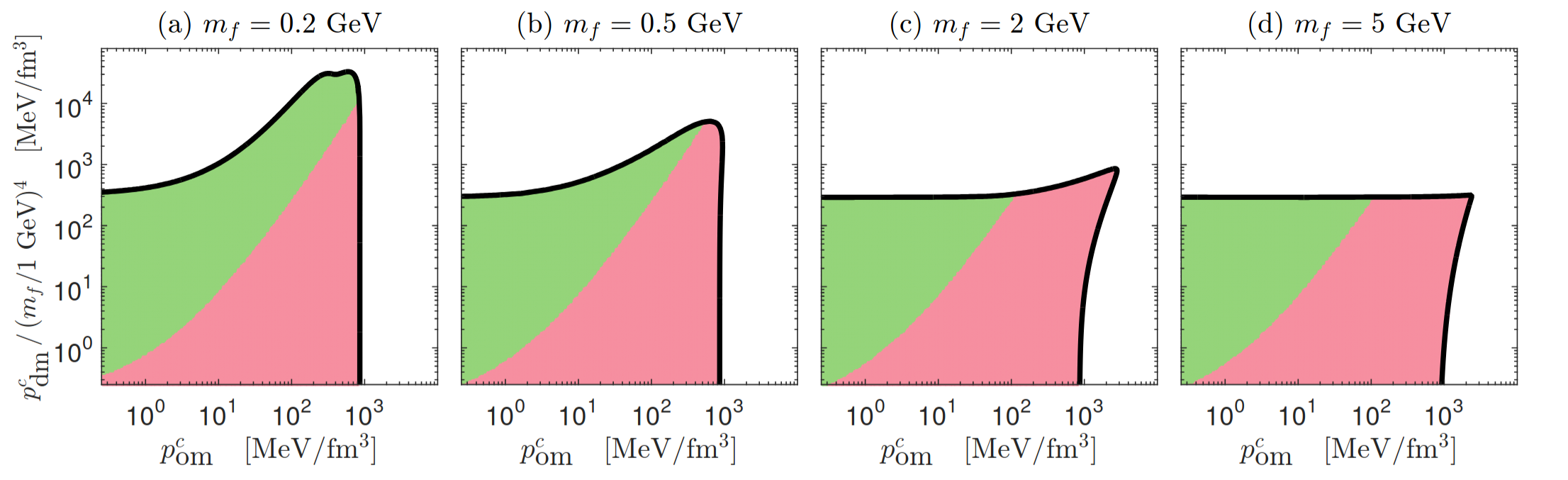}
    \caption{The region of stability of a DANS as a function of the central pressure of ordinary (i.e. baryonic) matter, $p_\mathrm{om}^c$, and the central DM pressure, $p_\mathrm{dm}^c$, by Kain~\cite{Kain:2021hpk}. They modelled DANS using the two-fluid formalism and modelled DM as a Fermi gas of particles with mass $m_f$. The colored region below the black curve marks the parameter space where DANS are stable. The colored red (lower) region marks DANS with a DM core, while the colored green (upper) region marks DANS with a DM halo/cloud. Adapted from Ref.~\cite{Kain:2021hpk}.}
    \label{fig:DANS-stability-region-Kain}
\end{figure}

Furthermore, the question of the stability of DANS to small radial perturbations has been under investigation since the early works by Henriques et al.~\cite{Henriques:1989ar,Henriques:1990xg}. The DM component adds complexity to the problem since the additional DM content might stabilize the baryonic NS component in some cases, while destabilizing it in others. For pure NSs, it is found that they are stable until the mass as a function of central density reaches a peak. A similar criterion can be found also for DANS, but this time the stable configurations span a region in the plane of central baryonic density and central DM density (or central scalar field amplitude in the case of field-theoretic treatments). We show some exemplary results from Kain~\cite{Kain:2021hpk} in Fig.~\ref{fig:DANS-stability-region-Kain}. Kain studied DANS with fermionic DM in the GeV mass-range and modeled it as a Fermi gas. They then used the two-fluid formalism to construct equilibrium solutions and subsequently analyzed them regarding stability. Kain found that the mixed DM-baryonic matter system is stable over a wide range of central densities. It is noteworthy that the addition of a DM component (or a baryonic component) allows for mixed systems which have much higher central densities compared to the pure DM or pure baryonic matter scenario. The shape of the stable region also depends on the DM mass and model, which could be used to constrain DM properties, if DANS are observed. We also refer to the numerical studies by DiGiovanni et al.~\cite{DiGiovanni:2021ejn,DiGiovanni:2022mkn,DiGiovanni:2020frc,DiGiovanni:2021vlu,DiGiovanni:2022xcc} of DANS with a scalar field, and to the studies by Diedrichs et al.~\cite{Diedrichs:2023trk} and Jockel et al.~\cite{Jockel:2023rrm}, who computed the stability regions for scalar and vector field DM, respectively.

The considerations regarding stability can also be interesting in the context of merging and/or accreting DANS. As was discussed by Hippert et al.~\cite{Hippert:2022snq}, DANS which exist in a binary system could eventually merge. During the merger process, if the DM exists as a cloud around the NS, some of the DM could be expelled from the system. If the remaining object forms again a stable DANS, then some of the formerly expelled matter could later be re-accreted onto the DANS over time. If sufficient matter is accreted this way, the DANS could then slowly migrate into the unstable region, which could lead to a delayed collapse of the remnant DANS to a BH.

The additional DM component also adds more complexity to rotating DANS configurations. This is because the DM and the baryonic component can rotate independently and at different rotational rates. Cipriani et al.~\cite{Cipriani:2025tga} and Issifu~\cite{Issifu:2025gsq} investigated this scenario using rotating two-fluid configurations. Cipriani et al. in particular examined DANS with an effective scalar-field DM EoS and considered rapidly rotating DANS configurations with DM cores or clouds, which are either co-rotating or counter-rotating with the baryonic component. They found that incorporating a DM component preserved the overall qualitative behavior of pure baryonic stars. If the DM core/cloud is non-rotating, it is only weakly affected by the NS rotation and may become slightly oblate by $<2\,\%$ (relative difference of polar radius and equatorial radius). For cases which are marginally DM core/cloud configurations, a rapid NS rotation might lead to the DM core having a larger polar radius than the NS, but a smaller equatorial radius. When the DM component rotates as well, there are some differences between co- and counter-rotating DM cases. Co-rotating cases lead to higher rotation rates with smaller deformation of the DM and NS component compared to the purely baryonic case. Counter-rotating configurations, however, lead to highly deformed DM cores. One effect of DM is that it increased the compactness of the NS component, which decreases the moment of inertia compared to a pure NS configuration.

\paragraph{binary systems}

Following the study of isolated DANS, a number of authors began to turn their attention towards binary systems and the GW emissions that they radiate. The importance of GW measurements has increased greatly since the detection of the first GW event in 2015, which is also reflected in the increasing number of studies of possible GW signatures of DM inside of NSs.

One of the first investigations into the GW signatures in DANS mergers was done by Ellis et al.~\cite{Ellis:2017jgp}. They used a mechanical toy model to obtain the qualitative spectral energy density of two merging equal-mass DANS with significant DM fractions of around $10\,\%$ concentrated in a DM core. They then computed the power spectral density (PSD) of equal-mass DANS at the time right after the merger. They identified three frequency peaks related to the hypermassive NS (HMNS) remnant and one additional peak, which comes from the DM cores orbiting each other inside the HMNS. This is possible because the DM cores have smaller radii than the NSs and can therefore complete additional orbits before merging. After the merger, in~\cite{Ellis:2017jgp}, the authors found that the additional DM peak decays faster than the NS baryonic matter contributions, which they explain with the DM cores relaxing faster due to the lack of repulsive forces. However, even after the DM peak weakens, the DM can still affect the PSD because the presence of a DM core can effectively soften the EoS and make the HMNS more compact, which can change the peak frequency of the HMNS modes through its gravitational effect. Through this effect, one could constrain DM only using the NS modes in the PSD. For the unequal mass case, Ellis et al.~\cite{Ellis:2017jgp} find that the DM cores now create two PSD peaks. This characteristic signal would be independent of the nuclear EoS and change for each merger, and could thus be identified by observing multiple DANS mergers.
In a similar vein, Horowitz and Reddy~\cite{Horowitz:2019aim} investigated the possible GW signals produced by a DM core which orbits/oscillates inside of a NS after being captured. For the case where the DM core has a sub-solar mass around $10^{-8}\,M_\odot$, the GW frequency can be multiple kHz, and the characteristic orbital lifetime is on the order of a few days. The resulting high-frequency GW signal could be detected with next-generation GW detectors if such an event occurs within our galaxy.

These semi-analytic works were later followed by increasingly complex numerical simulations, from which we highlight a few in the following. For a more complete list of works, see the previous sub-chapter and the references within the cited papers. 

One early simulation came by Bauswein et al.~\cite{Bauswein:2020kor}, who simulated the merger of binary NSs, which both host a DM core inside them. They considered the case where the DM cores are small in mass $<0.1\,M_\odot$ and radius smaller than about one km, and they subsequently simulated the DM cores as point particles which are affected by the gravitational potential. They found that during the inspiral, the DM particles closely follow the center of mass of the respective binary components. During the merger, the DM cores are then injected into a binary orbit inside of the newly formed common gravitational well and their movement decouples from the baryonic fluid motion. The orbital radius is very small at only a few km and is typically larger (up to $8$km) for unequal mass-ratio mergers. The eccentricity of the DM particle orbit lies in the range $\sim 0.2-0.8$, depending on the binary mass, mass ratio and nuclear matter EoS. The tight orbit means that the DM cores would be quickly swallowed if the merger remnant collapses to a BH. Due to the small orbital radii, the orbital frequency can be high and lead to GW emissions at $3-5\,$kHz, which can last days up to one year. Based on the expected sensitivity of GW detectors, Bauswein et al. finally estimate that only DM components as massive as $0.1\,M_\odot$ are detectable in advanced LIGO and $0.01\,M_\odot$ in the Einstein telescope.

\begin{figure}
    \centering
    \includegraphics[width=0.95\textwidth]{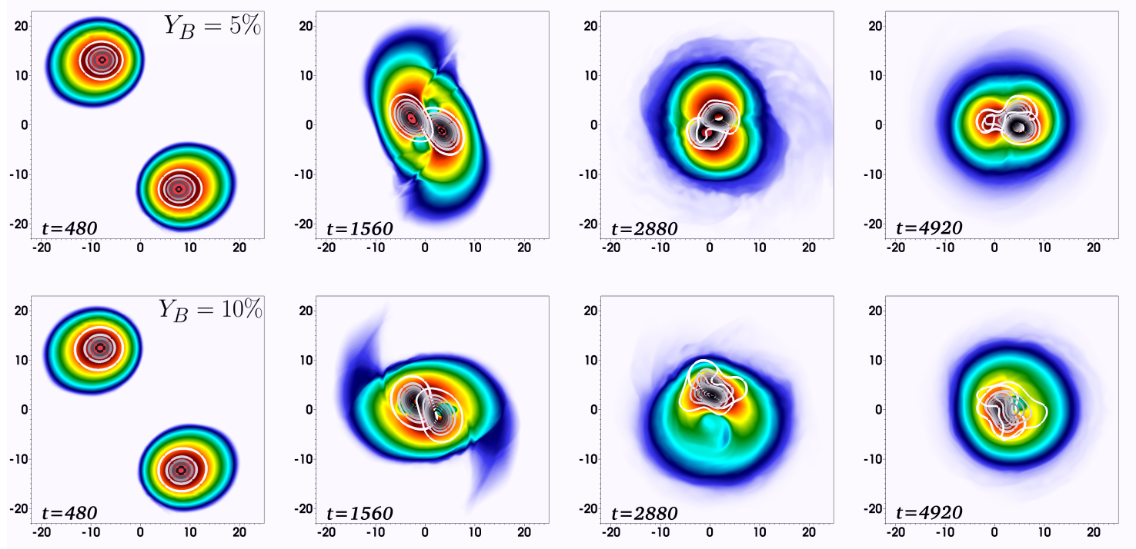}
    \caption{Snapshots of two binary DANS merger simulations, in the equatorial plane, by Bezares et al.~\cite{Bezares:2019jcb}, where the DANS have different DM fraction ($Y_B$ in their notation) of $5\,\%$ (upper row) and $10\,\%$ (lower row), respectively. The colored gradient shows the baryonic restmass density, and the white to black contours show the scalar field density amplitude. The first column shows the state in the early inspiral, the second one is roughly at merger time, the third one is during the post-merger phase, and the fourth shows the end of their simulation. In the $Y_B=10\,\%$ case, the development of the one-arm instability and loss of symmetry can be clearly seen, while for the $Y_B=5\,\%$ case the quadrupolar shape is retained. Adapted from Ref.~\cite{Bezares:2019jcb}.}
    \label{fig:DANS-merger-snapshot-Bezares}
\end{figure}
Bezares et al.~\cite{Bezares:2019jcb} later performed fully relativistic numerical simulations of DANS with scalar field DM, solving the scalar field alongside of the hydrodynamics of the baryonic matter and taking into account the gravitational back-reaction of the DM. As initial conditions they used two equal-mass Lorentz-boosted solutions of isolated scalar-field DANS with DM fractions of up to $10\,\%$. Due to the lack of self-consistent initial data, they could only simulate DANS with DM cores, as DM clouds would interfere non-trivially with the space between the two DANS. They then followed the inspiral and merger of the DANS numerically. While they found no noticeable differences in the inspiral phase, new phenomenology emerged in the merger phase. In agreement with previous works, they found that the DM cores continue to orbit each other for longer time after the baryonic components already merged. While in the low DM-fraction cases, the remnant NS has a strong quadrupolar shape, in the case with DM fraction of $10\,\%$ the DM back-reacts on the nuclear matter and breaks the quadrupolar symmetry. Instead, an overdense region is formed through the one-arm instability, which is excited by the asymmetry of the gravitational three-body interaction between the baryonic core and the two DM cores. We show this behavior in Fig. \ref{fig:DANS-merger-snapshot-Bezares}, which was reprinted from the paper of Bezares. This one-arm instability leads to an additional $l=m=1$ mode in the resulting GW signal, which weakens the quadrupolar $l=m=2$ mode and makes it decay faster. This additional lower-frequency $l=m=1$ mode could not be anticipated by earlier toy model of Ellis et al.. The simulations by Bezares thus show the purely emergent dynamics of DM within NSs. Although the one-arm instability has also been observed in unequal-mass NS mergers, in the DM case, it is much stronger and also appears for equal-mass binaries.

\begin{figure}
    \centering
    \includegraphics[width=0.9\textwidth]{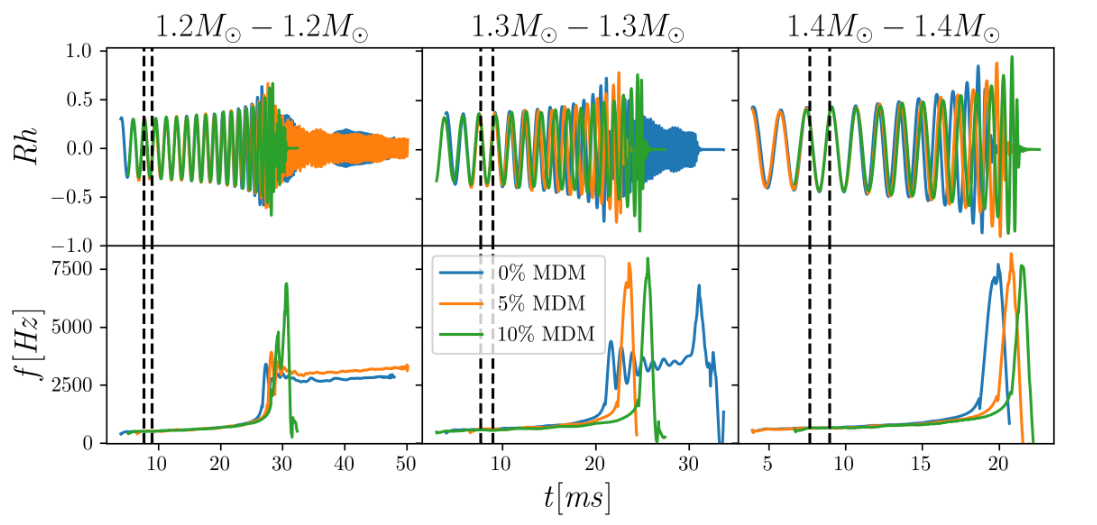}
    \caption{GW strain, $Rh$ (upper row), and frequency of the $l=m=2$ mode (lower row) for three different equal-mass binary DANS mergers with different DM fractions (denoted by the fraction of mirror DM (MDM)), as a function of time, as simulated by Emma et al.~\cite{Emma:2022xjs}. The vertical black lines denote the time range where the GW waveforms have been matched to each other to make the comparison easier. As can be seen, the presence of DM cores facilitate the prompt collapse of the merger remnant, all else being equal. Adapted from ref.~\cite{Emma:2022xjs}.}
    \label{fig:DANS-merger-waveform-Emmma}
\end{figure}

Another binary DANS merger simulation was performed by Emma et al.~\cite{Emma:2022xjs}, who investigated DANS admixed with mirror-DM and their possible multi-messenger observational signatures using the two-fluid formalism. They investigated DM core configurations for a range of DM fractions and binary masses and computed the resulting GW signal and mass ejecta properties. They found that the inspiral is slower in the case of DM cores, which they suspect is due to the reduced tidal deformability of DANS with DM cores compared to pure NS. During the merger phase, they observe that high-mass binaries of equal mass ($1.4\,M_\odot$ each) promptly collapse to a BH, but for lighter binaries, the collapse time depends on the DM content. For a less massive DANS binary ($1.3\,M_\odot$ each), Emma et al. observed the formation of a longer-lived (over 10 ms) HMNS in the case of no DM, but for the DM admixed cases (DM fraction of $5\,\%$ and $10\,\%$), they found the remnant to collapse promptly. The presence of DM cores thus facilitates the prompt collapse of the remnant to a BH. This can be seen in Fig. \ref{fig:DANS-merger-waveform-Emmma}, which is taken from Emma et al.~\cite{Emma:2022xjs}. Emma et al. also computed the mass of the ejecta and disk mass after BH formation, which is a proxy for the brightness of the resulting kilonova. While they found no clear correlation of ejecta mass with the DM fraction, they found that a larger DM fraction leads to a less massive disk. They suspect this being the case because DANS with DM cores are generally more compact with higher DM fraction, which makes it more difficult for matter to be separated from the matter bulk that falls into the BH. They finally conclude that, since the disk mass correlated with the kilonova brightness, the smaller disk size will lead to a less bright kilonova compared to the DM free case.

The first case of fully self-consistent initial data for binary DANS was finally achieved by Rüter et al.~\cite{Ruter:2023uzc} in 2023. They modified the code used by Emma et al.~\cite{Emma:2022xjs} to allow for arbitrary combinations of two-fluid DANS with cores or halos, where the DM is modeled as a fermi gas. In their paper, they demonstrate the performance of their initial-data algorithm and show convergence, which was not achieved in previous studies. Building upon these results, and in collaboration with Rüter, Giangrandi et al.~\cite{Giangrandi:2025rko} then used the formalism for consistent binary DANS initial data to perform full general-relativistic merger simulations of two-fluid DANS. In their setup, they investigated equal-mass DANS mergers with Fermi-fluid DM with a total binary mass of $2.8\,M_\odot$ and $2.6\,M_\odot$. They then ran three simulations for each binary mass: one without DM, one with a $3\,\%$ DM fraction ($m_\mathrm{DM}=1\,$GeV) concentrated in a core, and one with cloud configurations and a DM fraction of $0.5\,\%$ ($m_\mathrm{DM}=0.17\,$GeV). For the cores, they find a similar behavior to previous studies, that the DM cores continue to orbit each other even after the baryonic components merge and tend to reduce the lifetime of the remnant due to the increased compactness of the total system. The final BH mass is also increased in the presence of DM cores. In the case of DM clouds, however, the DM is seen to form a common envelope around the merger remnant HMNS, but has less impact on the merger remnant. Additionally, Giangrandi et al. found that the tidal deformability, that is obtained from the numerically extracted GW signal through matching with GW templates can be significantly miss-estimated (factor of $\sim 3$). Instead, better results are obtained if one only used the tidal deformability which is obtained by considering a pure baryonic or DM core. This result also suggests, as argued by Giangrandi et al., that previously obtained exclusion bounds for the DM fraction and particle mass, using GW parameter estimations, could be too strict and should be revisited.

In parallel, Srikanth et al.~\cite{Srikanth:2025lic} performed simulations of binary NSs that are immersed in a common envelope of ultralight DM particles, modeled as a scalar field. Their main objectives were to investigate how the presence of DM affects the inspiral and merger of the binaries. A common DM envelope could form in the case where the DM particle is too light or the repulsive self-interaction is too strong to form a core. In this case it might be difficult to probe the DM properties using isolated NSs, but NS binaries can probe this range. Srikanth et al. found that the scalar field does not disperse but  forms quasi-stable co-rotating waves that peak at the NS positions. This additional scalar field then leads to a dephasing  of up to $\sim0.5\,$rad of the binary compared to the vacuum case, and the effect gets stronger for larger scalar field mass. During the merger, they found a reduction in the ejecta mass because the presence of the scalar field suppresses shock driven ejecta. Also, the scalar field cloud reduces the baryonic central density of the remnant, which leads to a delayed collapse to a BH.

While numerical simulations are useful to study the dynamics of NSs admixed with DM, they remain computationally expensive and it is difficult to run sufficiently many simulations to exhaustively cover the parameter space for the sake of estimating parameters or producing exclusion bounds. We therefore highlight here some first works by Koehn et al.~\cite{Koehn:2024gal} and Rutherford et al.~\cite{Rutherford:2024uix} that tackle the problem of parameter estimation of the DM properties using NS mass-radius and GW (tidal deformability) measurements. Koehn et al.~\cite{Koehn:2024gal} studied whether the nuclear matter EoS and the DM properties can be extracted from GW observations of the tidal deformability using the ET. For this, they created a synthetic catalog of $500$ high-SNR binary NS merger events admixed with up to $1\,\%$ Fermi-fluid DM. Using this catalog, they then simulated mock-observations in ET and then used Bayesian inference to recover the nuclear matter and DM properties. They found that the presence can bias the recovered baryonic EOS towards softer EOS if DM is not accounted for, even though the effect is small. But, even for events which have DM, they find that the Bayes factor disfavors DM and is in general not able to distinguish between a DANS and a DM-free NS. Likewise, it was found that the Bayesian analysis is not able to break the degeneracy between DM fraction and DM particle mass.
Rutherford et al.~\cite{Rutherford:2024uix} instead focused on measurements of NS mass and radius, instead of the tidal deformability. They similarly considered DANS with fermionic mirror-DM and DM fractions below $2\,\%$. They then used Bayesian parameter estimations for the baryonic nuclear EoS and the DM parameters (mass and DM-baryon interaction-strength). They used two MR datasets for isolated NSs, one based on real measurements by NICER and another dataset based on a future mission which could reach $1\sigma$ radius-uncertainties of $\pm0.25\,$km. Similarly to Koehn et al., they found that the measurements are consistent with no DM and that the posteriors are nearly identical between the DM admixed DM free cases. This is true for the current NICER constraints and the synthetic measurements with smaller error bars. For the case with smaller error bars, they found only a little shift in the DM-baryon interaction-strength posterior.

In conclusion, both works find that it proved very difficult to estimate the DM parameters for DANS with DM fractions $<2\,\%$. Current methods are essentially unable to distinguish between a pure NS or a DANS, which implies that DANS are equally as consistent with NS data as purely baryonic NS. Our inability to meaningfully constrain DM properties using global NS properties remains strongly limited for current data, including the data that might be obtained in the foreseeable future.

\subsection{Axion-Like Particles in Stellar and High-Energy Astrophysical Contexts}

In this section, rather than focusing on DM accumulated inside NSs and its impact on their intrinsic properties, we concentrate on ALPs and Majorons as viable DM candidates. We review how their production and propagation leave characteristic spectral signatures in supernovae, the CMB, and other astrophysical and cosmological sources, and how such observations can be used to place constraints on these particles. ALPs can be produced in stellar interiors through interactions with photons~\cite{Dine:1982ah,CAST:2004gzq}, charged leptons~\cite{Raffelt:1985nk,Barth:2013sma,Bollig:2020xdr}, and nucleons~\cite{Chang:1993gm, DiLuzio:2020wdo}. Once they are generated, their weak coupling allows them to escape the stellar medium, leading to potentially observable energy losses. Therefore, the cooling of stars, particularly supernovae (SNe)~\cite{Raffelt:1987yt, Raffelt:1990yz} and the Sun~\cite{Raffelt:1999tx} put constraint on the mass and the coupling of ALPs to SM particles. The coupling between ALPs and photons can be also constrain from the magnetic resonance around NSs~\cite{Lai:2006af, Pshirkov:2007st, Safdi:2018oeu, Foster:2020pgt, Darling:2020uyo, Nurmi:2021xds, Foster:2022fxn, Noordhuis:2022ljw, Edwards:2020afl, Witte:2021arp, Witte:2022cjj, Gines:2024ekm, Long:2024qvd, Walters:2024vaw}.

The coupling of ALPs to nucleons and hadrons are intensively investigated so far, since one of the most investigated models of ALPs, the QCD axions, chirally couple to quark sector to solve strong CP problem~\cite{Peccei:1977hh, Weinberg:1977ma, Wilczek:1977pj}. We particularly focus on the constraint on ALPs-nucleon coupling from supernovae in Sec.~\ref{sec:SN ALPs-nucleons}.

We also highlight another window to search the properties of ALPs; the coupling to neutrinos. Neutrinos are uniquely positioned as messengers of DM: they are electrically neutral, weakly interacting, and possess sub-eV masses, enabling them to escape from dense astrophysical environments such as NSs and supernovae cores. This makes them not only valuable observational probes, but also key participants in high-energy astrophysical and cosmological processes where DM may leave imprints (see e.g.~Refs.~\cite{Arguelles:2019ouk,Arguelles:2022nbl} for review).

Moreover, the discovery of nonzero neutrino masses is itself a landmark indication of new physics beyond the SM. Many theoretical frameworks developed to explain neutrino mass generation naturally predict additional particles that could also serve as DM candidates. These include one of ALPs, {\it Majoron}-pseudo Nambu-Goldstone boson associated with the spontaneous breaking of global lepton number symmetries U(1)$_\text{L}$ or U(1)$_\text{B-L}$\cite{Chikashige:1980qk,Chikashige:1980ui}-as well as fermionic states like sterile (right-handed) neutrinos (see e.g.~Ref.~\cite{Drewes:2016upu} for review), and vector bosons such as the $Z'$ boson associated with gauged U(1)$_\text{B-L}$ (see e.g.~Ref.~\cite{Langacker:2008yv}). In this broader context, interactions between ALPs DM and neutrinos may arise naturally and offer a unique window into the interplay between DM physics and the origin of neutrino masses. We particularly focus on Majoron DM in Sec.~\ref{sec:Majoron DM}, to show the current status and future prospects of the detectability of Majoron DM based on Ref.~\cite{Akita:2023qiz}.

\subsubsection{The constraints on ALP-nucleon coupling from supernovae}
\label{sec:SN ALPs-nucleons}

One of the most efficient laboratories to test the phenomenology of exotic particle lighter than $\mathcal{O}(100\,{\rm MeV})$ is particle emission from the core of supernovae (SNe), and ALPs are not exception~\cite{Raffelt:1987yt, Raffelt:1990yz}. 
Particularly, the measured neutrino pulse from SN1987A~\cite{Kamiokande-II:1987idp,Hirata:1988ad,Bionta:1987qt} severely constrains the operation of an ``exotic'' cooling mechanism of SNe core. As ALPs are pseudo scalar bosons, they may have derivative coupling to the SM fermions. The interactions between axion and nucleons are given by~\cite{Chang:1993gm,DiLuzio:2020wdo},
\begin{align}
    \begin{split}\label{eq:aN interaction}
        &\mathcal{L}_{\rm int}= \frac{g_{a}}{2m_{N}}\partial_{\mu}a\\
        &\times\left\{
            C_{ap}\bar{p}\gamma^{\mu}\gamma_{5}p + C_{an}\bar{n}\gamma^{\mu}\gamma_{5}n 
            + \frac{C_{a\pi N}}{f_{\pi}}\left(i\pi^{+}\bar{p}\gamma^{\mu}n - i\bar{n}\gamma^{\mu}p\right) + C_{aN\Delta}\left(\bar{p}\Delta_{\mu}^{+} +\overline{\Delta_{\mu}^{+}}p + \bar{n}\Delta_{\mu}^{0}+ \overline{\Delta_{\mu}^{0}n} \right)
        \right\}
        \,,
    \end{split}
\end{align}
where the dimension-less coupling constant $g_{a}=m_{N}/f_{a}$ is given by the nucleon mass $m_{N}=938\,{\rm MeV}$ and the PQ scale $f_{a}$. The first two terms describe the direct interaction of ALPs to nucleons with model-dependent coupling constants $C_{aN}$ where $N=p,n$. The following terms represent axion-pion-nucleon couplings and axion-nucleon-$\Delta$ baryon couplings with $C_{a\pi N} = (C_{ap}-C_{an})/\sqrt{2}g_{A}$~\cite{Choi:2021ign} and $C_{aN\Delta} =-\sqrt{3}/2(C_{ap}-C_{an})$  where $f_{\pi}=92.4\,{\rm MeV}$ is the pion decay constant and $g_{A}\sim 1.28$~\cite{ParticleDataGroup:2022pth} is the axial coupling constant. We use the conventional coupling constant $g_{aN}=g_{a}C_{aN}$ instead of $g_{a}$ to present the strength of the axion-nucleon interaction.

There are two emission processes of ALPs inside the SN core; the bremsstrahlung process $NN\rightarrow NNa$ driven by $C_{aN}$ coupling and the pionic Compton-like scattering $\pi^{-} N\rightarrow aN$ driven by $C_{aN\pi}$ and $C_{aN\Delta}$. Initially the former process has been intensively studied neglecting the latter process. The first numerical estimation of the massless ALPs luminosity under One-Pion-Exchange (OPE) approximation was done in Ref.~\cite{Turner:1987by,Brinkmann:1988vi,Carena:1988kr}, which is now found to be reduced one order of magnitude by accounting several corrections shown by the follow-up works investigating the effect of mass of pions~\cite{Stoica:2009zh}, multiple pion exchanges~\cite{Ericson:1988wr}, contribution of medium to the mass of nucleons and multiple scattering process~\cite{Raffelt:1991pw,Janka:1995ir}. 

The pionic Compton-like scattering has been widely ignored after Refs.~\cite{Turner:1991ax,Raffelt:1993ix,Keil:1996ju}. Recently, however, Ref.~\cite{Carenza:2020cis} pointed out that it may dominate the bremsstrahlung process by revisiting negatively charged pion abundance in the core~\cite{Fore:2019wib} and the reduction of axion emission from the bremsstrahlung process beyond OPE approximations. 

The references above were mainly focused on the ALPs mass $m_{a}\ll 3T\sim 90\,{\rm MeV}$ where $T$ is the temperature of the environment of ALPs emission. In such case, the Boltzmann suppression $e^{-m_{a}/T}$ of ALPs production process can be neglected. Recent works derived comprehensive analysis by relaxing the condition of mass to $m_{a}<m_{N}$ for the bremsstrahlung process~\cite{Giannotti:2005tn}, and the pionic Compton-like scattering in addition to the correction of the bremsstrahlung process beyond OPE approximation~\cite{Lella:2022uwi,Carenza:2023lci}.

Once the scattering amplitude $|\mathcal{M}|^{2}$ is identified, the spectrum of ALPs per unit volume is obtained by usual integral over the phase space, 
\begin{align}
    \begin{split}
        \frac{\D ^{2}n_{a}}{\D E_{a}\D t}=\int \left(\frac{4\pi E_{a}|\vec{p}_{a}|}{(2\pi)^{3}2E_{a}}\right)\left(\Pi_{i}\frac{\D^{3}p_{i}}{(2\pi)^{3}2E_{i}}\right)\delta^{(3)}\left(\sum \vec{p}\right)\delta\left(\sum E\right)|\mathcal{M}|^{2}F_{\rm stat}\,,
    \end{split}
\end{align}
where $\vec{p}_{i}$ and $E_{i}$ are the momentum and energy of relevant particles in ALPs emission; $i=N_{1},\cdots, N_{4}$ for $N_{1}N_{2}\rightarrow aN_{3}N_{4}$ and $i=\pi,N_{1},N_{2}$ for $\pi N\rightarrow aN$, respectively. Delta functions confirm the momentum and energy conservation laws. $F_{\rm stat}$ is the statistical factors, namely $f_{N_{1}}f_{N_{2}}(1-f_{N_{3}})(1-f_{N_{4}})(1-f_{a}) - f_{N_{3}}f_{N_{4}}f_{a}(1-f_{N_{1}})(f_{N_{2}})$ for the former process and $f_{\pi}f_{N_{1}}(1-f_{a})(1-f_{N_{2}}) - f_{N_{2}}f_{a}(1-f_{\pi})(1-f_{N_{2}})$ for the latter process, respectively, where $f_{i} $ is the distribution function of the $i$-particle. See Ref.~\cite{Carenza:2023lci} for comprehensive review on the detail of the computation of the spectrum as well as the scattering amplitude of each ALPs emission processes.

The energy of the ALPs measured by an observer at infinity $E_{\infty}$ is shifted with respect to the local energy at the position of emission $E_{\rm e}$ as $E_{\infty}=E_{\rm e}\alpha(r)$, where the lapse factor $\alpha(r)\leq 1$ which encodes the gravitational redshift effects obtained from SN numerical simulation. Similarly, the local time coordinate at the particle emission $\D t_{\rm e}$ is slowed-down by the gravitational potential and related to the distant coordinate $\D t_{\infty}$ as $\D t_{\rm e}=\alpha(r) \D t_{\infty}$. 
Therefore, the energy spectrum of the ALPs for a distant observer is given by
\begin{align}
    \begin{split}
        \frac{\D N_{a}}{\D E_{\infty}}=\int \frac{\D ^{2}n_{a}}{\D E_{\rm e}\D t_{\rm e}}\frac{\D E_{\rm e}}{\D E_{\infty}}\D t_{\rm e}\D^{3}r = \int \frac{\D ^{2}n_{a}}{\D E_{\infty}\D t_{\infty}}\D t_{\infty} \D^{3}r\,,\label{eq:energy spectrum of ALPs from SN}
    \end{split}
\end{align}
where we used $\D E_{\infty}\D t_{\infty}=\D E_{\rm e}\D t_{\rm e}$.

Note that emitted ALPs escape from the SN gravitational field only if kinetic energy is sufficiently large. For relativistic ALPs, the condition is given by $E_{\infty}> m_{a}$ thus $E_{\rm e}\geq m_{a}\alpha(r)^{-1}$, while for non-relativistic ALPs, $m_{a}v^{2}/2> m_{a}M(r)/r$ thus $v^{2}> 2M(r)/r$ where $v$ is the speed of produced ALPs and $M(r)$ is the mass enclosed in the radius $r$ and we take $G_{N}=1$. Particles which satisfy those conditions can be detected by a distant observer.

If we take into account the absorption of ALPs at the radius $r$ during SN cooling, we multiply the integrand in Eq.~\eqref{eq:energy spectrum of ALPs from SN} by~\cite{Chang:2016ntp, Chang:2018rso,Caputo:2021rux, Caputo:2022rca, Lella:2023bfb}
\begin{align}
    \begin{split}
        \langle \exp\left\{-\tau(E_{\rm ab},r)\right\}\rangle
        =
        \frac{1}{2}\int_{-1}^{1} \D \mu \exp\left\{\int _{0}^{\infty} \D s \Gamma_{\rm ab}\left(E_{\rm ab}\sqrt{r^{2}+s^{2}+2rs\mu}\right)\right\}\,,
    \end{split}
\end{align}
where $\Gamma_{\rm ab}(E,r)$ is the (reduced) absorption rate of ALPs with the energy $E$ at the radius $r$, defined in Ref.~\cite{Caputo:2021rux}. The energy of the ALPs at absorption is given by $E_{\rm ab}=\alpha(r)E_{\rm e}/\alpha(\sqrt{r^{2}+s^{2}+2rs\mu})$, where $s$ is the distance that ALPs traveled after its emission at $r$ and $\mu$ is the cosine of the angle of the motion.

The observed luminosity is then given by 
\begin{align}
    \begin{split}
        L_{a}=
        \int  \left(E_{\infty}\frac{\D}{\D t_{\infty}}\frac{\D N_{a}}{\D E_{\infty}}\right) \D E_{\infty}
        = \int \alpha^{2}(r)E_{\rm e}\frac{\D ^{2}n_{a}}{\D E_{\rm e}\D t_{\rm e}}\D E_{\rm e}\D^{3}r\,,
    \end{split}
\end{align}
and the constraint on the coupling $g_{aN}$ is found from $L_{a}\lesssim L_{\nu}$; otherwise, the exotic energy losses of the core of the SN makes neutrino burst shorter than the standard expectation. 

\begin{figure}
    \centering
    \includegraphics[width=8cm]{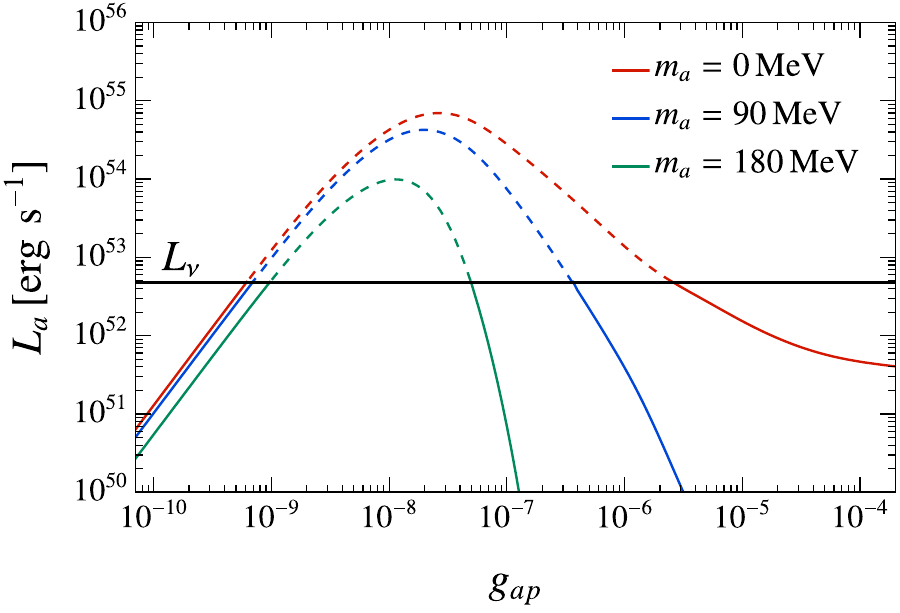}
    \caption{The luminosity of ALPs at the postbounce time $t_{\rm pb} = 1 \,{\rm s}$ as a function of the ALP-proton coupling $g_{ap}$ in KSVZ model where $(g_{ap},g_{an})=(-0.47 g_{a},0)$, reproduced from Ref.~\cite{Lella:2023bfb}, \textcopyright  Lella et al., CC-BY 4.0. Different curves correspond to three cases of ALPs mass; $m_{a}=0\,{\rm MeV}$ in red, $m_{a}=90\,{\rm MeV}$ in blue, and $m_{a}=180\,{\rm MeV}$ in green, respectively. The luminosity of neutrinos $L_{\nu}\simeq 5\times 10^{52}\,{\rm erg\,s^{-1}}$ is also shown in black, above which $L_{a}$ is depicted in dashed curves.}
    \label{fig:ALPs-luminosity}
\end{figure}

Figure \ref{fig:ALPs-luminosity}, reproduced from~\cite{Lella:2023bfb} shows the ALP luminosity at the postbounce time $t_{\rm pb}=1~\rm s$ as a function of ALP-proton coupling $g_{ap}$ with different mass of ALPs $m_{a}$, adapting the coupling $(C_{ap},C_{an})=(-0.47,0)$ as in the Kim-Shifman-Vainshtein-Zakharov (KSVZ) model~\cite{GrillidiCortona:2015jxo}. 
The luminosity has a peak around $g_{ap} \sim 10^{-8}$~\cite{Lella:2022uwi}; for sufficiently small $g_{ap}$, ALPs couple to nucleons weakly and thus escape from the core of collapse once produced. 
Therefore, the luminosity simply increases as the efficiency of ALPs production, i.e.~$L_{a}\propto g_{aN}^{2}$. 
For higher coupling, on the other hand, the absorption processes of ALPs to the medium in the core become relevant. In the regime, therefore, ALPs are trapped in the core and form ``ALPs spheres''~\cite{Caputo:2022rca}, similarly to neutrinos. Hence, the luminosity is approximated by that of the radiation from blackbody, i.e.~$L_{a}\propto r^{2}T_{a}(r)^{4}$ where $r$ is the emission radius and $T_{a}(r)$ is the local temperature of ALPs at the emission. Note that ALPs luminosity with the mass $m_{a}\gtrsim 10\,{\rm MeV}$ is exponentially suppressed inside the axion sphere because of the Boltzmann suppression $e^{-m_{a}/T_{a}}$.

\begin{figure}
    \centering
    \includegraphics[width=8cm]{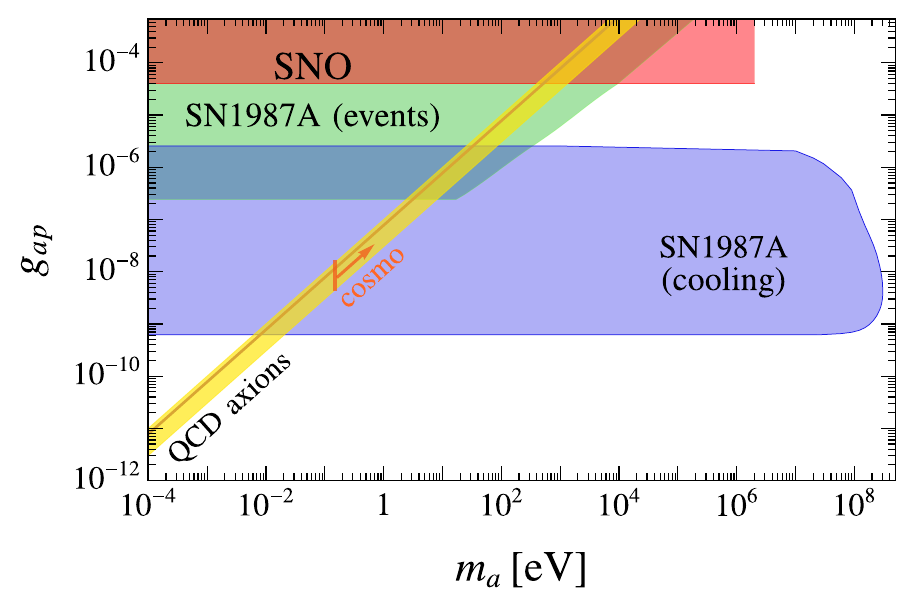}
    \caption{Summary plot reproduced from Ref.~\cite{Lella:2023bfb}, \textcopyright  Lella et al., CC-BY 4.0, showing the constraints on ALPs-proton coupling $g_{ap}$ as a function of ALPs mass $m_{a}$, in KSVZ model where $g_{an}=0$. The bound from SN cooling discussed in this section is shown in blue. The QCD axion band is in yellow with the orange line showing the current cosmological constraint on its mass $m_{a}\gtrsim 0.15\,{\rm eV}$~\cite{Archidiacono:2015mda, Caloni:2022uya, DEramo:2022nvb}. The result of the solar ALPs search in SNO experiment~\cite{Bhusal:2020bvx} is shown in red, and the constraint from the absence of ALP-induced events in Kamiokande-II experiment~\cite{Carenza:2023wsm} is shown in green labeled as SN1987A(events).}
    \label{fig:ALPs-SN-constraint}
\end{figure}

The condition $L_{a}\lesssim L_{\nu}\simeq 5\times 10^{52}\,{\rm erg\,s^{-1}}$ is translated to the constraint on $g_{ap}$, which is shown in blue region in Figure \ref{fig:ALPs-SN-constraint}, reproduced from~\cite{Lella:2023bfb}. The upper bound on the coupling $g_{ap}\lesssim 5\times 10^{-10}$, corresponds to the free-streaming ALPs without significant effect of absorption. On the other hand, lower bound $g_{ap}\gtrsim 2.5\times 10^{-6}$ is comes from absorption, dominantly in the inverse bremsstrahlung process. Those constraints may get altered to $g_{ap}\lesssim 8.5\times 10^{-10}$ and $g_{ap}\gtrsim 1.5\times 10^{-6}$, respectively, due to uncertainty of SN cooling simulation. 
Other regions in Figure \ref{fig:ALPs-SN-constraint} are the following; QCD axion band in yellow with the orange line showing the current cosmological constraint on its mass $m_{a}\gtrsim 0.15\,{\rm eV}$~\cite{Archidiacono:2015mda, Caloni:2022uya, DEramo:2022nvb}. The result of the solar ALPs search in SNO experiment~\cite{Bhusal:2020bvx} is shown in red, and the constraint from the absence of ALP-induced events in Kamiokande-II experiment~\cite{Carenza:2023wsm} is shown in green labeled as SN1987A(events). 

\subsubsection{Constraints on neutrino lines from ALPs dark matter decay}
\label{sec:Majoron DM}

Majoron is the Nambu-Goldstone boson associated with global $U(1)_{\rm L}$ (or $U(1)_{\rm B-L}$) symmetry~\cite{Chikashige:1980qk,Chikashige:1980ui}, where its spontaneous breaking generates Majorana neutrino mass in type-I seesaw mechanism. Perhaps the simplest model is the singlet Majoron model, where the Lagrangian is given by
\begin{align}
    \begin{split}
        \mathcal{L}=\mathcal{L}_{\rm SM} + iN_{R}\slashed{\partial}N_{R} -\mathcal{V}_{\rm Yukawa}-\mathcal{V}_{\Phi}\,,
    \end{split}
\end{align}
where $\mathcal{L}_{\rm SM}$ is the Lagrangian of minimal SM without right-handed neutrino, and $N_{R}$ is the SM singlet right-handed neutrino. We introduce three generation of right-handed neutrinos while the different number of species of right-handed neutrinos do not change the following discussion.
The Yukawa coupling terms to induce neutrino mass terms are given by 
\begin{align}
    \mathcal{V}_{\rm Yukawa} = \lambda_{D} H^{*}\overline{E_{L}}N_{R}+ \frac{\lambda_{R}}{2}\overline{N_{R}^{c}}\Phi N_{R}+{\rm h.c.}\,,
    \label{Lsingletmajoron}
\end{align}
where $\lambda_{D}$ and $\lambda_{R}$ are Yukawa couplings, $H$ is the SM Higgs doublet, $E_{L}$ is SM lepton doublet and $\Phi$ is a SM singlet complex scalar we introduce. 
The self-coupling of the newly introduced scalar $\Phi$ is given by
\begin{align}
    \mathcal{V}_{\Phi} = \lambda_{\Phi}\left(|\Phi|^{2}-\frac{f^{2}}{2}\right)^{2} + \mathcal{V}_{\rm sym.\,bkg.}\,,
\end{align}
where $\lambda_{\Phi}$ is the coupling constant and $f$ is the energy scale of spontaneous global $U(1)_{\rm L}$ symmetry breaking. 
By assigning the lepton number of $L(\Phi)=-2$ to $\Phi$, the whole Lagrangian respects global $U(1)_{L}$ symmetry at the tree-level. 

After the global $U(1)_{\rm L}$ symmetry is spontaneously broken, the Yukawa coupling between right-handed neutrinos and $\Phi=(f+\sigma(x) + iJ(x))/\sqrt{2}$ induces the Majorana mass $M_{R}=\lambda_{M}f/\sqrt{2}$. 
As the Dirac neutrino mass $m_{D}=\lambda_{D}v/\sqrt{2}$ is induced below the electroweak symmetry breaking $H=(0,v+h(x))^{T}/\sqrt{2}$ at the energy scale $v$, we take $v\ll f$ to let the type-I seesaw mechanism works even if $\lambda_{D} \sim\lambda_{M}\sim\mathcal{O}(1)$. 
Here we introduced two real scalars $\sigma(x)$ and $J(x)$. The former field corresponds to the radial mode of $\Phi(x)$ which decouples to the low-energy phenomenology below electroweak scale since it obtains the mass $m_{\sigma}=\lambda_{\Phi}f/\sqrt{2}\gg v$. The latter field is the Nambu-Goldstone boson associated to the global $U(1)_{\rm L}$ symmetry called Majoron, which may become massive, once the symmetry is explicitly broken by some additional term $\mathcal{V}_{\rm sym.\,bkg.}$.

One may introduce a SM triplet scalar to build a Yukawa coupling with SM lepton doublet to induce Majorana mass for left-handed neutrinos. The simplest class of such triplet Majoron models has been ruled out by the invisible $Z$-boson decay in the LEP experiment~\cite{Berezhiani:1992cd}. The combination of triplet and singlet Majoron models is an interesting possibility as discussed in literature (for example, see Refs.~\cite{Biggio:2023gtm,Lattanzi:2014zla}), while we focus on the simplest singlet Majoron model without introducing triplet in this article.

Once Majoron obtains the mass from explicit breaking of global $U(1)_{\rm L}$ symmetry, it becomes a viable DM candidate as other class of ALPs particles, since the decay rate is suppressed by the (inverse of) symmetry breaking scale $1/f$. There are variety of explicit symmetry breaking terms $\mathcal{V}_{\rm sym.\,bkg.}$ to explain Majoron DM production mechanism to satisfy the relic abundance. 
Several production mechanisms of Majoron DM appear in the literature that are consistent with the DM abundance in the Universe, including both thermal~\cite{Frigerio:2011in, Queiroz:2014yna} and non-thermal scenarios~\cite{Rothstein:1992rh,Berezinsky:1993fm,Frigerio:2011in,Abe:2020dut,Manna:2022gwn,Queiroz:2014yna}. 
In order to open the DM mass as a free parameter to be determined by observation, we do not specify either of the production mechanism or the form of $\mathcal{V}_{\rm sym.\,bkg.}$ but examine the observational constraints from the decay of Majoron DM, as investigated in Refs.~\cite{Garcia-Cely:2017oco,Akita:2023qiz}. This neglect of $\mathcal{V}_{\rm sym.\,bkg.}$ could give the conservative constraint on the Majoron decay. However, the additional decay through $\mathcal{V}_{\rm sym.\,bkg.}$ may mimic background and weaken the constraints on the Majoron DM in general.

We further assume the Majoron mass satisfies $m_{1},m_{2},m_{3}\ll m_{J}\ll m_{4},m_{5},m_{6}$ where $m_{i}(i=1,\cdots 6)$ are the eigenvalues of the mass matrix,
\begin{align}
    \begin{pmatrix}
        0&m_{D}\\
        m_{D}^{T}&M_{R}
    \end{pmatrix}=V^{*}{\rm diag}(m_{1}\cdots m_{6})V^{\dagger}\,,
\end{align}
with a unitary matrix $V$. 
Under this assumption, Majoron DM decays into neutrinos $J\rightarrow 2\nu_{i}(i=1,2,3)$ at tree-level, where the rate is given by
\begin{align}
    \Gamma(J\rightarrow 2\nu) &\simeq \frac{m_{J}}{16\pi f^{2}}\sum_{i=1}^{3}m_{i}^{2}
    \sim \frac{1}{3\times 10^{19}\,{\rm sec}}\left(\frac{m_{J}}{1\,{\rm MeV}}\right)\left(\frac{10^{9}\,{\rm GeV}}{f}\right)^{2}\left(\frac{\sum m_{i}^{2}}{10^{-3}{\rm eV^{2}}}\right)\,, \label{eq:total decay rate into neutrinos}
\end{align}
where we took the seesaw limit $M_{R}\gg m_{D}$ thus $m_{4,5,6}\gg m_{1,2,3}$. 
Due to the suppression from the (inverse of) decay constant $1/f$ and (the squared of) the mass of induced neutrinos $m_{i}^{2}$, Majoron obtains the lifetime which is long enough to make it DM candidate.

\begin{figure}
    \centering
    \includegraphics[width=10cm]{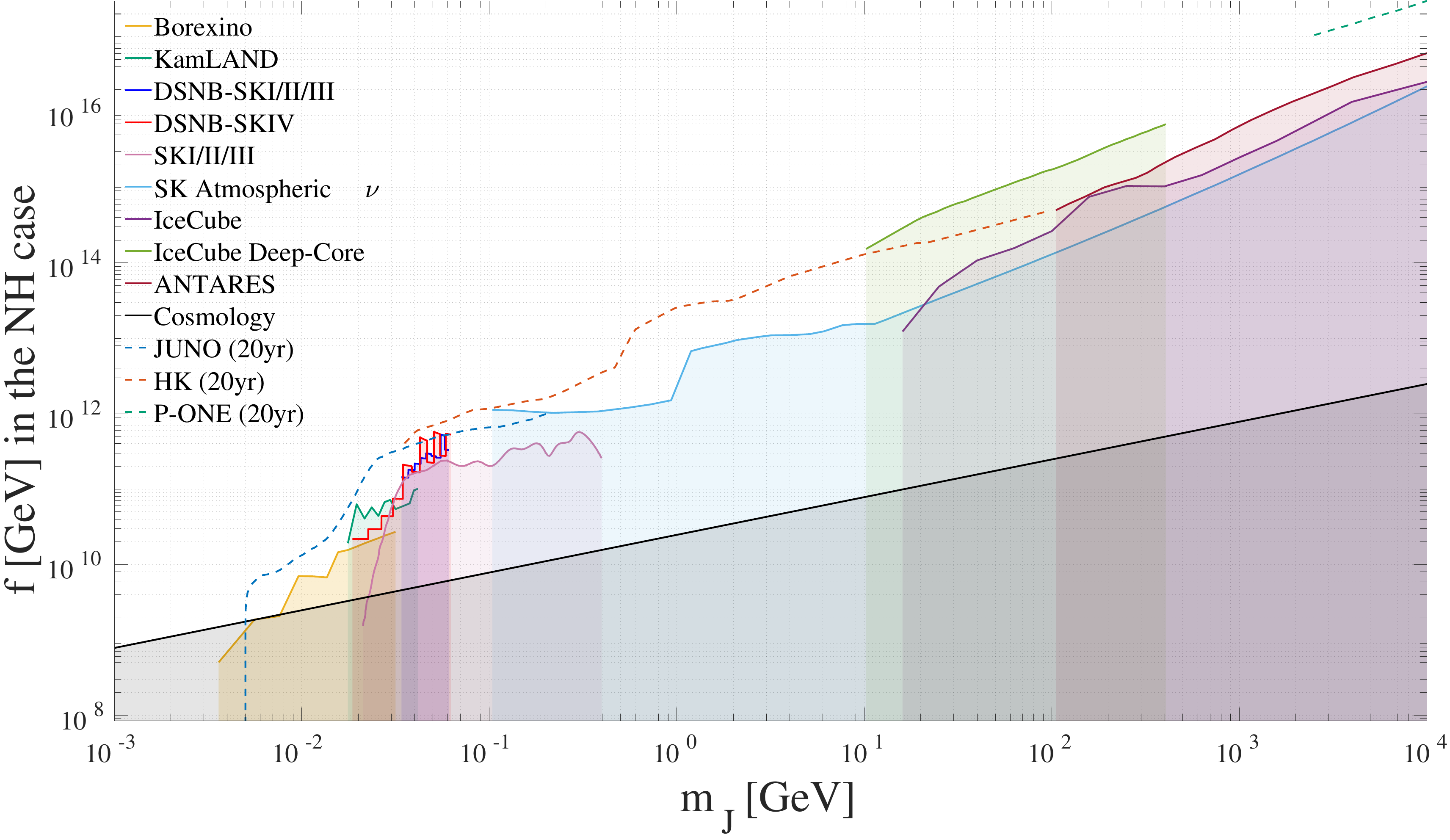}
    \caption{The current lower bounds and future prospects of the spontaneous symmetry breaking scale $f$, reproduced from Ref.~\cite{Akita:2023qiz}, \textcopyright  Akita et al., CC-BY 4.0. The black solid curve shows the model-independent constraint on the DM lifetime from CMB+BAO $\Gamma^{-1}\leq 250\,{\rm Gyr}$ and other colored regions with solid curves show the constraints from neutrino telescopes, while the dashed curves describe the expected sensitivities of future neutrino telescopes. The data used to produce the plot as well as the relevant experiments are explained in detail in Ref.~\cite{Akita:2023qiz}.}
    \label{fig:NH2}
\end{figure}

In order to estimate the observable, we need to determine the branching ratio of neutrino flux in the flavor basis, because neutrino telescopes detect neutrinos through weak interactions. The branching ratio is given by
\begin{align}
    \alpha_{\beta} = \frac{\sum_{i=1}^{3}|U_{\beta i}|^{2}m_{i}^{2}}{\sum_{i=1}^{3}m_{i}^{2}}\,,
\end{align}
where $\beta=e,\mu,\tau$ is the flavor index and $U$ is Pontecorvo-Maki-
Nakagawa–Sakata (PMNS) matrix. 
Figure \ref{fig:NH2} shows the current constraint (colored regions) and future prospects (dashed curves) on the spontaneous symmetry breaking scale $f$ within ${\rm MeV}\leq m_{J}\leq 10\,{\rm TeV}$, focusing on the line spectrum of neutrinos from Majoron DM inside Milky Way, assuming the NFW profile~\cite{Navarro:1995iw}.
Above $m_{J}\gtrsim 10\,{\rm TeV}$, the decay channel $J\rightarrow \nu\nu h(h)$ becomes dominant at tree-level~\cite{Dudas:2014bca} which we neglect in the review since the continuous spectrum of neutrino weakens the constraints. We also neglect continuous neutrino flux from extragalactic DM decays. We also assume normal-hierarchy (NH) of neutrino mass with one massless species, $0=m_{1}<m_{2}<m_{3}$, which leads $\alpha_{e}=0.03$, $\alpha_{\mu}=0.55$ and $\alpha_{\tau}=0.42$~\cite{Esteban:2020cvm}. The different mass hierarchies and different value of neutrino absolute mass change the value of $\alpha_{\beta}$ and shift the constraint on $f$.

\begin{figure}
    \centering
    \includegraphics[width=10cm]{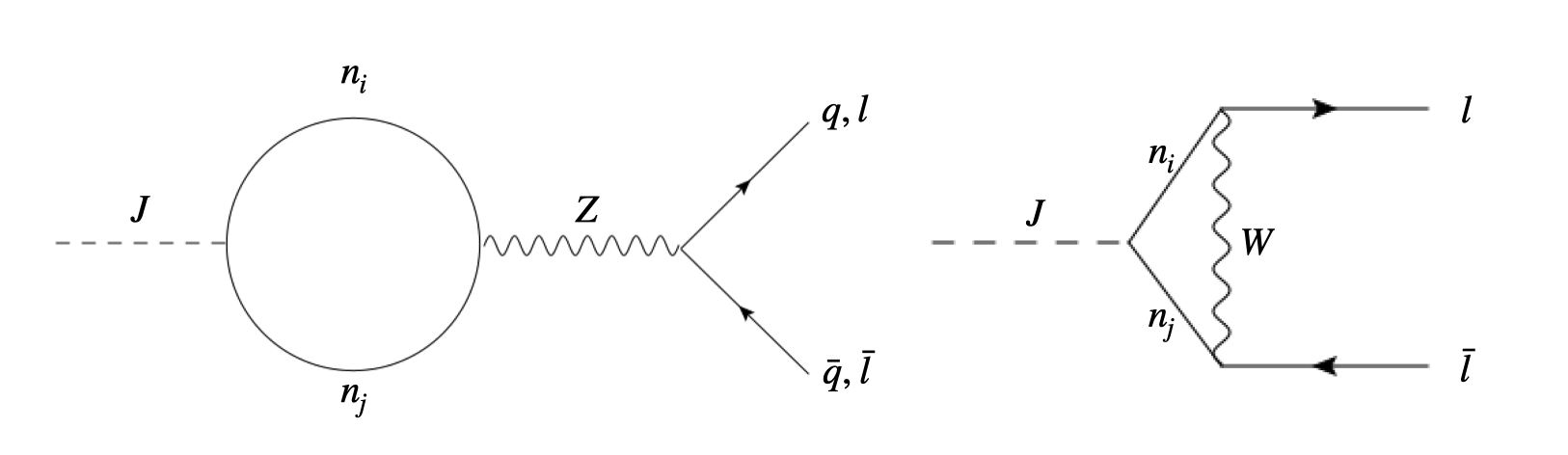}
    \caption{1-loop Feynman diagram for $J\rightarrow q\bar{q}, \,l\bar{l}$, reproduced from Ref.~\cite{Akita:2023qiz}.}
    \label{fig:Jqq}
\end{figure}

\begin{figure}
    \centering
    \includegraphics[width=15cm]{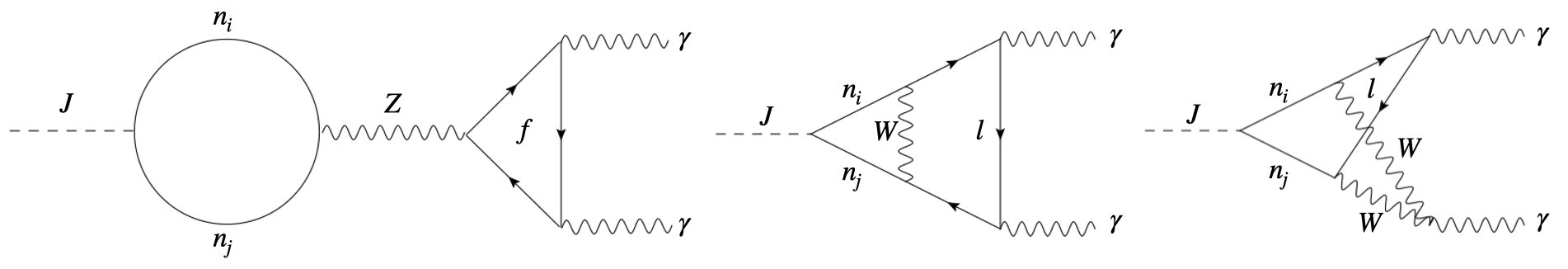}
    \caption{2-loop Feynman diagram for $J\rightarrow \gamma\gamma$, reproduced from Ref.~\cite{Akita:2023qiz}.}
    \label{fig:Jgammagamma1}
\end{figure}

At loop-level, the Majoron also has two-body decay channels into charged fermions and photons. The Feynman diagrams of those channels are given in Figs.~\ref{fig:Jqq} and \ref{fig:Jgammagamma1}, respectively, reproduced from Ref.~\cite{Akita:2023qiz} \footnote{There are similar diagrams to the left panel of Fig.~\ref{fig:Jgammagamma1} with $W$-boson triangle loops, which 
cancel with the diagrams with Faddeev-Popov ghosts and do not contribute~\cite{Garcia-Cely:2016hsk}.}. 
The decay rates into quarks and charged leptons in the seesaw limit are given by,
\begin{align}
    \Gamma(J\rightarrow q\bar{q})
    &\simeq\frac{3m_{J}}{8\pi}\left|\frac{m_{q}}{8\pi^{2}v}T_{3}^{q}{\rm tr}K\right|^{2}\,,\\
    \Gamma(J\rightarrow l\bar{l'})
    &\simeq\frac{m_{J}}{8\pi}\left(\left|\frac{m_{l}+m_{l'}}{16\pi^{2}v}(\delta_{ll'}T_{3}^{l}\,{\rm tr}K+K_{ll'})\right|^{2}+\left|\frac{m_{l}-m_{l'}}{16\pi^{2}v}K_{ll'}\right|^{2}\right)\,,
\end{align}
respectively, where $T_{3}^{d,l}=-1/2=-T_{3}^{u}$ and $K\equiv m_{D}m_{D}^{\dagger}/vf$.


The decay rate of diphoton decay is~\cite{Heeck:2019guh},
\begin{align}
    \Gamma(J\rightarrow 2\gamma)
    &\simeq\frac{\alpha^{2}}{4096\pi^{7}}\frac{m_{J}^{3}}{v^{2}} |K'|^2\,,\\
    \quad {\rm where} \quad K'&\equiv{\rm tr}K\sum_{f} N_{c}^{f}T_{3}^{f}Q_{f}^{2}h\left(\frac{m_{J}^{2}}{4m_{f}^{2}}\right)+\sum_l K_{ll}h\left(\frac{m_{J}^{2}}{4m_{l}^{2}}\right)\,.\label{eq:decay rate into photons}
\end{align}
Note that $\alpha$ is the fine-structure constant and $Q_{c}^{f}$ is the electric charge of particle species $f$. The factors $N_{c}^{q}=3,\, N_{c}^{l}=1$ are accounting the number of colors degree of freedoms. The loop function $h(x)$ is given by,
\begin{align}
    h(x) &=-\frac{1}{4x}\left({\rm log}(1-2x+2\sqrt{x(x-1)})\right)^{2}-1\,.
\end{align}

\begin{figure}
    \centering
    \includegraphics[width=13cm]{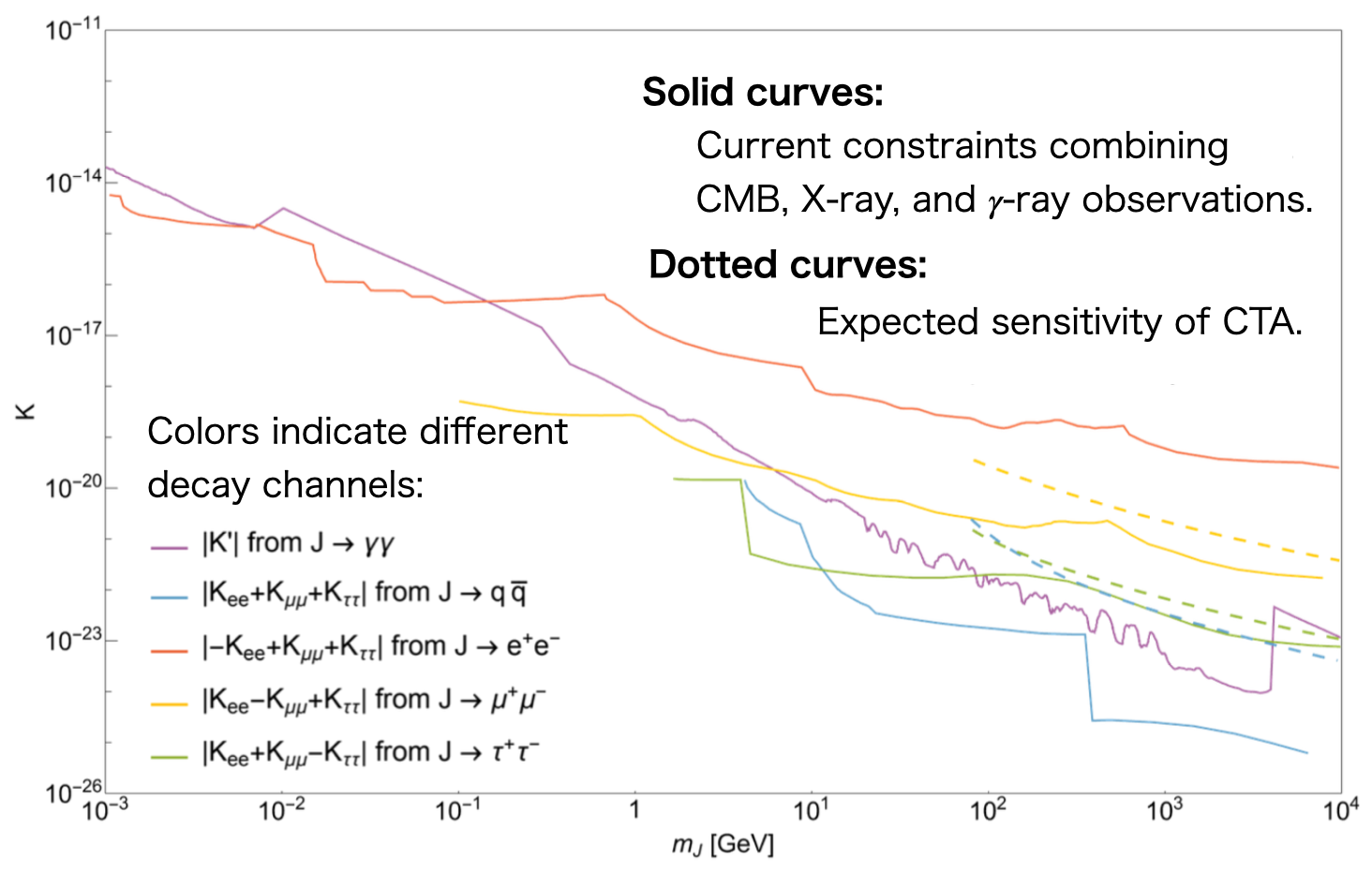} 
    \caption{The current upper bounds (solid) and future prospect (dashed) of the linear combination of the components of $K$ and $|K'|$, adopted from Ref.~\cite{Akita:2023qiz}, \textcopyright  Akita et al., CC-BY 4.0. The different colors correspond to the different daughter particles; photons (purple), quarks (blue), electron (red), muon (yellow) and tau (green). The detail of the data used to produce the plot is discussed in Ref.~\cite{Akita:2023qiz}.
    } 
    \label{fig:result_other}
\end{figure}

Figure \ref{fig:result_other}, reproduced from Ref.~\cite{Akita:2023qiz}, shows the current constraint (solid curves) and future prospects (dashed curve) on the linear combinations of the components of matrix $K$ and $|K'|$. 
Note that the dashed curves depict solely future effectivity of CTA~\cite{Pierre:2014tra} while the solid curves show the combined constraints using multiple datasets of CMB, X-rays, $\gamma$-rays and cosmic ray observations, where detail is given in Ref.~\cite{Akita:2023qiz}. The former is weaker, which may be simply because the latter deals the larger set of the data.
The constraints on the spontaneous symmetry breaking scale $f$ can be found once the Yukawa couplings $\lambda_{D}$ and $\lambda_{M}$ are determined.
In the figure, we neglect the decay of Majoron into $q=u,d,s,c$, since those channels are effectively replaced by Hadron-inducing Majoron decays which are not well understood. We also neglect the decay channel to $Z,W,h$-bosons and gluons, considering the fact that those channels are suppressed by either of (squared) Majorana neutrino mass $M_{R}^{2}$ or coupling constants $\alpha$ (or QCD corresponding, $\alpha_{S}$) compared to $J\rightarrow f\bar{f}$. 

Beyond axions and Majorons, vector gauge bosons associated with a $U_{B-L}$ symmetry can also be constrained through the energetics of GRBs, via their interactions with neutrinos emitted in these events \cite{Poddar:2022svz}. Ultralight scalar or vector particles that may constitute DM can similarly interact with neutrinos produced in supernova cores, potentially affecting the recoil (kick) velocities of pulsars; corresponding spectral observations can therefore be used to place limits on boson-neutrino couplings \cite{Lambiase:2023hpq}. In addition, forward-physics facilities \cite{Feng:2022inv} such as FASER and SHiP are designed to search for $\mathcal{O}(\mathrm{GeV})$-scale dark sector particles, including vector gauge bosons, sterile neutrinos, and heavy neutral leptons.

\newpage

%% file: 8_Discussion.tex
\section{Conclusions and Outlook}
\label{sec:discussion}

In this review, we have surveyed a broad class of extensions to GR and the SM motivated by recent developments in multi-messenger astronomy, including GW, EM, and neutrino observations. These observations increasingly point toward a ``dark" sector that cannot be explained within standard physics alone. Using data ranging from Solar System measurements to galactic and cosmological observations, a wide variety of DM scenarios and modified gravity models can be tested and constrained.

In particular, scalar-tensor theories such as Brans-Dicke gravity introduce additional scalar degrees of freedom, and many particle physics models predict light scalar fields that may constitute DM. Such fields can generate new long-range ``fifth forces" in addition to standard gravity. Precision Solar System tests including perihelion precession, light bending, and timing measurements as well as GW observations are in excellent agreement with standard theory; consequently, the absence of deviations places stringent limits on these models.

Primordial black holes, although not particle DM candidates, may also contribute to the DM abundance and can produce stochastic GW backgrounds potentially observable by current and future detectors. Ultralight bosons, another well-motivated DM candidate, can interact with compact astrophysical systems such as pulsars and magnetized stars, contributing to spin-down rates, orbital period decay in binaries, and timing delays. High-precision pulsar timing measurements therefore provide sensitive constraints on such particles. Modified gravity and massive gravity scenarios are similarly constrained by these observations, and future improvements in timing precision and GW detector sensitivity will further strengthen these bounds.

Ultralight bosons may also accumulate around BHs and extract rotational energy through superradiance, reducing the BH spin. Transitions between bosonic bound states can produce continuous GW signals, which next-generation detectors such as the Einstein Telescope may detect or constrain. In addition, heavier DM particles can be captured by NSs, altering their intrinsic properties including mass, radius, and tidal deformability which can be probed through X-ray and GW observations. DM decay into SM particles can also leave detectable signatures in high energy astrophysical spectra.

Overall, the study of DM and modified gravity in the multi-messenger era has become a rapidly advancing and highly interconnected field. The synergy between GW astronomy, EM observations, and particle physics has already yielded significant constraints on the dark sector. Future facilities including next-generation GW detectors, improved spectroscopic measurements, and precision astrophysical surveys promise to further illuminate the nature of DM and deepen our understanding of physics beyond the SM and GR.

%% file: 9_Acknowledgements.tex
\begin{acknowledgments}
GL, TKP, LV acknowledge support by Istituto Nazionale di Fisica Nucleare (INFN) through the Commissione Scientifica Nazionale 4 (CSN4) Iniziativa Specifica ``Quantum Universe'' (QGSKY). The work of MU is supported by IBS under the project code, IBS-R018-D3. MK is grateful to Carlos Herdeiro for reading the section on ``Superradiance'' and for insightful discussions along with introductions to some relevant references. It is also necessary to thank and appreciate the hosting and support of the School of Physics, Institute of Physics (IPM). LV acknowledges support by the National Natural Science Foundation of China (NSFC) through the grant No.\ 12350610240 ``Astrophysical Axion Laboratories'' and the State Key Laboratory of Dark Matter Physics at the Shanghai Jiao Tong University. LV additionally thanks the Tsung-Dao Lee Institute and the Xplorer Symposia Organization Committee of the New Cornerstone Science Foundation for hospitality during the final stages of this work. This publication is based upon work from the COST Actions ``COSMIC WISPers'' (CA21106), BridgeQG (CA23130) and ``Addressing observational tensions in cosmology with systematics and fundamental physics (CosmoVerse)'' (CA21136), all are supported by COST (European Cooperation in Science and Technology).
\end{acknowledgments}

\newpage

%% file: 10_Appendix.tex

%% file: 11_Acronyms_Conventions.tex
\appendix

\newpage

\section{List of conventions used}
\begin{table*}[ht]
\footnotesize
\begin{center}
\begin{tabular}{|l|l|}
\hline
\hspace{0.5cm} Greek small letters $\mu, \nu,$... & \hspace{0.5cm} Spacetime coordinates indices\\
\hspace{0.5cm} Latin small letters $i,j,k$... & \hspace{0.5cm} Space coordinates indices\\
\hspace{0.5cm} $g_{\mu\nu}$ & \hspace{0.5cm} Metric tensor \\
\hspace{0.5cm} $\nabla_{\mu}$ & \hspace{0.5cm} Covariant derivative \\
\hspace{0.5cm} $\left(-, +, +, +\right)$ & \hspace{0.5cm} Metric signature \\
\hspace{0.5cm} $\Gamma^{\mu}_{\phantom{\mu} \nu \rho}$ & \hspace{0.5cm} Levi-Civita connection\\
\hspace{0.5cm} $\mathcal{R}^{\mu}_{\phantom{\mu} \nu \alpha \beta}$ & \hspace{0.5cm} Riemann curvature tensor\\
\hspace{0.5cm} $\mathcal{R}_{\mu \nu} = \mathcal{R}^{\alpha}_{\phantom{\alpha} \mu \alpha \nu}$ & \hspace{0.5cm} Ricci tensor\\
\hspace{0.5cm} $\mathcal{R}=\mathcal{R}^{\alpha}_{\phantom{\alpha} \alpha}$ & \hspace{0.5cm} Ricci scalar\\
\hspace{0.5cm} $G_{\mu \nu} = \mathcal{R}_{\mu \nu} - \frac{1}{2} g_{\mu \nu} \mathcal{R}$ & \hspace{0.5cm} Einstein tensor\\
\hspace{0.5cm} $T^{\mu \nu}$ & \hspace{0.5cm} Energy-momentum tensor \\
\hspace{0.5cm} $a(t)$ & \hspace{0.5cm} Scale factor as a function of cosmic time $t$ \\
\hspace{0.5cm} $H(t) \equiv \frac{1}{a}\frac{\mathrm{d}a}{\mathrm{d}t}$ & \hspace{0.5cm} Hubble expansion rate at cosmic time $t$ \\
\hspace{0.5cm} $\tau = \int \frac{\mathrm{d}t}{a(t)}$ & \hspace{0.5cm} Conformal time\\
\hspace{0.5cm} $\dot{v} \equiv \frac{\mathrm{d}v}{\mathrm{d}\tau}$ & \hspace{0.5cm} Conformal time derivative of $v$\\
\hspace{0.5cm} ${\cal H}(\tau) \equiv \frac{1}{a}\frac{\mathrm{d}a}{\mathrm{d}\tau}$ & \hspace{0.5cm} Conformal Hubble expansion rate\\
\hspace{0.5cm} $\rho_m$, $\rho_{\rm DM}$, $\rho_b$ & \hspace{0.5cm} Energy density of matter, dark matter, baryons\\
\hspace{0.5cm} $\rho_r$, $\rho_\nu$ & \hspace{0.5cm} Energy density of radiation and neutrinos\\
\hspace{0.5cm} $\rho_{\rm DE}$, $p_{\rm DE}$ & \hspace{0.5cm} Energy density and pressure of dark energy\\
\hspace{0.5cm} $w_0$ & \hspace{0.5cm} Equation of state with a constant value\\
\hspace{0.5cm} $w_{\rm DE} \equiv p_{\rm DE}/\rho_{\rm DE}$ & \hspace{0.5cm} Equation of state for dark energy ($z$-dependent)\\
\hspace{0.5cm} $M_{\rm Pl} \equiv 1/\sqrt{8\pi G_N}$ & \hspace{0.5cm} Reduced Planck mass\\
\hspace{0.5cm} $\kappa \equiv \sqrt{8\pi G_N}$ & \hspace{0.5cm} Gravitational constant\\
\hspace{0.5cm} $\rho_{{\rm crit},0} \equiv 3H_0^2M_{\rm Pl}^2$ & \hspace{0.5cm} Present critical energy density\\
\hspace{0.5cm} $\mathcal{T}^{\mu}{}_{\nu\rho}$; $\mathcal{T}$ & \hspace{0.5cm} Torsion tensor; torsion scalar\\ 
\hspace{0.5cm} $\mathcal{Q}_{\alpha\mu\nu} \equiv \nabla_{\alpha}g_{\mu\nu}$; $\mathcal{Q}$ & \hspace{0.5cm} Non-metricity tensor; non-metricity scalar\\ 
\hspace{0.5cm} $N_{\rm eff}$; $N_{\rm eff}^{\rm SM} = 3.046$ & \hspace{0.5cm} Effective number of neutrino species; SM value of $N_{\rm eff}$ used here~\cite{Mangano:2005cc,deSalas:2016ztq,Akita:2020szl}\\
\hspace{0.5cm} $r_{\rm s}^*$ & \hspace{0.5cm} Comoving sound horizon at CMB last scattering\\
\hspace{0.5cm} $r_d$ & \hspace{0.5cm} Comoving sound horizon at the end of baryon-drag epoch\\
\hline
\end{tabular}
\end{center}
\caption{List of conventions used in the review.}
\label{tabnotation}
\end{table*}
\newpage

\section{List of acronyms used}
\begin{table*}[ht]
\footnotesize
\begin{center}
\begin{tabular}{|l|l|}
\hline
\hspace{0.5cm} ACTPol & \hspace{0.5cm} Atacama Cosmology Telescope Polarimeter\\
\hspace{0.5cm} (A)dS    & \hspace{0.5cm} (Anti-)de Sitter \\
\hspace{0.5cm} AGN    & \hspace{0.5cm} Active Galactic Nuclei\\
\hspace{0.5cm} ALPs   & \hspace{0.5cm} Axion like particles\\
\hspace{0.5cm} BAO    & \hspace{0.5cm} Baryon acoustic oscillations\\
\hspace{0.5cm} BBN    & \hspace{0.5cm} Big bang nucleosynthesis\\
\hspace{0.5cm} (P)BH  & \hspace{0.5cm} (Primordial) Black hole\\
\hspace{0.5cm} BM  & \hspace{0.5cm} Baryonic matter\\
\hspace{0.5cm} BOSS   & \hspace{0.5cm} Baryon Oscillation Spectroscopic Survey\\
\hspace{0.5cm} (B)SM  & \hspace{0.5cm} (Beyond) Standard Model\\
\hspace{0.5cm} CMB    & \hspace{0.5cm} Cosmic microwave background\\
\hspace{0.5cm} (C)DM  & \hspace{0.5cm} (Cold) dark matter\\
\hspace{0.5cm} CFT    & \hspace{0.5cm} Conformal Field Theory\\
\hspace{0.5cm} DANS   & \hspace{0.5cm} Dark matter admixed neutron star \\
\hspace{0.5cm} DE     & \hspace{0.5cm} Dark energy\\
\hspace{0.5cm} DES    & \hspace{0.5cm} Dark Energy Survey\\
\hspace{0.5cm} EM     & \hspace{0.5cm} Electromagnetic\\
\hspace{0.5cm} FLRW   & \hspace{0.5cm} Friedmann--Lemaitre--Robertson--Walker \\
\hspace{0.5cm} GR     & \hspace{0.5cm} General relativity\\
\hspace{0.5cm} GW     & \hspace{0.5cm} Gravitational wave\\
\hspace{0.5cm} HST    & \hspace{0.5cm} Hubble Space Telescope\\
\hspace{0.5cm} KGE    & \hspace{0.5cm} Klein-Gordon equation \\
\hspace{0.5cm} LMC    & \hspace{0.5cm} Large Magellanic Cloud\\
\hspace{0.5cm} LVK    & \hspace{0.5cm} Ligo-Virgo-Kagra \\
\hspace{0.5cm} {\it Planck} 2015/2018 TT & \hspace{0.5cm} {\it Planck} 2015/2018 temperature power spectrum at high-$\ell$\\
\hspace{0.5cm} {\it Planck} 2015/2018 & \hspace{0.5cm} {\it Planck} 2015/2018 temperature and polarization power spectra at high-$\ell$\\
\hspace{0.5cm} QNM    & \hspace{0.5cm} Quasi Normal Mode\\
\hspace{0.5cm} RD     & \hspace{0.5cm} Radiation domination\\
\hspace{0.5cm} SDSS   & \hspace{0.5cm} Sloan Digital Sky Survey\\
\hspace{0.5cm} SIMPs  & \hspace{0.5cm} Strongly Interacting Massive Particles\\
\hspace{0.5cm} SMBH   & \hspace{0.5cm} Supermassive black hole \\
\hspace{0.5cm} SPTPol & \hspace{0.5cm} South Pole Telescope Polarimeter\\
\hspace{0.5cm} SR     & \hspace{0.5cm} Superradiance/Superradiant\\
\hspace{0.5cm} SRI    & \hspace{0.5cm} Superradiant Instability\\
\hspace{0.5cm} TOV    & \hspace{0.5cm} Tolman-Oppenheimer-Volkoff\\ 
\hspace{0.5cm} ISCO   & \hspace{0.5cm} Innermost Stable Circular Orbit\\
\hspace{0.5cm} QCD    & \hspace{0.5cm} Quantum Chromodynamics \\
\hspace{0.5cm} NS     & \hspace{0.5cm} Neutron Star \\
\hspace{0.5cm} WD     & \hspace{0.5cm} White Dwarf \\
\hspace{0.5cm} WIMPs & \hspace{0.5cm} Weakly Interacting Massive Particles\\
\hspace{0.5cm} WMAP  & \hspace{0.5cm} Wilkinson Microwave Anisotropy Probe\\
\hline 
\end{tabular}
\end{center}
\caption{List of acronyms used in the review.}
\label{tab:acronyms}
\end{table*}

\newpage